\documentclass[aps,preprint,floats,epsf,epsfig,nofootinbib,letter]{revtex4}
\usepackage{epsfig}
\usepackage{graphicx}
\usepackage{dcolumn}
\usepackage{bm}
\usepackage{subfigure}
\usepackage{hyperref}                 
\usepackage{amsmath}
\usepackage{amsfonts}
\usepackage{xcolor}

\begin{document}
\def\be{\begin{eqnarray}}
\def\en{\end{eqnarray}}
\def\non{\nonumber}
\def\la{\langle}
\def\ra{\rangle}
\def\ov{\overline}
\def\pp{{\prime\prime}}
\def\vp{\varepsilon}
\def\Br{{\mathcal B}}
\def\A{{\mathcal A}}
\def\B{{\cal B}}
\def\D{{\cal D}}
\def\S{{\bf S}}
\def\T{{\bf T}}
\def\Bbar{\overline{\cal B}}
\def\bfB{{\rm\bf B}}
\def\bfBB{{\rm\bf B}\overline{\rm\bf B}}
\def\BB{{{\cal B} \overline {\cal B}}}
\def\BD{{{\cal B} \overline {\cal D}}}
\def\DB{{{\cal D} \overline {\cal B}}}
\def\DD{{{\cal D} \overline {\cal D}}}
\def\sq{\sqrt}


\title{Heavy Mesons to Open-Charmed Tetraquark Decays}

\author{Chun-Khiang Chua}
\affiliation{Department of Physics and Center for High Energy Physics,
Chung Yuan Christian University,
Chung-Li, Taiwan 320, Republic of China}

\date{\today}

\begin{abstract}
Motivated by the recent observations of $T^*_{cs0}(2870)^0$, $T^*_{c\bar s 0}(2900)^0$ and $T^*_{c\bar s 0}(2900)^{++}$ open-charmed tetraquark states by LHCb, we study the decays of heavy mesons to these open-charmed tetraquark states ($T$) using a topological amplitude approach. We first obtain the $T\to DP$ and $DS$ strong decay amplitudes by decomposing them into several topological amplitudes, where $P$ is a light pseudo-scalar particle and $S$ is a low-lying scalar particle. 
Subsequently, weak decay amplitudes of $\overline B\to D\overline T$, $\overline D T$ and $\overline B\to TP$, $TS$ decays are decomposed topologically.  
In addition, $B^-_c\to T\overline T$ decays are also discussed. Using these results, modes with an unambiguous exotic interpretation in flavor are highlighted.   

\end{abstract}


\maketitle

\tableofcontents


\vfill\eject

\section{Introduction}

In 2020, LHCb reported the observation of an open-charmed tetraquark $T^*_{cs 0} (2870)^0$ 
in $B^-\to D^- D^+ K^-$ decay with $B^-\to D^- T^*_{cs 0} (2870)^0, T^*_{cs 0} (2870)^0\to D^+ K^-$ decay~\cite{LHCb:2020pxc}. 
From $T^*_{cs 0} (2870)^0\to D^+ K^-$ decay, the minimal quark content of $cs\bar u\bar d$ for $T^*_{cs 0} (2870)^0$ is infered.
Several years later, the observations of two other open-charmed tetraquarks, $T^*_{c\bar s 0}(2900)^0$ and $T^*_{c\bar s 0}(2900)^{++}$, were reported in 
$\overline B{}^0\to D^0 D_s^-\pi^+$ and $B^-\to D^+ D_s^- \pi^-$ decays with 
$\overline B{}^0\to D^0 \overline {T^*_{c\bar s 0}} (2900)^{0}, \overline {T^*_{c\bar s 0}} (2900)^{0}\to D_s^-\pi^+$ 
and
$B^-\to D^+ \overline {T^*_{c\bar s 0}} (2900)^{\,--},  \overline {T^*_{c\bar s 0}} (2900)^{\,--}\to D_s^- \pi^-$ decays 
\cite{LHCb:2022sfr, LHCb:2022lzp}.
From $ {T^*_{c\bar s 0}} (2900)^{0}\to D_s^+\pi^-$ and ${T^*_{c\bar s 0}} (2900)^{\,++}\to D_s^+ \pi^+$ decays the minimal quark contents of these states are $c\bar s \bar u d$ and $c\bar s u \bar d$, respectively. 
The masses and widths of these states are~\cite{ParticleDataGroup:2024cfk}
\be
m(T^*_{cs 0}(2870)^0)&=&2.872\pm 0.016 {\rm GeV},
\quad
\Gamma(T^*_{cs 0}(2870)^0)=0.067\pm0.024 {\rm GeV},
\non\\
m(T^*_{c\bar s 0}(2900)^0)&=&2.892\pm 0.021 {\rm GeV},
\quad
\Gamma(T^*_{cs 0}(2900)^0)=0.119\pm0.029 {\rm GeV},
\non\\
m(T^*_{c\bar s 0}(2900)^{++})&=&2.921\pm 0.026 {\rm GeV},
\quad
\Gamma(T^*_{cs 0}(2900)^{++})=0.140\pm0.040 {\rm GeV}.
\en
Very recently, LHCb also reports the observation of a new decay mode of $T_{cs 0} (2870)^0$, namely  $T_{cs 0} (2870)^0\to D^0 K^0_S$,
in $B^-\to D^- D^0 K^0_S$ decay, 
and the  relative rates of $T_{cs 0} (2870)^0\to D^0 \overline K^0$ and $T^*_{cs 0} (2870)^0\to D^+ K^-$ decays is consistent with isospin symmetry 
within the experimental uncertainty~\cite{LHCb:2024xyx}.

There are many theoretical interpretations of these states; see, for example, refs.
\cite{Karliner:2020vsi,
He:2020jna,
Wang:2020xyc,
Molina:2020hde,
Huang:2020ptc,
Liu:2020nil,
Wang:2021lwy,
Yue:2022mnf,
Burns:2020epm,
Liu:2020orv},
see also ref.~\cite{PDGreview}.
In this work, we will study the decays of heavy mesons to these open-charmed tetraquark states ($T$). 
As a direct calculation of the decay amplitudes is too complicated, we will make use of a topological amplitude approach,
which have been applied to many heavy meson decays, see, for example, 
refs. \cite{Zeppenfeld:1980ex,
Chau:tk,
Chau:1990ay,
Gronau:1994rj,
Gronau:1995hn,
Rosner:1999xd,
Cheng:2002ai,
Chiang:2004nm,
Cheng:2014rfa,
Savage:ub,
Li:2012cfa,
Qin:2013tje,
Chua:2003it,
Chua:2013zga,
Cheng:2015cca,
Chua:2016aqy,
Chua:2022wmr,
Chua:2023qyj}. 
In particular, this approach can also be applied to decays involving exotic states.
For example, this method was employed to study heavy baryons to pentaquark decays in ref. \cite{Cheng:2015cca}, 
and the results obtained were confirmed by LHCb subsequently \cite{LHCb:2016lve}.
Note that a study on charmed tetraquark productions in $B$ decays using diagrammatic analysis was given in ref. \cite{Qin:2022nof}. 
We will compare our results with those obtained in that work in later sections.
 
In this work, we will consider two scenarios where the light antiquarks in the open-charmed tetraquarks are either antisymmetric or symmetric.
It will be interesting to see which scenario the above tetraquarks, namely $T_{cs 0} (2870)^0$, $T^*_{c\bar s 0}(2900)^0$ and $T^*_{c\bar s 0}(2900)^{++}$, fit.
We will consider the strong decays of open-charmed tetraquarks and weak decays involving these tetraquarks. 
Attentions are paid to predictions where all $T_{cs 0} (2870)^0$, $T^*_{c\bar s 0}(2900)^0$ and $T^*_{c\bar s 0}(2900)^{++}$ are involved.
We will try to identify other flavor exotic states as well.
It will be useful to identify modes involving flavor exotic states  as much as possible. 
We will be in a better position to understand these tetraquarks with this information.
   
The layout of this paper is as follows. 
We introduce two different scenarios of the light quark pair $\bar q \bar q$ in the open-charmed tetraquarks $T_{cq\bar q\bar q}$ in section II. 
All flavor exotic states are identified.
We consider 
$T\to DP$ and $DS$ decays in Sec. III, 
$\overline B\to D\overline T$ and $\overline D T$ decays in Sec.~IV,
$\overline B\to TP$ and $TS$ decays in Sec. V
and, finally, $B_c\to T\overline T$ decays in Sec. VI.
Decay amplitudes of these decays are given in terms of topological amplitudes.
These results can be easily generalized to modes with $D^*$ and $V$, instead of $D$ and $P$, respectively. 
Modes with unambiguous exotic interpretations in flavor are highlighted.
Sec. VII comes to the conclusions. 
We end this paper with three appendices.

\section{Two different configurations of the light anti-quarks in $T_{cq \bar q' \bar q''}$}

\subsection{Scenarios I and II}

The open-charmed tetraquark $T_{cq \bar q' \bar q''}$ has two light antiquarks, $\bar q' \bar q''$.
We can decompose the anti-quark pair $\bar q'\bar q''$ into anti-symmetric $[\bar q'\bar q'']$ and symmetric $\{\bar q'\bar q''\}$ at the outset,
since SU(3) transformation does not mix up these two sets. 
Hence, in this work, we consider two different configurations for the light anti-quarks in a $T_{cq \bar q' \bar q''}$ state, namely an antisymmetric $\bar q' \bar q''$ configuration and a symmetric $\bar q' \bar q''$ configuration. 
They can be further separated into traceless and traceful parts, which do not mix under SU(3) transformation, see Appendix~\ref{app: SU(3)}.
 
\begin{figure}[t]
\centering
 \subfigure[]{
  \includegraphics[width=0.41\textwidth]{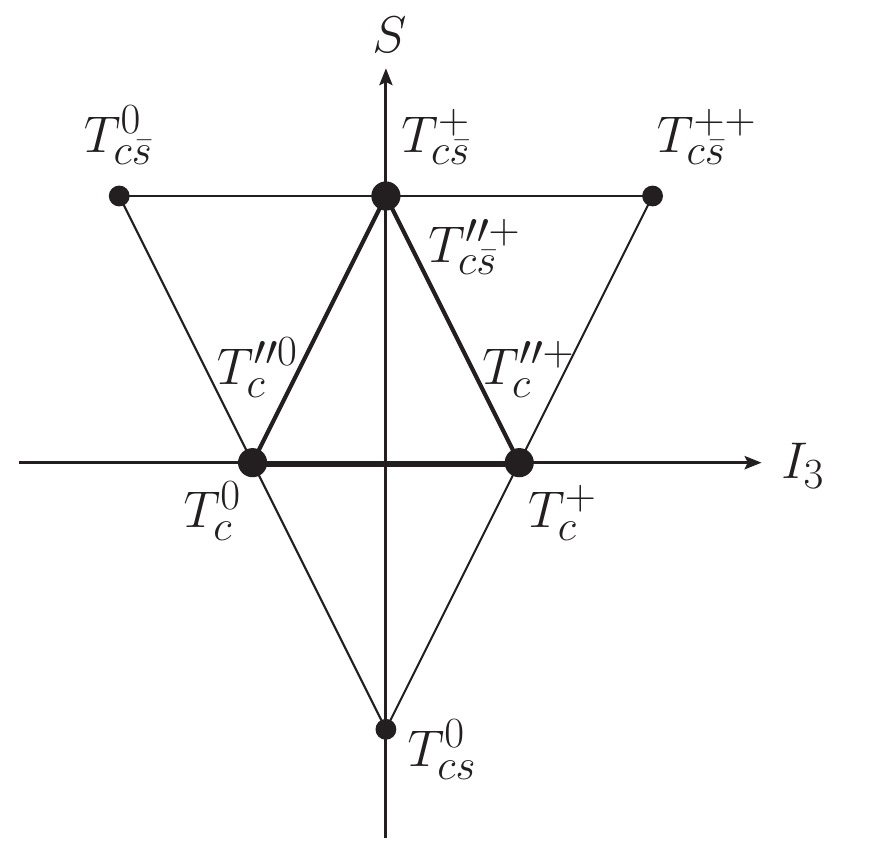}
}
\subfigure[]{
  \includegraphics[width=0.54\textwidth]{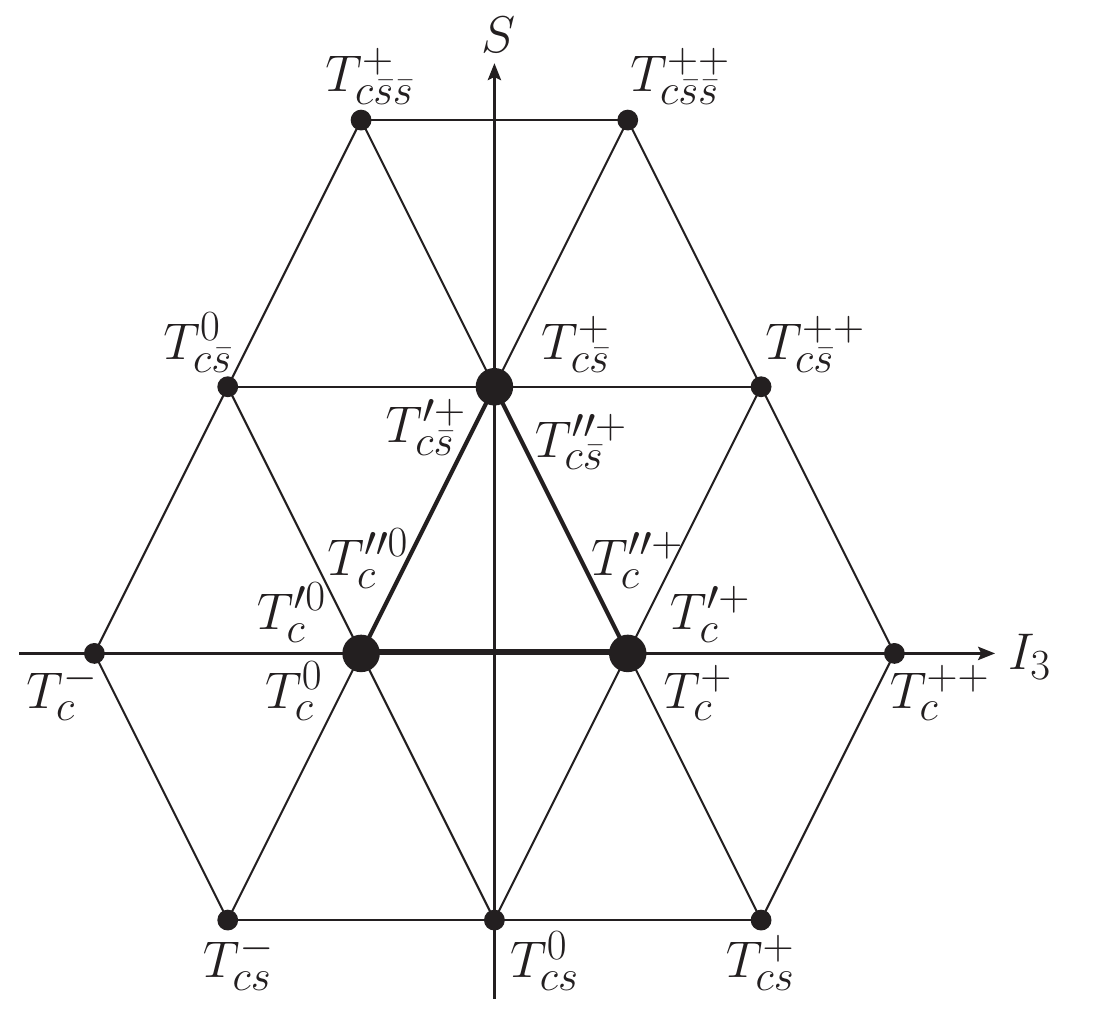}
}
\hspace{12pt}
\caption{
  (a) The ${\bf 6}$ and $\bar{\bf 3}$ in scenario I and (b) the $\overline{\bf 15}$ and $\bar{\bf 3}'$ in scenario II, see also Tables \ref{tab: 6+3bar} and \ref{tab: 15bar+3bar}. 
} \label{fig: SU(3)}
\end{figure}

For the antisymmetric $\bar q' \bar q''$ configuration, the nine $q[\bar q' \bar q'']$ are SU(3) ${\bf 3}\otimes {\bf 3}$ states, which can be expressed in a $3\times 3$ matrix given by
\be
\T= \left(
\begin{array}{ccc}
T_{cu [\bar d\bar s]}^{++}
    & T_{cu [\bar s\bar u]}^+
    & T_{cu [\bar u\bar d]}^+
    \\
T_{cd [\bar d\bar s]}^+
    & T_{cd [\bar s\bar u]}^0
    & T_{cd [\bar u\bar d]}^0
    \\
T_{cs [\bar d\bar s]}^+
    & T_{cs [\bar s\bar u]}^0
    & T_{cs [\bar u\bar d]}^0
\end{array}
\right),
\en
where the $[\bar q' \bar q'']$ notation indicates the antisymmetric configuration of these light anti-quarks.
These nine ${\bf 3}\times {\bf 3}$ states can be grouped into a ${\bf 6}$ and a $\bar{\bf 3}$, see Eq. (\ref{eq: SU(3) decompositions 0}),
where the ${\bf 6}$ is the traceless part, 
and it consists of an $S=+1$ isotriplet, $(T^{++}_{c\bar s}, T^{+}_{c\bar s}, T^{0}_{c\bar s})$, an $S=-1$ isosinglet, $T^0_{cs}$, 
and an $S=0$ isodoublet, $(T_c^{+},T_c^{0})$;
while the $\bar{\bf 3}$ is the traceful part,
and it has an $S=0$ isodoublet, $(T^{\prime\prime +}_c, T^{\prime\prime 0}_c)$, and an $S=+1$ isosinglet, $T^{\prime\prime +}_{c\bar s}$, see Fig.~\ref{fig: SU(3)} (a).
The quark contents and quantum numbers of these states are given in Table \ref{tab: 6+3bar}.
Consequently, the above $3\times 3$ matrix $\T$ can be re-expressed as
\be
\T= \left(
\begin{array}{ccc}
T_{c\bar s}^{++}
    & -\frac{T_{c\bar s}^+ + T_{c\bar s}^{\prime\prime +}}{\sqrt2}
    & \frac{T_{c}^+ + T_c^{\prime\prime +}}{\sqrt2}
    \\
-\frac{T_{c\bar s}^+ - T_{c\bar s}^{\prime\prime +}}{\sqrt2}
    & -T_{c\bar s}^0
    & \frac{T_{c}^0- T_c^{\prime\prime 0}}{\sqrt2}
    \\
\frac{T_{c}^+ - T_c^{\prime\prime +}}{\sqrt2}
    & \frac{T_{c}^0+ T_c^{\prime\prime 0}}{\sqrt2}
    & T_{cs}^0
\end{array}
\right).
\en
Note that for convenience we number these $T$ in Table \ref{tab: 6+3bar} and add asterisks to flavor exotic states.
\begin{table}[t!]
\caption{\label{tab: 6+3bar}
Quark contents and quantum numbers of open-charmed tetraquark in scenario I with antisymmetric $\bar q \bar q'$. 
We have a ${\bf 6}$ and a $\bar{\bf 3}$.
We numbered these states and states with flavor exotic are labeled by asterisks.
In this scenario, $T^*_{c\bar s0}(2900)^{++}$, $T_{c\bar s0}(2900)^0$ and $T^*_{cs0}(2870)^0$ are in the ${\bf 6}$.
}
\footnotesize{
\begin{ruledtabular}
\begin{tabular}{llcrcc}
\#
          & $T_c$
          & quark content
          & (multiplet,$I, I_z,S)$
          & remarks
          \\
\hline 
$1^{*}$
          & $T^{++}_{c\bar s}$
          & $\frac{1}{\sqrt2}(cu\bar d\bar s-c u\bar s \bar d)$
          & $({\bf 6},1,+1,+1)$
          & $T^*_{c\bar s 0}(2900)^{++}$
          \\
$2^*$
          & $T^{+}_{c\bar s}$
          & $\frac{1}{2}(-cu\bar s\bar u+c u\bar u \bar s-c d\bar d\bar s+c d\bar s\bar d)$
          & $({\bf 6},1,0,+1)$
          &
          \\          
$3^{*}$
          & $T^{0}_{c\bar s}$
          & $\frac{1}{\sqrt 2}(-c d\bar s\bar u+c d\bar u \bar s)$
          & $({\bf 6},1,-1,+1)$
          & $T^*_{c\bar s 0}(2900)^{0}$
          \\     
\hline
$4^{*}$
          & $T^{0}_{cs}$
          & $\frac{1}{\sqrt 2} (c s\bar u\bar d-cs\bar d\bar u)$
          & $({\bf 6},0,0,-1)$
          & $T^*_{cs0}(2870)^0$
          \\
\hline
$5$
          & $T^{+}_c$
          & $\frac{1}{2} (cu\bar u\bar d-cu\bar d\bar u+cs\bar d \bar s-cs\bar s\bar d)$
          & $({\bf 6},\frac{1}{2},+\frac{1}{2},0)$
          \\ 
$6$
          & $T^{0}_c$
          & $\frac{1}{2} (cd\bar u\bar d-cd\bar d\bar u+cs\bar s \bar u-cs\bar u\bar s)$
          & $({\bf 6},\frac{1}{2},-\frac{1}{2},0)$
          \\       
\hline
\hline
$ 7$
          & $T^{\prime\prime +}_{c}$
          & $\frac{1}{2} (cu\bar u\bar d-cu\bar d\bar u-cs\bar d \bar s+cs\bar s\bar d)$
          & $(\bar{\bf 3},\frac{1}{2},+\frac{1}{2}, 0)$
          \\                                                                                  
$ 8$
          & $T^{\prime \prime 0}_{c}$
          & $\frac{1}{2} (-cd\bar u\bar d+cd\bar d\bar u+cs\bar s \bar u-cs\bar u\bar s)$
          & $(\bar{\bf 3},\frac{1}{2},-\frac{1}{2}, 0)$
          \\ 
\hline                                       
$ 9$
          & $T^{\prime \prime+}_{c\bar s}$
          & $\frac{1}{2} (-cu\bar s \bar u+cu\bar u\bar s+cd\bar d\bar s-cd\bar s\bar d)$
          & $(\bar{\bf 3},0,0, +1)$
          \\  
\end{tabular}
\end{ruledtabular}
}
\end{table}

In this scenario, namely scenario I, $T^*_{c\bar s0}(2900)^{++}$, $T_{c\bar s0}(2900)^0$ and $T^*_{cs0}(2870)^0$ are in the ${\bf 6}$.
In particular,
$T^*_{c\bar s0}(2900)^{++}$ and $T_{c\bar s0}(2900)^0$ are $T^{++}_{c\bar s}$ and $T^0_{c\bar s}$, respectively, belonging to the $S=1$ isotriplet,
while $T^*_{cs0}(2870)^0$ is $T^0_{cs}$ in the $S=-1$ isosinglet as shown in Table \ref{tab: 6+3bar}. 
The three flavor exotic states are labeled by asterisks in the table.  

Mixing can happen for particles in different SU(3) multiplets, but mixing for particles in different isospin multiplets is expected to be more difficult. 
Consequently, $T^{+}_{c\bar s}$ in the $S=1$ isotripet does not mix with $T^{\prime\prime +}_{c\bar s}$ or $D_s$. 
Therefore, it should also be considered a flavor exotic particle and be labeled with an asterisk in the table. 
To our knowledge, such a particle has not been considered a flavor exotic particle in the literature.   

To match the flavor of these states, we can define the following tensor field,
\be
T^i_{[jk]}\equiv \epsilon_{jkl} \T^{il}.
\en
Note that from the above equations, we have
\be
T^i_{[i 1]}=-T^i_{[1i]}=\sqrt2 T^{\prime\prime 0}_c,
\quad
T^i_{[i 2]}=-T^i_{[2i]}=\sqrt2 T^{\prime\prime +}_c,
\quad
T^i_{[i 3]}=-T^i_{[3i]}=\sqrt2 T^{\prime\prime +}_{c\bar s},
\label{eq: trace T I}
\en
where only the $\bar {\bf 3}$ states remain after taking the trace $T^i_{[i k]}$, as the ${\bf 6}$ is traceless, which can also be inferred from Table \ref{tab: 6+3bar}.

\begin{table}[t!]
\caption{\label{tab: 15bar+3bar}
Quark contents and quantum numbers of open-charmed tetraquark in scenario II with symmetric $\bar q \bar q'$. 
We have a $\overline{\bf 15}$ and a $\bar{\bf 3}$.
We numbered these states, and states with flavor exotic are labeled by asterisks.
In this scenario, $T^*_{c\bar s0}(2900)^{++}$, $T_{c\bar s0}(2900)^0$ and $T^*_{cs0}(2870)^0$ are in the $\overline{\bf 15}$.
}
\footnotesize{
\begin{ruledtabular}
\begin{tabular}{llcrcc}
\#
          & $T_c$
          & quark content
          & (multiplet,$I, I_z,S)$
          & remarks
          & $T_c$ in ref. \cite{Qin:2022nof}
          \\
\hline 
$1'^{*}$
          & $T^{++}_{c\bar s}$
          & $\frac{1}{\sqrt2}(cu\bar d\bar s+c u\bar s \bar d)$
          & $(\overline{\bf 15},1,+1,+1)$
          & $T^*_{c\bar s 0}(2900)^{++}$
          & $T_{cu\bar d\bar s}$
          \\
$2'^*$
          & $T^{+}_{c\bar s}$
          & $\frac{1}{2}(cu\bar s\bar u+c u\bar u \bar s-c d\bar d\bar s-c d\bar s\bar d)$
          & $(\overline{\bf 15},1,0,+1)$
          & 
          \\          
$3'^{*}$
          & $T^{0}_{c\bar s}$
          & $\frac{1}{\sqrt 2}(c d\bar s\bar u+c d\bar u \bar s)$
          & $(\overline{\bf 15},1,-1,+1)$
          & $T^*_{c\bar s 0}(2900)^{0}$
          & $T_{cd\bar u\bar s}$
          \\     
\hline
$4'^{*}$
          & $T^{+}_{cs}$
          & $cs\bar d\bar d$
          & $(\overline{\bf 15},1,+1,-1)$
          &
          & $T_{cs\bar d\bar d}$
          \\
$5'^{*}$
          & $T^{0}_{cs}$
          & $\frac{1}{\sqrt 2} (c s\bar u\bar d+cs\bar d\bar u)$
          & $(\overline{\bf 15},1,0,-1)$
          & $T^*_{cs0}(2870)^0$
          & $T_{cs \bar u\bar d}$
          \\
$6'^{*}$
          & $T^{-}_{cs}$
          & $cs\bar u\bar u$
          & $(\overline{\bf 15},1,-1,-1)$
          &
          & $T_{cs\bar u\bar u}$
          \\
\hline
$7'^{*}$
          & $T^{++}_{c\bar s\bar s}$
          & $c u\bar s\bar s$
          & $(\overline{\bf 15},\frac{1}{2},+\frac{1}{2},+2)$
          & 
          & $T_{cu\bar s\bar s}$
          \\
$8'^{*}$
          & $T^{+}_{c\bar s\bar s}$
          & $c d\bar s\bar s$
          & $(\overline{\bf 15},\frac{1}{2},-\frac{1}{2},+2)$
          &
          & $T_{cd\bar s\bar s}$
          \\
\hline
$9'^{*}$
          & $T^{++}_c$
          & $cu\bar d\bar d$
          & $(\overline{\bf 15},\frac{3}{2},+\frac{3}{2},0)$
          &
          & $T_{cu\bar d\bar d}$
          \\
$10'^*$
          & $T^{+}_c$
          & $\frac{1}{\sqrt 3} (cu\bar u\bar d+cu\bar d\bar u-c d\bar d\bar d)$
          & $(\overline{\bf 15},\frac{3}{2},+\frac{1}{2},0)$
          & 
          \\ 
$11'^*$
          & $T^{0}_c$
          & $\frac{1}{\sqrt 3} (c u\bar u\bar u-cd\bar u\bar d-cd\bar d\bar u)$
          & $(\overline{\bf 15},\frac{3}{2},-\frac{1}{2},0)$
          & 
          \\       
$12'^{*}$
          &$T^{-}_c$
          & $c d\bar u\bar u$
          & $(\overline{\bf 15},\frac{3}{2},-\frac{3}{2},0)$
          &
          & $T_{cd\bar u\bar u}$
          \\  
\hline  
$13'$
          & $T^{\prime +}_c$
          & $\frac{1}{2\sqrt {6}} (cu\bar u\bar d+
             cu\bar d\bar u
             +2 c d\bar d\bar d-3 cs\bar d\bar s-3 c s\bar s\bar d)$
          & $(\overline{\bf 15},\frac{1}{2},+\frac{1}{2},0)$
          \\   
$14'$
          &$T^{\prime 0}_c$
          & $\frac{1}{2\sqrt {6}} (2 c u\bar u\bar u+cd\bar u\bar d+
            cd\bar d\bar u
             -3 cs\bar s\bar u-3 c s\bar u\bar s)$
          & $(\overline{\bf 15},\frac{1}{2},-\frac{1}{2},0)$
           \\ 
\hline
$15'$ 
          & $T^{\prime +}_{c\bar s}$
          & $\frac{1}{2\sqrt 2} (c u\bar s\bar u+cu\bar u\bar s+cd\bar d\bar s+cd\bar s\bar d-2 c s\bar s\bar s)$
          & $(\overline{\bf 15},0,0,+1)$
          \\ 
\hline
\hline
$16'$
          & $T^{\prime\prime +}_{c}$
          & $\frac{1}{2\sqrt 2} (cu\bar u\bar d+
             cu\bar d\bar u
            +2 c d\bar d\bar d+cs\bar d\bar s+c s\bar s\bar d)$
          & $(\bar{\bf 3}^\prime,\frac{1}{2},+\frac{1}{2}, 0)$
          \\                                                                                  
$17'$
          & $T^{\prime \prime 0}_{c}$
          & $\frac{1}{2\sqrt 2} (2 c u\bar u\bar u+cd\bar u\bar d+cd\bar d\bar u+cs\bar s\bar u+c s\bar u\bar s)$
          & $(\bar{\bf 3}^\prime,\frac{1}{2},-\frac{1}{2}, 0)$
          \\ 
\hline                                       
$18'$
          & $T^{\prime \prime+}_{c\bar s}$
          & $\frac{1}{2\sqrt 2} (2 c s\bar s\bar s+cd\bar d\bar s+cd\bar s\bar d+c u\bar s\bar u+cu\bar u\bar s)$
          & $(\bar{\bf 3}^\prime,0,0, +1)$
          \\  
\end{tabular}
\end{ruledtabular}
}
\end{table}

We now turn to the second scenario, scenario II, where the light anti-quarks are symmetric.
For the $q\{\bar q' \bar q''\}$ configuration, we have eighteen SU(3) ${\bf 3}\otimes \bar{\bf 6}$ states.
They form a $\overline{\bf 15}$ and another $\bar {\bf 3}$, see Eq. (\ref{eq: SU(3) decompositions 0}).
The $\overline{\bf 15}$ is the traceless part, which consists of six iso-multiplets:
an $S=+1$ isotriplet, $(T^{++}_{c\bar s},T^{+}_{c\bar s},T^{0}_{c\bar s})$, 
an $S=-1$ isotriplet, $(T^{+}_{c s},T^{0}_{c s},T^{-}_{c s})$,
an $S=+2$ isodoublet, $(T^{++}_{c\bar s\bar s},T^{+}_{c\bar s\bar s})$,
an $S=0$ iso-quarplet, $(T_c^{++},T_c^{+},T_c^{0},T_c^{-})$, 
an $S=0$ isodoublet $(T_c^{\prime +},T_c^{\prime 0})$
and an $S=+1$ isosinglet, $T_{c\bar s}^{\prime +}$;
while the $\bar{\bf 3}$ is the traceful part, and it has an $S=0$ isodoublet, $(T^{\prime\prime +}_c, T^{\prime\prime 0}_c)$, and an $S=+1$ isosinglet, $T^{\prime\prime +}_{c\bar s}$, see Fig.~\ref{fig: SU(3)} (b).
The quark contents and quantum numbers of these states are given in Table \ref{tab: 15bar+3bar}, 
and as in scenario I, we label these $T$ in Table~\ref{tab: 15bar+3bar} and add asterisks to flavor exotic states.

Note that in this scenario, 
$T^*_{c\bar s0}(2900)^{++}$, $T_{c\bar s0}(2900)^0$ and $T^*_{cs0}(2870)^0$ are in the $\overline{\bf 15}$.
In particular,
$T^*_{c\bar s0}(2900)^{++}$ and $T_{c\bar s0}(2900)^0$ are $T^{++}_{c\bar s}$ and $T^0_{c\bar s}$, respectively, belonging to the $S=1$ isotriplet,
while $T^*_{cs0}(2870)^0$ is $T^0_{cs}$ belonging to another isotriplet with $S=-1$. 
In this scenario, however, there are six other flavor exotic states similar to the above states, namely, 
$T^{+}_{cs}$.
$T^{-}_{cs}$,
$T^{++}_{c\bar s \bar s}$,
$T^{+}_{c\bar s \bar s}$,
$T^{++}_{c}$,
and
$T^{-}_{c}$.
In fact, $T^{+}_{cs}$, $T^{-}_{cs}$ and $T^*_{cs0}(2870)^0$ belong to the same isotriplet (with $S=-1$),
while $T^{++}_{c\bar s \bar s}$,  
$T^{+}_{c\bar s \bar s}$, 
$T^{++}_{c}$ 
and
$T^{-}_{c}$ 
belong to the $S=-2$ isodouble and the iso-quarplet with $S=0$, respectively, as shown in Table \ref{tab: 15bar+3bar}. These nine flavor exotic states are labeled with asterisks in the table. They have been identified in the literature, see, for example, ref. \cite{Qin:2022nof}, where a different notation was used.
For convenience, we also show their notation for these states in Table~\ref{tab: 15bar+3bar}.

Furthermore, as mentioned previously, mixing for particles in different isospin multiplets is expected to be more difficult. 
Consequently, $T^{+}_{c\bar s}$ in the $S=+1$ isotripet, 
$T^+_c$ and $T^0_c$ in the $S=0$ iso-quarplet
do not mix with $T^{\prime\prime +}_{c\bar s}$, $D_s$, $T'^+_c$, $T'^0_c$ or $D^{+,0}$. 
They should also be considered flavor exotic particles and be labeled with asterisks in the table. 
To our knowledge, these particles have not been considered flavor exotic particles in the literature. 
Altogether, we have twelve flavor exotic states in this scenario.  

The tensor fields corresponding to states in scenario II are defined as follows,
\be
T^1_{\{23\}}
=
T^{++}_{cu \{\bar d \bar s\}}
=T^{++}_{cs}, 
&&
T^1_{\{31\}}
=T^{+}_{cu \{\bar s \bar u\}}
=\frac{1}{\sqrt2}T^{+}_{c\bar s}+\frac{1}{2}T^{\prime +}_{c\bar s}+\frac{1}{2}T^{\prime\prime+}_{c\bar s},
\non\\
T^1_{\{22\}}=\sqrt2 T^{++}_{cu \{\bar d \bar d\}}
=\sqrt{2} T^{++}_c, 
&&
T^1_{\{11\}}
=\sqrt2 T^{0}_{cu \{\bar u \bar u\}}
=\sqrt{\frac{2}{3}} T^{0}_{c}+\frac{1}{\sqrt3} T^{\prime 0}_{c}+T^{\prime\prime 0}_{c},
\non\\
T^1_{\{33\}}
=\sqrt2 T^{++}_{cu \{\bar s \bar s\}}
=\sqrt2 T^{++}_{c\bar s \bar s},
&&
T^1_{\{12\}}
=T^{+}_{cu \{\bar u \bar d\}}
=\sqrt\frac{2}{3}T^{+}_{c}+\frac{1}{2\sqrt{3}}T^{\prime +}_{c}+\frac{1}{2}T^{\prime\prime+}_{c},
\en
\be
T^2_{\{23\}}
=
T^{+}_{cd \{\bar d \bar s\}}
=-\frac{1}{\sqrt 2}T^{+}_{c \bar s}+\frac{1}{2}T^{\prime +}_{c \bar s}+\frac{1}{2}T^{\prime\prime+}_{c \bar s},
&&
T^2_{\{31\}}
=
T^{0}_{cd \{\bar s \bar u\}}
=T^{0}_{c \bar s},
\non\\
T^2_{\{12\}}
=
T^{0}_{cd \{\bar u \bar d\}}
=-\sqrt{\frac{2}{3}} T^{0}_{c}+\frac{1}{2\sqrt3} T^{\prime 0}_{c}+\frac{1}{2}T^{\prime\prime 0}_{c},
&&
T^2_{\{11\}}
=
\sqrt2 T^{-}_{cd \{\bar u \bar u\}}
=\sqrt2 T^{-}_{c}, 
\non\\
T^2_{\{22\}}
=
\sqrt2 T^{+}_{cd \{\bar d \bar d\}}
=-\sqrt\frac{2}{3}T^{+}_{c}+\frac{1}{\sqrt{3}}T^{\prime +}_{c}+T^{\prime\prime+}_{c},
&&
T^2_{\{33\}}
=
\sqrt2 T^{+}_{cd \{\bar s \bar s\}}
=\sqrt2 T^{+}_{c\bar s \bar s}, 
\en
and
\be
T^3_{\{23\}}
=
T^{+}_{cs \{\bar d \bar s\}}
=-\frac{\sqrt3}{2}T^{\prime +}_{c}+\frac{1}{2} T^{\prime\prime+}_{c},
&&
T^3_{\{31\}}
=
T^{0}_{cs \{\bar s \bar u\}}
=-\frac{\sqrt3}{2}T^{\prime 0}_{c}+\frac{1}{2} T^{\prime\prime 0}_{c},
\non\\
T^3_{\{12\}}
=
T^{0}_{cs \{\bar u \bar d\}}
=T^0_{cs}, 
&&
T^3_{\{11\}}
=
\sqrt2 T^{-}_{cs \{\bar u \bar u\}}
=\sqrt2 T^{-}_{cs},
\non\\
T^3_{\{22\}}
=
\sqrt2 T^{+}_{cs \{\bar d \bar d\}}
=\sqrt2 T^{+}_{cs}, 
&&
T^3_{\{33\}}
=
\sqrt2 T^{+}_{cs \{\bar s \bar s\}}
= -T^{\prime +}_{c \bar s}+T^{\prime\prime+}_{c \bar s},
\label{eq: T123...}
\en
where we use $\{\bar q' \bar q'\}$ to indicates the symmetric configuration of these light anti-quarks.
Note that from the above equations, we have
\be
T^i_{\{i 1\}}=T^i_{\{1 i \}}=2 T^{\prime\prime 0}_c,
\quad
T^i_{\{i 2\}}=T^i_{\{2 i \}}=2 T^{\prime\prime +}_c,
\quad
T^i_{\{i 3\}}=T^i_{\{3 i \}}=2 T^{\prime\prime +}_{c\bar s},
\label{eq: trace T II}
\en
where only the $\bar {\bf 3}'$ states remain after taking the trace, $T^i_{\{i k\}}$, as the $\overline{\bf 15}$ is traceless, which can also be seen from Table \ref{tab: 15bar+3bar}.

\subsection{Remarks}

In this work, we consider $T_{cs 0} (2870)^0$, $T^*_{c\bar s 0}(2900)^0$ and $T^*_{c\bar s 0}(2900)^{++}$ either in the ${\bf 6}$ as in scenario I or in the $\overline{\bf 15}$ as in scenario II.
In principle, they may belong to different multiplets. 
For example, we may have $T_{cs 0} (2870)^0$ and $T^*_{c\bar s 0}(2900)^0$ belonging to the ${\bf 6}$, while $T_{cs 0} (2870)^0$ belonging to the $\overline{\bf 15}$, or vice verse. 
It is our working assumption that they only belong to one of the multiplets, and we work out the consequences accordingly.
It is useful to verify these predictions experimentally to check the working assumption. 
Therefore, we should pay attention to predictions where all $T_{cs 0} (2870)^0$, $T^*_{c\bar s 0}(2900)^0$ and $T^*_{c\bar s 0}(2900)^{++}$ are involved, or, at least,  
$T_{cs 0} (2870)^0$ and either one of $T^*_{c\bar s 0}(2900)^0$ and $T^*_{c\bar s 0}(2900)^{++}$ are involved.

As mentioned, it is our working assumption that particles in scenarios I and II do not mix. 
Even if they do, the amplitudes obtained in this work are still useful, since amplitudes for mixed states can be built upon them. 
Nevertheless, not all flavor exotic states can mix.
Although $T^{++}_{c\bar s}$, $T^0_{c\bar s}$ and $T^0_{c\bar s}$ in scenarios I and II can,
$T_{cs}^0$s cannot, as they have different isospin quantum numbers.
In addition, the rest of the flavor exotic states in scenario II cannot, as they do not have any counterparts in scenario I, see Fig.~\ref{fig: SU(3)}.
Furthermore, all flavor exotic states identified in this section do not mix with $D$ or any other $c\bar q$ states in any case. 
They are exotic.

\section{$T\to DP$ and $T\to DS$ decays} \label{sec: T to DP, DS}

\subsection{$T\to DP$ decays}

We consider the strong decays of $T\to DP$ decays in this sub-section.
We will first discuss how to decompose these decay amplitudes into topological amplitudes. 
Subsequently, $T\to DP$ decay amplitudes will be decomposed accordingly.

\subsubsection{Topological amplitudes in $T\to DP$ decays}

In scenario I the Hamiltonian governing $T(cq [\bar q'\bar q''])\to DP$ decays is given by
\be
H_{\rm eff}&=& 
        F_{DP} \,T^i_{[jk]} \overline D^j \Pi^k_i
     +AC_P\, T^{i}_{[ik]} \overline D^r \Pi^k_r
     +AC_D\, T^{i}_{[ik]} \overline D^k \Pi^r_r,
\label{eq: H T DP I}
\en
while in scenario II the Hamiltonian for $T(cq \{\bar q'\bar q''\})\to DP$ decays is given by
\be
H_{\rm eff}&=&
             F_{DP} \,T^i_{\{jk\}} \overline D^j \Pi^k_i
          + AC_P\, T^{i}_{\{ik\}} \overline D^r \Pi^k_r
          + AC_D\, T^{i}_{\{ik\}} \overline D^k \Pi^r_r,
 \label{eq: H T DP II}
\en
with
\be
D= \left(
\begin{array}{ccc}
D^0 &D^+ &D^+_s
\end{array}
\right),
\en
\be
\Pi= \left(
\begin{array}{ccc}
\frac{\pi^0}{\sqrt2}+\frac{\cos\theta\, \eta+\sin\theta \, \eta'}{\sqrt2}
    & \pi^+ 
    & K^+
    \\
\pi^-
    & -\frac{\pi^0}{\sqrt2}+\frac{\cos\theta\, \eta+\sin\theta \, \eta'}{\sqrt2}
    & K^0
    \\
K^-  
    & \overline K^0
    & \cos\theta\, \eta'-\sin\theta\, \eta 
\end{array}
\right),
\en
and
\be
\theta=(39.3\pm1.0)^\circ,
\en
the $\eta$--$\eta'$ mixing angle \cite{Feldmann:1998sh}.
Note that although we use the same notations, 
namely $F_{DP}$, $AC_P$ and $AC_D$, in Eqs. (\ref{eq: H T DP I}) and (\ref{eq: H T DP II}) for simplicity of notation, they are, in principle, different.

\begin{figure}[t]
\centering
 \subfigure[]{
  \includegraphics[width=0.45\textwidth]{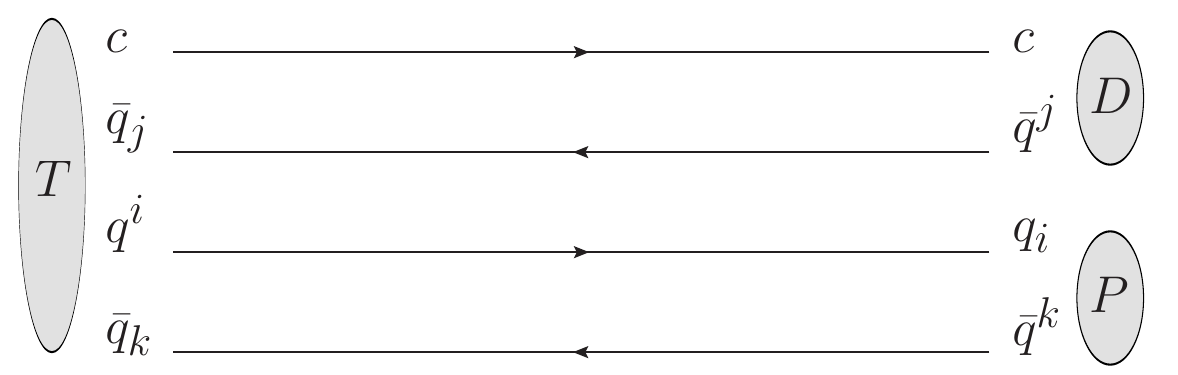}
}
\hspace{12pt}
\subfigure[]{
  \includegraphics[width=0.45\textwidth]{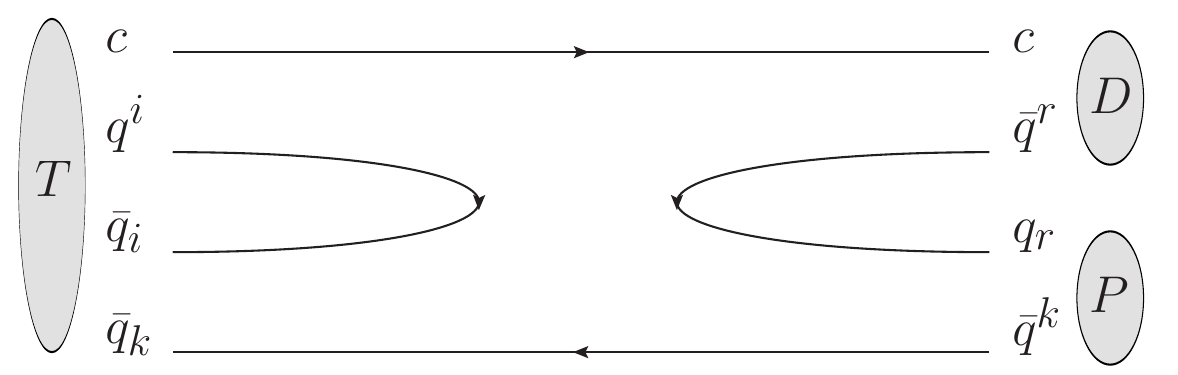}
}\\\subfigure[]{
  \includegraphics[width=0.45\textwidth]{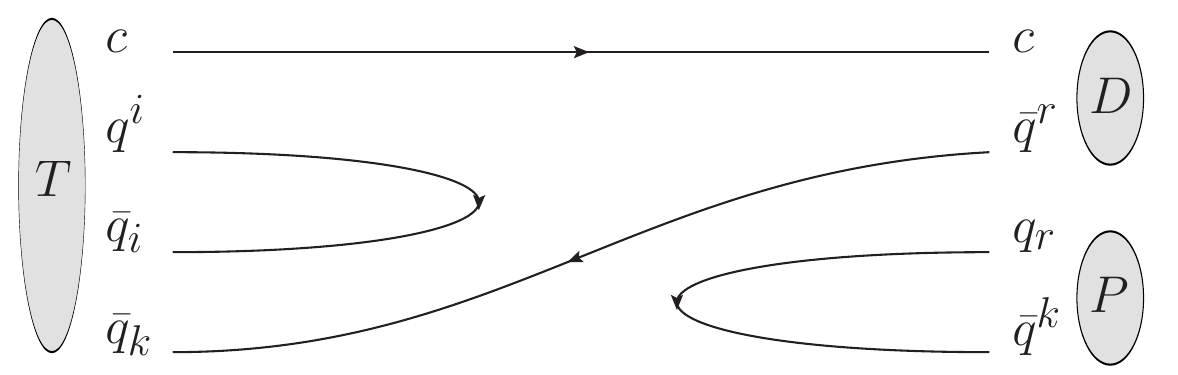}
}
\hspace{12pt}
\caption{Topological diagrams of 
  (a) $F_{DP}$ (fall-apart), (b) $AC_P$ (annihilation-creation) and (c) $AC_D$ (annihilation-creation)
   amplitudes in $T_{cq \bar q' \bar q'}\to DP$ decays. 
   The subscripts indicate the final state meson(s) receiving the light quark and/or the anti-quarks of $T_{cq \bar q' \bar q''}$.
  These are flavor flow diagrams. 
  The decay amplitudes of tetraquarks in the multiplet containing flavor exotic states can only have contributions from diagram (a).
} \label{fig: TA T2DP}
\end{figure}

As shown in the above effective Hamiltonian, there are three topological amplitudes in $T\to DP$ decays,
where
$F_{DP}$ is a fall-apart amplitude, 
$AC_P$ and $AC_D$ are annihilation-creation amplitudes.
The topological amplitudes are depicted in Fig.~\ref{fig: TA T2DP}. 
The subscripts indicate the final state meson(s) receiving the light quark and/or the anti-quarks of $T_{cq \bar q' \bar q''}$.
For example, for $AC_P$, an antiquark of $T$ goes into $P$, see the diagram in Fig.~\ref{fig: TA T2DP}(b). 
According to Okubo–Zweig–Iizuka  (OZI) rule \cite{Okubo:1963fa, Zweig:1964ruk, Iizuka:1966fk},
the $F_{DP}$ diagram is OZI superallowed, while $AC_P$ and$AC_D$ diagrams are OZI allowed.
It is reasonable to expect that the fall-apart amplitude is dominating, i.e.
\be
|F_{DP}|
>
 |AC_P|, |AC_D|.
\label{eq: T to DP}
\en

The situation is similar to the strong decay of low-lying scalar particles, where a tetraquark configuration is favored \cite{Jaffe:1976ig, Jaffe:1976ih}, 
see also ref. \cite{Cheng:2002ai}. 
For example, in the tetraquark picture, we have the following quark contents for $\sigma$ and $f_0$, $\sigma=ud \bar u \bar d$ and $f_0=(su\bar s\bar u+d s\bar d \bar s)/\sqrt 2$. 
The $\sigma\to\pi\pi$ decay amplitude is governed by a OZI superallowed fall-apart diagram, similar to $F_{DP}$ in  Fig.~\ref{fig: TA T2DP} (a), while the $f_0\to \pi\pi$ decay amplitude is by a OZI allowed diagram, similar to $AC_P$ in Fig.~\ref{fig: TA T2DP} (b).
Consequently, the $\sigma\to\pi\pi$ rate is larger than the $f_0\to \pi\pi$ rate, a fact supported by data~\cite{ParticleDataGroup:2024cfk}.

It is known that final state interaction can affect the size of a suppressed amplitude, see, for example, ref. \cite{Chua:2001br,Chua:2005dt,Chua:2007qw}.
Although $AC_P$ and $AC_D$ are expected to be smaller than $F_{DP}$,
they can be enhanced to some extent through final state interaction.
Their sizes remained to be determined experimentally.
Hence, one should keep them in the $T\to DP$ decay amplitudes.   
On the other hand, from the above discussion, one may have an impression that in the presence of final state interaction, the above result, i.e., Eq. (\ref{eq: T to DP}), based on the OZI rule, can be violated. 
This is, however, not true. 
For example, in $\overline B\to DP$ decays, one has exchange rescattering (governed by $r'_e$) and annihilation rescattering (governed by $r'_a$) in the presence of quasi-elastic rescattering~\cite{Chua:2001br,Chua:2005dt,Chua:2007qw}.
The former is OZI allowed, while the latter is OZI suppressed. 
By fitting to data, it was found that $|r'_e|>|r'_a|$ \cite{Chua:2005dt,Chua:2007qw}, hence, in line with the OZI rule. 
Therefore, final state interaction will not alter Eq. (\ref{eq: T to DP}), as they both respect the OZI rule.

The decay amplitudes of tetraquarks in the ${\bf 6}$ in scenario I and in the $\overline{\bf 15}$ in scenario II can only have contributions from diagram (a), the fall-apart amplitude, $F_{DP}$. This can be easily seen, as the $AC_P$ and $AC_D$ terms in Eqs. (\ref{eq: H T DP I}) and (\ref{eq: H T DP II}) necessary involve the trace $T^{i}_{[ik]}$ or $T^{i}_{\{ik\}}$, where only the states in the $\bar {\bf 3}$ in scenario I  and the $\bar{\bf 3}'$ in scenario II remain, see Eqs.(\ref{eq: trace T I}) and (\ref{eq: trace T II}).  
In fact, the reason for the above facts can be easily understood by noting $\bar {\bf 3}\otimes {\bf 8}=\overline{\bf 15}\oplus{\bf 6}\oplus\bar{\bf 3}$, see Eq.~(\ref{eq: SU(3) decompositions 1}).
Hence there is only one $\overline{\bf 15}$ and one ${\bf 6}$ in $D(\bar{\bf 3})\Pi({\bf 8})$ final states, which can match the $\overline{\bf 15}$ of $T$ in scenario II and ${\bf 6}$ of $T$ in scenario I, respectively. Adding $\eta_1$ in $P$ does not affect the above matching.

\subsubsection{$T\to DP$ decay amplitudes}

\begin{table}[t!]
\caption{\label{tab: TtoDP1}
$T\to DP$ decay amplitudes in scenario I, with $T=T_{cq[\bar q'\bar q'']}$.}
\footnotesize{
\begin{ruledtabular}
\begin{tabular}{llclc}
\#
          &Mode
          & $A (T\to D P)$
          & Mode
          & $A (T\to D P)$
          \\
\hline 
$1^{*}$
          & $T^{++}_{c\bar s}\to D^+ K^+$
          & $F_{DP}$
          & $T^{++}_{c\bar s}\to D_s^+ \pi^+$
          & $-F_{DP}$
          \\
$2^*$
          & $T^{+}_{c\bar s}\to D^0 K^+$
          & $\frac{1}{\sqrt2} F_{DP}$ 
          & $T^{+}_{c\bar s}\to D^+ K^0$
          & $-\frac{1}{\sqrt2} F_{DP}$
          \\
          & $T^{+}_{c\bar s}\to D_s^+ \pi^0$
          & $F_{DP}$
          \\                                                                
$3^{*}$
          & $T^{0}_{c\bar s}\to D^0 K^0$
          & $F_{DP}$
          & $T^{0}_{c\bar s}\to D_s^+ \pi^-$
          & $-F_{DP}$
          \\
\hline 
$4^{*}$ 
          & $T^{0}_{cs}\to D^0 \ov K^0$
          & $F_{DP}$
          & $T^{0}_{cs}\to D^+ K^-$
          & $-F_{DP}$
          \\
\hline 
5
          &$T^{+}_{c}\to D^0 \pi^+$
          & $\frac{1}{\sqrt2} F_{DP}$
          & $T^{+}_{c}\to D^+ \pi^0$
          & $-\frac{1}{2} F_{DP}$
          \\
          &$T^{+}_{c}\to D_s^+ \ov K^0$
          & $-\frac{1}{\sqrt2} F_{DP}$
          & $T^{+}_{c}\to D^+ \eta$
          & $-\frac{c_{\theta} +\sqrt2 s_{\theta} }{2} F_{DP} $           
          \\
          & $T^{+}_{c}\to D^+ \eta'$
          & $\frac{\sqrt2 c_{\theta} - s_{\theta} }{2} F_{DP}$ 
          \\ 
6
          &$T^{0}_{c}\to D^0 \pi^0$
          & $-\frac{1}{2} F_{DP}$
          & $T^{0}_{c}\to D^+ \pi^-$
          & $-\frac{F_{DP}}{\sqrt2}$
          \\
          &$T^{0}_{c}\to D_s^+ K^-$
          & $\frac{1}{\sqrt2} F_{DP}$
          & $T^{0}_{c}
          \to D^+ \eta$
          &  $\frac{c_{\theta} +\sqrt2 s_{\theta} }{2} F_{DP} $           
          \\
          & $T^{0}_{c}\to D^+ \eta'$
          & $- \frac{\sqrt2 c_{\theta} - s_{\theta}}{2} F_{DP}$ 
          \\    
\hline
\hline
7 
          & $T^{\prime\prime +}_{c}\to D^0 \pi^+$
          & $\frac{1}{\sqrt2}(F_{DP}+2AC_P)$
          & $T^{\prime\prime +}_{c}\to D^+ \pi^0$
          & $-\frac{1}{2}(F_{DP}+2AC_P)$ 
          \\
          & $T^{\prime\prime +}_{c}\to D_s^+ \ov K^0$
          & $\frac{1}{\sqrt2} (F_{DP}+2AC_P)$
          \\
          & $T^{\prime\prime +}_{c}\to D^+ \eta$
          & $\frac{c_{\theta}+\sqrt2 s_{\theta}}{6} (F_{DP}+2 AC_P)$        
          & $T^{\prime\prime +}_{c}\to D^+ \eta'$
          & $\frac{s_{\theta}-\sqrt2 c_{\theta}}{6} (F_{DP}+2 AC_P)$ 
          \\ 
          & 
          & $+\frac{\sqrt2 s_{\theta}-2c_{\theta}}{3}(F_{DP}-AC_P-3AC_D)$          
          & 
          & $ -\frac{\sqrt2 c_{\theta}+2s_{\theta}}{3}(F_{DP}-AC_P-3AC_D)$  
          \\ 
8
          & $T^{\prime\prime 0}_{c}\to D^0 \pi^0$
          & $\frac{1}{2}(F_{DP}+2AC_P)$
          & $T^{\prime\prime 0}_{c}\to D^+ \pi^-$
          & $\frac{1}{\sqrt2}(F_{DP}+2AC_P)$
          \\
          &$T^{\prime\prime 0}_{c}\to D_s^+ K^-$
          & $\frac{1}{\sqrt2}(F_{DP}+2AC_P)$
          \\
          & $T^{\prime\prime 0}_{c}\to D^0 \eta$
          & $\frac{c_{\theta}+\sqrt2 s_{\theta}}{6} (F_{DP}+2 AC_P)$           
          & $T^{\prime\prime 0}_{c}\to D^+ \eta'$
          & $\frac{s_{\theta}-\sqrt2 c_{\theta}}{6} (F_{DP}+2 AC_P)$  
          \\  
          & 
          & $+\frac{\sqrt2 s_{\theta}-2c_{\theta}}{3}(F_{DP}-AC_P-3AC_D)$          
          & 
          & $ -\frac{\sqrt2 c_{\theta}+2s_{\theta}}{3}(F_{DP}-AC_P-3AC_D)$  
          \\  
\hline  
9
          & $T^{\prime\prime +}_{c\bar s}\to D^0 K^+$
          & $\frac{1}{\sqrt2}(F_{DP}+2AC_P)$
          & $T^{\prime\prime +}_{c\bar s}\to D^+ K^0$
          & $\frac{1}{\sqrt2}(F_{DP}+2AC_P)$
          \\
          & $T^{\prime\prime +}_{c\bar s}\to D_s^+ \eta$
          & $-\frac{c_{\theta}+\sqrt2 s_{\theta}}{3} (F_{DP}+2 AC_P)$        
          & $T^{\prime\prime +}_{c\bar s}\to D_s^+ \eta'$
          & $\frac{\sqrt2 c_{\theta}-s_{\theta}}{3} (F_{DP}+2 AC_P)$  
          \\                              
          & 
          & $ +\frac{\sqrt2 s_{\theta}-2c_{\theta}}{3}(F_{DP}-AC_P-3AC_D)$          
          & 
          & $ -\frac{\sqrt2 c_{\theta}+2s_{\theta}}{3}(F_{DP}-AC_P-3AC_D)$  
          \\                              
\end{tabular}
\end{ruledtabular}
}
\end{table}

\begin{table}[t!]
\caption{\label{tab: TtoDP2}
$T\to DP$ decay amplitudes in scenario II, with $T=T_{cq\{\bar q'\bar q''\}}$.} 
\scriptsize{
\begin{ruledtabular}
\begin{tabular}{llclc}
\#
          & Mode
          & $A (T\to D P)$
          & Mode
          & $A (T\to D P)$
          \\
\hline 
$1'^{*}$
          &$T^{++}_{c\bar s}\to D^+ K^+$
          & $F_{DP}$
          & $T^{++}_{c\bar s}\to D_s^+ \pi^+$
          & $F_{DP}$
          \\
$2'^*$
          &$T^{+}_{c\bar s}\to D^0 K^+$
          & $\frac{1}{\sqrt2} F_{DP}$ 
          & $T^{+}_{c\bar s}\to D^+ K^0$
          & $-\frac{1}{\sqrt2} F_{DP}$ 
          \\
          &$T^{+}_{c\bar s}\to D_s^+ \pi^0$
          & $F_{DP}$
          \\
$3'^{*}$
          &$T^{0}_{c\bar s}\to D^0 K^0$
          & $F_{DP}$
          & $T^{0}_{c\bar s}\to D_s^+ \pi^-$
          & $F_{DP}$
          \\
\hline 
$4'^{*}$
          &$T^{+}_{cs}\to D^+ \ov K^0$
          & $\sqrt2 F_{DP}$
          \\ 
$5'^{*}$
          &$T^{0}_{cs}\to D^0 \ov K^0$
          & $F_{DP}$
          & $T^{0}_{cs}\to D^+ K^-$
          & $F_{DP}$
          \\         
$6'^{*}$       
          &$T^{-}_{cs}\to D^0 K^-$
          & $\sqrt2 F_{DP}$
          \\
\hline          
$7'^{*}$
          & $T^{++}_{c\bar s\bar s}\to D_s^+ K^+$
          & $\sqrt2 F_{DP}$
          \\
$8'^{*}$
          & $T^{+}_{c\bar s\bar s}\to D_s^+ K^0$
          & $\sqrt2 F_{DP}$
          \\
\hline
$9'^{*}$
          & $T^{++}_{c}\to D^+ \pi^+$
          & $\sqrt2 F_{DP}$
          \\
$10'^*$
          &$T^{+}_{c}\to D^0 \pi^+$
          & $\sqrt{\frac{2}{3}}F_{DP}$
          & $T^{+}_{c}\to D^+ \pi^0$
          & $\frac{2}{\sqrt3} F_{DP}$
          \\
$11'^*$
          &$T^{0}_{c}\to D^0 \pi^0$
          & $\frac{2}{\sqrt 3}F_{DP}$
          & $T^{0}_{c}\to D^+ \pi^-$
          & $-\sqrt{\frac{2}{3}}F_{DP}$
          \\  
$12'^{*}$
          & $T^{-}_{c}\to D^0 \pi^-$
          & $\sqrt2 F_{DP}$
          \\
\hline 
$13'$
          &$T^{\prime +}_{c}\to D^0 \pi^+$
          & $\frac{1}{2\sqrt3}F_{DP}$
          & $T^{\prime +}_{c}\to D^+ \pi^0$
          & $-\frac{1}{2\sqrt 6}F_{DP}$
          \\
          & $T^{\prime +}_{c}\to D^+ \eta$
          & $\frac{\sqrt 6 }{4} (c_{\theta}+\sqrt2 s_{\theta}) F_{DP}$           
          & $T^{\prime +}_{c}\to D^+ \eta'$
          & $\frac{\sqrt 6 }{4} (s_{\theta}-\sqrt2 c_{\theta}) F_{DP}$ 
          \\ 
          &$T^{\prime +}_{c}\to D_s^+ \ov K^0$
          & $-\frac{\sqrt 3}{2} F_{DP}$
          \\
$14'$
          &$T^{\prime 0}_{c}\to D^0 \pi^0$
          &  $\frac{1}{2\sqrt 6}F_{DP}$
          & $T^{\prime 0}_{c}\to D^0 \eta$
          & $\frac{\sqrt 6 }{4} (c_{\theta}+\sqrt2 s_{\theta}) F_{DP}$     
          \\       
          & $T^{\prime 0}_{c}\to D^0 \eta'$
          & $\frac{\sqrt 6 }{4} (s_{\theta}-\sqrt2 c_{\theta}) F_{DP}$ 
          & $T^{\prime 0}_{c}\to D^+ \pi^-$
          & $\frac{1}{2\sqrt 3} F_{DP}$
          \\
          &$T^{\prime 0}_{c}\to D_s^+ K^-$
          & $-\frac{\sqrt 3}{2} F_{DP}$
          \\            
\hline  
$15'$
          &$T^{\prime +}_{c\bar s}\to D^0 K^+$
          & $\frac{1}{2} F_{DP}$
          &$T^{\prime +}_{c\bar s}\to D^+ K^0$
          & $\frac{1}{2} F_{DP}$  
          \\
          & $T^{\prime +}_{c\bar s}\to D_s^+ \eta$
          & $\frac{1}{\sqrt2} (c_{\theta}+\sqrt2 s_{\theta}) F_{DP}$
          & $T^{\prime +}_{c\bar s}\to D_s^+ \eta'$
          & $\frac{1}{\sqrt2} (s_{\theta}-\sqrt2 c_{\theta}) F_{DP}$
          \\            
\hline
\hline
$16'$
          &$T^{\prime \prime +}_{c}\to D^0 \pi^+$
          & $\frac{1}{2} (F_{DP}+4 AC_P)$
          & $T^{\prime \prime +}_{c}\to D^+ \pi^0$
          & $-\frac{1}{2\sqrt2} (F_{DP}+4 AC_P)$
          \\
          &$T^{\prime \prime +}_{c}\to D_s^+ \ov K^0$
          & $\frac{1}{2} (F_{DP}+4 AC_P)$
          \\
          & $T^{\prime \prime +}_{c}\to D^0 \eta$
          & $\frac{\sqrt2 c_{\theta}+2 s_{\theta}}{12} (F_{DP}+4 AC_P)$          
          & $T^{\prime \prime +}_{c}\to D^+ \eta'$
          & $\frac{\sqrt2 s_{\theta}-2 c_{\theta}}{12} (F_{DP}+4 AC_P) $
          \\  
          & 
          & $+\frac{2 c_{\theta}-\sqrt2 s_{\theta}}{3} [3AC_D+\sqrt2(F_{DP}+AC_P)] $          
          & 
          & $+\frac{\sqrt2 c_{\theta}+2 s_{\theta}}{3} [3AC_D+\sqrt2(F_{DP}+AC_P)] $
          \\  
$17'$   
          &$T^{\prime \prime 0}_{c}\to D^0 \pi^0$
          & $\frac{1}{2\sqrt2} (F_{DP}+4 AC_P)$
          & $T^{\prime \prime 0}_{c}\to D^+ \pi^-$
          & $\frac{1}{2}(F_{DP}+4 AC_P)$ 
          \\
         & $T^{\prime \prime 0}_{c}\to D_s^+ K^-$
         & $\frac{1}{2} (F_{DP}+4 AC_P)$
          \\
         & $T^{0}_{c}\to D^0 \eta$
         & $\frac{\sqrt2 c_{\theta}+2 s_{\theta}}{12} (F_{DP}+4 AC_P)$          
         & $T^{0}_{c}\to D^+ \eta'$
         & $\frac{\sqrt2 s_{\theta}-2 c_{\theta}}{12} (F_{DP}+4 AC_P) $
         \\   
         & 
          & $+\frac{2 c_{\theta}-\sqrt2 s_{\theta}}{3} [3AC_D+\sqrt2(F_{DP}+AC_P)] $          
         & 
          & $+\frac{\sqrt2 c_{\theta}+2 s_{\theta}}{3} [3AC_D+\sqrt2(F_{DP}+AC_P)] $
         \\   
\hline
$18'$         
          &$T^{\prime \prime +}_{c\bar s}\to D^0 K^+$
          & $\frac{1}{2} (F_{DP}+4 AC_P)$
          & $T^{\prime \prime +}_{c\bar s}\to D^+ K^0$
          &$\frac{1}{2} (F_{DP}+4 AC_P)$
          \\
          & $T^{\prime \prime +}_{c\bar s}\to D_s^+ \eta$
          & $-\frac{\sqrt2 c_{\theta}+2 s_{\theta}}{6} (F_{DP}+4 AC_P)$          
          & $T^{+}_{c\bar s}\to D_s^+ \eta'$
          & $\frac{2 c_{\theta}-\sqrt2 s_{\theta}}{6} (F_{DP}+4 AC_P) $
          \\
         & 
         & $+\frac{2 c_{\theta}-\sqrt2 s_{\theta}}{3} [3AC_D+\sqrt2(F_{DP}+AC_P)] $          
         & 
         & $+\frac{\sqrt2 c_{\theta}+2 s_{\theta}}{3} [3AC_D+\sqrt2(F_{DP}+AC_P)] $
         \\                                               
\end{tabular}
\end{ruledtabular}
}
\end{table}

With the above effective Hamiltonians, we can decompose $T\to DP$ decay amplitudes in terms of these topological amplitudes. 
The $T\to DP$ decay amplitudes for $T=T(cq [\bar q'\bar q''])$ in scenario I and $T=T(cq \{\bar q'\bar q''\})$ in scenario II are shown in Table \ref{tab: TtoDP1} and \ref{tab: TtoDP2}, respectively.
Note that some of the modes may be kinematically prohibited, but the information on these amplitudes can still be useful, as the open-charmed tetraquark can be a virtual particle in some other processes.

From Table \ref{tab: TtoDP1} and \ref{tab: TtoDP2}, we see that for flavor exotic states and states in the same multiplets, namely ${\bf 6}$ in scenario I and $\overline{\bf 15}$ in scenario II, the $T\to DP$ decay amplitudes only consist of the fall-apart amplitude, i.e. $F_{DP}$. 
Their decay amplitudes are all highly related.

In fact, for flavor exotic states in scenario I, most of their decay amplitudes are identical up to a sign. 
As one can infer from Table \ref{tab: TtoDP1}, the squares of the amplitudes of the following six modes,
namely 
$T^{++}_{c\bar s}\to D^+ K^+, D_s^+ \pi^+$,
$T^{+}_{c\bar s}\to D_s^+ \pi^0$,
$T^{0}_{c\bar s}\to D^0 K^0, D_s^+ \pi^-$ 
and 
$T^{0}_{cs}\to D^0 \ov K^0, D^+ K^-$ decays,
should be identical,
while those in $T^{+}_{c\bar s}\to D^0 K^+$ and $T^{+}_{c\bar s}\to D^+ K^0$ decays are half the above.
The flavor exotic nature of most of these states is self-evident from their decay modes shown in the table.

The three other states in ${\bf 6}$, namely, $T^+_{c\bar s}$, $T^+_c$ and $T^0_c$, have $T\to DP$ decay rates that are proportional to the previous ones with simple factors.
Note that in $T^{+}_{c\bar s}\to D_s^+ \pi^0$ decay the exotic quantum number of $T^{+}_{c\bar s}$ can also be easily identified,
as $\pi^0$ is an iso-vector, $T^{+}_{c\bar s}$ cannot be a $c\bar s$ state. 

The situation is different for the three states in $\bar{\bf 3}$, namely $T^{\prime\prime +}_{c}$, $T^{\prime\prime 0}_{c}$ and $T^{\prime\prime +}_{cs}$, as their amplitudes contain $F_{DP}$, $AC_P$ and $AC_D$ as well.
Nevertheless, the $T\to DP$ decay amplitudes of these states for $P=\pi$, $K$ and $\overline K$, are proportional to $F_{DP}+ 2AC_P$, giving 
$\Gamma(T\to D\pi^+)=\Gamma(T\to D\overline K)=\Gamma(T\to DK)=2\Gamma(T\to D\pi^0)$ for these states.
As noted previously we have $\bar {\bf 3}\otimes {\bf 8}=\overline{\bf 15}\oplus{\bf 6}\oplus\bar{\bf 3}$, see Eq.~(\ref{eq: SU(3) decompositions 1}).
The $\bar{\bf 3}$ in the $D\Pi$ final states can match the $\bar{\bf 3}$ of $T$ in scenario I, but the addition of $\eta_1$ in $P$ provides another $\bar{\bf 3}'$ in $DP$ and another combination, 
namely $F_{DP}-AC_P-3AC_D$, 
is needed for the $T\to D\eta, D\eta'$ decay amplitudes.

Indeed, the above form on $T\to DP$ decay amplitudes still holds in the presence of rescattering,
as strong interaction does not affect the flavor structure of the decay amplitudes.
For illustration, we consider the effect of final state interaction in the $DP$ system with quasi-elastic strong rescattering \cite{Chua:2001br,Chua:2005dt,Chua:2007qw}. 
Before rescattering the topological amplitudes in the $T\to DP$ decay amplitudes are given by Table~\ref{tab: TtoDP1}, 
but with $F_{DP}$, $F_{DP}+2AC_P$ and $F_{DP}-AC_P-3AC_D$ replaced with 
$F^0_{DP}$, $F^0_{DP}+2AC^0_P$ and $F^0_{DP}-AC^0_P-3AC^0_D$ (superscript $0$ are added to indicate amplitudes before rescattering), respectively. 
The question is whether the form of $T\to DP$ decay amplitudes, in terms of topological amplitudes, holds in the presence of the rescattering.
As shown in Appendix~\ref{sec: FSI}, after rescattering, the form of the decay ampltudes remain unchanged, 
but with topological amplitudes modified and redefined as follows,  
\be
F_{DP}&=& e^{i\delta_{6}} F^0_{DP},
\non\\
\left(
\begin{array}{c}
F_{DP}+2AC_P\\
F_{DP}-AC_P-3AC_D
\end{array}
\right)
&=&
\left(
\begin{array}{cc}
e^{i\delta_{\bar 3}}\cos^2\tau+e^{i\delta_{\bar 3'}}\sin^2\tau
& \frac{e^{i\delta_{\bar 3}}-e^{i\delta_{\bar 3'}}}{\sqrt2}\sin \tau\cos\tau
\\
\sqrt2 (e^{i\delta_{\bar 3}}-e^{i\delta_{\bar 3'}})\sin \tau\cos\tau
& e^{i\delta_{\bar 3'}}\cos^2\tau+e^{i\delta_{\bar 3}}\sin^2\tau
\end{array}
\right)
\non\\
&&\times
\left(
\begin{array}{c}
F^0_{DP}+2AC^0_P\\
F^0_{DP}-AC^0_P-3AC^0_D
\end{array}
\right),
\label{eq: FSI I}
\en
where $\delta_{6}$, $\delta_{\bar 3}$ and $\delta_{\bar 3'}$ are strong phases and $\tau$ is the mixing angle between the $\bar{\bf 3}$ and $\bar{\bf 3}'$ in the $DP$ system.
For example, according to Table~\ref{tab: TtoDP1}, before rescattering the $T^{++}_{c\bar s}\to D^+ K^+$ and $D_s^+\pi^+$ amplitudes are $F^0_{DP}$ and $-F^0_{DP}$, respectively.
As shown in Appendix~\ref{sec: FSI}, after quasi-elastic strong rescattering, they become $e^{i\delta_{6}} F^0_{DP}$ and $-e^{i\delta_{6}} F^0_{DP}$, respectively, which agree to those shown in Table \ref{tab: TtoDP1} by taking $F_{DP}=e^{i\delta_{6}} F^0_{DP}$.
Therefore, the form of the amplitudes in terms of topological amplitudes is unchanged, but the topological amplitude itself is affected by rescattering.
This agrees with the common understanding that topological amplitudes can include long-distance final state interaction effects, 
see, for example, \cite{Chau:tk, Chau:1990ay,Rosner:1999xd, Cheng:2002ai}.

Similarly, for scenario II, as one can infer from Table \ref{tab: TtoDP2}, the following modes,
namely,
$T^{++}_{c\bar s}\to D^+ K^+, D_s^+ \pi^+$,
$T^{+}_{c\bar s}\to D_s^+ \pi^0$,
$T^{0}_{c\bar s}\to D^0 K^0, D_s^+ \pi^-$
and
$T^{0}_{cs}\to D^0 \ov K^0, D^+ K^-$ decays,
should have identical amplitudes squared.
The amplitudes squared of
$T^{+}_{cs}\to D^+ \ov K^0$,
$T^{-}_{cs}\to D^0 K^-$,
$T^{++}_{c\bar s\bar s}\to D_s^+ K^+$,
$T^{+}_{c\bar s\bar s}\to D_s^+ K^0$,
$T^{++}_{c}\to D^+ \pi^+$
and
$T^{-}_{c}\to D^0 \pi^-$ decays are doubled compared to the former modes.
In addition, for the other two flavor exotic states, we have $T^{+}_{c}\to D^0 \pi^+, D^+ \pi^0$, $T^{0}_{c}\to D^0 \pi^0$ and $D^+ \pi^-$ decays with amplitudes squared porpotianal to the aboves.
The flavor exotic nature of most of these states is self-evident from their decay modes shown in the table.
For $T^{+}_{c\bar s}$, the $T^{+}_{c\bar s}\to D_s^+ \pi^0$ decay reveals the exotic nature of $T^{+}_{c\bar s}$,
but for $T^{+}_{c}$ and $T^{0}_{c}$, elebrated works are required, 
as one needs to verify that they do not decay to $D\eta$ and $D\eta'$.

Furthermore, four other states in $\overline{\bf 15}$, namely, 
$T^+_{c\bar s}$,
$T^{\prime +}_c$,
$T^{\prime 0}_c$
and
$T^{\prime +}_{c\bar s}$,    
have $T\to DP$ decay amplitudes squared proportional to the previous ones with some simple factors,
while the situation is slightly different for the three states in $\bar{\bf 3}'$, 
namely $T^{\prime\prime +}_{c}$, $T^{\prime\prime 0}_{c}$ and $T^{\prime\prime +}_{cs}$, as their amplitudes contain $F_{DP}$, $AC_P$ and $AC_D$ as well. 
Nevertheless, as shown in Table~\ref{tab: TtoDP2}, the $T\to DP$ decay amplitudes of these states for $P=\pi$, $K$ and $\overline K$, are proportional to the combination of $F_{DP}+ 4AC_P$, giving $\Gamma(T\to D\pi^+)=\Gamma(T\to D\overline K)=\Gamma(T\to DK)=2\Gamma(T\to D\pi^0)$ for these states.
The $\bar{\bf 3}$ in the $D\Pi$ final states can match the $\bar{\bf 3}'$ of $T$ in scenario II, but the addition of $\eta_1$ in $P$ provide another $\bar{\bf 3}'$ in $DP$, 
where another combination, namely $3AC_D+\sqrt2 (F_{DP}+AC_P)$, is needed for the $T\to D\eta, D\eta'$ decay amplitudes.

Similarly, under quasi-elastic scattering  \cite{Chua:2001br,Chua:2005dt,Chua:2007qw}, the topological amplitudes are given by, see Appendix~\ref{sec: FSI},
\be
F_{DP}&=& e^{i\delta_{\overline {15}}} F^0_{DP},
\non\\
\left(
\begin{array}{c}
F_{DP}+4AC_P\\
F_{DP}+AC_P+\frac{3}{\sqrt2}AC_D
\end{array}
\right)
&=&
\left(
\begin{array}{cc}
e^{i\delta_{\bar 3}}\cos^2\tau+e^{i\delta_{\bar 3'}}\sin^2\tau
& \sqrt2 (e^{i\delta_{\bar 3'}}-e^{i\delta_{\bar 3}})\sin \tau\cos\tau
\\
\frac{e^{i\delta_{\bar 3'}}-e^{i\delta_{\bar 3}}}{\sqrt2}\sin \tau\cos\tau
& e^{i\delta_{\bar 3'}}\cos^2\tau+e^{i\delta_{\bar 3}}\sin^2\tau
\end{array}
\right)
\non\\
&&\times
\left(
\begin{array}{c}
F^0_{DP}+4AC^0_P\\
F^0_{DP}+AC^0_P+\frac{3}{\sqrt2}AC^0_D
\end{array}
\right),
\label{eq: FSI II}
\en
where $\delta_{\overline{15}}$, $\delta_{\bar 3}$ and $\delta_{\bar 3'}$ are strong phases and $\tau$ is the mixing angle between the $\bar{\bf 3}$ and $\bar{\bf 3}'$ in the $DP$ system,
and $F^0_{DP}$, $F^0_{DP}+4AC^0_P$ and $F^0_{DP}+AC^0_P+\frac{3}{\sqrt2}AC^0_D$ are the topological amplitudes before rescattering.
The form of $T\to DP$ decay amplitudes in terms of the topological amplitudes does not change under rescattering. 
For example, according to Table \ref{tab: TtoDP2}, before rescattering the $T^{++}_{c\bar s}\to D^+ K^+$ and $D_s^+\pi^+$ amplitudes are $F^0_{DP}$ and $F^0_{DP}$, respectively.
As shown in Appendix~\ref{sec: FSI}, after quasi-elastic strong rescattering, they become $e^{i\delta_{\overline {15}}} F^0_{DP}$ and $e^{i\delta_{\overline {15}}} F^0_{DP}$, respectively, which agree to those shown in Table \ref{tab: TtoDP2} by taking $F_{DP}=e^{i\delta_{\overline {15}}} F^0_{DP}$.
Therefore, the form of the amplitudes in terms of topological amplitudes is unchanged, but the topological amplitude itself is affected by rescattering.

Recently, LHCb found that at the current experimental precision, the relative rates of 
$T^*_{cs0}(2870)\to D^+ K^-$ and $T^*_{cs0}(2870)\to D^0 \ov K{}^0$ decays are consistent with an isospin invariant~\cite{LHCb:2024xyx}.
This result supports the above results, where in the
$T^{0}_{cs}\to D^0 \ov K^0, D^+ K^-$ modes we have 
$|A(T^{0}_{cs}\to D^0 \ov K^0)|=|A(T^{0}_{cs}\to D^+ K^-)|$ in both scenarios.

\subsection{$T\to DS$ decays}

\subsubsection{Topological amplitudes in $T\to DS$ decays}

The Hamiltonian in scenario I governing $T(cq [\bar q\bar q'])\to DS$ decays is given by
\be
H_{\rm eff}&=& 
       C_{S} \,T^i_{[jk]} \overline D^p S^{kj}_{pi}
     +C_{DS} \,T^i_{[jk]} \overline D^j S^{kp}_{pi}
     +ACC_{S}\, T^{i}_{[ik]} \overline D^r S^{kp}_{pr}
     +ACC_{D}\, T^{i}_{[ik]} \overline D^k S^{rp}_{pr},
\en
while the Hamiltonian in scenario II for $T(cq \{\bar q\bar q'\})\to DS$ decays is given by,
\be
H_{\rm eff}&=&
             C_{DS} \,T^i_{\{jk\}} \overline D^j S^{kp}_{pi}
     +ACC_{S}\, T^{i}_{\{ik\}} \overline D^r S^{kp}_{pr}
     +ACC_{D}\, T^{i}_{\{ik\}} \overline D^k S^{rp}_{pr}.
\en
Note that in the above equations, $S^{ij}_{kl}$ denotes the field for low-lying scalar particles with 
\be
S^{ij}_{kl}=\epsilon^{ijp} \S_p^q \epsilon_{qkl}=S( [q^i q^j] [\bar q_k \bar q_l]),
\en
with
\be
\S= \left(
\begin{array}{ccc}
\frac{a^0_0}{\sqrt2}+\frac{\cos\phi\, f_0-\sin\phi \,\sigma}{\sqrt2}
    & a^+_0 
    & \kappa^+
    \\
a^-_0
    & -\frac{a^0_0}{\sqrt2}+\frac{\cos\phi\, f_0-\sin\phi \,\sigma}{\sqrt2}
    & \kappa^0
    \\
\kappa^-  
    & \bar\kappa^0
    & \sin\phi\, f_0+\cos\phi\, \sigma 
\end{array}
\right),
\en
the mixing angle is taken to be $\phi=174.6^\circ$\cite{Maiani:2004uc}.
For simplicity of notation, we do not distinguish the topological amplitudes in scenario I and scenario II, which, however, do not imply that they are identical.
We follow refs.~\cite{Jaffe:1976ig, Jaffe:1976ih} on the quark contents of these low-lying scalar particles. 
For more information on the flavor structure of these particles, see Appendix~\ref{sec: scalar}.

\begin{figure}[t]
\centering
 \subfigure[]{
  \includegraphics[width=0.45\textwidth]{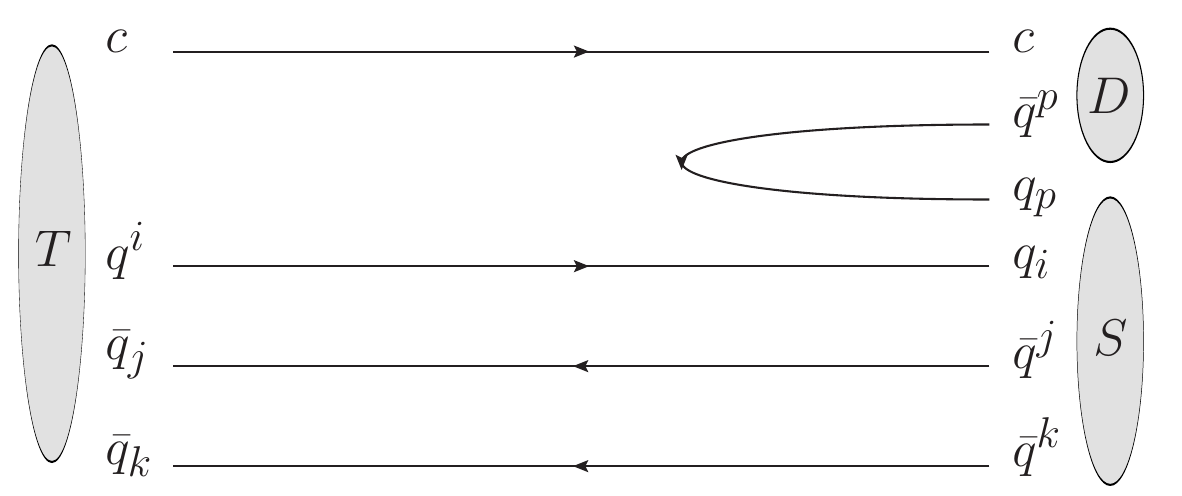}
}
\hspace{12pt}
\subfigure[]{
  \includegraphics[width=0.45\textwidth]{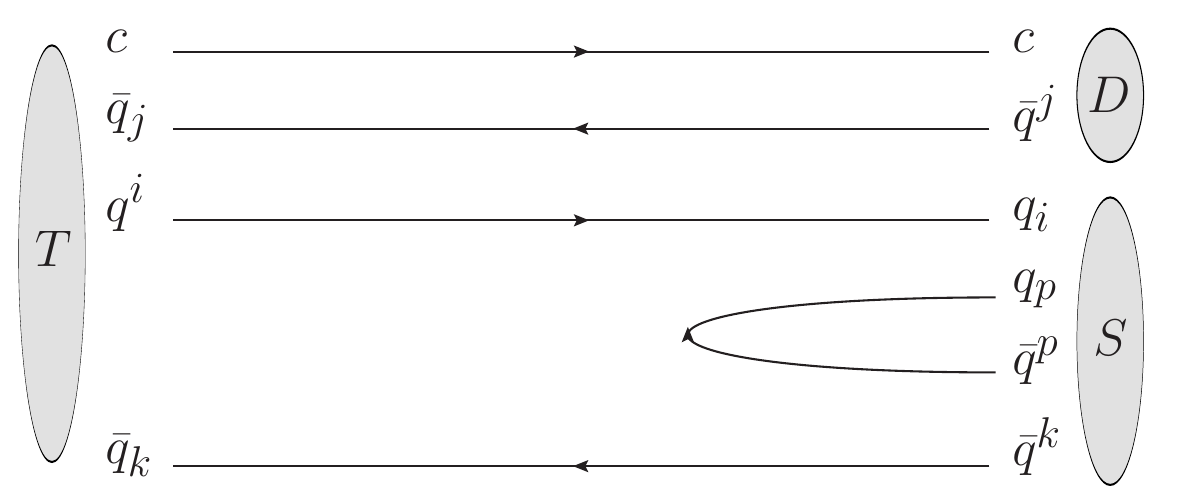}
}\\\subfigure[]{
  \includegraphics[width=0.45\textwidth]{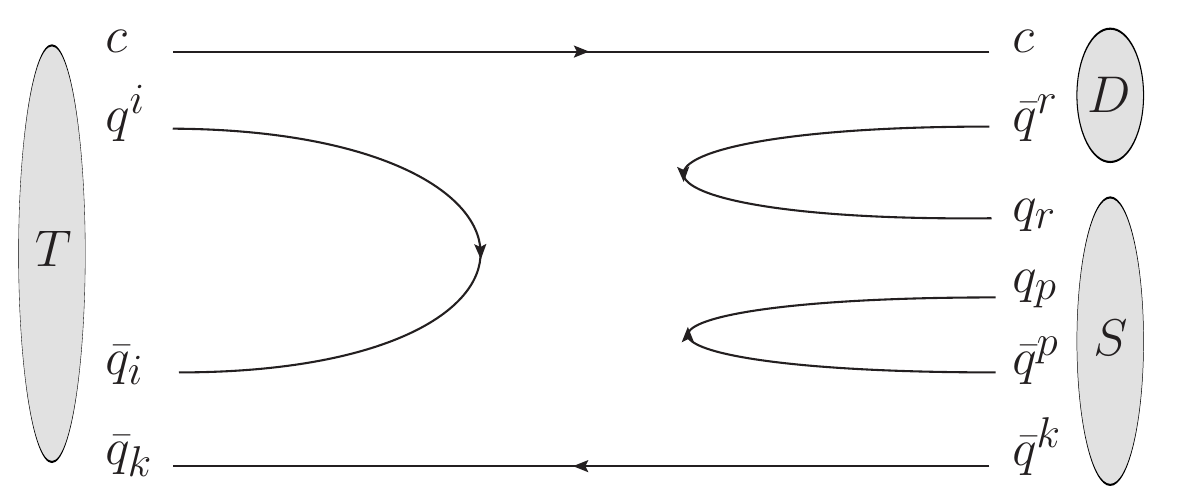}
}
\hspace{12pt}
\subfigure[]{
  \includegraphics[width=0.45\textwidth]{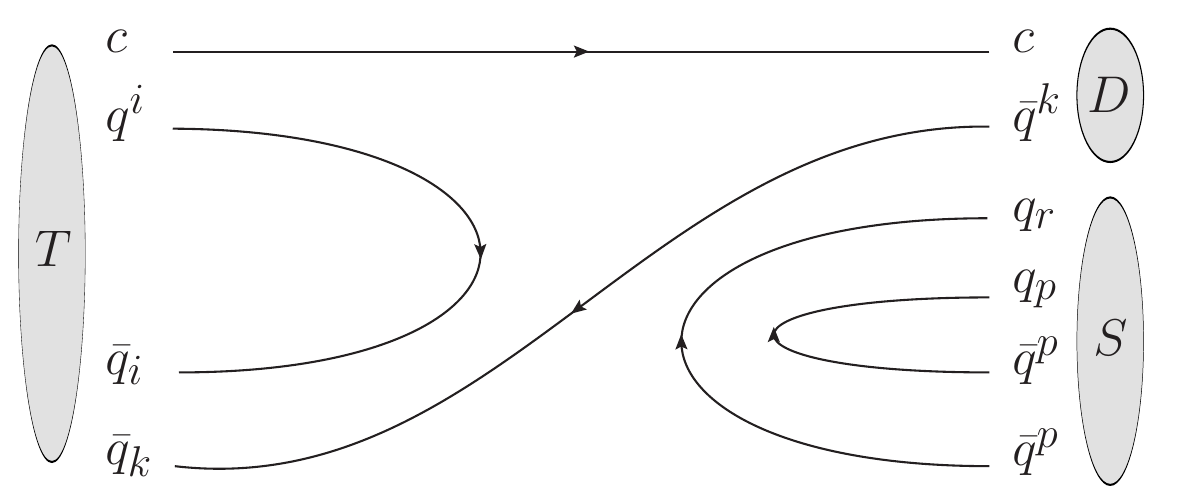}
}
\caption{Topological diagrams of 
  (a) $C_S$ (creation) (only for $T_{cq [\bar q' \bar q'']}$ decays), (b) $C_{DS}$ (creation), (c) $ACC_S$ (annihilation-double creation)
  and (d) $ACC_D$ (annihilation-double creation)
   amplitudes in $T_{cq \bar q' \bar q'}\to DS$ decays. 
   The subscripts indicate the final state meson(s) receiving the light quark and/or the anti-quarks of $T_{cq \bar q' \bar q''}$.
  The decays of flavor exotic states can only have diagrams (a) and (b) in scenario I, 
  but for scenario II only diagram (b) is allowed for flavor exotic states.
} \label{fig: TA T2DS}
\end{figure}

There are four topological amplitudes for $T\to DS$ decays in scenario I ($T=T(cq [\bar q\bar q'])$),
where $C_S$, $C_{DS}$ are creation amplitudes with a $q\bar q$ pair created in the decay processes, see Fig.~\ref{fig: TA T2DS} (a) and (b),
and $ACC_S$ and $ACC_D$ are annihilation-double-creation amplitudes with a $q\bar q$ pair annihilated and two $q\bar q$ pairs created, see Fig.~\ref{fig: TA T2DS}(c) and (d).
The subscripts of these topological amplitudes denote the final state mesons to which the quarks and antiquark from $T$ go.
In scenario II ($T=T(cq \{\bar q\bar q'\})$),
it is evident that a $T(cq \{\bar q\bar q'\})\to DS$ decay cannot have a $C_{S}$ amplitude, as their flavors cannot match, 
since the anti-quark pair in $T(cq \{\bar q\bar q'\})$ is symmetric, while those in $S$ is antisymmetric.

As shown in Fig.~\ref{fig: TA T2DS},
we do not have fall-apart amplitude $F$ in this $T\to DS$ decays.
However, we expect creation amplitudes $C_{S,DS}$ to dominate over annihilation-double creation amplitudes $ACC_{S,D}$ in
these decays, as the former involves only $q\bar q$ creation, while the latter needs a $q\bar q$ pair annihilated and two $q\bar q$ pairs created.

Furthermore, diagrams (c) and (d) in Fig.~\ref{fig: TA T2DS} involve the traces $T^i_{[ik]}$ and $T^i_{\{i k\}}$,
where only $\bar{\bf 3}$ and $\bar{\bf 3}'$ remain.
Therefore, in scenario I the decays of flavor exotic states and all other states in ${\bf 6}$ can only have diagrams (a) and (b), 
while for scenario II only diagram (b) is allowed for flavor exotic states and all other states in $\overline{\bf 15}$.

\subsubsection{$T\to DS$ decay amplitudes}

\begin{table}[t!]
\caption{\label{tab: TtoDS1}
$T\to DS$ decay amplitudes in scenario I, with $T=T_{cq[\bar q'\bar q'']}$.} 
\footnotesize{
\begin{ruledtabular}
\begin{tabular}{llclc}
\#
          & Mode
          & $A (T\to D S)$
          & Mode
          & $A (T\to D S)$
          \\
\hline 
$1^{*}$
          & $T^{++}_{c\bar s}\to D^+ \kappa^+$
          & $2C_{S}+C_{DS}$
          & $T^{++}_{c\bar s}\to D_s^+ a_0^+$
          & $-(2 C_{S}+C_{DS})$
          \\
$2^*$
          & $T^{+}_{c\bar s}\to D^0 \ov \kappa^+$
          & $\frac{1}{\sqrt2}(2C_{S}+C_{DS})$ 
          & $T^{+}_{c\bar s}\to D^+ \kappa^0$
          & $-\frac{1}{\sqrt2}(2C_{S}+C_{DS})$ 
          \\
          & $T^{+}_{c\bar s}\to D_s^+ a_0^0$
          & $-2(C_{S}+C_{DS})$
          \\                                                                        
$3^{*}$
          &$T^{0}_{c\bar s}\to D^0 \kappa^0$
          & $2 C_{S}+C_{DS}$
          & $T^{0}_{c\bar s}\to D_s^+ a_0^-$
          & $-(2C_{S}+C_{DS})$
          \\
\hline 
$4^{*}$
          & $T^{0}_{cs}\to D^0 \ov \kappa^0$
          & $2C_{S}+C_{DS}$
          & $T^{0}_{cs}\to D^+ \kappa^-$
          & $-(2C_{S}+C_{DS})$
          \\
\hline  
5
          & $T^{+}_{c}\to D^0 a_0^+$
          & $\frac{1}{\sqrt2}(2C_{S}+C_{DS})$
          & $T^{+}_{c}\to D^+ a_0^0$
          & $-\frac{1}{2}(2C_{S}+C_{DS})$
          \\
          & $T^{+}_{c}\to D_s^+ \ov \kappa^0$
          & $-\frac{1}{\sqrt2}(2 C_{S}+C_{DS})$
          & $T^{+}_{c}\to D^+ \sigma$
          & $ \frac{1}{2} (\sqrt2 c_\phi + s_\phi) (2C_{S}+C_{DS})$         
          \\
          & $T^{+}_{c}\to D^+ f_0$
          & $ -\frac{1}{2} (c_\phi - \sqrt2 s_\phi) (2C_{S}+C_{DS})$ 
          \\ 
6
          & $T^{0}_{c}\to D^0 a_0^0$
          & $-\frac{1}{2}(2C_{S}+C_{DS})$
          & $T^{0}_{c}\to D^+ a_0^-$
          & $-\frac{1}{\sqrt2}(2C_{S}+C_{DS})$
          \\
          & $T^{0}_{c}\to D_s^+ \kappa^-$
          & $\frac{1}{\sqrt2}(2C_{S}+C_{DS})$
          & $T^{0}_{c}\to D^+ \sigma$
          & $ -\frac{1}{2} (\sqrt2 c_\phi + s_\phi) (2C_{S}+C_{DS})$             
          \\
          & $T^{0}_{c}\to D^+ f_0$
          & $ \frac{1}{2} (c_\phi - \sqrt2 s_\phi) (2C_{S}+C_{DS})$
          \\  
\hline \hline 
7
          & $T^{\prime\prime +}_{c}\to D^0 a_0^+$
          & $\frac{1}{\sqrt2}(C_{DS}-2C_{S}+2ACC_{S})$
          & $T^{\prime\prime +}_{c}\to D^+ a_0^0$
          & $-\frac{1}{2}(C_{DS}-2C_{S}+2ACC_{S})$
          \\
          & $T^{\prime\prime+}_{c}\to D_s^+ \ov \kappa^0$
          & $\frac{1}{\sqrt2}(C_{DS}-2C_{S}+2ACC_{S})$
          \\
          & $T^{\prime\prime +}_{c}\to D^0 f_0$
          & $\frac{(c_{\phi}-\sqrt2 s_\phi)}{6} (C_{DS}-2C_{S}+2ACC_S)$
          & $T^{\prime\prime +}_{c}\to D^+ \sigma$
          & $-\frac{(s_{\phi}+\sqrt2 c_\phi)}{6} (C_{DS}-2C_{S}+2ACC_S)$
           \\ 
          & 
          & $ +\frac{2(\sqrt2 s_{\phi}+2c_\phi)}{3} (C_{DS}+C_S$           
          & 
          &  $ +\frac{2(\sqrt2 c_{\phi}-2s_\phi)}{3} (C_{DS}+C_S$
           \\ 
          & 
          & $   -ACC_S-3ACC_D)$            
          & 
          &  $  -ACC_S-3ACC_D)$
           \\ 
8
          & $T^{\prime\prime 0}_{c}\to D^0 a_0^0$
          & $\frac{1}{2}(C_{DS}-2C_{S}+2ACC_{S})$
          & $T^{\prime\prime 0}_{c}\to D^+ a_0^-$
          & $\frac{1}{\sqrt2}(C_{DS}-2C_{S}+2ACC_{S})$
          \\
          & $T^{\prime\prime 0}_{c}\to D_s^+ \kappa^-$
          & $\frac{1}{\sqrt2}(C_{DS}-2C_{S}+2ACC_{S})$
          \\
          & $T^{\prime\prime 0}_{c}\to D^0 f_0$
          & $\frac{(c_{\phi}-\sqrt2 s_\phi)}{6} (C_{DS}-2C_{S}+2ACC_S)$
          & $T^{\prime\prime 0}_{c}\to D^+ \sigma$
          & $-\frac{(s_{\phi}+\sqrt2 c_\phi)}{6} (C_{DS}-2C_{S}+2ACC_S)$
           \\ 
          & 
          & $ +\frac{2(\sqrt2 s_{\phi}+2c_\phi)}{3} (C_{DS}+C_S$         
          & 
          & $ +\frac{2(\sqrt2 c_{\phi}-2s_\phi)}{3} (C_{DS}+C_S$
          \\  
          & 
          & $    -ACC_S-3ACC_D)$         
          & 
          & $  -ACC_S-3ACC_D)$
          \\  
 \hline 
9        
          & $T^{\prime\prime +}_{c\bar s}\to D^0 \kappa^+$
          & $\frac{1}{\sqrt2}(C_{DS}-2C_{S}+2ACC_{S})$
          & $T^{\prime\prime +}_{c\bar s}\to D^+ \kappa^0$
          & $\frac{1}{\sqrt2}(C_{DS}-2C_{S}+2ACC_{S})$
          \\
          & $T^{\prime\prime +}_{c\bar s}\to D_s^+ f_0$
          & $-\frac{(c_{\phi}-\sqrt2 s_\phi)}{3} (C_{DS}-2C_{S}+2ACC_S)$
          & $T^{\prime\prime +}_{c\bar s}\to D_s^+ \sigma$
          & $\frac{(s_{\phi}+\sqrt2 c_\phi)}{3} (C_{DS}-2C_{S}+2ACC_S)$
          \\                      
          & 
          & $+ \frac{2(\sqrt2 s_{\phi}+2c_\phi)}{3} (C_{DS}+C_S$  
          & 
          & $ +\frac{2(\sqrt2 c_{\phi}-2s_\phi)}{3} (C_{DS}+C_S$
          \\ 
          & 
          & $   -ACC_S-3ACC_D)$  
          & 
          & $  -ACC_S-3ACC_D)$
          \\ 
\end{tabular}
\end{ruledtabular}
}
\end{table}

The $T_{cq \bar q' \bar q''}\to DS$ decay amplitudes decomposed in these topological amplitudes are shown in Table \ref{tab: TtoDS1} and \ref{tab: TtoDS2} for scenarios I and II, respectively.
where we have $T=T(cq [\bar q'\bar q''])$ and $T(cq \{\bar q'\bar q''\})$, respectively. 
Note that although some of the decay modes listed in these tables are phase space limited if not kinematically prohibited,
the information of their amplitudes can still be useful, as the $T$ in these modes can be a virtual particle involved in a decay of a heavy particle, such as a $B$ meson, to multi-particle final states.

As noted previously for flavor exotic states and all other states in ${\bf 6}$ the $T\to DS$ decay amplitudes in the first scenario can only 
have topological amplitudes $C_S$ and $C_{DS}$, as depicted in Fig.~\ref{fig: TA T2DS} (a) and (b). 
Although we have two different topological amplitudes contributing to the decay amplitudes of $T$ in ${\bf 6}$, 
the amplitudes are proportional to a single combination $2C_{S}+C_{DS}$ as shown in Table \ref{tab: TtoDS1}.
Consequently, their decay amplitudes are highly related.
This can be easily understood as in the $T\to D P$ case,
since there is only one ${\bf 6}$ in $\bar{\bf 3}\otimes ({\bf 8}\oplus {\bf 1})$, see Eq.~(\ref{eq: SU(3) decompositions 1}), 
from the final state $DS$ that can match the SU(3) quantum number of these $T$.

\begin{table}[t!]
\caption{\label{tab: TtoDS2}
$T\to DS$ decay amplitudes in scenario II, with $T=T_{cq\{\bar q'\bar q''\}}$.} 
\scriptsize{
\begin{ruledtabular}
\begin{tabular}{llclc}
\#
          & Mode
          & $A (T\to D S)$
          & Mode
          & $A (T\to D S)$
          \\
\hline
$1'^{*}$
          & $T^{++}_{c\bar s}\to D^+ \kappa^+$
          & $C_{DS}$
          & $T^{++}_{c\bar s}\to D_s^+ a_0^+$
          & $C_{DS}$
          \\
$2'^*$
          & $T^{+}_{c\bar s}\to D^0 \kappa^+$
          & $\frac{1}{\sqrt2}C_{DS}$ 
          & $T^{+}_{c\bar s}\to D^+ \kappa^0$
          & $-\frac{1}{\sqrt2}C_{DS}$
          \\
          & $T^{+}_{c\bar s}\to D_s^+ a_0^0$
          & $C_{DS}$
           \\                        
$3'^{*}$
          & $T^{0}_{c\bar s}\to D^0 \kappa^0$
          & $C_{DS}$
          & $T^{0}_{c\bar s}\to D_s^+ a_0^-$
          & $C_{DS}$
          \\
\hline          
$4'^{*}$         
          & $T^{+}_{cs}\to D^+ \ov \kappa^0$ 
          & $\sqrt2 C_{DS}$
          \\ 
$5'^{*}$
          & $T^{0}_{cs}\to D^0 \ov \kappa^0$
          & $C_{DS}$
          & $T^{0}_{cs}\to D^+ \kappa^-$
          & $C_{DS}$
          \\
$6'^{*}$       
          & $T^{-}_{cs}\to D^0 \kappa^-$
          & $\sqrt2 C_{DS}$
          \\
\hline
$7'^{*}$          
          & $T^{++}_{c\bar s\bar s}\to D_s^+ \kappa^+$
          & $\sqrt2 C_{DS}$
          \\
$8'^{*}$          
          & $T^{+}_{c\bar s\bar s}\to D_s^+ \kappa^0$ 
          & $\sqrt2 C_{DS}$
          \\
\hline 
$9'^{*}$
          & $T^{++}_{c}\to D^+ a_0^+$
          & $\sqrt2 C_{DS}$
          \\
$10'^*$
          & $T^{+}_{c}\to D^0 a_0^+$
          & $\sqrt{\frac{2}{3}} C_{DS}$
          & $T^{+}_{c}\to D^+ a_0^0$
          & $\frac{2}{\sqrt3} C_{DS}$
          \\
$11'^*$
          & $T^{0}_{c}\to D^0 a_0^0$
          & $\frac{2}{\sqrt3} C_{DS}$
          & $T^{0}_{c}\to D^+ a_0^-$
          & $-\sqrt{\frac{2}{3}} C_{DS}$
          \\
$12'^{*}$       
          & $T^{-}_{c}\to D^0 a_0^-$
          & $\sqrt2 C_{DS}$
          \\
\hline 
$13'$
          & $T^{\prime +}_{c}\to D^0 a_0^+$
          & $\frac{1}{2 \sqrt3} C_{DS}$
          & $T^{\prime +}_{c}\to D^+ a_0^0$
          & $-\frac{1}{2 \sqrt6} C_{DS}$
          \\
          & $T^{\prime +}_{c}\to D^+ \sigma$
          & $-\frac{\sqrt 6}{4}  (\sqrt2 c_{\phi}+s_{\phi})C_{DS}$           
          & $T^{\prime +}_{c}\to D^+ f_0$
          &  $-\frac{\sqrt 6}{4}  (\sqrt2 s_{\phi}-c_{\phi})C_{DS}$ 
          \\ 
          & $T^{\prime +}_{c}\to D_s^+ \ov \kappa^0$
          & $-\frac{\sqrt3}{2} C_{DS}$
          \\
$14'$
          & $T^{\prime 0}_{c}\to D^0 a_0^0$
          & $\frac{1}{2 \sqrt6} C_{DS}$
          & $T^{\prime 0}_{c}\to D^+ a_0^-$
          & $\frac{1}{2 \sqrt3} C_{DS}$
          \\
          & $T^{\prime 0}_{c}\to D^+ \sigma$
          & $-\frac{\sqrt 6}{4}  (\sqrt2 c_{\phi}+s_{\phi})C_{DS}$            
          & $T^{\prime 0}_{c}\to D^+ f_0$
          & $-\frac{\sqrt 6}{4}  (\sqrt2 s_{\phi}-c_{\phi})C_{DS}$ 
          \\ 
          & $T^{\prime 0}_{c}\to D_s^+ \kappa^-$
          & $-\frac{\sqrt3}{2} C_{DS}$
          \\           
\hline 
$15'$
          & $T^{\prime +}_{c\bar s}\to D^0 \kappa^+$
          & $\frac{1}{2} C_{DS}$
          & $T^{\prime +}_{c\bar s}\to D^+ \kappa^0$
          & $\frac{1}{2} C_{DS}$
          \\
          & $T^{\prime +}_{c\bar s}\to D_s^+ \sigma$
          & $-\frac{1}{\sqrt 2}  (\sqrt2 c_{\phi}+s_{\phi})C_{DS}$ 
          & $T^{\prime +}_{c\bar s}\to D_s^+ f_0$
          & $-\frac{1}{\sqrt 2}   (\sqrt2 s_{\phi}-c_{\phi})C_{DS}$
          \\ 
\hline
\hline
$16'$
          & $T^{\prime \prime +}_{c}\to D^0 a_0^+$
          & $\frac{1}{2}(C_{DS}+4ACC_{S})$
          & $T^{\prime \prime +}_{c}\to D^+ a_0^0$
          & $-\frac{1}{2\sqrt2 }(C_{DS}+4ACC_{S})$
          \\
          & $T^{\prime \prime +}_{c}\to D_s^+ \ov \kappa^0$
          & $\frac{1}{2}(C_{DS}+4ACC_{S})$
          \\
          & $T^{\prime \prime +}_{c}\to D^+ f_0$
          & $\frac{\sqrt2 c_\phi-2s_\phi}{12}(C_{DS}+4ACC_{S})$     
          & $T^{\prime \prime +}_{c}\to D^+ \sigma$
          & $-\frac{\sqrt2 s_\phi+2c_\phi}{12}(C_{DS}+4ACC_{S})$
          \\ 
          & 
          & $-\frac{4(\sqrt2 c_\phi+s_\phi) }{3}  (C_{DS}+ACC_S+3ACC_D)$     
          & 
          & $-\frac{4(c_\phi-\sqrt2 s_\phi) }{3}  (C_{DS}+ACC_S+3ACC_D)$
          \\            
$17'$
          & $T^{\prime \prime 0}_{c}\to D^0 a_0^0$
          & $\frac{1}{2\sqrt2 }(C_{DS}+4ACC_{S})$
          & $T^{\prime \prime 0}_{c}\to D^+ a_0^-$
          & $\frac{1}{2}(C_{DS}+4ACC_{S})$
          \\
          & $T^{\prime \prime 0}_{c}\to D_s^+ \kappa^-$
          & $\frac{1}{2}(C_{DS}+4ACC_{S})$
          \\
          & $T^{\prime \prime 0}_{c}\to D^0 f_0$
          & $\frac{\sqrt2 c_\phi-2s_\phi}{12}(C_{DS}+4ACC_{S})$           
          & $T^{\prime \prime 0}_{c}\to D^+ \sigma$
          & $-\frac{\sqrt2 s_\phi+2c_\phi}{12}(C_{DS}+4ACC_{S})$
          \\
          & 
          & $-\frac{4(\sqrt2 c_\phi+s_\phi) }{3}  (C_{DS}+ACC_S+3ACC_D)$             
          & 
          & $-\frac{4(c_\phi-\sqrt2 s_\phi) }{3}  (C_{DS}+ACC_S+3ACC_D)$
          \\                       
\hline  
$18'$
          & $T^{\prime \prime +}_{c\bar s}\to D^0 \kappa^+$
          & $\frac{1}{2}(C_{DS}+4ACC_{S})$
          & $T^{\prime \prime +}_{c\bar s}\to D^+ \kappa^0$
          & $\frac{1}{2}(C_{DS}+4ACC_{S})$
          \\
          & $T^{\prime \prime +}_{c\bar s}\to D_s^+ f_0$
          & $-\frac{\sqrt2 c_\phi-2s_\phi}{6}(C_{DS}+4ACC_{S})$ 
          & $T^{\prime \prime +}_{c\bar s}\to D_s^+ \sigma$
          & $\frac{\sqrt2 s_\phi+2c_\phi}{6}(C_{DS}+4ACC_{S})$
          \\
          & 
          & $-\frac{4(\sqrt2 c_\phi+s_\phi) }{3}  (C_{DS}+ACC_S+3ACC_D)$
          & 
          & $-\frac{4(c_\phi-\sqrt2 s_\phi) }{3}  (C_{DS}+ACC_S+3ACC_D)$
          \\                              
\end{tabular}
\end{ruledtabular}
}
\end{table}

The sizes of the amplitudes of the following decay modes of flavor exotic states,
$T^{++}_{c\bar s}\to D^+ \kappa^+, D_s^+ a_0^+$,
$T^{+}_{c\bar s}\to D_s^+ a_0^0$,
$T^{0}_{c\bar s}\to D^0 \kappa^0, D_s^+ a_0^-$
and
$T^{0}_{cs}\to D^0 \ov \kappa^0, D^+ \kappa^-$,
are identical, giving $|A|=|2C_{S}+C_{DS}|$,
and factor $1/\sqrt2$ are needed for $|A|$ of $T^{+}_{c\bar s}\to D^0 \ov \kappa^+$ and $T^{+}_{c\bar s}\to D^+ \kappa^0$ decays.
The flavor exotic nature of most of these states is self-evident from their decay modes shown in the table.
For example, the $T^{+}_{c\bar s}\to D_s^+ a_0^0$ decay reveals the exotic quantum number of $T^{+}_{c\bar s}$,
as $a_0$ is an iso-vector.

For $T$ in ${\bf 3}$, namely $T^{\prime\prime +}_c$, $T^{\prime\prime 0}_c$ and $T^{\prime\prime +}_{c\bar s}$, 
there is only one combination, namely $C_{DS}-2C_{S}+2ACC_{S}$, in the amplitudes of $T\to Da_0, D \kappa, D\bar\kappa$ decays,
while another combination, $C_S+C_{DS}-3ACC_{D}-ACC_{S}$, is needed in $T\to D f_0, D\sigma$ decays
as there are two $\bar{\bf 3}$ in $\bar{\bf 3}\otimes ({\bf 8}\oplus {\bf 1})$, see Eq. (\ref{eq: SU(3) decompositions 1}),
from the final states that can match the SU(3) quantum number of these $T^{\prime\prime}_c$ and $T^{\prime\prime}_{c\bar s}$.

The situation in scenario II is slightly different.
In this case, we only have one topological amplitude, $C_{DS}$, contributing to the multiplet containing flavor exotic states, namely $\overline{\bf 15}$,
as noted previously, $C_S$ is not allowed in this scenario.
Here, the SU(3) quantum number of the final state $DS$ is $\bar{\bf 3}\otimes ({\bf 8}\oplus {\bf 1})$,
but these is only one $\overline{\bf 15}$ in $\bar{\bf 3}\otimes ({\bf 8}\oplus {\bf 1})$, see Eq. (\ref{eq: SU(3) decompositions 1}), that can match the SU(3) quantum number of these $T$.
Consequently, the $T\to DS$ amplitudes of these states are highly related.
In particular, as shown in Table \ref{tab: TtoDS2}, the following decay modes of flavor exotic states,
$T^{++}_{c\bar s}\to D^+ \kappa^+$,
$D_s^+ a_0^+$,
$T^{+}_{c\bar s}\to D_s^+ a_0^0$,
$T^{0}_{c\bar s}\to D^0 \kappa^0$,
$D_s^+ a_0^-$,
$T^{0}_{cs}\to D^0 \ov \kappa^0$,
$D^+ \kappa^-$,
all have identical amplitude, $A=C_{DS}$.
Furthermore, the decay amplitudes of flavor exotic states,
$T^{+}_{cs}\to D^+ \ov \kappa^0$, 
$T^{-}_{cs}\to D^0 \kappa^-$,
$T^{++}_{c\bar s\bar s}\to D_s^+ \kappa^+$,
$T^{+}_{c\bar s\bar s}\to D_s^+ \kappa^0$,
$T^{++}_{c}\to D^+ a_0^+$
and
$T^{-}_{c}\to D^0 a_0^-$ decays have $A=\sqrt2 C_{DS}$,
and the amplitudes of
$T^{+}_{c}\to D^0 a_0^+$,
$T^{+}_{c}\to D^+ a_0^0$,
$T^{0}_{c}\to D^0 a_0^0$
and
$T^{0}_{c}\to D^+ a_0^-$
decays have $A\propto C_{DS}$ with some simple proportional constants.
The flavor exotic nature of most of these states is self-evident from the table.
For $T^{+}_{c\bar s}$ the $T^{+}_{c\bar s}\to D_s^+ a_0^0$ decay reveals the exotic nature of $T^{+}_{c\bar s}$,
while for $T^{+}_{c}$ and $T^{0}_{c}$, more works are required.
For example, one needs to verify that they do not decay to $Df_0$ and $D\sigma$.

For $T$ in ${\bf 3}$, namely $T^{\prime\prime +}_c$, $T^{\prime\prime 0}_c$ and $T^{\prime\prime +}_{c\bar s}$, 
there is only one combination, namely $C_{DS}+4ACC_{S}$, in the amplitudes of $T\to Da_0, D \kappa, D\bar\kappa$ decays,
while another combination, $C_{DS}+ACC_{S}+3ACC_{D}$, is needed in $T\to D f_0, D\sigma$ decays
as there are two $\bar{\bf 3}$ in $\bar{\bf 3}\otimes ({\bf 8}\oplus {\bf 1})$ from the final states that can match the SU(3) quantum number of these $T^{\prime\prime}_c$ and $T^{\prime\prime}_{c\bar s}$.

\section{$\overline B\to D \overline T$ and $\overline B\to \overline D T$ decays}

\subsection{Topological amplitudes in $\overline B \to D \overline T$ and $\overline D T$ decays}

\subsubsection{Topological amplitudes in $\overline B \to D \overline T$ decays}

\begin{figure}[t]
\centering
 \subfigure[]{
  \includegraphics[width=0.45\textwidth]{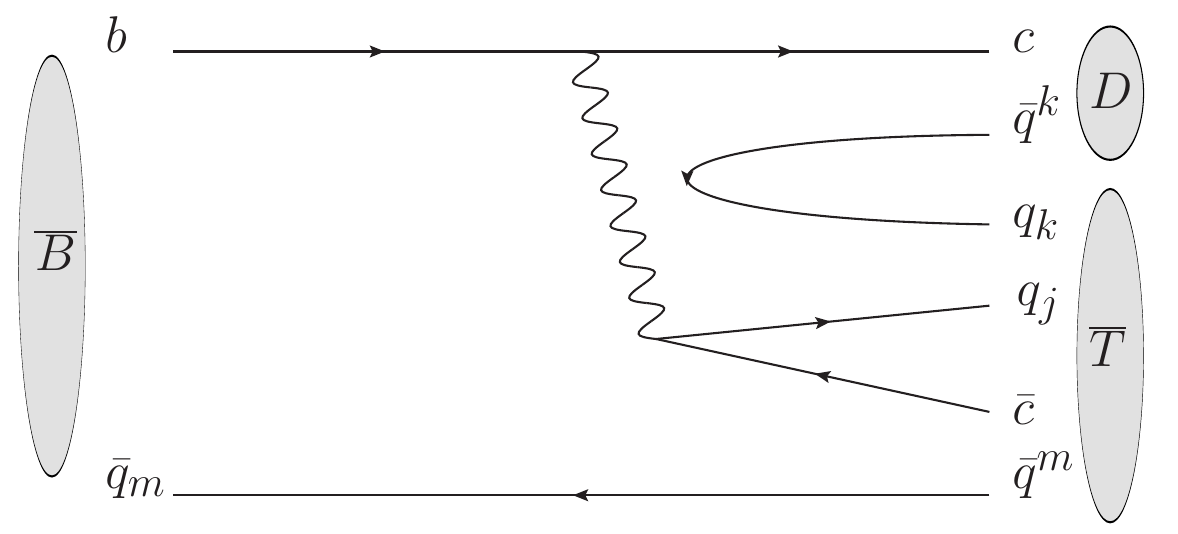}
}
\hspace{12pt}
\subfigure[]{
  \includegraphics[width=0.45\textwidth]{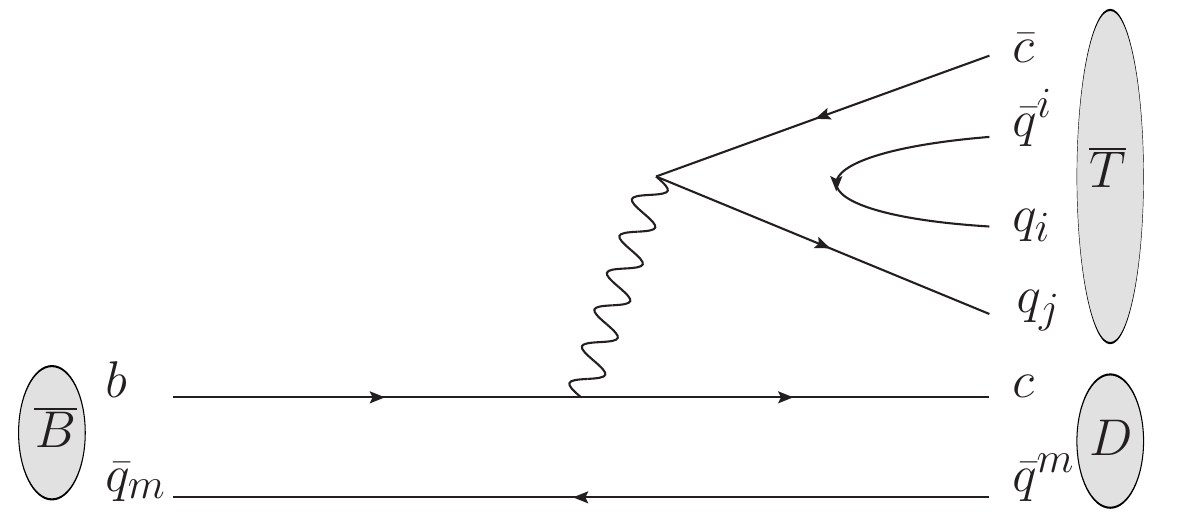}
}\\
\subfigure[]{
  \includegraphics[width=0.45\textwidth]{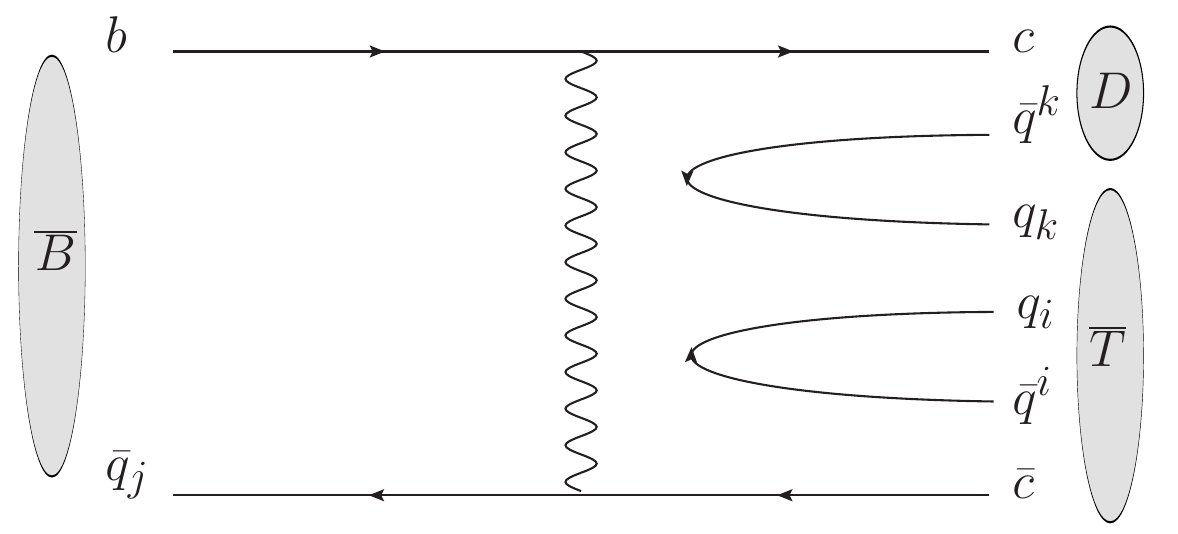}
}
\hspace{12pt}
\subfigure[]{
  \includegraphics[width=0.45\textwidth]{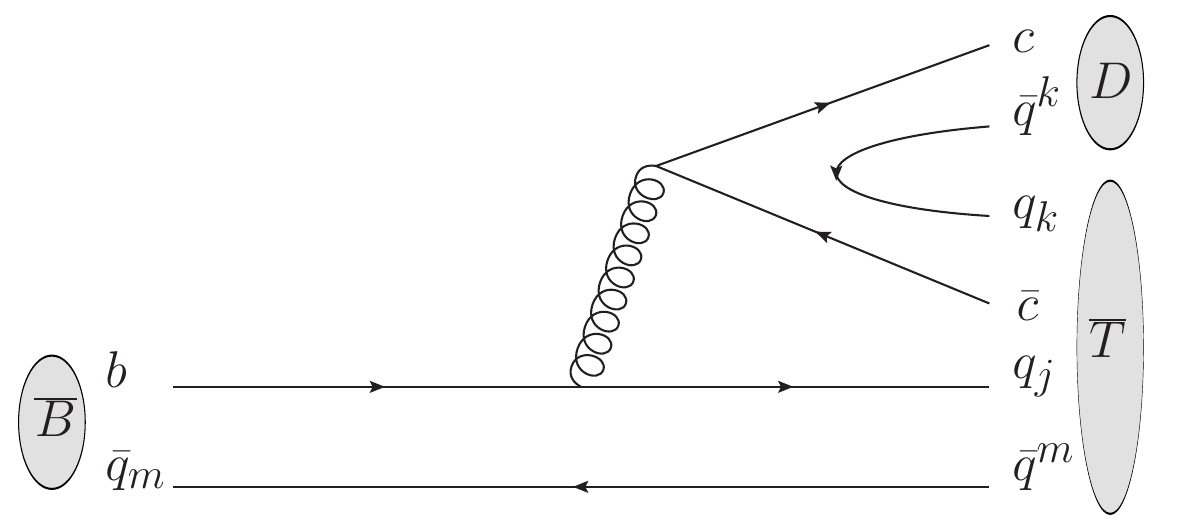}
}\\
\subfigure[]{
  \includegraphics[width=0.45\textwidth]{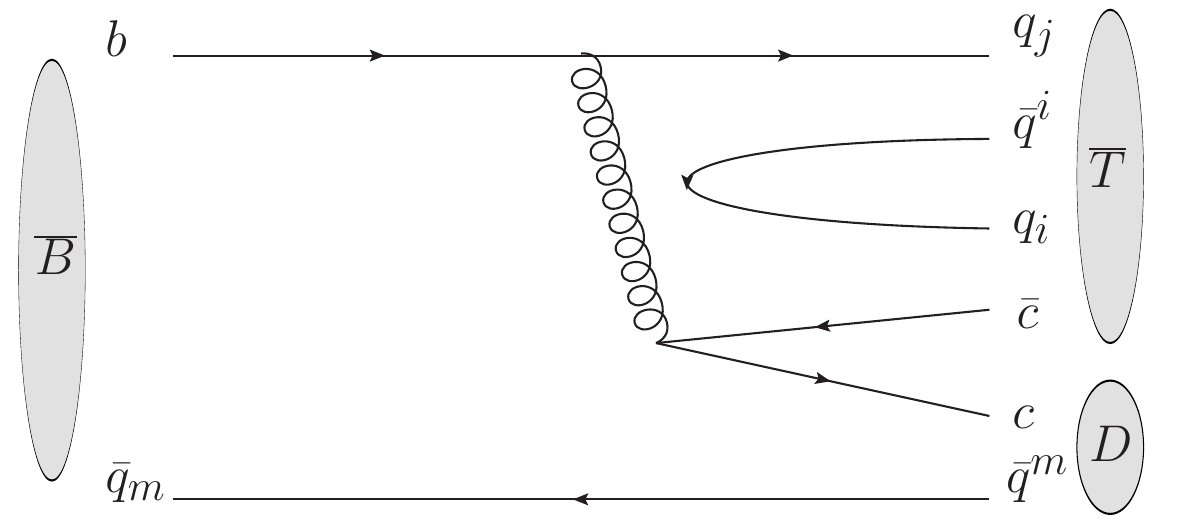}
}
\hspace{12pt}  
\subfigure[]{
  \includegraphics[width=0.45\textwidth]{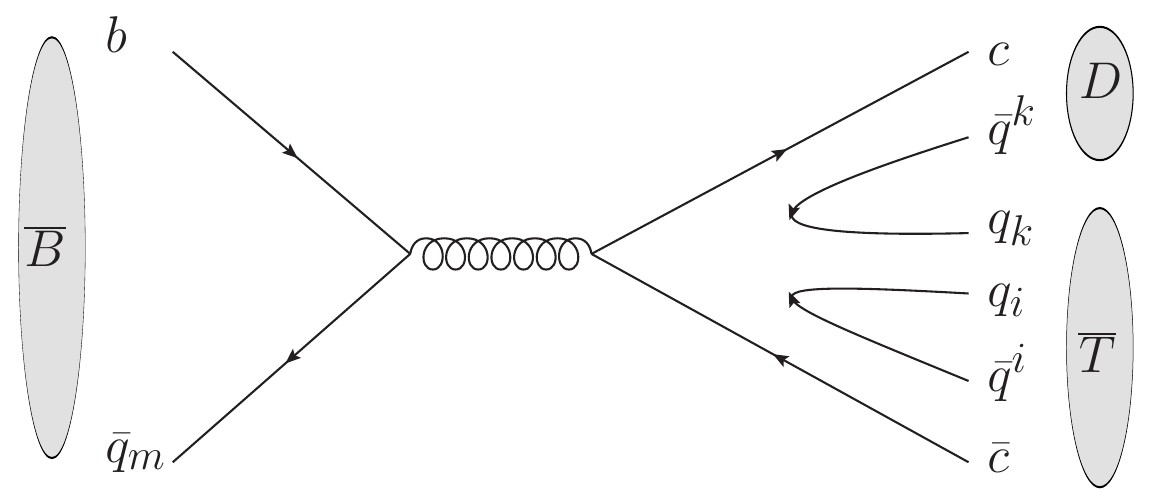}
}  
\hspace{12pt}
\caption{Topological diagrams of 
  (a) $C_{\bar T}$ (internal $W$), (b) $T_{\bar T}$ (external $W$), (c) $E_{\bar T\bar B}$ (exchange),
  (d) $P_{\bar T,1}$ (penguin), (e) $P_{\bar T,2}$ (penguin)and (f) $PA_{\bar B}$ (penguin annihilation)
   amplitudes in $\overline B\to D \overline T$ decays. 
  These are flavor flow diagrams. Only (a) and (d) contribute to modes with flavor exotic states or other states in the same multiplets.
} \label{fig: TA B2DTbar}
\end{figure}

The $\overline B \to D \overline T$ decays in $\Delta S=-1$ transitions are governed by the a tree operator $(\bar s c)_{V-A}(\bar c b)_{V-A}$ and a penguin operator $(\bar c c)(\bar s b)$.
These operators have the same flavor structure that can be expressed as $(\bar q_j H^j c)\,(\bar c b)$ with $\bar q_j=(\bar u,\,\bar d,\,\bar s)$ and
\begin{equation}
H= \left(
\begin{array}{c}
0 \\
0 \\
1 
\end{array}
\right).
\label{eq: Hj}
\end{equation}
Consequently, in scenario I the Hamiltonian governing the $\overline B \to D \overline T(\bar c\bar q [q'q''])$ decays is given by
\be
H_{\rm eff}&=& 
       T_{\bar T} \, \overline B_m \overline D^m  H^j T^i_{[ji]}
      +C_{\bar T} \, \overline B_m T^m_{[jk]} H^j \overline D^k
      +E_{\bar T\bar B} \, \overline B_j H^j T^i_{[ik]} \overline D^k
\non\\      
      &&+P_{\bar T,1} \, \overline B_m T^m_{[jk]} H^j \overline D^k
        +P_{\bar T,2} \, \overline B_m \overline D^m  H^j T^i_{[ji]}
      +PA_{\bar B} \, \overline B_j H^j T^i_{[ik]} \overline D^k,
\label{eq: Heff B to DTbar I}      
\en
while in scenario II the Hamiltonian for the $\overline B \to D \overline T(\bar c\bar q \{q'q''\})$ decays is given by,
\be
H_{\rm eff}&=& 
       T_{\bar T} \, \overline B_m \overline D^m  H^j T^i_{\{ji\}}
      +C_{\bar T} \, \overline B_m T^m_{\{jk\}} H^j \overline D^k
      +E_{\bar T\bar B} \, \overline B_j H^j T^i_{\{ik\}} \overline D^k
\non\\
&&      +P_{\bar T,1} \, \overline B_m T^m_{\{jk\}} H^j \overline D^k
       +P_{\bar T,2} \, \overline B_m \overline D^m  H^j T^i_{\{ji\}}
      +PA_{\bar B} \, \overline B_j H^j T^i_{\{ik\}} \overline D^k.
\label{eq: Heff B to DTbar II}        
\en
For notational simplicity, we use the same notations on topological amplitudes in both scenarios, without any assumption on their equality.  
There are three tree topological amplitudes, $T_{\bar T}$, $C_{\bar T}$ and $E_{\bar T\bar B}$
and three penguin topological amplitudes, $P_{\bar T,1}$, $P_{\bar T,2}$ and $PA_{\bar B}$, 
as depicted in Fig.~\ref{fig: TA B2DTbar}.
These $T_{\bar T}$, $C_{\bar T}$ and $E_{\bar T\bar B}$ amplitudes denote external $W$, internal $W$ and exchange amplitudes, respectively, 
where the subscripts denote the particles receiving the $\bar cs$ quarks from the $W$ line in the $b\to c W^-, W^-\to \bar c s$ transition. 
For $P_{\bar T,1}$, $P_{\bar T,2}$ and $PA_{\bar B}$ amplitudes, they are penguin and penguin-annihilation amplitudes, 
respectively,
where the subscripts denote the particles receiving the $s$ quark from the $b\to s \bar c c$ transition. 

The relative sizes of these topological amplitudes are expected to be
\be
|T_{\bar T}|
>|C_{\bar T}|
>|P_{\bar T, 1}|, |P_{\bar T,2}|,
|E_{\bar T\bar B}|, 
|PA_{\bar B}|.
\en
Although the external $W$ diagram is expected to dominate, it is unable to produce a flavor exotic $\overline T$ in the final state, as the $q\bar q$ pair is flavor connected, see Fig. \ref{fig: TA B2DTbar} (b).
These flavor exotic $\overline T$ and other states in the same multiplets, namely ${\bf 6}$ in scenario I and $\overline{\bf 15}$ in scenario II, only have contribution from $(C_{\bar T}+P_{\bar T,1})$,\footnote{These two tpological amplitudes can only occur in this combination, see Eqs. (\ref{eq: Heff B to DTbar I}) and (\ref{eq: Heff B to DTbar II}).} and, consequently, their $\overline B\to D\overline T$ decay amplitudes are highly related.

The above effective Hamiltonians are for the $\Delta S=-1$ transition, 
but they can be easily transformed into those for the $\Delta S=0$ transition, by simply replacing $H$ in Eq. (\ref{eq: Hj}) with
\begin{equation}
H'= \left(
\begin{array}{c}
0 \\
1 \\
0 
\end{array}
\right),
\label{eq: Hj'}
\end{equation}
and with 
$T_{\bar T}$, 
$C_{\bar T}$, 
$E_{\bar T\bar B}$,
$P_{\bar T,1}$,  
$P_{\bar T,2}$,
$PA_{\bar B}$
 replaced with 
$T'_{\bar T}$, 
$C'_{\bar T}$, 
$E'_{\bar T\bar B}$,
$P'_{\bar T,1}$,  
$P'_{\bar T,2}$, 
$PA'_{\bar B}$
respectively. 
It is understood that these amplitudes are related by Cabibbo–Kobayashi–Maskawa (CKM) matrix elements, giving
$T'_{\bar T}=\frac{V^*_{cd}}{V^*_{cs}} T_{\bar T}$, 
$C'_{\bar T}=\frac{V^*_{cd}}{V^*_{cs}} C_{\bar T}$, 
$E'_{\bar T\bar B}=\frac{V^*_{cd}}{V^*_{cs}} E_{\bar T\bar B}$,
$P'_{\bar T,1}=\frac{V^*_{td}}{V^*_{ts}} P_{\bar T,1}$,  
$P'_{\bar T,2}=\frac{V^*_{td}}{V^*_{ts}} P_{\bar T,2}$
and
$PA'_{\bar B}=\frac{V^*_{td}}{V^*_{ts}} PA_{\bar B}$.

\begin{figure}[t]
\centering
 \subfigure[]{
  \includegraphics[width=0.45\textwidth]{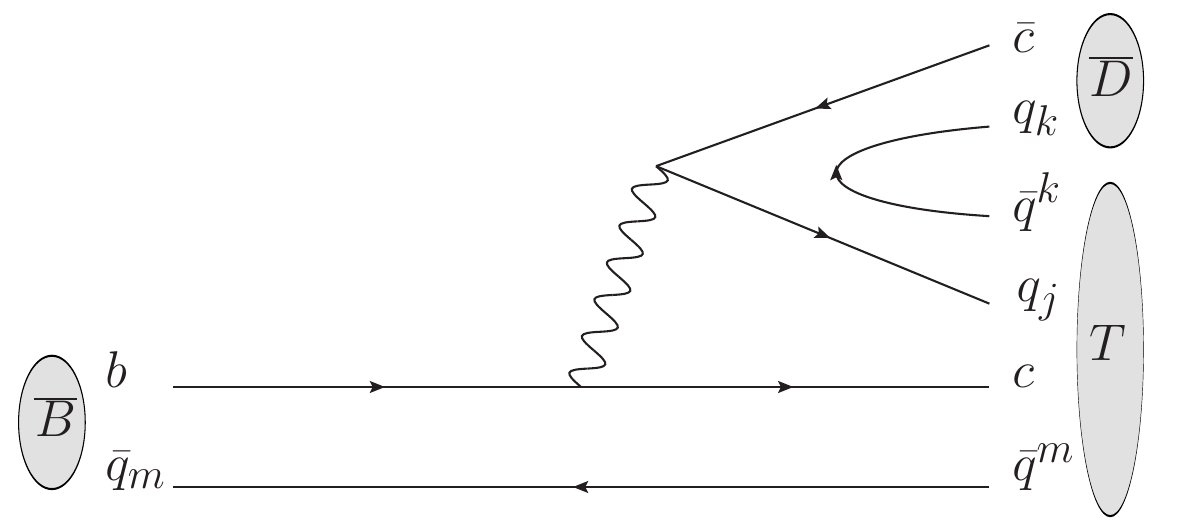}
}
\hspace{12pt}
\subfigure[]{
  \includegraphics[width=0.45\textwidth]{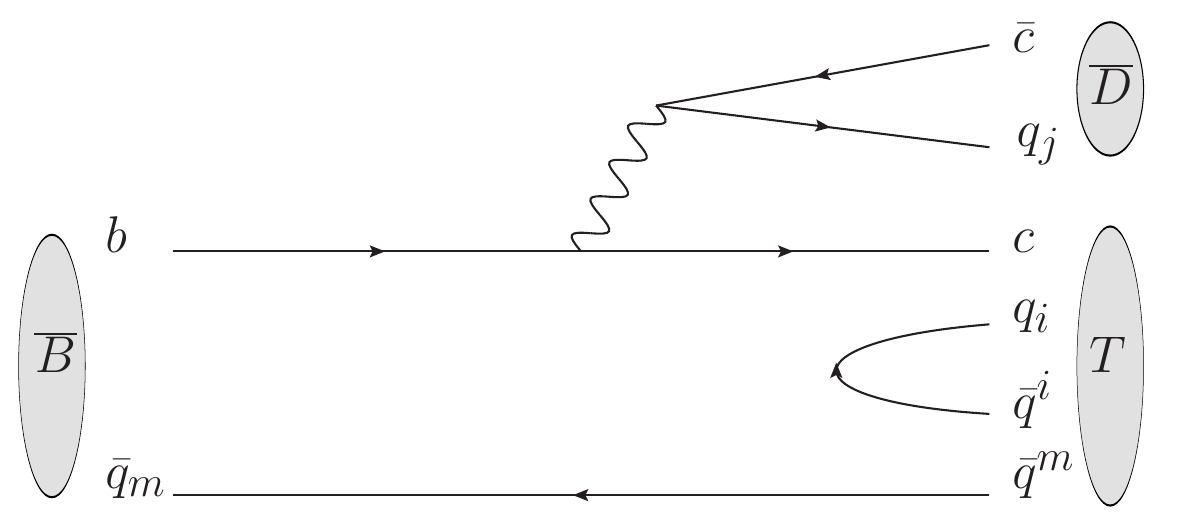}
}\\\subfigure[]{
  \includegraphics[width=0.45\textwidth]{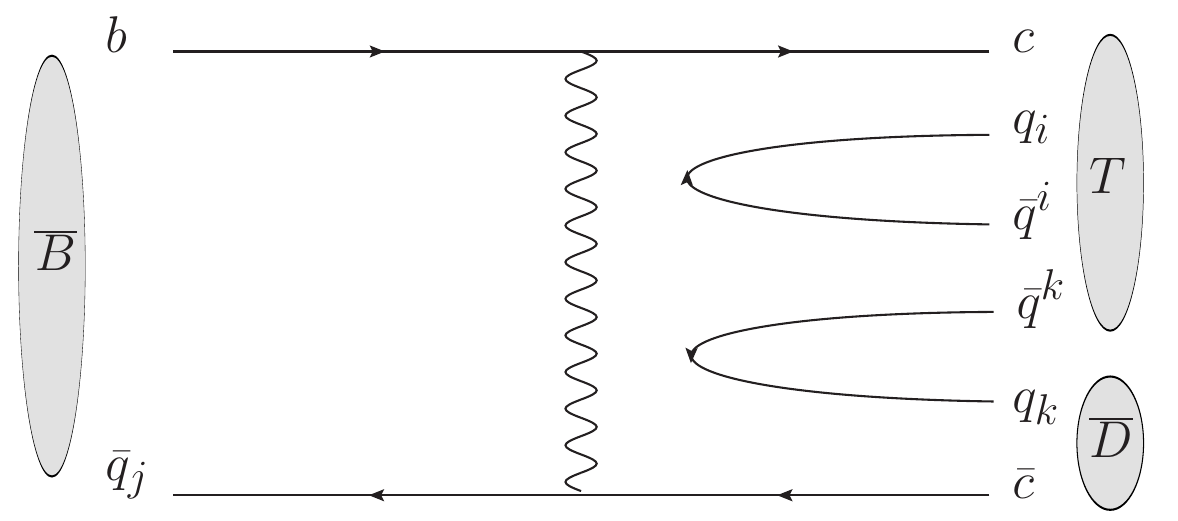}
}
\hspace{12pt}
\subfigure[]{
  \includegraphics[width=0.45\textwidth]{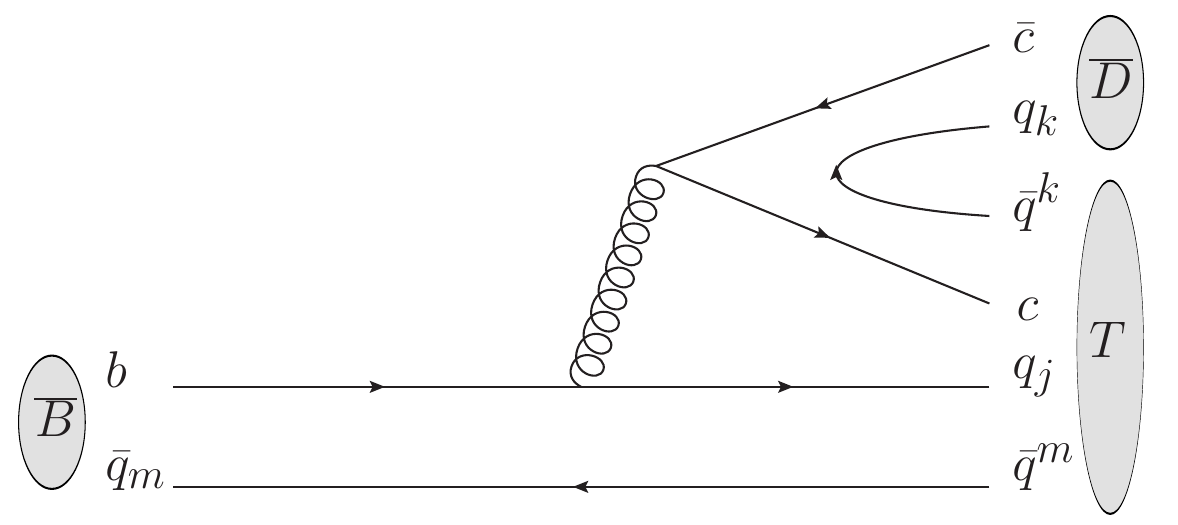}
}\\
\subfigure[]{
  \includegraphics[width=0.45\textwidth]{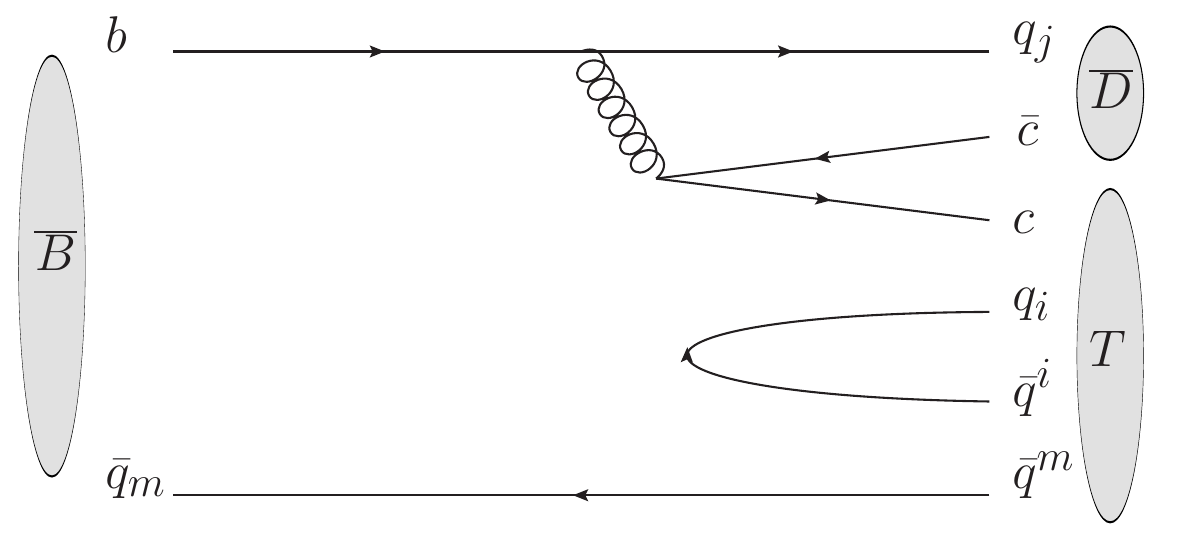}
}
\hspace{12pt}  
\subfigure[]{
  \includegraphics[width=0.45\textwidth]{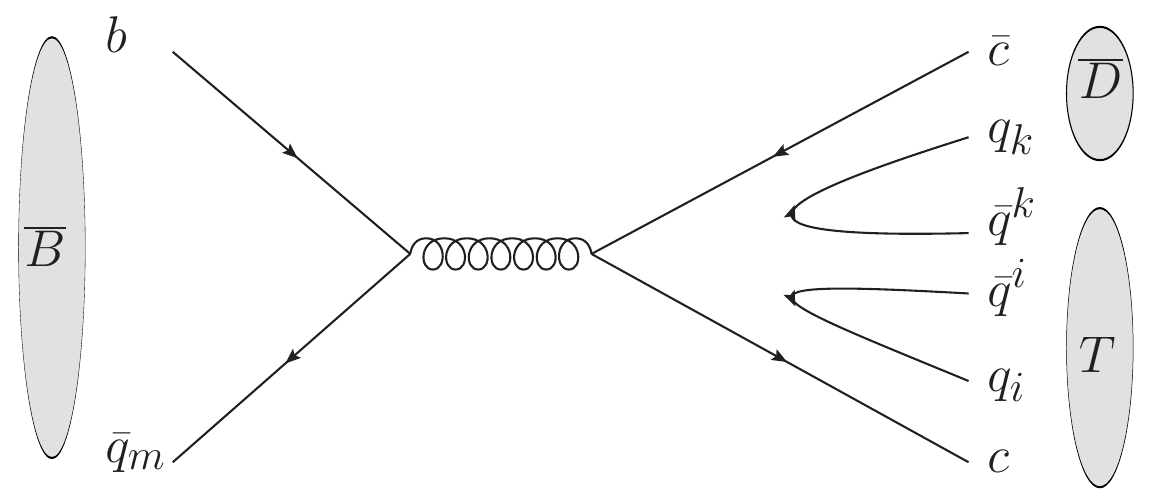}
}  
\hspace{12pt}
\caption{Topological diagrams of 
   (a) $C_{\bar D T}$ (internal $W$), (b) $T_{\bar D}$ (external $W$), (c) $E_{\bar D\bar B}$ (exchange),
   (d) $P_{T}$ (penguin), (e) $P_{\bar D}$ (penguin) and (f) $PA_{\bar B}$ (penguin annihilation) 
   amplitudes in $\overline B\to \overline D T$ decays. 
  These are flavor flow diagrams. Only (a) and (d) contribute to modes with flavor exotic states or other states in the same multiplets.
} \label{fig: TA B2DbarT}
\end{figure}

\subsubsection{Topological amplitudes in $\overline B \to \overline D T$ decays}

We turn to $\overline B_q\to \overline D T$ decays.
The tree $b\to c W^-, W^-\to \bar c s$ and the penguin $b\to s \bar c c$ transitions can also induce $\overline B \to \overline D T(c q [\bar q'\bar q''])$ and $\overline B \to \overline D T(c q \{\bar q'\bar q''\})$ decays as well.
The Hamiltonian governing the $\overline B \to \overline D T(c q [\bar q'\bar q''])$ decays in $\Delta S=-1$ transitions in scenario I is given by
\be
H_{\rm eff}&=& 
       T_{\bar D} \, \overline B_m \overline T_i^{[mi]} H^j D_j  
      +C_{\bar D T} \, \overline B_m  H^j \overline T_j^{[mk]} D_k 
      +E_{\bar D\bar B} \, \overline B_i H^i \overline T_j^{[jk]}  D_k
\non\\
&& +P_{T} \, \overline B_m  H^j \overline T_j^{[mk]} D_k 
       +P_{\bar D} \, \overline B_m \overline T_i^{[mi]} H^j D_j  
      +PA_{\bar B} \, \overline B_i H^i \overline T_j^{[jk]}  D_k,
\label{eq: Heff B to DbarT I}    
\en
while the one for the $\overline B \to D \overline T(\bar c\bar q \{q'q''\})$ decays in scenario II is given by,
\be
H_{\rm eff}&=& 
       T_{\bar D} \, \overline B_m \overline T_i^{\{mi\}} H^j D_j  
      +C_{\bar D T} \,\overline B_m  H^j \overline T_j^{\{mk\}} D_k 
      +E_{\bar D\bar B} \,  \overline B_i H^i \overline T_j^{\{jk\}}  D_k
\non\\
&& +P_{T} \,\overline B_m  H^j \overline T_j^{\{mk\}} D_k 
      +P_{\bar D} \, \overline B_m \overline T_i^{\{mi\}} H^j D_j  
      +PA_{\bar B} \,  \overline B_i H^i \overline T_j^{\{jk\}}  D_k.
\label{eq: Heff B to DbarT II}     
\en
For notational simplicity, we use the same notations on topological amplitudes in both scenarios.
There are three tree topological amplitudes, $T_{\bar D}$, $C_{\bar D T}$ and $E_{\bar D\bar B}$
and three penguin topological amplitudes, $P_{T}$, $P_{\bar D}$ and $PA_{\bar B}$, 
and
they are depicted in Fig. \ref{fig: TA B2DbarT}.
These $T_{\bar D}$, $C_{\bar D T}$ and $E_{\bar D\bar B}$ amplitudes denote external $W$, internal $W$ and exchange amplitudes, respectively, and the subscripts denote the particle(s) receiving the $\bar c s$ quarks from the $W$ line in the $b\to c W^-, W^-\to \bar c s$ transition.
For $P_{T}$, $P_{\bar D}$ and $PA_{\bar B}$ amplitudes, they are penguin and penguin-annihilation amplitudes, 
where the subscripts denote the particles receiving the $s$ quark from the $b\to s \bar c c$ transition. 
The effective Hamiltonians governing $\Delta S=0$ transitions can be obtained from the above equations with $H$ replaced with $H'$ given in Eq.~(\ref{eq: Hj'}), 
and with $T_{\bar D}, C_{\bar D T}, E_{\bar D\bar B}$, $P_{T}$, $P_{\bar D}$ and $PA_{\bar B}$ replaced with 
$T'_{\bar D},C'_{\bar D},E'_{\bar D\bar B}$, $P'_{T}$, $P'_{\bar D}$ and $PA'_{\bar B}$, respectively. 
It is understood that these amplitudes are related by CKM matrix elements, giving
$T'_{\bar D}=\frac{V^*_{cd}}{V^*_{cs}} T_{\bar D}$, 
$C'_{\bar D T}=\frac{V^*_{cd}}{V^*_{cs}} C_{\bar D T}$,
$E'_{\bar D\bar B}=\frac{V^*_{cd}}{V^*_{cs}} E_{\bar D\bar B}$,
$P'_{\bar D}=\frac{V^*_{td}}{V^*_{ts}} P_{\bar D}$, 
$P'_{T}=\frac{V^*_{td}}{V^*_{ts}} P_{T}$
and
$PA'_{\bar B}=\frac{V^*_{td}}{V^*_{ts}} PA_{\bar B}$.

The relative sizes of the tree topological amplitudes are expected to be
\be
|T_{\bar D}|>|C_{\bar D T}|>|P_{T}|, |P_{\bar D}|,|E_{\bar D\bar B}|, |PA_{\bar B}|.
\en
Although the external $W$ diagram is expected to dominate, it is unable to produce a flavor exotic $T$, as the $q\bar q$ pair in the diagram, see Fig. \ref{fig: TA B2DbarT} (b), is flavor connected.
The $\overline B_q\to \overline D T$ decay amplitudes of modes involving $T$ in ${\bf 6}$ in scenario I or $\overline{\bf 15}$ in scenario II only have contribution from $(C_{\bar D T}+P_{T})$,\footnote{These two tpological amplitudes can only occur in this combination, see Eqs. (\ref{eq: Heff B to DbarT I}) and (\ref{eq: Heff B to DbarT II}).} i.e. the diagrams depicted in \ref{fig: TA B2DbarT} (a) and (d). Consequently, the amplitudes of these states are highly related.

From the above discussion, we may wrongly have the impression that a $\overline B \to \overline D T$ decay is similar to a  $\overline B \to D \overline T$ decay.
In fact, by comparing Fig. \ref{fig: TA B2DbarT} to Fig. \ref{fig: TA B2DTbar}, one can easily see the differences. 
For example, 
in $\overline B \to D \overline T$ decays, the external $W$ diagram, $T_{\bar T}$, has a $\overline T$ emitted from $W^-$, as shown in Fig. \ref{fig: TA B2DTbar} (b), while
in $\overline B \to \overline D T$ decays, the emitted particle is $\overline D$ instead, see Fig. \ref{fig: TA B2DbarT} (b).
Furthermore, there is no counterpart of the internal $W$ diagram $C_{\bar D T}$, Fig.~\ref{fig: TA B2DbarT}~(a), in $\overline B \to D \overline T$ decays, 
as the  $C_{\bar T}$ in Fig.~\ref{fig: TA B2DTbar}~(a) is different from $C_{\bar D T}$.
Explicitly, the former has $\bar D T$ sharing the $c\bar q$ from $W$, while in the latter, both $c\bar q$ go to $\overline T$.
Consequently, one cannot identify the topological amplitudes in $\overline B \to D \overline T$ to those in $\overline B \to \overline D T$ decays.

\subsection{$\overline B \to D \overline T$ and $\overline D T$ decay amplitudes}

\subsubsection{$\overline B \to D \overline T$ decay amplitudes}

\begin{table}[t!]
\caption{\label{tab: BtoDTbar I}
$\overline B_q\to D\overline T_{\bar c\bar q[ q' q'']} $ decay amplitudes in $\Delta S=-1$ and  $\Delta S=0$ transitions in scenario I.}
\footnotesize{
\begin{ruledtabular}
\begin{tabular}{llcllccccr}
\#
&Mode
          & $A (\overline B_q\to D\overline T_{\bar c\bar q[ q' q'']})$
          & \#
          & Mode
          & $A (\overline B_q\to D\overline T_{\bar c\bar q [q' q'']})$
          \\
\hline
$\bar 1^{*}$ 
          & $B^-\to D^+ \overline T^{\,--}_{\bar c s}$
          & $-(C_{\bar T}+P_{\bar T,1})$
&$\bar 2^*$ 
          & $B^-\to D^0 \overline T^{\,-}_{\bar c  s }$
          & $-\frac{1}{\sqrt2} (C_{\bar T}+P_{\bar T,1})$
           \\
$\bar 2^*$ 
          & $\overline B{}^0\to D^+ \overline T^{\,-}_{\bar c  s }$
          & $\frac{1}{\sqrt2} (C_{\bar T}+P_{\bar T,1})$
&$\bar 3^{*}$ 
          & $\overline B{}^0\to D^0 \overline T^0_{\bar c s}$
          & $-(C_{\bar T}+P_{\bar T,1})$
          \\
$\bar 5$ 
          & $\overline B{}_s^0\to D^+ \overline T^{\,-}_{\bar c}$
          & $-\frac{1}{\sqrt2} (C_{\bar T}+P_{\bar T,1})$
& $\bar  6$ 
          & $\overline B{}_s^0\to D^0 \overline T^0_{\bar c}$
          & $\frac{1}{\sqrt2} (C_{\bar T}+P_{\bar T,1})$
          \\
\hline
$\bar 7$
          & $\overline B{}_s^0\to D^+ \overline T^{\,\prime\prime -}_{\bar c}$
          & $\frac{1}{\sqrt2}((C_{\bar T}+P_{\bar T,1})+2(E_{\bar T\bar B}+PA_{\bar B}))$
&$\bar 8$ 
          & $\overline B{}_s^0\to D^0 \overline T^{\,\prime\prime 0}_{\bar c}$
          & $\frac{1}{\sqrt2}((C_{\bar T}+P_{\bar T,1})+2(E_{\bar T\bar B}+PA_{\bar B}))$
          \\
$\bar 9$          
          & $B^-\to D^0 \overline T^{\,\prime\prime -}_{\bar c s}$
          & $-\frac{1}{\sqrt2}((C_{\bar T}+P_{\bar T,1})+2(T_{\bar T}+P_{\bar T,2}))$
&$\bar 9$          
          & $\overline B{}^0\to D^+ \overline T^{\,\prime\prime -}_{\bar c s}$
          & $-\frac{1}{\sqrt2}((C_{\bar T}+P_{\bar T,1})+2(T_{\bar T}+P_{\bar T,2}))$
           \\                                                     
$\bar 9$
          & $\overline B{}_s^0\to D_s^+ \overline T^{\,\prime\prime -}_{\bar c s}$
          & $\sqrt2(-T_{\bar T}-P_{\bar T,2}+E_{\bar T\bar B}+PA_{\bar B})$
           \\ 
           \hline
           \hline
\#
&Mode
          & $A'(\overline B_q\to D\overline T_{\bar c\bar q[ q' q'']})$
          & \#
          & Mode
          & $A' (\overline B_q\to D\overline T_{\bar c\bar q\{q' q''\}})$
          \\
\hline
$\bar 1^{*}$ 
          & $B^-\to D_s^+ \overline T^{\,--}_{\bar c s}$
          & $(C'_{\bar T}+P'_{\bar T,1})$
&$\bar 2^*$ 
          & $\overline B{}^0\to D_s^+ \overline T^{\,-}_{\bar c s }$
          & $-\frac{(C'_{\bar T}+P'_{\bar T,1})}{\sqrt2}$
          \\
$\bar 4^{*}$ 
          & $\overline B{}_s^0\to D^0 \overline T^0_{\bar c\bar s}$
          & $-(C'_{\bar T}+P'_{\bar T,1})$
          \\
$\bar 5$ 
          & $B^-\to D^0 \overline T^{\,-}_{\bar c}$
          & $-\frac{1}{\sqrt2} (C'_{\bar T}+P'_{\bar T,1})$
&$\bar 5$ 
          & $\overline B{}_s^0\to D_s^+ \overline T^{\,-}_{\bar c}$
          & $\frac{1}{\sqrt2} (C'_{\bar T}+P'_{\bar T,1})$
          \\
$\bar 6$ 
          & $\overline B{}^0\to D^0 \overline T^0_{\bar c}$
          & $-\frac{1}{\sqrt2} (C'_{\bar T}+P'_{\bar T,1})$
          \\
\hline
$\bar 7$          
          & $B^-\to D^0 \overline T^{\,\prime\prime -}_{\bar c}$
          & $-\frac{1}{\sqrt2} ((C'_{\bar T}+P'_{\bar T,1})+2(T'_{\bar T}+P'_{\bar T,2}))$
&$\bar 7$          
          & $\overline B{}^0\to D^+ \overline T^{\,\prime\prime -}_{\bar c}$
          & $\sqrt2((E'_{\bar T\bar B}+PE'_{\bar B})-(T'_{\bar T}+P'_{\bar T,2}))$
          \\          
$\bar 7$
          & $\overline B{}_s^0\to D_s^+ \overline T^{\,\prime\prime -}_{\bar c}$
          & $-\frac{1}{\sqrt2} ((C'_{\bar T}+P'_{\bar T,1})+2(T'_{\bar T}+P'_{\bar T,2}))$
&$\bar 8$
          & $\overline B{}^0\to D^0 \overline T^{\,\prime\prime 0}_{\bar c}$
          & $\frac{1}{\sqrt2} ((C'_{\bar T}+P'_{\bar T,1})+2(E'_{\bar T\bar B}+PE'_{\bar B}))$
           \\                              
$\bar 9$          
          & $\overline B{}^0\to D_s^+ \overline T^{\,\prime\prime -}_{\bar c s}$
          & $\frac{1}{\sqrt2}((C'_{\bar T}+P'_{\bar T,1})+2 (E'_{\bar T\bar B}+PA'_{\bar B}))$
           \\                                                                                                                
\end{tabular}
\end{ruledtabular}
}
\end{table}

\begin{table}[t!]
\caption{\label{tab: BtoDTbar II}
$\overline B_q\to D\overline T_{\bar c\bar q\{q' q''\}}$ decay amplitudes in $\Delta S=-1$ and $\Delta S=0$ transitions in scenario~II.}
\scriptsize{
\begin{ruledtabular}
\begin{tabular}{llcllccccr}
\#
&Mode
          & $A (\overline B_q\to D\overline T_{\bar c\bar q[ q' q'']})$
          & \#
          & Mode
          & $A (\overline B_q\to D\overline T_{\bar c\bar q\{q' q''\}})$
          \\
\hline
$\bar 1'^{*}$ 
          & $B^-\to D^+ \overline T^{\,--}_{\bar c s}$
          & $(C_{\bar T}+P_{\bar T,1})$
&$\bar 2'^*$        
          & $B^-\to D^0 \overline T^{\,-}_{\bar c s}$
          & $\frac{1}{\sqrt2} (C_{\bar T}+P_{\bar T,1})$
          \\
$\bar 2'^*$          
          & $\overline B{}^0\to D^+ \overline T^{\,-}_{\bar c s }$
          & $-\frac{1}{\sqrt2} (C_{\bar T}+P_{\bar T,1})$
&$\bar 3'^{*}$
          & $\overline B{}^0\to D^0 \overline T^0_{\bar c s }$
          & $(C_{\bar T}+P_{\bar T,1})$
          \\
$\bar 7'^{*}$         
          & $B^-\to D_s^+ \overline T^{\,--}_{\bar c s s}$
          & $\sqrt2 (C_{\bar T}+P_{\bar T,1})$
&$\bar 8'^{*}$          
          & $\overline B{}^0\to D_s^+ \overline T^{\,-}_{\bar c s s}$
          & $\sqrt2 (C_{\bar T}+P_{\bar T,1})$
          \\  
$\overline{13'}$
          & $\overline B{}_s^0\to D^+ \overline T^{\prime -1}_{\bar c}$
          & $-\frac{\sqrt3}{2} (C_{\bar T}+P_{\bar T,1})$         
&$\overline{14'}$
          & $\overline B{}_s^0\to D^0 \overline T^{\prime 0}_{\bar c}$
          & $-\frac{\sqrt3}{2} (C_{\bar T}+P_{\bar T,1})$
          \\
$\overline{15'}$
          & $B^-\to D^0 \overline T^{\prime -}_{\bar c s}$
          & $\frac{1}{2} (C_{\bar T}+P_{\bar T,1})$
&$\overline{15'}$          
          & $\overline B{}^0\to  D^+\overline T^{\prime -}_{\bar c s}$
          & $\frac{1}{2} (C_{\bar T}+P_{\bar T,1})$
          \\          
$\overline{15'}$
          & $\overline B{}_s^0\to D_s^+ \overline T^{\prime -}_{\bar c s}$
          & $-(C_{\bar T}+P_{\bar T,1})$
          \\                                                                                                               
\hline
$\overline {16'}$
          & $\overline B{}_s^0\to D^+ \overline T^{\prime \prime -}_{\bar c}$
          & $\frac{1}{2} ((C_{\bar T}+P_{\bar T,1})+4(E_{\bar T\bar B}+PA_{\bar B}))$
&$\overline {17'}$
          & $\overline B{}_s^0\to D^0 \overline T^{\prime \prime 0}_{\bar c}$
          & $\frac{1}{2} ((C_{\bar T}+P_{\bar T,1})+4(E_{\bar T\bar B}+PA_{\bar B}))$
          \\
$\overline{18'}$          
          & $B^-\to D^0 \overline T^{\prime \prime -}_{\bar c s}$
          & $\frac{1}{2} (4T_{\bar T}+P_{\bar T,2}+(C_{\bar T}+P_{\bar T,1}))$
          &$\overline{18'}$
          & $\overline B{}_s^0\to D_s^+ \overline T^{\prime \prime -}_{\bar c s}$
          & $2(T_{\bar T}+P_{\bar T,2})+(C_{\bar T}+P_{\bar T,1})$
          \\ 
$\overline {18'}$          
          & $\overline B{}^0\to D^+ \overline T^{\prime \prime -}_{\bar c s}$
          & $\frac{1}{2} (4T_{\bar T}+P_{\bar T,2}+(C_{\bar T}+P_{\bar T,1}))$
          & 
          & 
          & $2E_{\bar T\bar B}+2PA_{\bar B}$
          \\ 
\hline\hline
\#
&Mode
          & $A' (\overline B_q\to D\overline T_{\bar c\bar q[ q' q'']})$
          & \#
          & Mode
          & $A' (\overline B_q\to D\overline T_{\bar c\bar q\{q' q''\}})$
          \\
\hline
$\bar 1'^{*}$ 
          & $B^-\to D_s^+ \overline T^{\,--}_{\bar c s}$
          & $(C'_{\bar T}+P'_{\bar T,1})$
&$\bar 2'^*$          
          & $\overline B{}^0\to D_s^+ \overline T^{\,-}_{\bar c s }$
          & $-\frac{1}{\sqrt2} (C'_{\bar T}+P'_{\bar T,1})$
          \\
$\bar 4'^{*}$
          & $\overline B{}_s^0\to D^+ \overline T^{\,-}_{\bar c\bar s }$
          & $\sqrt2 (C'_{\bar T}+P'_{\bar T,1})$
&$\bar 5'^{*}$
          & $\overline B{}_s^0\to D^0 \overline T^0_{\bar c\bar s }$
          & $(C'_{\bar T}+P'_{\bar T,1})$
          \\
$\bar 9'^{*}$         
          & $B^-\to D^+ \overline T^{\,--}_{\bar c }$
          & $\sqrt2  (C'_{\bar T}+P'_{\bar T,1})$
&$\overline{10'}^*$          
          & $B^-\to D^0 \overline T^{\,-}_{\bar c }$
          & $\sqrt{\frac{2}{3}} (C'_{\bar T}+P'_{\bar T,1})$
          \\
$\overline{10'}^*$          
          & $\overline B{}^0\to D^+ \overline T^{\,-}_{\bar c }$
          & $-\sqrt{\frac{2}{3}} (C'_{\bar T}+P'_{\bar T,1})$
&$\overline{11'}^*$
          & $\overline B{}^0\to D^0 \overline T^0_{\bar c}$
          & $-\sqrt{\frac{2}{3}} (C'_{\bar T}+P'_{\bar T,1})$
          \\
$\overline{13'}$
          & $B^-\to D^0 \overline T^{\,\prime -}_{\bar c }$
          & $\frac{1}{2\sqrt3}  (C'_{\bar T}+P'_{\bar T,1})$
&$\overline{13'}$         
          & $\overline B{}^0\to D^+ \overline T^{\,\prime -}_{\bar c}$
          & $\frac{1}{\sqrt3} (C'_{\bar T}+P'_{\bar T,1})$
          \\ 
$\overline{13'}$
          & $\overline B{}_s^0\to D_s^+ \overline T^{\,\prime -}_{\bar c}$
          & $-\frac{\sqrt3}{2} (C'_{\bar T}+P'_{\bar T,1})$
&$\overline{14'}$
          & $\overline B{}_s^0\to D^0 \overline T^{\,\prime 0}_{\bar c}$
          & $ \frac{1}{2\sqrt3} (C'_{\bar T}+P'_{\bar T,1})$         
          \\                                                                       
$\overline{15'}$          
          & $\overline B{}^0\to  D_s^+\overline T^{\,\prime -}_{\bar c  s}$
          & $\frac{1}{ 2} (C'_{\bar T}+P'_{\bar T,1})$
          \\ 
\hline 
$\overline{16'}$          
          & $B^-\to D^0 \overline T^{\,\prime \prime -}_{\bar c }$
          & $\frac{1}{2} (4(T'_{\bar T}+P'_{\bar T,2})+(C'_{\bar T}+P'_{\bar T,1}))$
&$\overline{16'}$          
          & $\overline B{}^0\to D^+ \overline T^{\,\prime \prime -}_{\bar c }$
          & $2 (T'_{\bar T}+P'_{\bar T,2})+(C'_{\bar T}+P'_{\bar T,1})+2(E'_{\bar T\bar B}+PE'_{\bar B})$
          \\          
$\overline{16'}$
          & $\overline B{}_s^0\to D_s^+ \overline T^{\,\prime \prime 0}_{\bar c}$
          & $\frac{1}{2} (4(T'_{\bar T}+P'_{\bar T,2})+(C'_{\bar T}+P'_{\bar T,1}))$
&$\overline{17'}$          
          & $\overline B{}^0\to D^0 \overline T^{\,\prime \prime 0}_{\bar c }$
          & $\frac{1}{2}((C'_{\bar T}+P'_{\bar T,1})+4(E'_{\bar T\bar B}+PE'_{\bar B}))$
          \\
\hline 
$\overline{18'}$          
          & $\overline B{}^0\to D_s^+ \overline T^{\,\prime \prime -}_{\bar c  s}$
          & $\frac{1}{2}((C'_{\bar T}+P'_{\bar T,1})+4(E'_{\bar T\bar B}+PE'_{\bar B}))$
          \\
\end{tabular}
\end{ruledtabular}
}
\end{table}

The $\overline B \to D \overline T$ decay amplitudes decomposed in these topological amplitudes are shown in Table \ref{tab: BtoDTbar I} for $\Delta S=-1$ and $\Delta S=0$  transitions in scenario I and in Table \ref{tab: BtoDTbar II} in scenario~II. 
Note that these results also hold by replacing $D$ by $D^*$.

From Table \ref{tab: BtoDTbar I} we see that decay modes involving flavor exotic $\overline T$ in scenario I are governed by $(C_{\bar T}+P_{\bar T,1})$,
and, consequently, their decay rates are highly related.
Furthermore, as shown in Table \ref{tab: BtoDTbar I}, all twelve modes with $T$ in ${\bf 6}$ are highly related, as they are either proportional to $(C_{\bar T}+P_{\bar T,1})$ or $(C'_{\bar T}+P'_{\bar T,1})$.
In particular, the following modes involving flavor exotic states have relations on rates,
\be
\Gamma_1
&\equiv&
\Gamma(B^-\to D^+ \overline T^{\,--}_{\bar c s})
= \Gamma(\overline B{}^0\to D^0 \overline T^0_{\bar c s })
=2\Gamma(\overline B{}^0\to D^+ \overline T^{\,-}_{\bar c  s })
=2\Gamma(B^-\to D^0 \overline T^{\,-}_{\bar c  s })
\non\\
\Gamma_2
&\equiv &
\Gamma(B^-\to D_s^+ \overline T^{\,--}_{\bar c s})
=2\Gamma(\overline B{}^0\to D_s^+ \overline T^{\,-}_{\bar c s })
=\Gamma(\overline B{}_s^0\to D^0 \overline T^0_{\bar c\bar s}),
\non\\
\Gamma_1
&\simeq&
 \left|\frac{V_{cb} V^*_{cs}}{V_{cb} V^*_{cd}}\right|^2 \Gamma_2,
\label{eq: Gamma C+P/C'+P' 1}
\en 
where the approximation in the last line neglects the penguin contributions.
Note that all four flavor exotic states in scenario I are involved in these $\overline B\to D\overline T$ decays, 
but we need to consider decay modes in $\Delta S=-1$ and $\Delta S=0$ transitions. 

One can roughly estimate the error in the approximation in Eq. (\ref{eq: Gamma C+P/C'+P' 1}) by using 
\be
\frac{|P_{\bar T,1}|}{|C_{\bar T}|}\sim \left|\frac{V_{tb} V^*_{ts}(c_4+\frac{c_3}{3})}{V_{cb} V^*_{cs}(c_2+\frac{c_1}{3})}\right|\simeq 0.19,
\label{eq: P/C 1}
\en
and
\be
\frac{|P'_{\bar T,1}|}{|C'_{\bar T}|}\sim \left|\frac{V_{tb} V^*_{td}(c_4+\frac{c_3}{3})}{V_{cb} V^*_{cd}(c_2+\frac{c_1}{3})}\right|\simeq 0.17,
\label{eq: P'/C' 1}
\en
where $c_i$ are the Wilson coefficients, see \cite{Beneke:2001ev}.
Therefore, the errors of neglecting penguin contributions are roughly estimated to be of the order $\lambda$.
To be more precise, we define the following ratio:
\be
R
&\equiv&
\left|\frac{V_{cb}V^*_{cd}}{V_{cb} V^*_{cs}}\right|^2 \frac{\Gamma_1}{\Gamma_2}
=\left|\frac{V_{cb} V^*_{cd}}{V_{cb} V^*_{cs}}\right|^2\frac{|A|^2+|\bar A|^2}{|A'|^2+|\bar A'|^2}
\non\\
&=&\left|\frac{V_{cb} V^*_{cd}}{V_{cb} V^*_{cs}}\right|^2
\frac{|C_{\bar T}+P_{\bar T,1}|^2+|\bar C_{\bar T}+\bar P_{\bar T,1}|^2}{|C'_{\bar T}+P'_{\bar T,1}|^2+|\bar C'_{\bar T}+\bar P'_{\bar T,1}|^2},
\en
where $A^{(\prime)}$ is the decay amplitude of $\overline B\to D\overline T$ decay in $\Delta S=-1 (0)$ transitions, and $\bar A^{(\prime)}$ is the decay amplitude in the conjugated $B\to \overline D T$ decay.
As the average rate (without $B$ tagging) is easier to determine experimentally, we defined $R$ as the ratio of averaged rates with CKM factor corrected. 
Note that $\Gamma$s in Eq. (\ref{eq: Gamma C+P/C'+P' 1}) should also be averaged rates.
We can estimate the value of $R$ using Eqs. (\ref{eq: P/C 1}) and (\ref{eq: P'/C' 1})
\be
R&=&
\left|\frac{V_{cb} V^*_{cd}}{V_{cb} V^*_{cs}}\right|^2
\non\\
&&\times
\frac
{|V_{cb} V^*_{cd}(c_2+\frac{c_1}{3})+\rho e^{i\phi} V_{tb} V^*_{ts}(c_4+\frac{c_3}{3}) |^2+|V^*_{cb} V_{cd}(c_2+\frac{c_1}{3})+\rho e^{i\phi} V^*_{tb} V_{ts}(c_4+\frac{c_3}{3}) |^2}
{|V_{cb} V^*_{cd}(c_2+\frac{c_1}{3})+\rho e^{i\phi}V_{tb} V^*_{td}(c_4+\frac{c_3}{3})|^2+|V^*_{cb} V_{cd}(c_2+\frac{c_1}{3})+\rho e^{i\phi}V^*_{tb} V_{td}(c_4+\frac{c_3}{3})|^2},
\non\\
\en
where we introduce a correction factor $\rho$ and a relative strong phase $\phi$ between the penguin and the tree amplitudes.
We take the value of $\rho$ from 0 to 2 to parameterize our ignorance of the actual size of the penguin-to-tree amplitude ratio.
By taking $\rho=0\sim 2$ and $\phi=0\sim 2\pi$, we obtain
\be
R-1=-0.20\sim 0.09.
\label{eq: R value 1}
\en
Hence, the correction on the approximation in the last line of Eq. (\ref{eq: Gamma C+P/C'+P' 1}) is estimated to be about 10 to 20\% in the presence of penguin contributions.
To effectively control the penguin pollution, further work is needed; see, for example, a study in a similar situation \cite{Fleischer:2007zn},
where information of direct CP violation is needed. 
With these data, the sizes of the penguin and tree amplitudes, as well as the relative strong phase, can be probed.

We turn to scenario II.
Similarly all twenty four modes with $T$ in $\overline {\bf 15}$ are highly related,
as they are either proportional to $(C_{\bar T}+P_{\bar T,1})$ or $(C'_{\bar T}+P'_{\bar T,1})$.
In particular, for the following fourteen modes involving flavor exotic states in scenario II, we have
\be
\Gamma_3
&\equiv&
2\Gamma(B^-\to D^+ \overline T^{\,--}_{\bar c s})
= 4\Gamma(B^-\to D^0 \overline T^{\,-}_{\bar c s})
= 4\Gamma(\overline B{}^0\to D^+ \overline T^{\,-}_{\bar c s })
=2\Gamma(\overline B{}^0\to D^0 \overline T^0_{\bar c s })
\non\\
&=&  \Gamma(B^-\to D_s^+ \overline T^{\,--}_{\bar c s s})
=\Gamma(\overline B{}^0\to D_s^+ \overline T^{\,-}_{\bar c s s})
\non\\
\Gamma_4&\equiv&2 \Gamma(B^-\to D_s^+ \overline T^{\,--}_{\bar c s})
=4 \Gamma(\overline B{}^0\to D_s^+ \overline T^{\,-}_{\bar c s })
=
\Gamma(\overline B{}_s^0\to D^+ \overline T^{\,-}_{\bar c\bar s })
=2 \Gamma(\overline B{}_s^0\to D^0 \overline T^0_{\bar c\bar s})
\non\\
&=& \Gamma(B^-\to D^+ \overline T^{\,--}_{\bar c })
=3 \Gamma(B^-\to D^0 \overline T^{\,-}_{\bar c })
=3 \Gamma(\overline B{}^0\to D^+ \overline T^{\,-}_{\bar c })
=3 \Gamma(\overline B{}^0\to D^0 \overline T^0_{\bar c}),
\non\\
\Gamma_3
&\simeq&
\left|\frac{V_{cs}}{V_{cd}}\right|^2 \Gamma_4,
\en 
where the approximation in the last line neglects the penguin contributions.
Similarly, in the presence of penguin contributions, a correction of 10\% to 20\% on the above approximation is called for, see Eq. (\ref{eq: R value 1}).
Note that even with both $\Delta S=-1$ and $\Delta S=0$ transitions, we only have nine out of twelve flavor exotic states involved in these $\overline B\to D\overline T$ decays.
On the other hand, the above relations impose strong restrictions on these rates.
The continue unobservation of some of these modes, such as $B^-\to D_s^+ \overline T^{\,--}_{\bar c s s}$ and $\overline B{}^0\to D_s^+ \overline T^{\,-}_{\bar c s s}$ decays, will certainly pose tension to scenario II.

The $B^-\to D^+ \overline T^{\,--}_{\bar c s}$ 
and $\overline B{}^0\to D^0 \overline T^0_{\bar c s}$ decays
correspond to the $B^-\to D^+ \overline {T^*_{c\bar s 0}} (2900)^{\,--}$
and $\overline B{}^0\to D^0 \overline {T^*_{c\bar s 0}} (2900)^{0}$ decays, respectively, in both scenarios.
These modes were 
reported by  LHCb in the observation of $\overline {T^*_{c\bar s 0}} (2900)^{\,--}$ and $\overline{ T^*_{c\bar s 0}} (2900)^{0}$
in $B^-\to D^+ D_s^- \pi^-$ and $\overline B{}^0\to D^0 D_s^-\pi^+$ decays \cite{LHCb:2022sfr, LHCb:2022lzp}.
Note that the $\overline B{}_s^0\to D^0\overline {T} ^0_{\bar c\bar s}$ decay involving $\overline{T^*_{c s 0}}(2870)^0$ has not been observed yet, as it is a CKM suppressed $\Delta S=-1$ transition.
It will be interesting to search for this mode to check whether these three states belong to the same multiplet or not.

Modes in $\overline{\bf 3}$ or $\overline{\bf 3}'$ are also related. 
From Table \ref{tab: BtoDTbar I} and Table \ref{tab: BtoDTbar II}, one can infer the following relations on rates in both scenarios,
\be
\Gamma(\overline B{}_s^0\to D^+ \overline T^{\,\prime\prime -}_{\bar c})
=\Gamma(\overline B{}_s^0\to D^0 \overline T^{\,\prime\prime 0}_{\bar c})
\simeq \left|\frac{V_{cs}}{V_{cd}}\right|^2 \Gamma(\overline B{}^0\to D_s^+ \overline T^{\,\prime\prime -}_{\bar c s}),
\label{eq: Bbar to D Tbar  3bar a}
\en
\be
\Gamma(B^-\to D^0 \overline T^{\,\prime\prime -}_{\bar c s})
&=&\Gamma(\overline B{}^0\to D^+ \overline T^{\,\prime\prime -}_{\bar c s})
\simeq \left|\frac{V_{cs}}{V_{cd}}\right|^2 \Gamma(B^-\to D^0 \overline T^{\,\prime\prime -}_{\bar c})
\non\\
&=&\left|\frac{V_{cs}}{V_{cd}}\right|^2 \Gamma(\overline B{}_s^0\to D_s^+ \overline T^{\,\prime\prime -}_{\bar c}),
\label{eq: Bbar to D Tbar  3bar b}
\en
and
\be
\Gamma(\overline B{}_s^0\to D_s^+ \overline T^{\,\prime\prime -}_{\bar c s})
\simeq\left|\frac{V_{cs}}{V_{cd}}\right|^2 \Gamma(\overline B{}^0\to D^+ \overline T^{\,\prime\prime -}_{\bar c}),
\label{eq: Bbar to D Tbar  3bar c}
\en
where the above approximations neglect penguin contributions with a 10\% to 20\% error, see Eq. (\ref{eq: R value 1}).
The rates in Eqs. (\ref{eq: Bbar to D Tbar  3bar b}) and (\ref{eq: Bbar to D Tbar  3bar c}) are expected to be larger than those in Eq. (\ref{eq: Bbar to D Tbar  3bar a}) in the same $\Delta S$ transition. as the last two sets have contributions from external $W$ amplitudes, $T_{\bar T}$ or $T'_{\bar T}$. 

Note that in ref. \cite{Qin:2022nof} the amplitudes of the following four modes in $\Delta S=-1$ transitions in scenario II, namely
$B^-\to D^+ \overline T^{\,--}_{\bar c s}$, 
$\overline B{}^0\to D^0 \overline T^0_{\bar c s }$,
$B^-\to D_s^+ \overline T^{\,--}_{\bar c s s}$, 
$\overline B{}^0\to D_s^+ \overline T^{\,-}_{\bar c s s}$,
were also given. 
Their results agree with those shown in Table \ref{tab: BtoDTbar II}, but with $C_{\bar T}$ replaced with $C V_{cb} V^*_{cs}$ and neglecting $P_{\bar T,1}$,
as they do not consider penguin contributions. 
The rest in Table \ref{tab: BtoDTbar II} and all in Table \ref{tab: BtoDTbar I} are new.  
Note that ref. \cite{Qin:2022nof} also used a different notation for the open-charmed tetraquark states, see Table~\ref{tab: 15bar+3bar} for a comparison.
For example, in Table 2 of ref. \cite{Qin:2022nof}, one finds
\be
A(B^-\to D^+T_{\bar c \bar u d s})=C V_{cb} V^*_{cs},
\en
which corresponds to 
\be
A(B^-\to D^+ \overline T^{\,--}_{\bar c s})=C_{\bar T}+P_{\bar T,1},
\en
in Table \ref{tab: BtoDTbar II}, but with $C_{\bar T}$ replaced with $C V_{cb} V^*_{cs}$ and neglecting $P_{\bar T,1}$.

The final state $\overline T$ can further decay to $\bar DP$ and $\bar DS$, or they can connect to these states virtually. 
One can obtain their full amplitudes using the results in this section and those in Sec.~\ref{sec: T to DP, DS}.

\subsubsection{$\overline B \to \overline D T$ decay amplitudes}

The $\overline B \to \overline D T$ decay amplitudes for $\Delta S=-1$ and $\Delta S=0$ transitions decomposed in these topological amplitudes are shown in Table \ref{tab: BtoDbarT I} for scenario I and in Table \ref{tab: BtoDbarT II} for scenario~II. 
Note that these results can easily generalize to $\overline B \to \overline D^* T$ decay by replacing $\overline D$ by $\overline D^*$ in the tables.

\begin{table}[t!]
\caption{\label{tab: BtoDbarT I}
$\overline B_q\to \overline D T_{c q[ \bar q' \bar q'']} $ decay amplitudes in $\Delta S=-1$ and $\Delta S=0$ transitions in scenario I.}
\footnotesize{
\begin{ruledtabular}
\begin{tabular}{llcllcccr}
\#
&Mode
          & $A (\overline B_q\to \overline D T_{c q[ \bar q' \bar q'']} )$
          &\#
          & Mode
          & $A (\overline B_q\to \overline D T_{c q [\bar q' \bar q'']})$
          \\
\hline
$4^{*}$ 
          & $B^-\to D^- T^0_{c s}$
          & $(C_{\bar D T}+P_{T})$
&$4^{*}$
          & $\overline B{}^0\to \overline D^0 T^0_{c s}$
          & $-(C_{\bar D T}+P_{T})$
          \\
5        
          & $\overline B{}^0\to D_s^- T^+_{c}$
          & $\frac{(C_{\bar D T}+P_{T})}{\sqrt2}$
&5
          & $\overline B{}_s^0\to D^- T^+_{c}$
          & $-\frac{1}{\sqrt2} (C_{\bar D T}+P_{T})$
           \\
6          
          & $B^-\to D_s^- T^0_{c}$
          & $-\frac{1}{\sqrt2} (C_{\bar D T}+P_{T})$
&6
          & $\overline B{}_s^0\to \overline D^0 T^0_{c}$
          & $\frac{1}{\sqrt2} (C_{\bar D T}+P_{T})$
          \\
\hline
7          
          & $\overline B{}^0\to D_s^- T^{\prime\prime +}_{c}$
          & $-\frac{1}{\sqrt2} (2 (T_{\bar D}+P_{\bar D})+(C_{\bar D T}+P_{T}))$
&7
          & $\overline B{}_s^0\to D^- T^{\prime\prime +}_{c}$
          & $\frac{1}{\sqrt2} ((C_{\bar D T}+P_{T})+2(E_{\bar D\bar B}+PA_{\bar B}))$
          \\                              
8          
          & $B^-\to D_s^- T^{\prime\prime 0}_{c}$
          & $-\frac{1}{\sqrt2}(2(T_{\bar D}+P_{\bar D})+(C_{\bar D T}+P_{T}))$
&8
          & $\overline B{}_s^0\to \overline D^0 T^{\prime\prime 0}_{c}$
          & $\frac{1}{\sqrt2}((C_{\bar D T}+P_{T})+2 (E_{\bar D\bar B}+PA_{\bar B}))$
           \\
9
          & $\overline B{}_s^0\to D_s^- T^{\prime\prime +}_{c \bar s}$
          & $-\sqrt2 ((T_{\bar D}+P_{\bar D})-(E_{\bar D\bar B}+PA_{\bar B}))$
          \\                                                          
\hline
\hline
\#
&Mode
          & $A' (\overline B_q\to \overline D T_{c q[ \bar q' \bar q'']} )$
          &\#
          & Mode
          & $A' (\overline B_q\to\overline D T_{c q [\bar q' \bar q'']})$
          \\
\hline
$2^*$         
          & $\overline B{}^0\to D_s^- T^+_{c\bar s}$
          & $-\frac{1}{\sqrt2} (C'_{\bar D T}+P'_{T})$
&$2^*$
          & $\overline B{}_s^0\to D^- T^+_{c \bar s} $
          & $\frac{1}{\sqrt2} (C'_{\bar D T}+P'_{T})$
          \\
$3^{*}$ 
          & $B^-\to D_s^- T^0_{c \bar s}$
          & $(C'_{\bar D T}+P'_{T})$
&$3^{*}$
          & $\overline B{}_s^0\to \overline D^0 T^0_{c \bar s}$
          & $-(C'_{\bar D T}+P'_{T})$
          \\                              
6          
          & $B^-\to D^- T^0_{c}$
          & $\frac{1}{\sqrt2} (C'_{\bar D T}+P'_{T})$
&6
          & $\overline B{}^0\to \overline D^0 T^0_{c}$
          & $-\frac{1}{\sqrt2} (C'_{\bar D T}+P'_{T})$
          \\
\hline
7          
          & $\overline B{}^0\to D^- T^{\prime\prime +}_{c}$
          & $-\sqrt2((T'_{\bar D}+P'_{\bar D})-(E'_{\bar D\bar B}+PA'_{\bar B}))$
&8          
          & $B^-\to D^- T^{\prime\prime 0}_{c}$
          & $-\frac{1}{\sqrt2} (2(T'_{\bar D}+P'_{\bar D})+(C'_{\bar D T}+P'_{T}))$
          \\
8          
          & $\overline B{}^0\to  \overline D^0 T^{\prime\prime 0}_{c}$
          & $\frac{1}{\sqrt2 } ((C'_{\bar D T}+P'_{T})+2(E'_{\bar D\bar B}+PA'_{\bar B}))$
          \\

9          
          & $\overline B{}^0\to D_s^- T^{\prime\prime +}_{c \bar s}$
          & $\frac{1}{\sqrt2 } ((C'_{\bar D T}+P'_{T})+2(E'_{\bar D\bar B}+PA'_{\bar B}))$
&9
          & $\overline B{}_s^0\to D^- T^{\prime\prime +}_{c \bar s}$
          & $-\frac{1}{\sqrt2} (2(T'_{\bar D}+P'_{\bar D})+(C'_{\bar D T}+P'_{T}))$
          \\                                             
\end{tabular}
\end{ruledtabular}
}
\end{table}

\begin{table}[t!]
\caption{\label{tab: BtoDbarT II}
$\overline B_q\to \overline D T_{c q\{\bar q' \bar q''\}}$ decay amplitudes in $\Delta S=-1$ and $\Delta S=0$ transitions in scenerio II.}
\footnotesize{
\begin{ruledtabular}
\begin{tabular}{llcllccccr}
\#
&Mode
          & $A (\overline B_q\to \overline D T_{c q\{ \bar q' \bar q''\} } )$
          &\#
          & Mode
          & $A (\overline B_q\to \overline D T_{c q\{\bar q' \bar q''\}})$
          \\
\hline
$4'^{*}$          
          & $\overline B{}^0\to D^- T^+_{c s}$
          & $\sqrt2 (C_{\bar D T}+P_{T})$
&$5'^{*}$ 
          & $B^-\to D^- T^0_{c s}$
          & $(C_{\bar D T}+P_{T})$
          \\
$5'^{*}$
          & $\overline B{}^0\to \overline D^0 T^0_{c s}$
          & $(C_{\bar D T}+P_{T})$
&$6'^{*}$         
          & $B^-\to \overline D^0 T^-_{c s}$
           & $\sqrt2 (C_{\bar D T}+P_{T})$
          \\ 
$13'$         
          & $\overline B{}^0\to  D_s^- T^{\prime +}_{c}$
          & $-\frac{\sqrt3}{2} (C_{\bar D T}+P_{T})$
&$13'$
          & $\overline B{}_s^0\to D^- T^{\prime +}_{c}$
          & $-\frac{\sqrt3}{2} (C_{\bar D T}+P_{T})$
          \\
$14'$         
          & $B^-\to D_s^- T^{\prime 0}_{c}$
          & $-\frac{\sqrt3}{2} (C_{\bar D T}+P_{T})$
&$14'$
          & $\overline B{}_s^0\to \overline D^0 T^{\prime 0}_{c}$
          & $-\frac{\sqrt3}{2} (C_{\bar D T}+P_{T})$
          \\
$15'$
          & $\overline B{}_s^0\to D_s^- T^{\prime +}_{c \bar s}$
          & $-(C_{\bar D T}+P_{T})$
          \\                                                    
\hline
$16'$          
          & $\overline B{}^0\to D_s^- T^{\prime\prime +}_{c}$
          & $\frac{1}{2}(4(T_{\bar D}+P_{\bar D})+(C_{\bar D T}+P_{T}))$
&$16'$
          & $\overline B{}_s^0\to D^- T^{\prime\prime +}_{c}$
          & $\frac{1}{2} ((C_{\bar D T}+P_{T})+4(E_{\bar D\bar B}+PA_{\bar B}))$
          \\                              
$17'$         
          & $B^-\to D_s^- T^{\prime\prime 0}_{c}$
          & $\frac{1}{2} (4 (T_{\bar D}+P_{\bar D})+(C_{\bar D T}+P_{T}))$
&$17'$
          & $\overline B{}_s^0\to \overline D^0 T^{\prime\prime 0}_{c}$
          & $\frac{1}{2} ((C_{\bar D T}+P_{T})+4(E_{\bar D\bar B}+PA_{\bar B}))$
          \\
$18'$
          & $\overline B{}_s^0\to D_s^- T^{\prime\prime +}_{c \bar s}$
          & $2(T_{\bar D}+P_{\bar D})+(C_{\bar D T}+P_{T})$
          \\ 
          & 
          & $+2(E_{\bar D\bar B}+PA_{\bar B})$
          \\           
\hline
\hline
\#
&Mode
          & $A' (\overline B_q\to \overline D T_{c q\{ \bar q' \bar q''\} } )$
          & \#
          & Mode
          & $A' (\overline B_q\to \overline D T_{c q\{\bar q' \bar q''\}})$
          \\
\hline
$2'^*$         
          & $\overline B{}^0\to D_s^- T^+_{c \bar s}$
          & $-\frac{ (C'_{\bar D T}+P'_{T})}{\sqrt2}$
&$2'^*$
          & $\overline B{}_s^0\to D^- T^+_{c \bar s} $
          & $-\frac{(C'_{\bar D T}+P'_{T})}{\sqrt2}$
          \\
$3'^{*}$ 
          & $B^-\to D_s^- T^0_{c \bar s}$
          & $(C'_{\bar D T}+P'_{T})$
&$3'^{*}$
          & $\overline B{}_s^0\to \overline D^0 T^0_{c \bar s}$
          & $(C'_{\bar D T}+P'_{T})$
          \\                              
$8'^{*}$
          & $\overline B{}_s^0\to D_s^- T^+_{c \bar s \bar s}$
          & $\sqrt2 (C'_{\bar D T}+P'_{T})$
          \\
$10'^*$        
          & $\overline B{}^0\to D^- T^+_{c}$
          & $-\sqrt{\frac{2}{3}} (C'_{\bar D T}+P'_{T})$
&$11'^*$         
          & $B^-\to D^- T^0_{c}$
          & $-\sqrt{\frac{2}{3}} (C'_{\bar D T}+P'_{T})$
          \\
$11'^*$
          & $\overline B{}^0\to \overline D^0 T^0_{c}$
          & $-\sqrt{\frac{2}{3}} (C'_{\bar D T}+P'_{T})$
&$12'^{*}$         
          & $B^-\to \overline D^0 T^-_{c}$
          & $\sqrt2 (C'_{\bar D T}+P'_{T})$
          \\ 
$13'$         
          & $\overline B{}^0\to  D^- T^{\prime +}_{c}$
          & $\frac{1}{\sqrt3} (C'_{\bar D T}+P'_{T})$
&$14'$        
          & $B^-\to D^- T^{\prime 0}_{c}$
          & $\frac{ 1}{2\sqrt3} (C'_{\bar D T}+P'_{T})$
          \\    
$14'$       
          & $\overline B{}^0\to \overline D^0 T^{\prime 0}_{c}$
          & $\frac{1}{2\sqrt3} (C'_{\bar D T}+P'_{T})$
          \\ 
$15'$         
          & $\overline B{}^0\to  D_s^- T^{\prime +}_{c \bar s}$
          & $\frac{1}{2} (C'_{\bar D T}+P'_{T})$
&$15'$
          & $\overline B{}_s^0\to D^- T^+_{c \bar s}$
          & $\frac{1}{2} (C'_{\bar D T}+P'_{T})$
          \\
\hline                                                              
$16'$          
          & $\overline B{}^0\to D^- T^{\prime\prime +}_{c}$
          & $2(T'_{\bar D}+P'_{\bar D})+(C'_{\bar D T}+P'_{T})$
&$17'$        
          & $B^-\to D^- T^{\prime\prime 0}_{c}$
          & $\frac{1}{2}(4 (T'_{\bar D}+P'_{\bar D})+(C'_{\bar D T}+P'_{T}))$
          \\
          & 
          & $+2(E'_{\bar D\bar B}+PA'_{\bar B})$
&$17'$        
          & $\overline B{}^0\to \overline D{}^0 T^{\prime\prime 0}_{c}$
          & $\frac{1}{2}((C'_{\bar D T}+P'_{T})+4 (E'_{\bar D\bar B}+PA'_{\bar B}))$
          \\
$18'$          
          & $\overline B{}^0\to D_s^- T^{\prime\prime +}_{c \bar s}$
          & $\frac{1}{2}(C'_{\bar D T}+P'_T+4(E'_{\bar D\bar B}+PA'_{\bar B}))$
&$18'$
          & $\overline B{}_s^0\to D^- T^{\prime\prime +}_{c \bar s}$
          & $\frac{1}{2}(4(T'_{\bar D}+P'_{\bar D})+(C'_{\bar D T}+P'_{T}))$
          \\ 
\end{tabular}
\end{ruledtabular}
}
\end{table}

From Table \ref{tab: BtoDbarT I} we see that the decay modes involving $T$ in ${\bf 6}$ in scenario I 
are governed by $(C_{\bar D T}+P_{T})$ or $(C'_{\bar D T}+P'_{T})$.
Consequently, their decay rates are highly related.
In particular, the modes involving flavor exotic states have the following relations on decay rates,
\be
\Gamma_5
&\equiv&\Gamma(B^-\to D^- T^0_{c s})
= \Gamma(\overline B{}^0\to \overline D^0 T^0_{c s}),
\non\\
\Gamma_6
&\equiv&
\Gamma(B^-\to D_s^- T^0_{c \bar s})
= \Gamma(\overline B{}_s^0\to \overline D^0 T^0_{c \bar s})
=2  \Gamma(\overline B{}^0\to D_s^- T^+_{c\bar s})
\non\\
&=&2 \Gamma(\overline B{}_s^0\to D^- T^+_{c \bar s} ),
\non\\
\Gamma_5
&\simeq&
\left|\frac{V_{cs}}{V_{cd}}\right|^2 \Gamma_6,
\label{eq: Bto Dbar T relations I}
\en
where the approximation in the last line neglects the penguin contributions.
Note that three out of four flavor exotic states, namely $T^0_{c s}$, $T^+_{c s}$  and $T^0_{c \bar s}$, are involved in these $\overline B\to \overline D T$ decays,
while the other flavor exotic state, $T^{++}_{c \bar s}$ is absent.
The absence of the doubly charged state in $\overline B \to \overline D T$ decays can be easily understood from charge conservation.
We need to consider both $\Delta S=-1$ and $\Delta S=0$ transitions in $\overline B\to \overline D T$ decays to have both $T^0_{c s}$ and $T^0_{c \bar s}$ final states.

One can roughly estimate the error in the above approximation by using 
\be
\frac{|P_{T}|}{|C_{\bar D T}|}\sim \left|\frac{V_{tb} V^*_{ts}(c_4+\frac{c_3}{3})}{V_{cb} V^*_{cs}(c_2+\frac{c_1}{3})}\right|\simeq 0.19,
\en
and
\be
\frac{|P'_{T}|}{|C'_{\bar D T}|}\sim \left|\frac{V_{tb} V^*_{td}(c_4+\frac{c_3}{3})}{V_{cb} V^*_{cd}(c_2+\frac{c_1}{3})}\right|\simeq 0.17,
\en
where $c_i$ are the Wilson coefficients, see \cite{Beneke:2001ev}.
Therefore, the errors of neglecting penguin contributions are roughly estimated to be of the order $\lambda$.
To be more precise, we can similarly define the following ratio:
\be
\tilde R
&\equiv& \left|\frac{V_{cb}V^*_{cd}}{V_{cb} V^*_{cs}}\right|^2\frac{\Gamma_5}{\Gamma_6}
\non\\
&=&
\left|\frac{V_{cb} V^*_{cd}}{V_{cb} V^*_{cs}}\right|^2
\frac{|C_{\bar D T}+P_{T}|^2+|\bar C_{\bar D T}+\bar P_{T}|^2}{|C'_{\bar D T}+P'_{T}|^2+|\bar C'_{\bar D T}+\bar P'_{T}|^2}
\non\\
&=&
\left|\frac{V_{cb} V^*_{cd}}{V_{cb} V^*_{cs}}\right|^2
\non\\
&&\times
\frac
{|V_{cb} V^*_{cd}(c_2+\frac{c_1}{3})+\tilde \rho e^{i\tilde \phi} V_{tb} V^*_{ts}(c_4+\frac{c_3}{3}) |^2+|V^*_{cb} V_{cd}(c_2+\frac{c_1}{3})+\tilde \rho e^{i\tilde \phi} V^*_{tb} V_{ts}(c_4+\frac{c_3}{3}) |^2}
{|V_{cb} V^*_{cd}(c_2+\frac{c_1}{3})+\tilde \rho e^{i\tilde \phi}V_{tb} V^*_{td}(c_4+\frac{c_3}{3})|^2+|V^*_{cb} V_{cd}(c_2+\frac{c_1}{3})+\tilde \rho e^{i\tilde \phi} V^*_{tb} V_{td}(c_4+\frac{c_3}{3})|^2},
\non\\
\en
where we introduce a correction factor $\tilde \rho$ and a relative strong phase $\phi$ between the penguin and the tree amplitudes.
Again by taking $\tilde \rho=0\sim 2$ and $\tilde\phi=0\sim 2\pi$, we obtain
\be
\tilde R-1=-0.20\sim 0.09.
\label{eq: R value 2}
\en
Hence, the correction on the above approximation is estimated to be about 10\% to 20\% in the presence of penguin contributions.
To effectively control the penguin pollution, further work is needed; see, for example, a study in a somewhat similar situation \cite{Fleischer:2007zn},
where information of direct CP violation is needed. 
With this information, the sizes of the penguin and tree amplitudes, as well as the relative strong phase, can be probed.

We turn to scenario II.
Likewise, from Table \ref{tab: BtoDbarT II}, 
the decay modes involving $T$ in $\overline {\bf 15}$ only have contributions from $(C_{\bar D T}+P_{T})$ or $(C'_{\bar D T}+P'_{T})$, and, consequently, their amplitudes are highly related.
In particular, the modes involving flavor exotic $T$ have the following relations on decay rates,
\be
 \Gamma_7
 &\equiv&\Gamma(\overline B{}^0\to D^- T^+_{c s})
 =
2\Gamma(B^-\to D^- T^0_{c s})
=2 \Gamma(\overline B{}^0\to \overline D^0 T^0_{c s})
=\Gamma(B^-\to \overline D^0 T^-_{c s})
\non\\
\Gamma_8
&\equiv&
2\Gamma(B^-\to D_s^- T^0_{c \bar s})
=2\Gamma(\overline B{}_s^0\to \overline D^0 T^0_{c \bar s})
\non\\
&=&4\Gamma(\overline B{}^0\to D_s^- T^+_{c \bar s})
=4 \Gamma(\overline B{}_s^0\to D^- T^+_{c \bar s} )
=\Gamma(\overline B{}_s^0\to D_s^- T^+_{c \bar s \bar s})
=3 \Gamma(\overline B{}^0\to D^- T^+_{c})
\non\\
&=&\Gamma(B^-\to D^- T^0_{c})
=3 \Gamma(\overline B{}^0\to \overline D^0 T^0_{c})
= \Gamma(B^-\to \overline D^0 T^-_{c}),
\non\\
\Gamma_7 &\simeq&
\left|\frac{V_{cs}}{V_{cd}}\right|^2 \Gamma_8,
\label{eq: Bto Dbar T relations II}
\en
where the approximation in the last line neglects the penguin contributions and the correction to the approximation is about 10\% to 20\%, see Eq. (\ref{eq: R value 2}).
We have nine out of twelve flavor exotic states involved in these $\overline B\to \overline D T$ decays,
while all doubly charged states, $T^{++}_{c \bar s}$, $T^{++}_{c\bar s\bar s}$ and $T^{++}_c$, are absent,
as $\overline B\to \overline D T^{++}$ decays are forbiden by charge conservation.

Note that the $B^-\to D^- T^0_{c s}$ decay
corresponds to the $B^-\to D^- T^*_{cs 0} (2870)^0$ decay
reported by LHCb in the observation of $T_{c s 0} (2870)^{0}$,
from $B^-\to D^- D^+ K^-$ decay with $B^-\to D^- T^*_{cs 0} (2870)^0, T^*_{cs 0} (2870)^0\to D^+ K^-$ decay \cite{LHCb:2020pxc}. 
Recently, LHCb also reports the observation of $B^-\to D^- T^*_{cs 0} (2870)^0, T^*_{cs 0} (2870)^0\to D^0 K^0_S$
in $B^-\to D^- D^0 K^0_S$ decay \cite{LHCb:2024xyx} agreeing with
\be
\Gamma(B^-\to D^- T^0_{c s},  T^0_{c s}\to D^+ K^-) 
=
\Gamma(B^-\to D^- T^0_{c s},  T^0_{c s}\to D^0 \overline K^0), 
\en
which can be inferred from Tables \ref{tab: TtoDP1} and \ref{tab: TtoDP2} in scenarios I and II, respectively.

Note that the $\overline B^0\to \overline D^0 T^0_{c s}$ decay has not been observed yet. It will be useful to search for this mode and check the relations in Eqs. (\ref{eq: Bto Dbar T relations I}) and (\ref{eq: Bto Dbar T relations II}). 

Modes in $\overline{\bf 3}$ or $\overline{\bf 3}'$ are also related. 
From Table \ref{tab: BtoDTbar I} and Table \ref{tab: BtoDTbar II}, one can infer the following relations on rates in both scenarios,
\be
\Gamma(\overline B{}^0\to D_s^- T^{\prime\prime +}_{c})
&=&\Gamma(B^-\to D_s^- T^{\prime\prime 0}_{c})
\simeq \left|\frac{V_{cs}}{V_{cd}}\right|^2 \Gamma(B^-\to D^- T^{\prime\prime 0}_{c})
\non\\
&=&\left|\frac{V_{cs}}{V_{cd}}\right|^2 \Gamma(\overline B{}_s^0\to D^- T^{\prime\prime +}_{c \bar s}),
\label{eq: Bbar to D barT  3bar a}
\en
\be
\Gamma(\overline B{}_s^0\to D_s^- T^{\prime\prime +}_{c \bar s})
\simeq\left|\frac{V_{cs}}{V_{cd}}\right|^2 \Gamma(\overline B{}^0\to D^- T^{\prime\prime +}_{c}),
\label{eq: Bbar to D barT  3bar b}
\en
and
\be
\Gamma(\overline B{}_s^0\to D^- T^{\prime\prime +}_{c})
&=&\Gamma(\overline B{}_s^0\to \overline D^0 T^{\prime\prime 0}_{c})
\simeq\left|\frac{V_{cs}}{V_{cd}}\right|^2 \Gamma(\overline B{}^0\to  \overline D^0 T^{\prime\prime 0}_{c})
\non\\
&=&\left|\frac{V_{cs}}{V_{cd}}\right|^2 \Gamma(\overline B{}^0\to D_s^- T^{\prime\prime +}_{c \bar s}),
\label{eq: Bbar to D barT  3bar c}
\en
where the approximation is neglecting penguin contributions and the correction to the approximation is about 10\% to 20\%, see Eq. (\ref{eq: R value 2}).
As they have contributions from external $W$ amplitudes, $T_{\bar D}$ or $T'_{\bar D}$,
the rates in Eqs. (\ref{eq: Bbar to D barT  3bar a}) and (\ref{eq: Bbar to D barT  3bar b}) are expected to be larger than those in Eq. (\ref{eq: Bbar to D barT  3bar c}) in the same $\Delta S$ transition.

In ref. \cite{Qin:2022nof} the amplitudes of the following four modes in $\Delta S=-1$ transitions in scenario~II, namely
$\overline B{}^0\to D^- T^+_{c s}$,
$B^-\to D^- T^0_{c s}$,
$\overline B{}^0\to \overline D^0 T^0_{c s}$
and
$B^-\to \overline D^0 T^-_{c s}$ decays,
were also given. 
Their results agree with those shown in Table \ref{tab: BtoDbarT II}, but with $C_{\bar D T}$ replaced with $T_3 V_{cb} V^*_{cs}$ and neglecting $P_{T}$, as they do not consider penguin contributions.
The rest in Table \ref{tab: BtoDbarT II} and all in Table \ref{tab: BtoDbarT I} are new.  
Note that ref. \cite{Qin:2022nof} also used a different notation for the open-charmed tetraquark states, see Table~\ref{tab: 15bar+3bar} for a comparison.
For example, in Table 2 of ref. \cite{Qin:2022nof}, one finds
\be
A(B^-\to D^- T_{c s \bar u\bar d})=T_3 V_{cb} V^*_{cs},
\en
which corresponds to 
\be 
A(B^-\to D^- T^0_{c s})=C_{\bar D T}+P_{T},
\en
in Table \ref{tab: BtoDbarT II}  but with $C_{\bar D T}$ replaced with $T_3 V_{cb} V^*_{cs}$ and neglecting $P_{T}$.

The final state $T$ can further decay to $DP$ and $DS$, or they can connect to those states virtually. 
One can obtain their full amplitudes using results in the section and those in Sec.~\ref{sec: T to DP, DS}. 

\begin{table}[t!]
\caption{\label{tab: BtoDTbar BtoDbarT}
Participation of $T^*_{c\bar s0}(2900)^{++}$, $T^*_{c\bar s0}(2900)^{0}$ and $T^*_{c s 0}(2870)^{0}$ in $\overline B\to D \overline T$ and $\overline B\to  \overline D T$ decays in $\Delta S=-1$ and $\Delta S=0$ transitions. TA stands for topological amplitude.
}
\begin{ruledtabular}
\begin{tabular}{lcccc}
Decays
          & $T^*_{c\bar s0}(2900)^{++}$
          & $T^*_{c\bar s0}(2900)^{0}$
          & $T^*_{c s 0}(2870)^{0}$
          & TA involved
          \\
\hline
$\overline B\to D \overline T$, $\Delta S=-1$
          & $\checkmark$
          & $\checkmark$
          & {\sffamily X}
          & $(C_{\bar T}+P_{\bar T,1})$
          \\
$\overline B\to  \overline D T$, $\Delta S=-1$
          & {\sffamily X}
          & {\sffamily X}
          & $\checkmark$
          & $(C_{\bar D T}+P_{T})$
          \\
\hline           
$\overline B\to D \overline T$, $\Delta S=0$
          & $\checkmark$
          & {\sffamily X}
          & $\checkmark$
          & $(C'_{\bar T}+P'_{\bar T,1})$
          \\
$\overline B\to  \overline D T$, $\Delta S=0$
          &  {\sffamily X}
          & $\checkmark$
          & {\sffamily X}
          & $(C'_{\bar D T}+P'_{T})$
          \\                  
\end{tabular}
\end{ruledtabular}
\end{table}

Before we end this section, 
it will be useful to compare the participation of $T^*_{c\bar s0}(2900)^{++}$, $T^*_{c\bar s0}(2900)^{0}$ and $T^*_{c s 0}(2870)^{0}$
in $\overline B\to D\overline T$ and $\overline B\to \overline D T$ decays in $\Delta S=-1$ and $\Delta S=0$ transitions.
We show in Table \ref{tab: BtoDTbar BtoDbarT} such a comparison with the relevant topological amplitudes listed.
It is evident that using only $\Delta S=-1$ transitions, one is unable to check if these three states belong to the same multiplet or not, as $(C_{\bar T}+P_{\bar T,1})$ and $(C_{\bar D T}+P_{T})$ are unrelated.
Therefore, to verify that, in any case, we need to consider the CKM suppressed $\Delta S=0$ transitions as well.

\section{$\overline B\to T P$ and $\overline B\to T S$ decays}

\subsection{Topological amplitudes in $\overline B\to T P$ and $TS$ decays}

\subsubsection{Topological amplitudes in $\overline B\to T P$ decays}

The $(\bar d u)(\bar c b)$ operator in the weak Hamiltonian
$H_{\rm W}$ can be expressed as $(\bar q_i H^i_j q^j)\,(\bar c
b)$, where $q^j=(u,\,d,\,s)$ and
\begin{equation}
H= \left(
\begin{array}{ccc}
0 &0 &0\\
1 &0 &0\\
0 &0 &0
\end{array}
\right).
\label{eq: H 3x3}
\end{equation}
Consequently, for $\overline B \to T_{c q [q' q'']} P$ decays in scenario I, we have the following effective Hamiltonian,
\be
H_{\rm eff}
&=&
         T_P\, \overline B_m  \overline T_k^{[mk]}\, H^i_j\, \Pi^j_i
         +C_{T}\, \overline B_m \Pi^m_k \, H^i_j\, \overline T_i^{[jk]}
\non\\
         && +C_{PT,1}\, \overline B_m  \overline T_k^{[mj]}\, H^i_j\, \Pi^k_i
              +C_{PT,2}\,\overline B_m \Pi^m_i\, H^i_j\, \overline T_k^{[jk]}
\non\\
         &&+C_{TP}\, \overline B_m  \overline T_i^{[mk]}\, H^i_j\, \Pi^j_k
         +E_{\bar B T}\,\overline B_i\,H^i_j\, \overline T_m^{[jk]}   \Pi^m_k
\non\\
        &&+E_{\bar B P}\,\overline B_i\,H^i_j\,\Pi^j_m  \overline T_k^{[mk]},      
\label{eq: B to TP 1}
\en
while for $\overline B \to T_{c q \{q' q''\}} P$ decays in scenario II, we have
\be
H_{\rm eff}
&=&
         T_P\, \overline B_m  \overline T_k^{\{mk\}}\, H^i_j\, \Pi^j_i
         +C_{T}\, \overline B_m \Pi^m_k \, H^i_j\, \overline T_i^{\{jk\}}
\non\\
         && +C_{PT,1}\, \overline B_m  \overline T_k^{\{mj\}}\, H^i_j\, \Pi^k_i
              +C_{PT,2}\,\overline B_m \Pi^m_i\, H^i_j\, \overline T_k^{\{jk\}}
\non\\
         &&+C_{TP}\, \overline B_m  \overline T_i^{\{mk\}}\, H^i_j\, \Pi^j_k
         +E_{\bar B T}\,\overline B_i\,H^i_j\, \overline T_m^{\{jk\}}   \Pi^m_k
\non\\         
          &&+E_{\bar B P}\,\overline B_i\,H^i_j\,\Pi^j_m  \overline T_k^{\{mk\}}.
\label{eq: B to TP 2}
\en
For notational simplicity, we do not distinguish the topological amplitudes in scenarios I and II.
The above construction also applies to $\overline B\to T V$ decays by simple substitution. 
Note that in the above effective Hamiltonians, we do not consider terms with $\Pi^i_i$. 
In principle, they can take place, but these diagrams are more detached and are expected to be suppressed. 
Furthermore, the addition of these terms only affects $\eta_1$. 
We will neglect them in this work.
The effect of this approximation is that we do not differentiate $\eta_1$ from $\eta_8$, and one may view this as a nonet symmetry. 

\begin{figure}[t]
\centering
\hspace{12pt}
\subfigure[]{
  \includegraphics[width=0.3\textwidth]{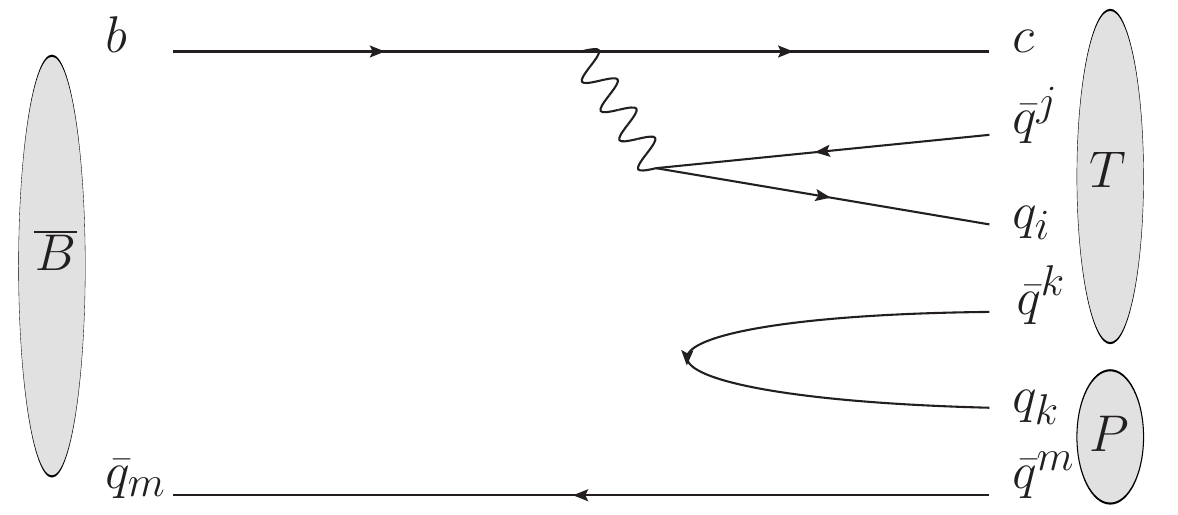}
}
\subfigure[]{
  \includegraphics[width=0.3\textwidth]{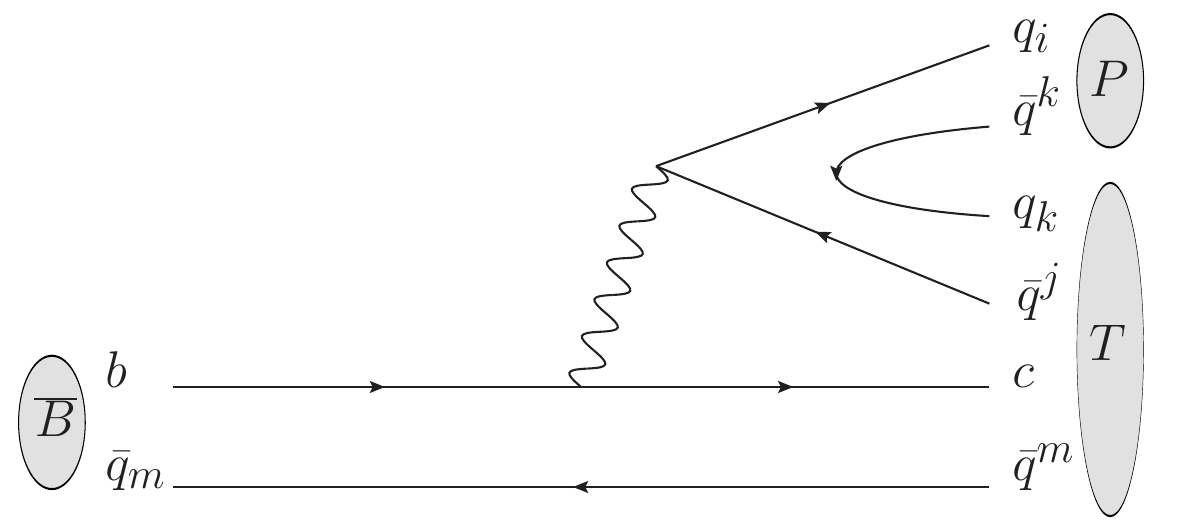}
}
\subfigure[]{
  \includegraphics[width=0.3\textwidth]{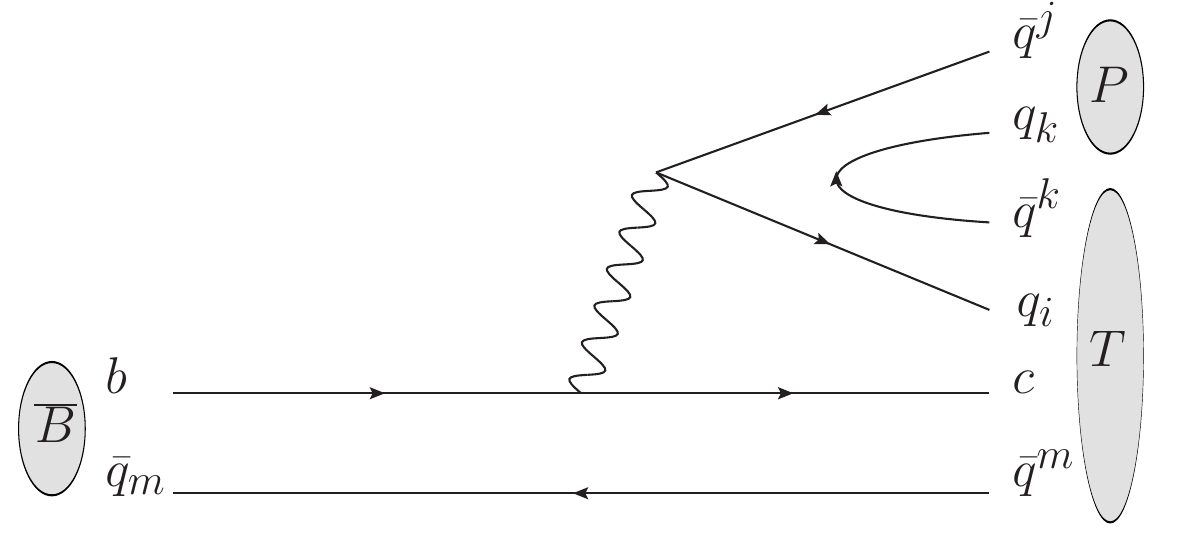}
}
\subfigure[]{
  \includegraphics[width=0.3\textwidth]{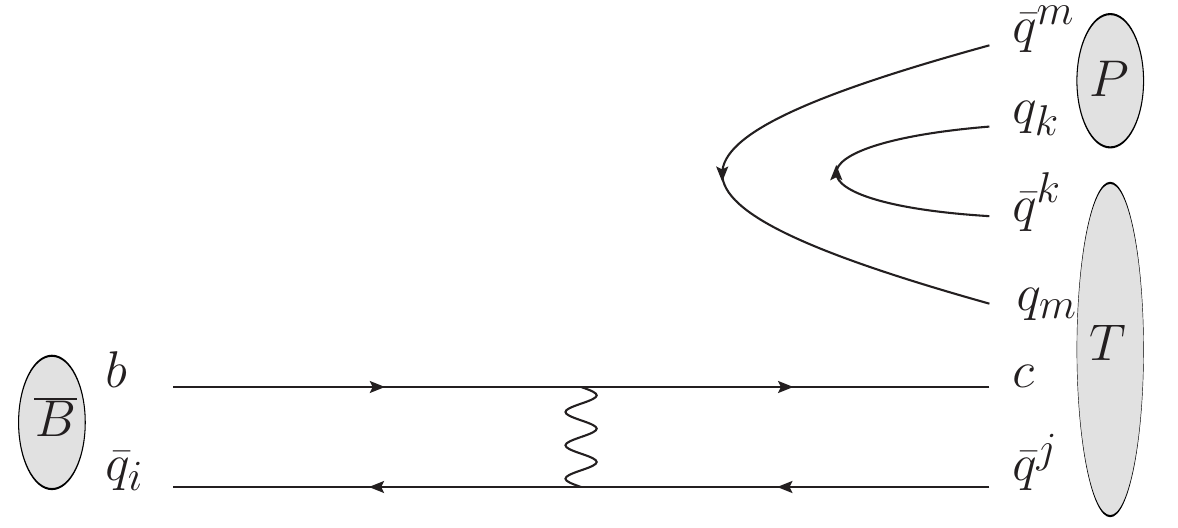}
}
\subfigure[]{
  \includegraphics[width=0.3\textwidth]{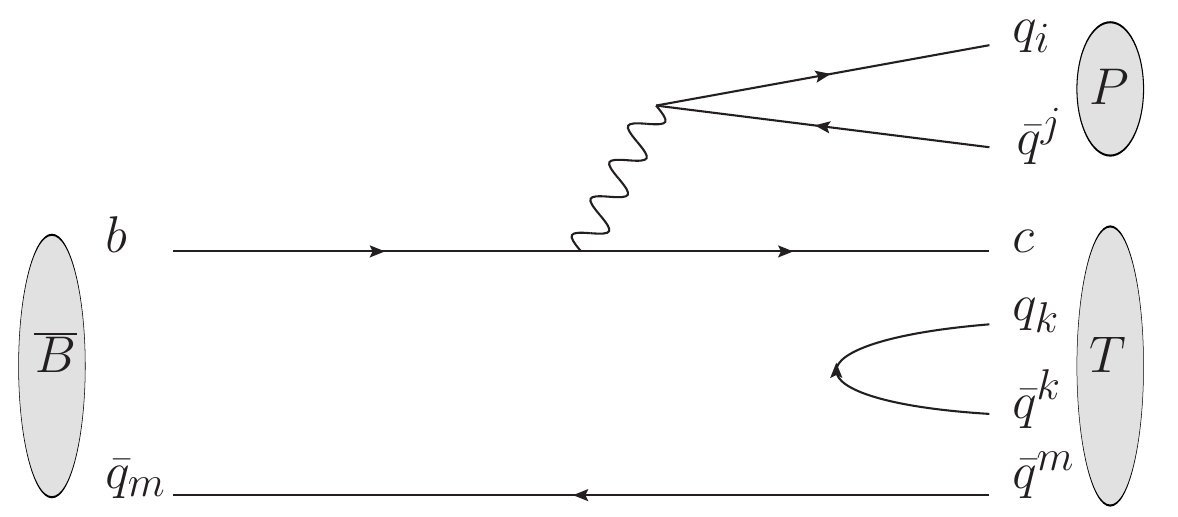}
}
\subfigure[]{
  \includegraphics[width=0.3\textwidth]{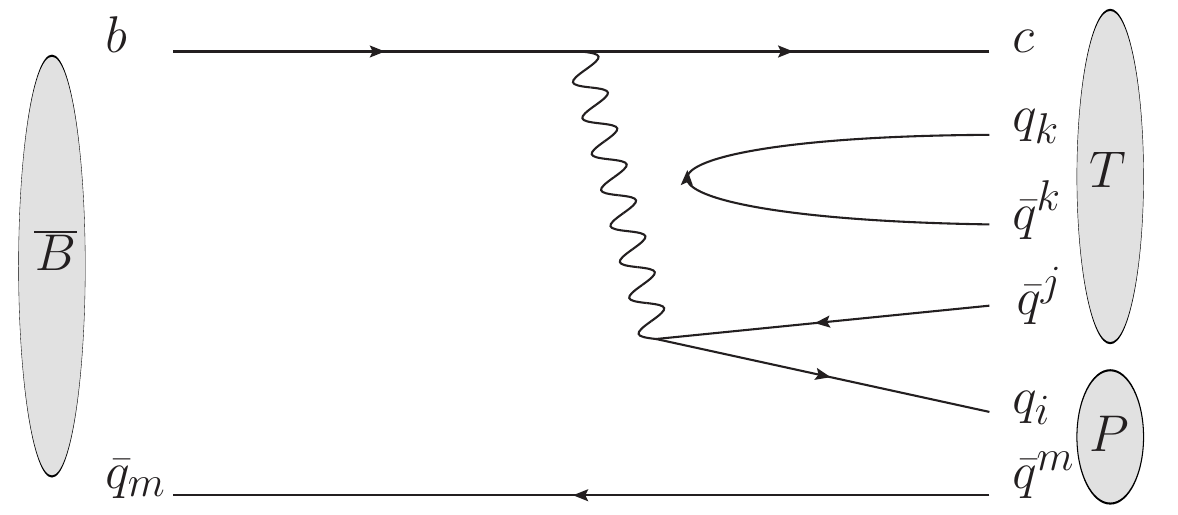}
  }
\subfigure[]{
  \includegraphics[width=0.3\textwidth]{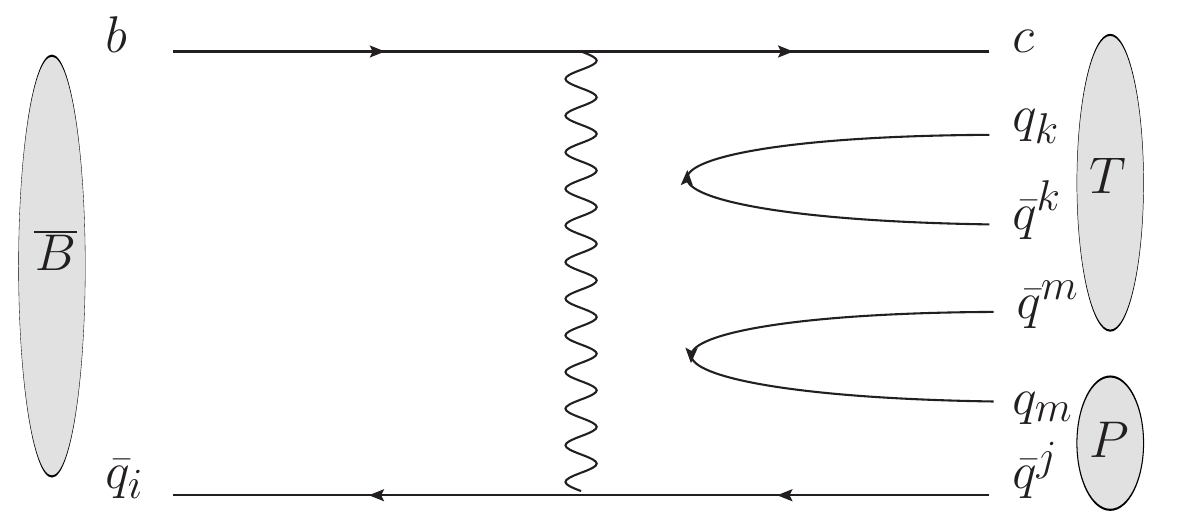}
}
\caption{Topological diagrams of 
(a) $C_T$,  
(b) $C_{PT,1}$, 
(c) $C_{TP}$, 
(d) $E_{\bar B T}$, 
(e) $T_P$,
(f) $C_{PT,2}$ 
and
(g)~$E_{\bar B P}$  
amplitudes in $\overline B\to T P$ decays, with $T$, $C$ and $E$ as external $W$,
  internal $W$ and exchange amplitudes, respectively. 
   The subscripts indicate the final state meson(s) receiving the $q_i \bar q^j$ quarks from the $W$ line in the $b\to c W^-, W^-\to q_i \bar q^j$ transition.
 Only diagrams (a) to (d) contribute to modes involving flavor exotic states.}
  \label{fig: TA B2TP}
\end{figure}

There are seven topological amplitudes in $\overline B\to T P$ decays, namely
$C_T$ (internal $W$), 
$C_{PT,1}$ (internal $W$), 
$C_{TP}$ (internal $W$),  
$E_{\bar B T}$ (exchange),
$T_P$ (external $W$), 
$C_{PT,2}$ (internal $W$)
and
$E_{\bar B P}$ (exchange) 
topological amplitudes.
They are depicted in Fig.~\ref{fig: TA B2TP}. 
The subscripts of these topological amplitudes indicate the final state meson(s) receiving the $q_i \bar q^j$ quarks 
from the $W$ line in the $b\to c W^-, W^-\to q_i \bar q^j$ transition.
For example. $C_{TP}$ indicates $T$ receiving the $q_i$ while $P$ receiving the $\bar q^j$, see Fig.~\ref{fig: TA B2TP} (c).
It is evident that when both are received by a single meson, as in Fig.~\ref{fig: TA B2TP} (a), one simply indicates that meson in the subscript, as in $C_T$.

The above effective Hamiltonians are for the $\Delta S=0$ transition, and they can be easily transformed into those for the $\Delta S=-1$ transition, by simply replacing $H$ in Eq. (\ref{eq: H 3x3}) with
\begin{equation}
H'= \left(
\begin{array}{ccc}
0 &0 &0\\
0 &0 &0\\
1 &0 &0
\end{array}
\right).
\label{eq: H' 3x3}
\end{equation}
and with 
$T_P$, $C_T$, $C_{PT,1}$, $C_{PT,2}$, $C_{TP}$, $E_{\bar B T}$, $E_{\bar B P}$ replaced with 
$T'_P$, $C'_T$, $C'_{PT,1}$, $C'_{PT,2}$, $C'_{TP}$, $E'_{\bar B T}$, $E'_{\bar B, P,2}$, respectively. 
It is understood that these amplitudes are related by CKM matrix elements giving
$T'_P=\frac{V^*_{us}}{V^*_{ud}} T_P$, 
$C'_T=\frac{V^*_{us}}{V^*_{ud}}  C_T$, 
$C'_{PT,1}=\frac{V^*_{us}}{V^*_{ud}}  C_{PT,1}$,
$C'_{PT,2}=\frac{V^*_{us}}{V^*_{ud}} C_{PT,2}$,
$C'_{TP}=\frac{V^*_{us}}{V^*_{ud}} C_{TP}$, 
$E'_{\bar B T}=\frac{V^*_{us}}{V^*_{ud}}  E_{\bar B T}$
and 
$E'_{\bar B P}=\frac{V^*_{us}}{V^*_{ud}}  E_{\bar B P}$.

Although we have seven topological amplitudes in $\overline B\to T P$ decays,
only four topological amplitudes, namely $C^{(\prime)}_T$, $C^{(\prime)}_{PT,1}$, $C^{(\prime)}_{TP}$ and $E^{(\prime)}_{\bar B P}$, 
contribute to modes involving flavor exotic states and other states of the same multiplets.
As one can see from Fig.~\ref{fig: TA B2TP} (a) to (d), only these four topological amplitudes do not have flavor-connected $q\bar q$ pairs in tetraquarks. 
All other diagrams necessarily involve flavor-connected $q\bar q$ pairs in tetraquarks, and they are ``traceful", which, as a result, cannot produce ``traceless" tetraquarks. 

\subsubsection{Topological amplitudes in $\overline B\to T S$ decays}

\begin{figure}[t]
\centering
\hspace{10pt}
\subfigure[]{
  \includegraphics[width=0.3\textwidth]{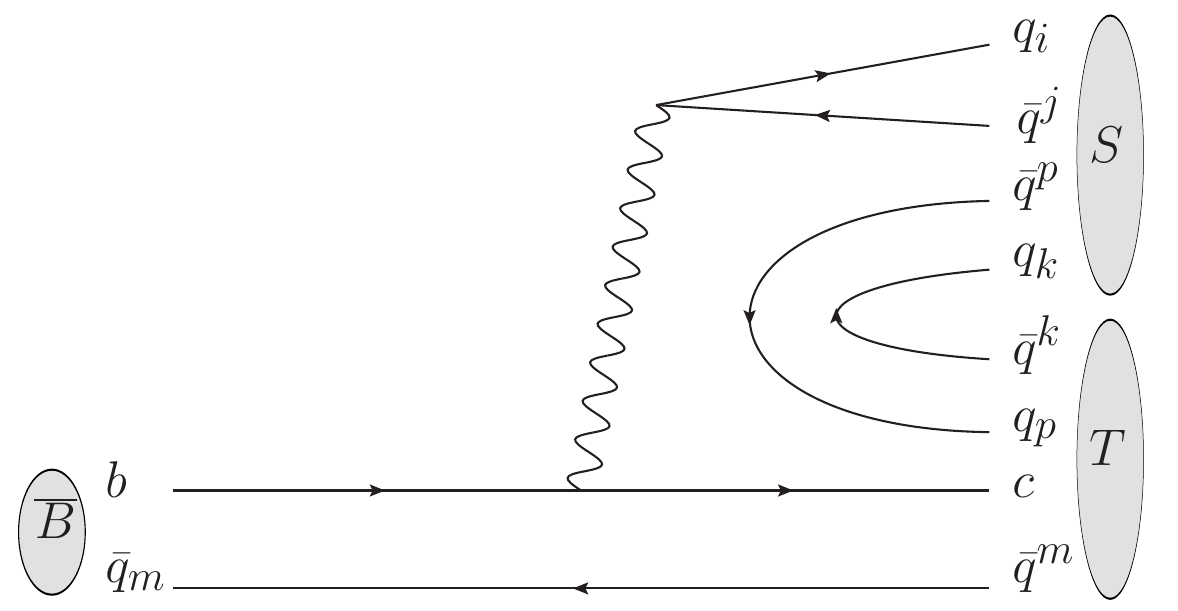}%
}
\hspace{10pt}
\subfigure[]{
  \includegraphics[width=0.3\textwidth]{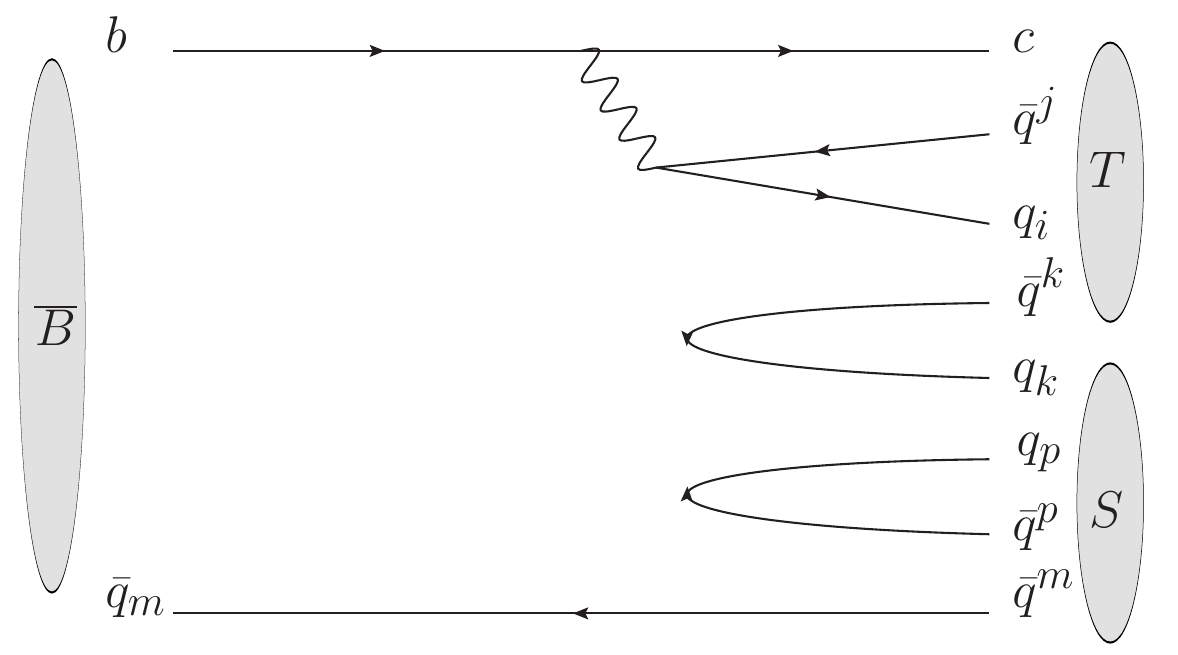}%
}
\subfigure[]{
  \includegraphics[width=0.3\textwidth]{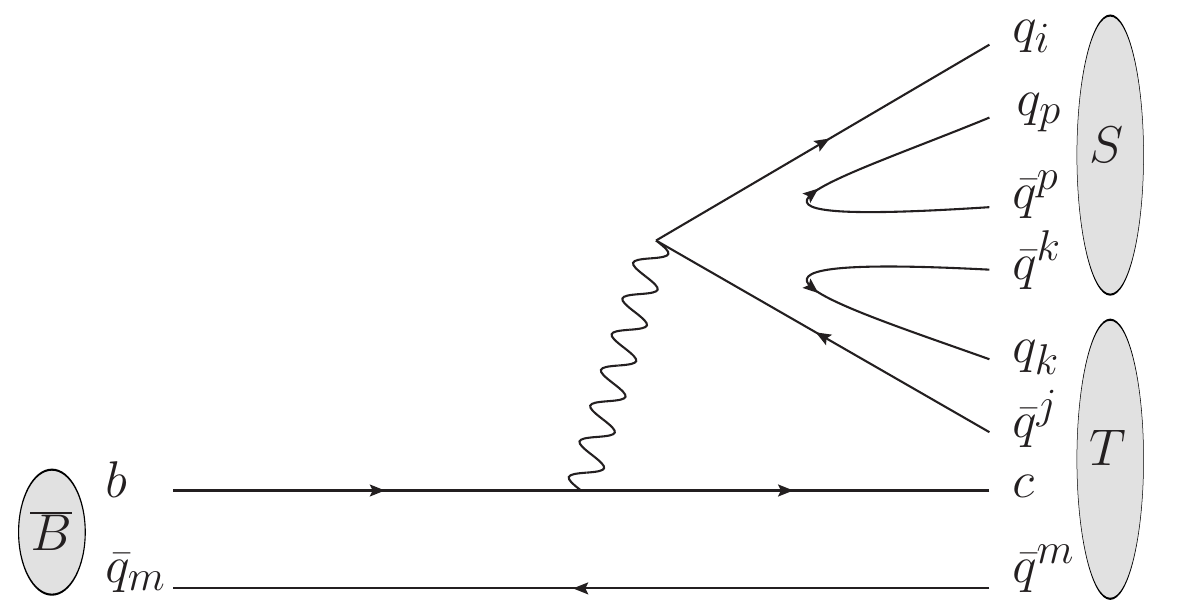}%
}
\hspace{10pt}
\subfigure[]{
  \includegraphics[width=0.3\textwidth]{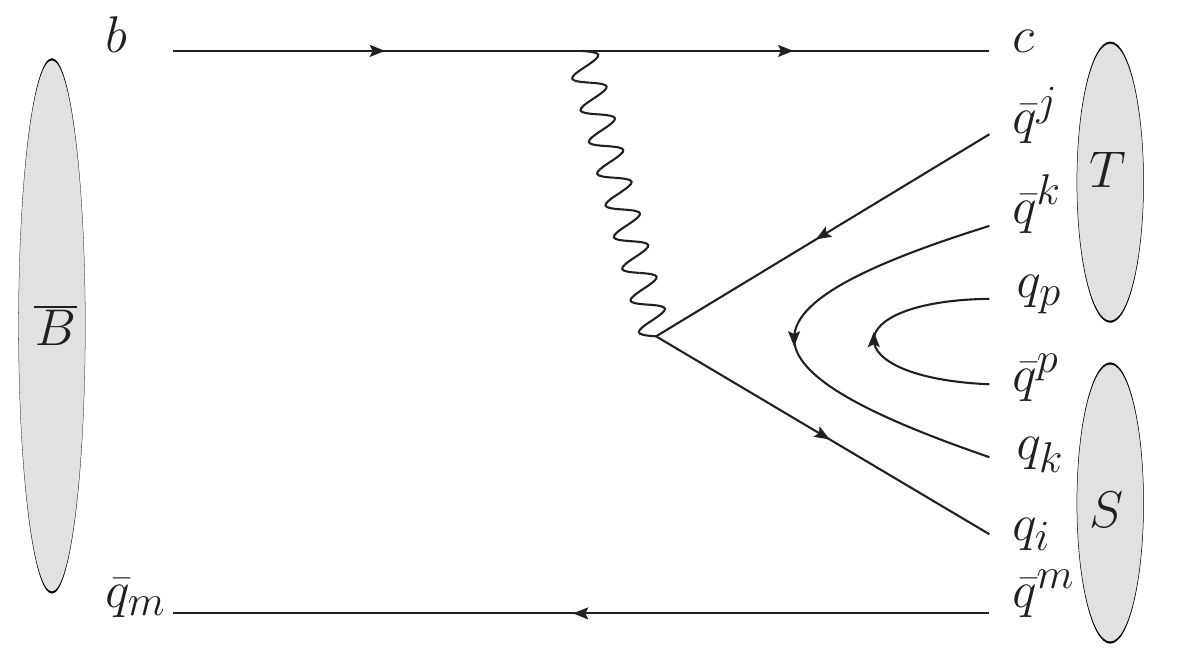}%
}
\hspace{10pt}
\subfigure[]{
  \includegraphics[width=0.3\textwidth]{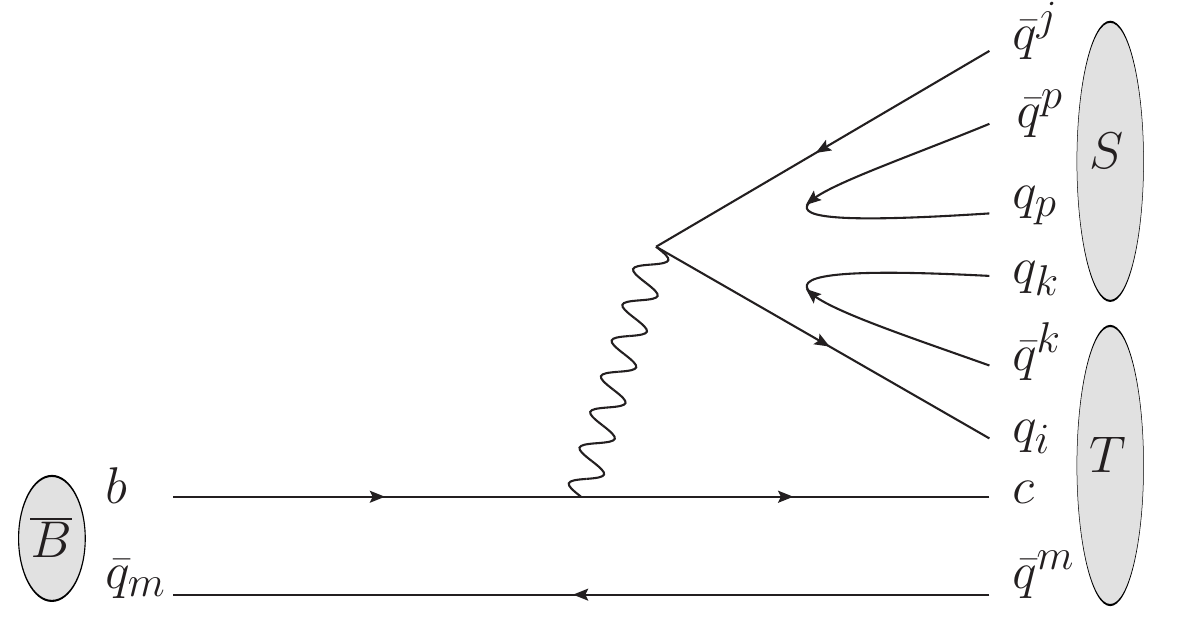}%
}
\hspace{10pt}
\subfigure[]{
  \includegraphics[width=0.3\textwidth]{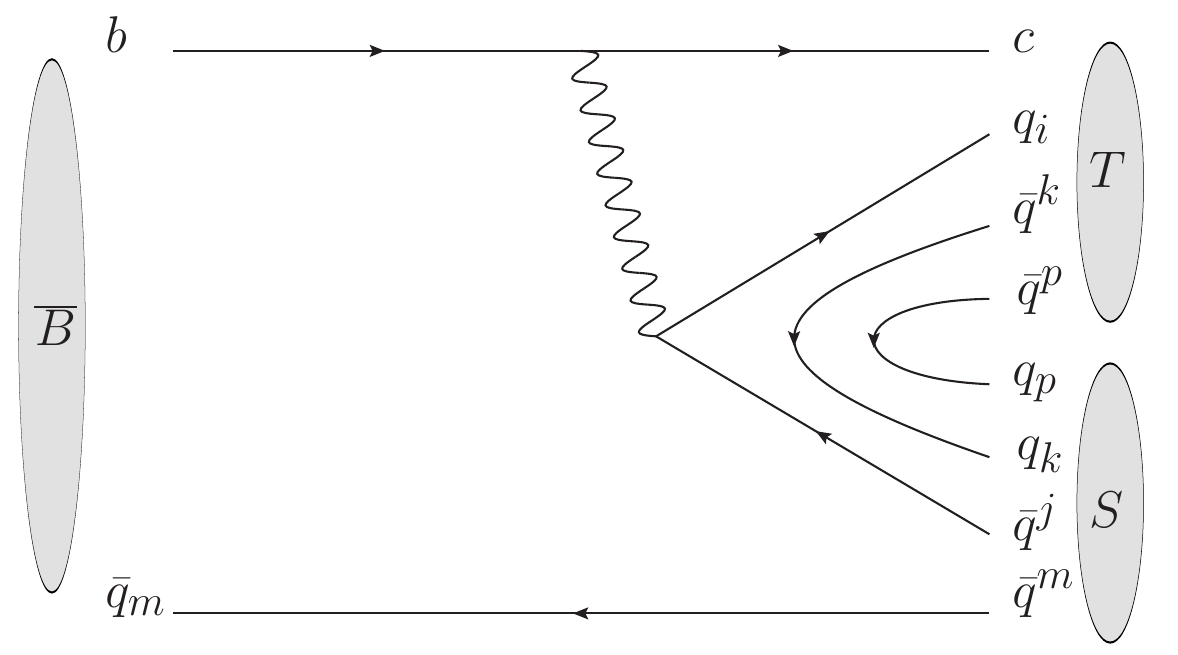}%
}
\subfigure[]{
  \includegraphics[width=0.3\textwidth]{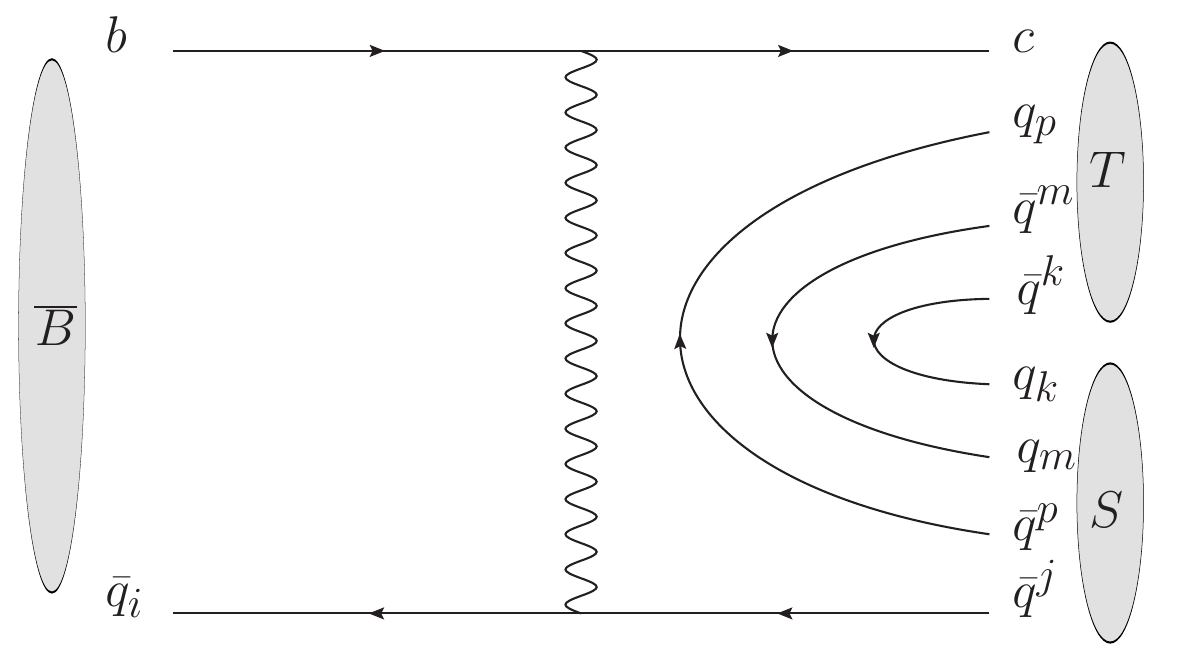}%
}
\hspace{10pt}
\subfigure[]{
  \includegraphics[width=0.3\textwidth]{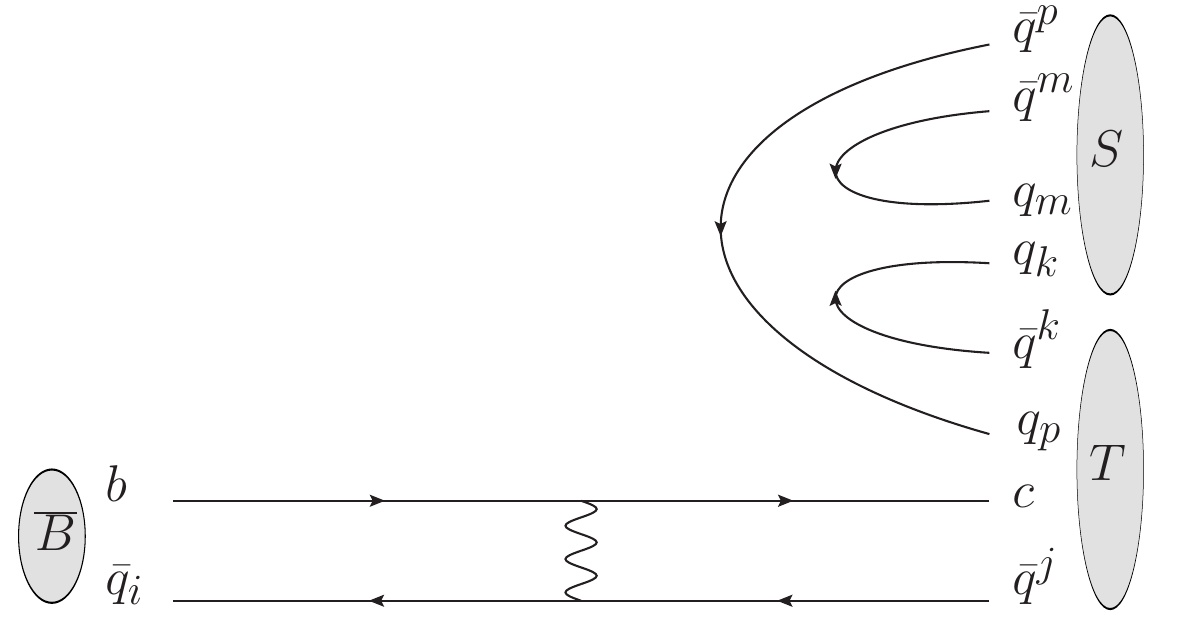}%
}
 \subfigure[]{
  \includegraphics[width=0.3\textwidth]{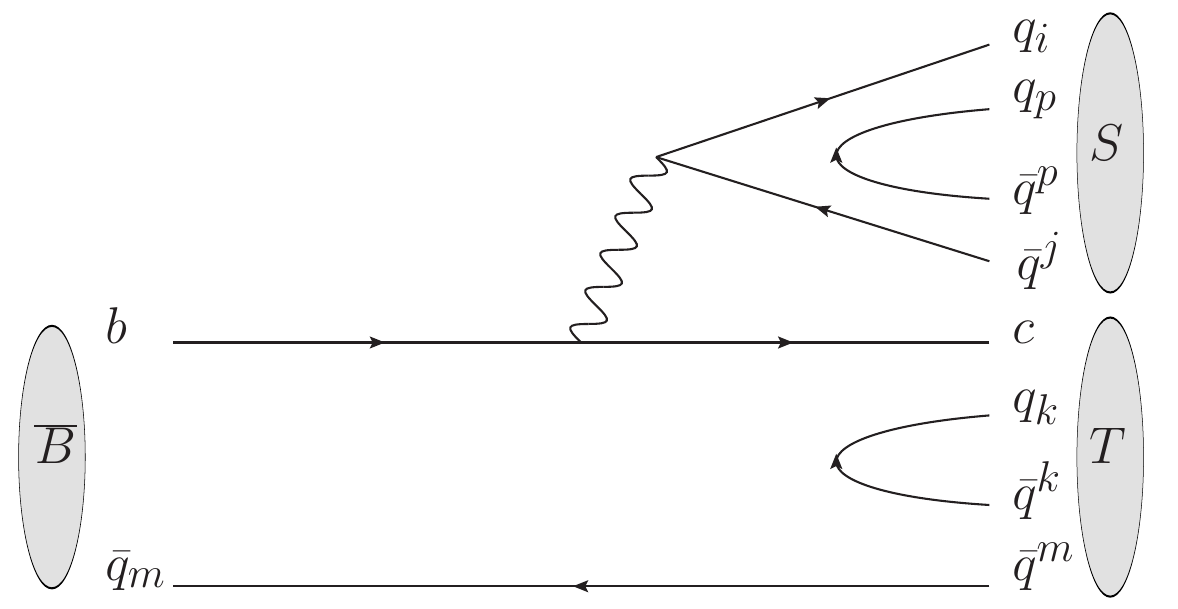}%
}
\subfigure[]{
  \includegraphics[width=0.3\textwidth]{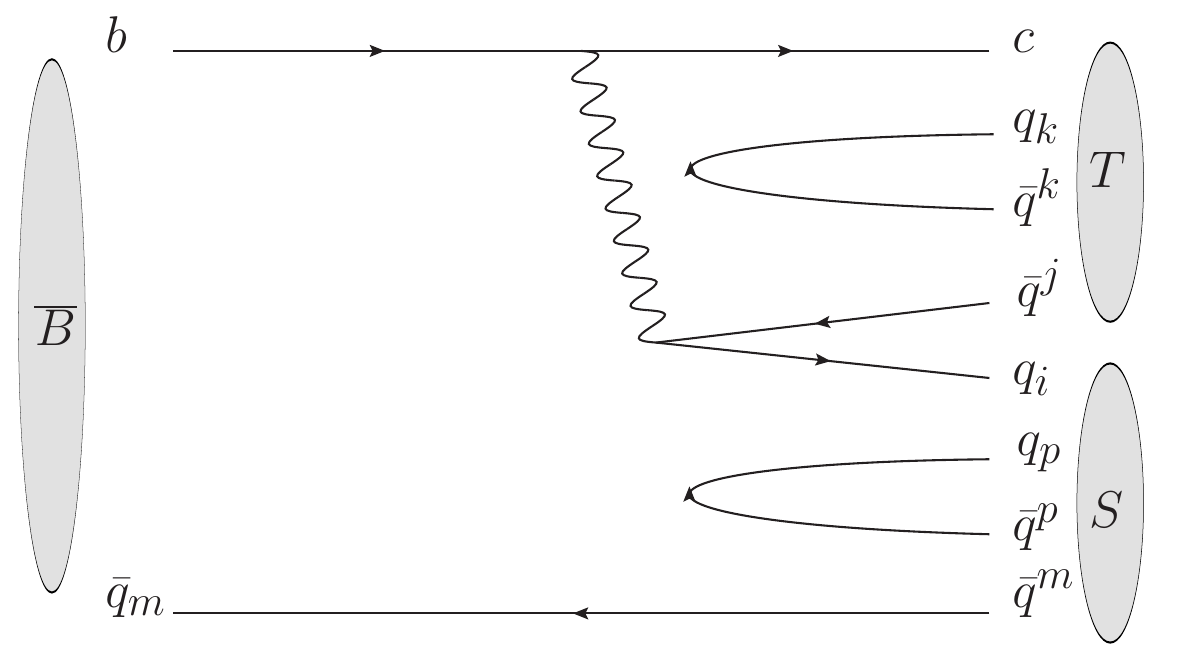}%
}
\hspace{10pt}
\subfigure[]{
  \includegraphics[width=0.3\textwidth]{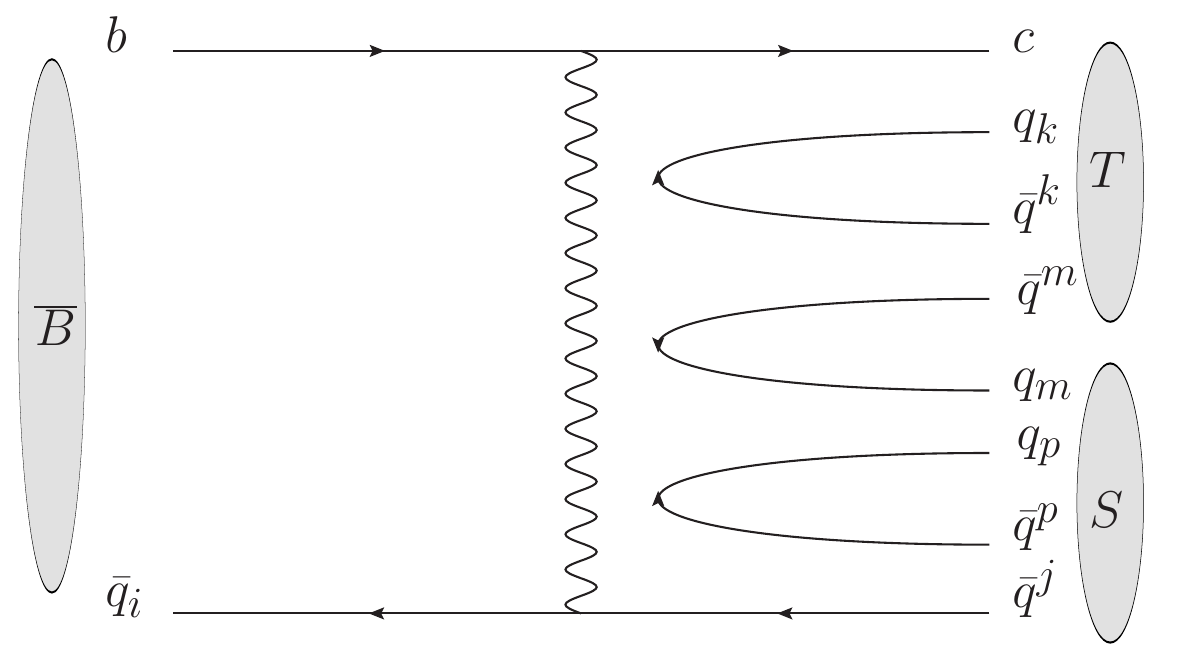}%
}
\caption{Topological diagrams of  
  (a) $C_S$, 
  (b) $C_T$, 
  (c) $C_{ST,1}$, 
  (d) $C_{ST,2}$, 
  (e) $C_{TS,1}$, 
  (f) $C_{TS,2}$
  (g) $E_{\bar B S, 1}$, 
  (h) $E_{\bar B T}$,
  (i) $T_S$, 
  (j) $C_{ST,3}$, 
and (k) $E_{\bar B S,2}$ 
in $\overline B\to T S$ decays, 
with $T$, $C$ and $E$ as external $W$,
internal $W$ and exchange amplitudes, respectively.
Only diagrams (a) to (h) contribute to modes involving flavor exotic states. 
Diagrams (f) and (g) are absent in scenario II. 
} \label{fig: TA B2TS}
\end{figure}

For $\overline B \to T_{c q [q' q'']} S$ decays in $\Delta S=0$ transitions in scenario I, we have
\be
H_{\rm eff}
&=&
         T_S\, \overline B_m  \overline T_k^{[mk]}\, H^i_j\, S^{jp}_{pi}
         +C_S\, \overline B_m  \overline T_p^{[mk]}\, H^i_j\, S^{jp}_{ki}
\non\\         
         &&+C_{T}\, \overline B_m S^{mp}_{pk} \, H^i_j\, \overline T_i^{[jk]}
         +C_{ST,1}\, \overline B_m  \overline T_k^{[mj]}\, H^i_j\, S^{kp}_{pi}
\non\\
         && +C_{ST,2}\,\overline B_m S^{mp}_{ki}\, H^i_j\, \overline T_p^{[jk]}
               +C_{ST,3}\,\overline B_m S^{mp}_{pi}\, H^i_j\, \overline T_k^{[jk]}   
\non\\
         && +C_{TS,1}\,\overline B_m  \overline T_i^{[mk]} \, H^i_j\, S^{jp}_{pk}
         +C_{TS,2}\,\overline B_m S^{mj}_{kp}\, H^i_j\, \overline T_i^{[pk]}
\non\\
         &&+E_{\bar B S,2}\,\overline B_i\,H^i_j\, S^{jp}_{pm}  \overline T_k^{[mk]}
         +E_{\bar B S,1}\,\overline B_i\,H^i_j\, S^{jp}_{km}  \overline T_p^{[mk]}
\non\\
         &&+E_{\bar B T}\,\overline B_i\,H^i_j\,  \overline T_p^{[jk]} S^{mp}_{km}             
\label{eq: B to TS 1}
\en
while for $\overline B \to T_{c q \{q' q''\}} S$ decays in scenario II, we have
\be
H_{\rm eff}
&=&
         T_S\, \overline B_m  \overline T_k^{\{mk\}}\, H^i_j\, S^{jp}_{pi}
         +C_S\, \overline B_m  \overline T_p^{\{mk\}}\, H^i_j\, S^{jp}_{ki}
\non\\         
         &&+C_{T}\, \overline B_m S^{mp}_{pk} \, H^i_j\, \overline T_i^{\{jk\}}
         +C_{ST,1}\, \overline B_m  \overline T_k^{\{mj\}}\, H^i_j\, S^{kp}_{pi}
\non\\
         && +C_{ST,2}\,\overline B_m S^{mp}_{ki}\, H^i_j\, \overline T_p^{\{jk\}} 
              +C_{ST,3}\,\overline B_m S^{mp}_{pi}\, H^i_j\, \overline T_k^{\{jk\}}  
\non\\
         && +C_{TS,1}\,\overline B_m  \overline T_i^{\{mk\}} \, H^i_j\, S^{jp}_{pk}
              +E_{\bar B S,2}\,\overline B_i\,H^i_j\, S^{jp}_{pm}  \overline T_k^{\{mk\}}
\non\\
         && +E_{\bar B T}\,\overline B_i\,H^i_j\,  \overline T_p^{\{jk\}} S^{mp}_{km}.     
\label{eq: B to TS 2}
\en
For notational simplicity, we do not distinguish the topological amplitudes in scenarios I and II.
Note that in the above effective Hamiltonians, we do not consider terms with $S^{mp}_{pm}$. 
In principle, they can take place, but these diagrams are more detached and are expected to be small. 
For practical purposes, we will neglect them in this work, and this approximation only affects modes with $\sigma$ and $f_0$.
One may view this as we use nonet symmetry. 

In scenario I we have eleven topological amplitudes, namely
$C_S$, 
$C_T$, 
$C_{ST,1}$, 
$C_{ST,2}$, 
$C_{TS,1}$, 
$C_{TS,2}$
$E_{\bar B S, 1}$, 
$E_{\bar B T}$,
$T_S$, 
$C_{ST,3}$, 
and $E_{\bar B S,2}$, 
in $\overline B\to T S$ decays, 
with $T$, $C$ and $E$ as external $W$,
internal $W$ and exchange amplitudes, respectively.
They are depicted in Fig.~\ref{fig: TA B2TS}.
This is the case where we have the largest set of topological amplitudes.
As in the $\overline B \to TP$ case, the subscript denotes the meson(s) where the $q^i \bar q_j$, from the $b\to c W^-\to c q^i \bar q_j$ transition, contribute to.
For example, $C_{ST,1}$ meaning $S$ receives the $q^i$, while $T$ receives the $\bar q_j$.
Note that there are only nine topological amplitudes in scenario II.
This can be understood as follows. 
The light quarks in $S$ are antisymmetric, while in scenario II the anti-quarks in $T$ are symmetric. 
Hence, when these pairs are fully connected, as in Fig.~\ref{fig: TA B2TS} (f) and (g), the corresponding topological amplitudes, namely $C_{TS,2}$
and $E_{\bar B S, 1}$, vanish.
Furthermore, only diagrams in Fig.~\ref{fig: TA B2TS} (a) to (h) contribute to modes with flavor exotic states in scenario I and with those in the same multiplets, while only those in Fig.~\ref{fig: TA B2TS} (a) to (e) and (h) contribute to those modes in scenario II.
However, as we shall see, the numbers of independent combinations of these topological amplitudes in these modes are smaller.  

Note that in general, topological diagrams in $\overline B \to TS$ decays, see Fig.~\ref{fig: TA B2TS}, have more $q\bar q$ line(s) produced than those in $\overline B\to TP$, see Fig~\ref{fig: TA B2TP}.
According to the OZI rule, we will expect topological amplitudes in $\overline B \to TS$ decays to receive some suppressions. 
However, since most of the diagrams in $\overline B \to TS$ decays are rather different from those in $\overline B\to TP$ decays,
it is hard to provide a reliable comparison on rates on this basis.

\subsection{$\overline B \to T P$ and $TS$ decay amplitudes}

\subsubsection{$\overline B \to T P$ decay amplitudes}

The $\overline B \to T P$ decay amplitudes, decomposed in these topological amplitudes, are shown in Table \ref{tab: BtoTP I} for $\Delta S=0$ and $-1$ transitions in scenario I, and in Tables \ref{tab: BtoTP II 0} and \ref{tab: BtoTP II -1} for $\Delta S=0$ and $\Delta S=-1$ transitions in scenario II, respectively. 
These results can be applied to $\overline B \to T V$ decays by replacing $P$ with $V$ suitably.

From Table \ref{tab: BtoTP I}, we see that in scenario I, there are eight modes in $\Delta S=0$ transitions and seven modes in $\Delta S=-1$ transitions that involve flavor exotic states. Except $T^{++}_{c\bar s}$, all three other flavor exotic states are included, as a $\overline B \to T^{++} P$ decay is prohibited by charge conservation. 

The amplitudes are more complicated than in the previous cases, and one expect complex relations between rates.
Nevertheless, concerning modes with $T^0_{c\bar s}$, $T^{+}_{c\bar s}$ and $T^0_{c s}$, we have several relations, 
\be
\Gamma(B^-\to T^0_{c\bar s} K^-)
&=&\left|\frac{V_{ud}}{V_{us}}\right|^2 \Gamma(B^-\to T^0_{c s} \pi^-)
=2\left|\frac{V_{ud}}{V_{us}}\right|^2 \Gamma(\overline B{}^0\to T^0_{c s} \pi^0),
\label{eq: TP I1}\\
\Gamma(\overline B{}^0\to T^0_{c \bar s} \overline K{}^0)
&=&\left|\frac{V_{ud}}{V_{us}}\right|^2 \Gamma(\overline B{}_s^0\to T^0_{c s} K^0),
\label{eq: TP I2}
\en
and
\be
\Gamma(\overline B{}^0\to T^0_{c s} K^0)
&=&2\left|\frac{V_{ud}}{V_{us}}\right|^2 \Gamma(\overline B{}_s^0\to T^{ +}_{c \bar s}  K^-)
=\left|\frac{V_{ud}}{V_{us}}\right|^2 \Gamma(\overline B{}_s^0\to T^0_{c \bar s} \overline K^0),
\label{eq: TP I3}\\
\Gamma(\overline B{}_s^0\to T^+_{c \bar s} \pi^-)
&=&\Gamma(\overline B{}_s^0\to T^0_{c \bar s} \pi^0).
\label{eq: TP I4}
\en 
These relations are useful in verifying the multiplet to which these three states belong.
In other words, if they belong to the same multiplet, the ${\bf 6}$ in this case, their $\overline B\to TP$ decays should satisfy the above relations.

The above relations can be understood as follows.
Since $\overline B$ is a $\bar {\bf 3}$, the operator $(\bar c b)(\bar d u)$ or $(\bar c b)(\bar s u)$ is an ${\bf 8}$,
and we have $\bar{\bf 3}\times {\bf 8}=\overline{\bf 15}\oplus{\bf 6}\oplus\bar{\bf 3}$, as shown in Eq. (\ref{eq: SU(3) decompositions 1}).
On the other hand, the final states consist of $T$ in ${\bf 6}$ and $P=\pi, K$ in ${\bf 8}$,
they form ${\bf 6}\otimes{\bf 8}$ and there are one $\overline{\bf 15}$, one ${\bf 6}$ and one ${\bf 3}$ in ${\bf 6}\otimes{\bf 8}$, see Eq. (\ref{eq: SU(3) decompositions 2}), 
that can match the SU(3) quantum numbers of $\overline B$ and the operator. 
As noted previously, only four out of eleven topological amplitudes can contribute to modes with flavor exotic states.
In fact, from the above argument we see that there are just only three independent combinations,
namely $C^{(\prime)}_{T}+C^{(\prime)}_{TP}$, $C^{(\prime)}_{T}+E^{(\prime)}_{\bar B T}$ and $C^{(\prime)}_{TP}-C^{(\prime)}_{PT,1}$,
that contibute to modes involving flavor exotic states and $P=\pi$ or $K$. 
These modes are related. 

\begin{table}[t!]
\caption{\label{tab: BtoTP I}
$\overline B_q\to T_{c q'[ \bar q'' \bar q''']} P$ decay amplitudes in $\Delta S=0$ and $\Delta S=-1$ transitions in scenario~I.}
\scriptsize{
\begin{ruledtabular}
\begin{tabular}{llcllcccr}
\#
&Mode
          & $A (\overline B_q\to T_{c q[ \bar q' \bar q'']} P)$
          & \#
          & Mode
          & $A (\overline B_q\to T_{c q [\bar q' \bar q'']} P)$
          \\
\hline
$2^*$         
          & $\overline B{}^0\to T^+_{c \bar s} K^-$
          & $-\frac{1}{\sqrt2} (C_{TP}-E_{\bar B T})$
&$2^*$
          & $\overline B{}_s^0\to T^+_{c \bar s} \pi^-$
          & $\frac{1}{\sqrt2}(C_{TP}-C_{PT,1})$
          \\
$3^{*}$ 
          & $B^-\to T^0_{c\bar s} K^-$
          & $C_{T}+C_{TP}$
&$3^{*}$
          & $\overline B{}^0\to T^0_{c \bar s} \overline K{}^0$
          & $C_{T}+E_{\bar B T}$
          \\  
$3^{*}$
          & $\overline B{}_s^0\to T^0_{c \bar s} \pi^0$
          & $-\frac{1}{\sqrt2} (C_{TP}-C_{PT,1})$
&$3^{*}$
          & $\overline B{}_s^0\to T^0_{c \bar s} \eta$
          & $-\frac{c_{\theta}}{\sqrt2}(C_{TP}+C_{PT,1})-s_{\theta} C_T$
          \\
$3^{*}$
          & $\overline B{}_s^0\to T^0_{c \bar s} \eta'$
          & $-\frac{s_{\theta}}{\sqrt2}(C_{TP}+C_{PT,1})+c_{\theta} C_T$
&$4^{*}$
          & $\overline B{}^0\to T^0_{c s} K^0$
          & $-C_{PT,1}+E_{\bar B T}$
          \\
$5$         
          & $\overline B{}^0\to T^+_{c}\pi^-$
          & $-\frac{1}{\sqrt2}(C_{PT,1}-E_{\bar B T})$
&$6$         
          & $B^-\to T^0_{c} \pi^-$
          & $\frac{1}{\sqrt2} (C_{T}+C_{TP})$
          \\
$6$         
          & $\overline B{}^0\to T^0_{c} \pi^0$
          & $\frac{1}{2}(-C_T-C_{TP}+C_{PT,1}-E_{\bar B T})$
&$6$         
          & $\overline B{}^0\to T^0_{c} \eta$
          & $\frac{c_{\theta}}{2}(C_T-C_{TP}-C_{PT,1}+E_{\bar B T})+\frac{s_{\theta}}{\sqrt2} E_{\bar B T}$
          \\   
$6$
          & $\overline B{}_s^0\to T^0_{c} K^0$
          & $\frac{1}{\sqrt2} (C_T+C_{PT,1})$
&$6$         
          & $\overline B{}^0\to T^0_{c} \eta'$
          & $\frac{s_{\theta}}{2}(C_T-C_{TP}-C_{PT,1}+E_{\bar B T})-\frac{c_{\theta}}{\sqrt2} E_{\bar B T}$
          \\ 
\hline            
$7$          
          & $\overline B{}^0\to T^{\prime\prime +}_{c} \pi^-$
          &  $-\frac{1}{\sqrt2} (2T_{P}+C_{PT,1}-E_{\bar B T}+2E_{\bar B P})$
&$8$          
          & $B^-\to T^{\prime\prime 0}_{c} \pi^-$
          & $-\frac{1}{\sqrt2} (2T_{P}+C_T+C_{TP}+2C_{PT,2})$
          \\
$8$         
          & $\overline B{}^{0}\to T^{\prime\prime 0}_{c} \eta$
          & $\frac{c_{\theta}}{2}(-C_T+C_{TP}+C_{PT,1}-2C_{PT,2}$
&$8$             
          & $\overline B{}^0\to T^{\prime\prime 0}_{c} \eta'$
          & $\frac{s_{\theta}}{2}(-C_T+C_{TP}+C_{PT,1}-2C_{PT,2}$
          \\
          &
          & $ -E_{\bar B T}-2E_{\bar B P})+\frac{s_{\theta}}{\sqrt2} E_{\bar BT}$
&
          & 
          & $ -E_{\bar B T}-2E_{\bar B P})-\frac{c_{\theta}}{\sqrt2} E_{\bar BT}$
          \\
$8$         
          & $\overline B{}^0\to T^{\prime\prime 0}_{c} \pi^0$
          & $\frac{1}{2}(C_T+C_{TP}-C_{PT,1}+2C_{PT,2}$
&$8$
          & $\overline B{}_s^0\to T^{\prime\prime 0}_{c} K^0$
          & $-\frac{1}{\sqrt2} (C_T-C_{PT,1}+2C_{PT,2})$
          \\                        
          & 
          & $   +E_{\bar B T}-2E_{\bar B P})$
          \\
$9$          
          & $\overline B{}^0\to T^{\prime\prime +}_{c \bar s} K^-$
          & $\frac{1}{\sqrt2} (C_{TP}+E_{\bar B T}-2 E_{\bar B P})$
&$9$
          & $\overline B{}_s^0\to T^{\prime\prime +}_{c \bar s} \pi^-$
          & $-\frac{1}{\sqrt2} (2T_{P}+C_{TP}+C_{PT,1})$
          \\                          
\hline\hline
\#
&Mode
          & $A' (\overline B_q\to T_{c q[ \bar q' \bar q'']} P)$
          & \#
          & Mode
          & $A' (\overline B_q\to T_{c q [\bar q' \bar q'']} P)$
          \\
\hline
$2^*$
          & $\overline B{}_s^0\to T^{ +}_{c \bar s}  K^-$
          & $-\frac{1}{\sqrt2}(C'_{PT,1}-E'_{\bar B T})$
&$3^{*}$
          & $\overline B{}_s^0\to T^0_{c \bar s} \overline K^0$
          & $-C'_{PT,1}+E'_{\bar B T}$
          \\
$4^{*}$ 
          & $B^-\to T^0_{c s} \pi^-$
          & $C'_{T}+C'_{TP}$
&$4^{*}$
          & $\overline B{}^0\to T^0_{c s} \pi^0$
          & $-\frac{1}{\sqrt2} (C'_T+C'_{TP})$
          \\
$4^{*}$
          & $\overline B{}^0\to T^0_{c s} \eta$
          & $\frac{c_{\theta}}{\sqrt2}(C'_T-C'_{TP})+s_{\theta} C'_{PT,1} $
&$4^{*}$
          & $\overline B{}^0\to T^0_{c s} \eta'$
          & $\frac{s_{\theta}}{\sqrt2}(C'_T-C'_{TP})-c_{\theta} C'_{PT,1} $
          \\
$4^{*}$
          & $\overline B{}_s^0\to T^0_{c s} K^0$
          & $C'_T+E'_{\bar B T}$
&$5$         
          & $\overline B{}^0\to T^+_{c } K^-$
          & $\frac{1}{\sqrt2}(C'_{TP}-C'_{PT,1})$
          \\
$5$         
          & $\overline B{}_s^0\to T^+_{c} \pi^-$
          & $-\frac{1}{\sqrt2}(C'_{TP}-E'_{\bar B T})$
&$6$          
          & $B^-\to T^0_{c} K^-$
          & $-\frac{1}{\sqrt2} (C'_T+C'_{TP})$
          \\
$6$        
          & $\overline B{}^0\to T^0_{c} \overline K^0$
          & $-\frac{1}{\sqrt2} (C'_{T}+C'_{PT,1})$
&$6$
          & $\overline B{}_s^0\to T^0_{c} \pi^0$
          & $\frac{1}{2} (C'_{TP}-E'_{\bar B T})$
          \\
$6$
          & $\overline B{}_s^0\to T^0_{c} \eta$
          & $\frac{c_{\theta}}{2} (C'_{TP}+E'_{\bar B T})$
&$6$
          & $\overline B{}_s^0\to T^0_{c} \eta'$
          & $\frac{s_{\theta}}{2} (C'_{TP}+E'_{\bar B T})$
          \\
          & 
          & $ +\frac{s_{\theta}}{\sqrt2} (C'_T-C'_{PT,1}+E'_{\bar B T})$
&
          & 
          & $  -\frac{c_{\theta}}{\sqrt2} (C'_T-C'_{PT,1}+E'_{\bar B T})$
          \\
\hline          
$7$         
          & $\overline B{}^0\to T^{\prime\prime +}_{c} K^-$
          & $-\frac{1}{\sqrt2} (2T'_{P}+C'_{TP}+C'_{PT,1})$
&$7$          
          & $\overline B_s{}^0\to T^{\prime\prime +}_{c} \pi^-$
          & $\frac{1}{\sqrt2} (C'_{TP}+E'_{\bar B T}-2 E'_{\bar BP})$
          \\
$8$          
          & $B^-\to T^{\prime\prime 0}_{c} K^-$
          & $-\frac{1}{\sqrt2} (2 T'_P+C'_T+C'_{TP}+2C'_{PT,2})$
&$8$         
          & $\overline B{}^0\to T^{\prime\prime 0}_{c} \overline K^0$
          & $-\frac{1}{\sqrt2} (C'_{T}-C'_{PT,1}+2C'_{PT,2})$
          \\                   
$8$
          & $\overline B{}_s^0\to T^{\prime\prime 0}_{c} \eta$
          & $\frac{c_{\theta}}{2} (C'_T-E'_{\bar B T}-2E'_{\bar B P})$
&$8$
          & $\overline B{}_s^0\to T^{\prime\prime 0}_{c} \eta'$
          & $\frac{s_{\theta}}{2} (C'_T-E'_{\bar B T}-2E'_{\bar B P})$
          \\     
          & 
          & $+\frac{s_{\theta}}{\sqrt2} (C'_T-C'_{PT,1}+2 C'_{PT,2}+E'_{\bar B T})$
&
          & 
          & $-\frac{c_{\theta}}{\sqrt2} (C'_T-C'_{PT,1}+2 C'_{PT,2}+E'_{\bar B T})$
          \\  
$8$
          & $\overline B{}_s^0\to T^{\prime\prime 0}_{c} \pi^0$
          & $\frac{1}{2} (C'_{TP}+E'_{\bar B T}-2 E'_{\bar B P})$
&$9$
          & $\overline B{}_s^0\to T^{\prime\prime +}_{c \bar s} K^-$
          & $-\frac{1}{\sqrt2}(2T'_P+C'_{PT,1}-E'_{\bar B T}+2 E'_{\bar B P})$
          \\  
\end{tabular}
\end{ruledtabular}
}
\end{table}

\begin{table}[t!]
\caption{\label{tab: BtoTP II 0}
$\overline B_q\to T_{c q'\{\bar q'' \bar q'''\}} P$ decay amplitudes in $\Delta S=0$ transitions in scenario II.}
\scriptsize{
\begin{ruledtabular}
\begin{tabular}{llccccr}
\#
&Mode
          & $A (\overline B_q\to T_{c q\{\bar q' \bar q''\}} P)$
          \\
\hline
$2'^*$          
          & $\overline B{}^0\to T^+_{c \bar s} K^-$
          & $-\frac{1}{\sqrt2}(C_{TP}-E_{\bar B T})$
          \\
$2'^*$
          & $\overline B{}_s^0\to T^+_{c \bar s} \pi^-$
          & $-\frac{1}{\sqrt2}(C_{TP}-C_{PT,1})$
          \\
$3'^{*}$ 
          & $B^-\to T^0_{c \bar s} K^-$
          & $C_{T}+C_{TP}$
          \\
$3'^{*}$
          & $\overline B{}_s^0\to T^0_{c \bar s} \pi^0$
          & $\frac{1}{\sqrt2}(C_{TP}-C_{PT,1})$
          \\
$3'^{*}$
          & $\overline B{}_s^0\to T^0_{c \bar s} \eta$
          & $-s_{\theta} C_T+\frac{c_{\theta}}{\sqrt2}(C_{PT,1}+C_{TP})$
          \\
$3'^{*}$
          & $\overline B{}_s^0\to T^0_{c \bar s} \eta'$
          & $c_{\theta} C_T+\frac{s_{\theta}}{\sqrt2} (C_{PT,1}+C_{TP})$
          \\
$3'^{*}$
          & $\overline B{}^0\to T^0_{c \bar s} \overline K{}^0$
          & $C_{T}+E_{\bar B T}$
          \\  
\hline
$5'^{*}$
          & $\overline B{}^0\to T^0_{c s} K^0$
          & $C_{PT,1}+E_{\bar B T}$
          \\
$6'^{*}$         
          & $B^-\to T^-_{c s} K^0$
          & $\sqrt2 C_{PT,1}$
          \\ 
$6'^{*}$          
          & $\overline B{}^0\to T^-_{c s} K^+$
          & $\sqrt2 E_{\bar B T}$
          \\ 
\hline
$8'^{*}$          
          & $\overline B{}^0_s\to T^+_{c \bar s\bar s} K^-$
          & $\sqrt2 C_{TP}$
          \\ 
\hline
$10'^*$         
          & $\overline B{}^0\to T^+_{c}\pi^-$
          & $\sqrt\frac{2}{3}(-C_{TP}+C_{PT,1}+E_{\bar B T})$
          \\
$11'^*$          
          & $B^-\to T^0_{c} \pi^-$
          & $-\sqrt\frac{2}{3} (C_{T}+C_{TP}-C_{PT,1})$
          \\
$11'^*$         
          & $\overline B{}^0\to T^0_{c} \pi^0$
          & $\frac{1}{\sqrt3}(C_T-C_{PT}+C_{PT,1}+2E_{\bar B T})$
          \\                    
$11'^*$        
          & $\overline B{}^0\to T^0_{c} \eta$
          & $-\frac{c_{\theta}}{\sqrt3}(C_T+C_{TP}+C_{PT,1})$
          \\   
$11'^*$         
          & $\overline B{}^0\to T^0_{c} \eta'$
          & $-\frac{s_{\theta}}{\sqrt3}(C_T+C_{TP}+C_{PT,1})$
          \\   
$11'^*$
          & $\overline B{}_s^0\to T^0_{c} K^0$
          & $-\sqrt\frac{2}{3} C_T$
          \\
$12'^{*}$         
          & $B^-\to T^-_{c} \pi^0$
          & $C_T+C_{TP}-C_{PT,1}$
          \\ 
$12'^{*}$         
          & $B^-\to T^-_{c} \eta$
          & $c_{\theta}(C_T+C_{TP}+C_{PT,1})$
          \\ 
$12'^{*}$         
          & $B^-\to T^-_{c} \eta'$
          & $s_{\theta}(C_T+C_{TP}+C_{PT,1})$
          \\ 
$12'^{*}$          
          & $\overline B{}^0\to T^-_{c} \pi^+$
          & $\sqrt2 (C_T+E_{\bar B T})$
          \\ 
$12'^{*}$
          & $\overline B{}_s^0\to T^-_{c} K^+$
          & $\sqrt2 C_T$
          \\          
\hline
$13'$         
          & $\overline B{}^0\to  T^{\prime +}_{c} \pi^-$
          & $\frac{1}{2\sqrt3}(2C_{TP}+C_{PT,1}+E_{\bar B T})$
          \\ 
$14'$         
          & $B^-\to T^{\prime -}_{c} \pi^-$
          & $\frac{1}{2\sqrt3} (C_T+C_{TP}+2C_{PT,1})$
          \\    
$14'$         
          & $\overline B{}^0\to T^{\prime 0}_{c} \pi^0$
          & $-\frac{1}{2\sqrt6}(C_T-C_{TP}+C_{PT,1}-E_{\bar B T})$
          \\ 
$14'$          
          & $\overline B{}^0\to T^{\prime 0}_{c} \eta$
          & $\frac{c_{\theta}}{2\sqrt6}(C_T+C_{TP}+C_{PT,1}+3E_{\bar B T})+\frac{\sqrt3 s_{\theta}}{2} E_{\bar B T}$
          \\ 
$14'$          
          & $\overline B{}^0\to T^{\prime 0}_{c} \eta'$
          & $\frac{s_{\theta}}{2\sqrt6}(C_T+C_{TP}+C_{PT,1}+3E_{\bar B T})-\frac{\sqrt3 c_{\theta}}{2} E_{\bar B T}$
          \\ 
$14'$
          & $\overline B{}_s^0\to T^{\prime 0}_{c} K^0$
          & $\frac{1}{2\sqrt3} (C_T-3 C_{PT,1})$
          \\
\hline
$15'$         
          & $\overline B{}^0\to  T^{\prime +}_{c \bar s} K^-$
          & $\frac{1}{2}(C_{TP}+E_{\bar B T})$
          \\ 
$15'$
          & $\overline B{}_s^0\to T^{\prime +}_{c \bar s} \pi^-$
          & $\frac{1}{2} (C_{TP}+C_{PT,1})$
          \\ 
\hline
\hline                                          
$16'$          
          & $\overline B{}^0\to T^{\prime\prime +}_{c} \pi^-$
          & $\frac{1}{2} (4T_{P}+2C_{TP}+C_{PT,1}+E_{\bar B T}+4 E_{\bar B P})$
          \\
$17'$          
          & $B^-\to T^{\prime\prime 0}_{c} \pi^-$
          & $\frac{1}{2}(4T_{P}+C_T+C_{TP}+2C_{PT,1}+4C_{PT,2})$
          \\
$17'$
          & $\overline B{}^0\to T^{\prime\prime 0}_{c} \pi^0$
          & $\frac{1}{2\sqrt2}(-C_T+C_{TP}-C_{PT,1}-4C_{PT,2}+E_{\bar B T}+4 E_{\bar B P})$
          \\
$17'$        
          & $\overline B{}^0\to T^{\prime\prime 0}_{c} \eta$
          & $\frac{c_{\theta}}{2\sqrt2}(C_T+C_{TP}+C_{PT,1}+4C_{PT,2}+3E_{\bar B T}+4 E_{\bar B P})-\frac{s_{\theta}}{2} E_{\bar B T}$
          \\
$17'$         
          & $\overline B{}^0\to T^{\prime\prime 0}_{c} \eta'$
          & $\frac{s_{\theta}}{2\sqrt2}(C_T+C_{TP}+C_{PT,1}+4C_{PT,2}+3E_{\bar B T}+4 E_{\bar B P})+\frac{c_{\theta}}{2} E_{\bar B T}$
          \\
$17'$
          & $\overline B{}_s^0\to T^{\prime\prime 0}_{c} K^0$
          & $\frac{1}{2} (C_T+C_{PT,1}+4C_{PT,2})$
          \\                        
\hline
$18'$          
          & $\overline B{}^0\to T^{\prime\prime +}_{c \bar s} K^-$
          & $\frac{1}{2}(C_{TP}+E_{\bar B T}+4 E_{\bar B P})$
          \\
$18'$
          & $\overline B{}_s^0\to T^{\prime\prime +}_{c \bar s} \pi^-$
          & $\frac{1}{2} (4T_{P}+C_{TP}+C_{PT,1})$
          \\                                     
\end{tabular}
\end{ruledtabular}
}
\end{table}

\begin{table}[t!]
\caption{\label{tab: BtoTP II -1}
$\overline B_q\to T_{c q'\{\bar q'' \bar q'''\}} P$ decay amplitudes in $\Delta S=-1$ transitions in scenario II.}
\scriptsize{
\begin{ruledtabular}
\begin{tabular}{llccccr}
\#
&Mode
          & $A' (\overline B_q\to T_{c q\{\bar q' \bar q''\}} P)$
          \\
\hline
$2'^*$
          & $\overline B{}_s^0\to T^+_{c \bar s}  K^-$
          & $\frac{1}{\sqrt2}(C'_{PT,1}+E'_{\bar B T})$
          \\
$3'^{*}$
          & $\overline B{}_s^0\to T^0_{c \bar s} \overline K^0$
          & $C'_{PT,1}+E'_{\bar B T}$
          \\
\hline
$4'^{*}$        
          & $\overline B{}^0\to T^+_{c s} \pi^-$
          &$\sqrt2 C'_{TP}$
          \\ 
$5'^{*}$ 
          & $B^-\to T^0_{c s} \pi^-$
          & $C'_{T}+C'_{TP}$
          \\          
$5'^{*}$
          & $\overline B{}^0\to T^0_{c s } \pi^0$
          & $\frac{1}{\sqrt2}(-C'_T+C'_{TP})$
          \\
$5'^{*}$
          & $\overline B{}^0\to T^0_{c s } \eta$
          & $\frac{c_{\theta}}{\sqrt2} (C'_T+C'_{TP})-s_{\theta} C'_{PT,1} $
          \\          
$5'^{*}$
          & $\overline B{}^0\to T^0_{c s} \eta'$
          & $\frac{s_{\theta}}{\sqrt2}(C'_T+C'_{TP})+c_{\theta} C'_{PT,1} $
          \\
$5'^{*}$
          & $\overline B{}_s^0\to T^0_{c s} K^0$
          & $C'_T+E'_{\bar B T}$
          \\
$6'^{*}$         
          & $B^-\to T^-_{c s} \pi^0$
          & $C'_T+C'_{TP}$
          \\ 
$6'^{*}$         
          & $B^-\to T^-_{c s} \eta$
          & $c_{\theta}(C'_T+C'_{TP})-\sqrt2 s_{\theta}C'_{PT,1}$
          \\ 
$6'^{*}$         
          & $B^-\to T^-_{c s} \eta'$
          &$s_{\theta} (C'_T+C'_{TP})+\sqrt2 c_{\theta}C'_{PT,1}$
          \\ 
$6'^{*}$         
          & $\overline B{}^0\to T^-_{c s} \pi^+$
          &$\sqrt2 C'_T$
          \\ 
$6'^{*}$         
          & $\overline B{}^0\to T^-_{c s} K^+$
          &$\sqrt2 (C'_T+E'_{\bar B T})$
          \\ 
\hline
$10'^*$         
          & $\overline B{}^0\to T^+_{c} K^-$
          & $\sqrt\frac{2}{3} C'_{PT,1}$
          \\
$10'^*$         
          & $\overline B{}_s^0\to T^+_{c} \pi^-$
          & $\sqrt\frac{2}{3}E'_{\bar B T}$
          \\
$11'^*$          
          & $B^-\to T^0_{c} K^-$
          & $\sqrt\frac{2}{3} C'_{PT,1}$
          \\
$11'^*$        
          & $\overline B{}^0\to T^0_{c} \overline K^0$
          & $-\sqrt\frac{2}{3} C'_{PT,1}$
          \\                    
$11'^*$
          & $\overline B{}_s^0\to T^0_{c} \pi^0$
          & $\frac{2}{\sqrt3} E'_{\bar B T}$
          \\
$12'^{*}$         
          & $B^-\to T^-_{c} \overline K^0$
          & $\sqrt2 C'_{PT,1}$
          \\ 
$12'^{*}$
          & $\overline B{}_s^0\to T^-_{c} \pi^+$
          & $\sqrt2 E'_{\bar B T}$
          \\
\hline
$13'$         
          & $\overline B{}^0\to  T^+_{c} K^-$
          & $-\frac{1}{2\sqrt3}(3C'_{TP}-C'_{PT,1})$
          \\ 
$13'$         
          & $\overline B{}_s^0\to  T^{\prime +}_{c} \pi^-$
          & $-\frac{1}{2\sqrt3} (3C'_{TP}-E'_{\bar B T})$
          \\ 
$14'$         
          & $B^-\to T^{\prime 0}_{c} K^-$
          & $-\frac{1}{2\sqrt3} (3C'_{T}+3C'_{TP}-2C'_{PT,1})$
          \\    
$14'$          
          & $\overline B{}^0\to T^0_{c} \overline K^0$
          & $-\frac{1}{2\sqrt3}(3C'_T-C'_{PT,1})$
          \\ 
$14'$
          & $\overline B{}_s^{\prime 0}\to T^{\prime 0}_{c} \pi^0$
          & $-\frac{1}{2\sqrt 6}(3 C'_{TP}-E'_{\bar B T})$
          \\
$14'$
          & $\overline B{}_s^0\to T^{\prime 0}_{c} \eta$
          & $\frac{\sqrt3 c_{\theta}}{2\sqrt2}(-C'_{TP}+E'_{\bar B T})+\frac{\sqrt 3 s_{\theta}}{2} (C'_T+C'_{PT,1}+E'_{\bar B T})$
          \\
$14'$
          & $\overline B{}_s^0\to T^{\prime 0}_{c} \eta'$
          & $\frac{\sqrt3 s_{\theta}}{2\sqrt2}(-C'_{TP}+E'_{\bar B T})-\frac{\sqrt 3 c_{\theta}}{2} (C'_T+C'_{PT,1}+E'_{\bar B T})$
          \\
\hline
$15'$
          & $\overline B{}_s^0\to T^{\prime +}_{c \bar s} K^-$
          & $-\frac{1}{2} (2 C'_{TP}-C'_{PT,1}-E'_{\bar B T})$
          \\ 
\hline 
\hline                                                                             
$16'$        
          & $\overline B{}^0\to T^{\prime\prime+}_{c} K^-$
          & $\frac{1}{2}(4T'_{P}+C'_{TP}+C'_{PT,1})$
          \\
$16'$         
          & $\overline B_s{}^0\to T^{\prime\prime +}_{c} \pi^-$
          & $\frac{1}{2}(C'_{TP}+E'_{\bar B T}+4 E'_{\bar B P})$
          \\
$17'$          
          & $B^-\to T^{\prime\prime 0}_{c} K^-$
          & $\frac{1}{2}(4T'_{P}+C'_T+C'_{TP}+2C'_{PT,1}+4C'_{PT,2})$
          \\
$17'$         
          & $\overline B{}^0\to T^{\prime\prime 0}_{c} \overline K^0$
          & $\frac{1}{2}(C'_{T}+C'_{PT,1}+4C'_{PT,2})$
          \\
$17'$
          & $\overline B{}_s^0\to T^{\prime\prime 0}_{c} \pi^0$
          & $\frac{1}{2\sqrt2}(C'_{TP}+E'_{\bar B T}+4E'_{\bar B P})$
          \\                        
$17'$
          & $\overline B{}_s^0\to T^{\prime\prime 0}_{c} \eta$
          & $\frac{c_{\theta}}{2\sqrt2}  (C'_{TP}+3E'_{\bar BP,1}+4 E'_{\bar BP,2})-\frac{s_{\theta}}{2} (C'_T+C'_{PT,1}+4C'_{PT,2}+E'_{\bar B T})$
          \\      
$17'$
          & $\overline B{}_s^0\to T^{\prime\prime 0}_{c} \eta'$
          & $\frac{s_{\theta}}{2\sqrt2}  (C'_{TP}+3E'_{\bar BP,1}+4 E'_{\bar BP,2})+\frac{c_{\theta}}{2} (C'_T+C'_{PT,1}+4C'_{PT,2}+E'_{\bar B T})$
          \\     
\hline
$18'$
          & $\overline B{}_s^0\to T^{\prime\prime +}_{c \bar s} K^-$
          & $\frac{1}{2} (4 T'_{P}+2 C'_{TP}+C'_{PT,1}+E'_{\bar B T}+4 E'_{\bar B P})$
          \\ 
\end{tabular}
\end{ruledtabular}
}
\end{table}

We turn to the discussion on $\overline B\to TP$ decays in scenario II.
As shown in Tables \ref{tab: BtoTP II 0} and \ref{tab: BtoTP II -1}, there are twenty-two modes involving flavor exotic states in $\Delta S=0$ transitions, 
while there are twenty such modes in $\Delta S=-1$ transitions.
In $\Delta S=0$ transitions, eight out of twelve flavor exotic states are involved, while in $\Delta S=-1$ transitions, we also have eight flavor exotic states involved.
Explicitly, three doubly charged states are forbidden by charge conservation, while $T^+_{c s}$ is absent from the $\Delta S=0$ transitions 
and $T^+_{c \bar s\bar s}$ from the $\Delta S=-1$ transitions.

Note that the amplitudes of the following ten $\overline B \to TP$ modes in $\Delta S=0$ transitions in scenario II, namely
$B^-\to T^0_{c \bar s} K^-$, 
$\overline B{}_s^0\to T^0_{c \bar s} \pi^0$,
$\overline B{}^0\to T^0_{c \bar s} \overline K{}^0$,
$\overline B{}^0\to T^0_{c s} K^0$,
$B^-\to T^-_{c s} K^0$,
$\overline B{}^0\to T^-_{c s} K^+$,
$\overline B{}^0_s\to T^+_{c \bar s\bar s} K^-$,
$B^-\to T^-_{c} \pi^0$,
 $\overline B{}^0\to T^-_{c} \pi^+$
and
$\overline B{}_s^0\to T^-_{c} K^+$ decays,
were also given in ref. \cite{Qin:2022nof}. 
Five instead of eight flavor exotic states were considered.
Their results agree with those shown in Table \ref{tab: BtoTP II 0}, but with $C_{T}$, $C_{PT,1}$, $C_{TP}$ and $E_{\bar B T}$ replaced with 
$T_1 V_{cb} V^*_{ud}$, $T_2 V_{cb} V^*_{ud}$, $T_3 V_{cb} V^*_{ud}$ and $E V_{cb} V^*_{ud}$, respectively.
Note that ref. \cite{Qin:2022nof} also used a different notation for the open-charmed tetraquark states, see Table~\ref{tab: 15bar+3bar} for a comparison.
For example, in Table 2 of ref. \cite{Qin:2022nof}, one finds
\be
A(\overline B{}_s^0\to T_{c d\bar u\bar s} \pi^0)=\frac{1}{\sqrt2}(T_3-T_2) V_{cb} V^*_{ud},
\en
which corresponds to 
\be 
A(\overline B{}_s^0\to T^0_{c \bar s} \pi^0)=\frac{1}{\sqrt2}(C_{TP}-C_{PT,1}),
\en
in Table \ref{tab: BtoTP II 0} but with $ T^0_{c \bar s}$ replaced with $T_{c d\bar u\bar s}$, 
$C_{TP}$ replaced with $T_3 V_{cb} V^*_{ud}$ 
and $C_{PT,1}$ replaced with $T_2 V_{cb} V^*_{ud}$.
Except for the above-mentioned modes, the rest in Table \ref{tab: BtoTP II 0} and all in Tables \ref{tab: BtoTP I} and \ref{tab: BtoTP II -1} are new.

One can find many relations on rates using Tables \ref{tab: BtoTP II 0} and \ref{tab: BtoTP II -1}.
In particular, concerning modes with $T^0_{c\bar s}$,  $T^+_{c\bar s}$ and $T^0_{c s}$, we have the following relations, 
\be
\Gamma(B^-\to T^0_{c\bar s} K^-)
&=&\left|\frac{V_{ud}}{V_{us}}\right|^2 \Gamma(B^-\to T^0_{c s} \pi^-)
\neq 2\left|\frac{V_{ud}}{V_{us}}\right|^2 \Gamma(\overline B{}^0\to T^0_{c s} \pi^0),
\label{eq: TP II1}\\
\Gamma(\overline B{}^0\to T^0_{c \bar s} \overline K{}^0)
&=&\left|\frac{V_{ud}}{V_{us}}\right|^2 \Gamma(\overline B{}_s^0\to T^0_{c s} K^0),
\label{eq: TP II2}
\en
and
\be
\Gamma(\overline B{}^0\to T^0_{c s} K^0)&=&2\left|\frac{V_{ud}}{V_{us}}\right|^2 \Gamma(\overline B{}_s^0\to T^{ +}_{c \bar s}  K^-)
=\left|\frac{V_{ud}}{V_{us}}\right|^2 \Gamma(\overline B{}_s^0\to T^0_{c \bar s} \overline K^0),
\label{eq: TP II3}\\
\Gamma(\overline B{}_s^0\to T^+_{c \bar s} \pi^-)
&=&\Gamma(\overline B{}_s^0\to T^0_{c \bar s} \pi^0).
\label{eq: TP II4}
\en 
We can compare the above relations to those in scenario I.
We find that relations in Eqs.~(\ref{eq: TP II2}), (\ref{eq: TP II3}) and (\ref{eq: TP II4}) agree with those in Eqs. (\ref{eq: TP I2}), (\ref{eq: TP I3}) and (\ref{eq: TP I4}),
but those in Eq. (\ref{eq: TP II1}) do not always agree with those in Eq. (\ref{eq: TP I1}).
In scenario I, as shown in Table~\ref{tab: BtoTP I}, 
$B^-\to T^0_{c\bar s} K^-$ decay is governed by $C_{T}+C_{TP}$ and $B^-\to T^0_{c s} \pi^-$ and $\overline B{}^0\to T^0_{c s} \pi^0$ decays are governed by $C'_{T}+C'_{TP}$.
However, things are slightly different in scenario II.
Although the first two modes are still governed by $C_{T}+C_{TP}$ or $C'_{T}+C'_{TP}$, the last one is governed by a different combination, $-C'_{T}+C'_{TP}$, see Tables~\ref{tab: BtoTP II 0} and \ref{tab: BtoTP II -1}.
The discrepancy can be traced to the different isospin quantum numbers of  $T^0_{c s}$ in these two scenarios. 
In scenario I, it is an isosinglet; there is only one way to match the isospin quantum numbers of the inintial state, $\overline B_{u,d}$, the operator $(\bar cb)(\bar s u)$ and the final states $T^0_{c s}$ $\pi$,
while in scenario II, it is an isovector, and, hence, there is more than one way to match.
Consequently, the $B^-\to T^0_{c s} \pi^-$ and $\overline B{}^0\to T^0_{c s} \pi^0$ amplitudes can be rather different in the latter case.

Besides the above relations, there are other relations between flavor exotic modes in scenario II.
For example, we can have
\be
\Gamma(B^-\to T^-_{c s} K^0)
&=&\Gamma(\overline B{}^0_s\to T^+_{c \bar s\bar s} K^-)
=3 \Gamma(\overline B{}_s^0\to T^0_{c} K^0)
=\Gamma(\overline B{}_s^0\to T^-_{c} K^+)
\non\\
&=&\left|\frac{V_{ud}}{V_{us}}\right|^2 \Gamma(\overline B{}^0\to T^+_{c s} \pi^-)
=\left|\frac{V_{ud}}{V_{us}}\right|^2 \Gamma(\overline B{}^0\to T^-_{c s} \pi^+)
\non\\
&=&3\left|\frac{V_{ud}}{V_{us}}\right|^2 \Gamma(\overline B{}^0\to T^+_{c} K^-)
=3\left|\frac{V_{ud}}{V_{us}}\right|^2 \Gamma(B^-\to T^0_{c} K^-)
\non\\
&=&3\left|\frac{V_{ud}}{V_{us}}\right|^2 \Gamma(\overline B{}^0\to T^0_{c} \overline K^0)
=3\left|\frac{V_{ud}}{V_{us}}\right|^2 \Gamma(B^-\to T^-_{c} \overline K^0).
\en
Note that the above relations involve nine flavor exotic states. 
It will be useful to check these relations and those inferred from the tables experimentally.

The above relations can be understood as in scenario I,
but now the final states consist of $T$ in $\overline{\bf 15}$, with $P=K,\pi$, they form $\overline{\bf 15}\otimes{\bf 8}$.
Furthermore, there are two $\overline{\bf 15}$, one ${\bf 6}$ and one ${\bf 3}$ in $\overline{\bf 15}\otimes{\bf 8}$, see Eq. (\ref{eq: SU(3) decompositions 2}), 
that can match the SU(3) quantum numbers of $\overline B$ and the operator. 
Although we have seven topological amplitudes, there are only four independent combinations,
namely $C^{(\prime)}_{T}+C^{(\prime)}_{TP}$, $-C^{(\prime)}_{T}+C^{(\prime)}_{TP}$, $C^{(\prime)}_{T}+E^{(\prime)}_{\bar B P,1}$ and $C^{(\prime)}_{TP}-C^{(\prime)}_{PT,1}$,
that contribute to modes with flavor exotic states.
Note that the number also matches the number of topological amplitudes that can contribute to these modes.
These modes are related. 

The final state $T$ can further decay to $DP$ and $DS$, or they can connect to these states virtually. 
One can obtain their full amplitudes using results in the section and those in Sec.~\ref{sec: T to DP, DS}.

\subsubsection{$\overline B \to T S$ decay amplitudes}

\begin{table}[t!]
\caption{\label{tab: BtoTS I 0}
$\overline B_q\to T_{c q'[ \bar q'' \bar q''']} S$ decay amplitudes in $\Delta S=0$ transitions in scenario I.}
\scriptsize{
\begin{ruledtabular}
\begin{tabular}{llccccr}
\#
&Mode
          & $A (\overline B_q\to T_{c q[ \bar q' \bar q'']} S)$
          \\
\hline
$2^*$         
          & $\overline B{}^0\to T^+_{c \bar s} \kappa^-$
          & $\frac{1}{\sqrt2} (C_S+C_{ST,2}-C_{TS,1}-2C_{TS,2}+2E_{\bar B S,1}+E_{\bar B T})$
          \\
$2^*$ 
          & $\overline B{}_s^0\to T^+_{c  \bar s} a_0^-$
          & $-\frac{1}{\sqrt2} (C_{ST,1}+C_{ST,2}-C_{TS,1}-2C_{TS,2})$
          \\
$3^{*}$ 
          & $B^-\to T^0_{c \bar s } \kappa^-$
          & $C_{T}-C_S-C_{ST,2}+C_{TS,1}$
          \\
$3^{*}$ 
          & $\overline B{}^0\to T^0_{c\bar s} \overline \kappa{}^0$
          & $C_{T}-2C_{TS,2}+2E_{\bar B S,1}+E_{\bar B T}$
          \\  
$3^{*}$
          & $\overline B{}_s^0\to T^0_{c \bar s} a_0^0$
          & $\frac{1}{\sqrt2}(C_{ST,1}+C_{ST,2}-C_{TS,1}-2C_{TS,2})$
          \\
$3^{*}$ 
          & $\overline B{}_s^0\to T^0_{c  \bar s} f_0$
          & $-\frac{c_{\phi}}{\sqrt2} (2C_T-C_{ST,1}-C_{ST,2}-C_{TS,1}-2C_{TS,2})$
          \\
          & 
          & $-s_{\phi} (C_S-C_{ST,1}-C_{TS,1})$
          \\
$3^{*}$
          & $\overline B{}_s^0\to T^0_{c  \bar s} \sigma$
          &  $\frac{s_{\phi}}{\sqrt2} (2C_T-C_{ST,1}-C_{ST,2}-C_{TS,1}-2C_{TS,2})$
          \\
          & 
          & $-c_{\phi} (C_S-C_{ST,1}-C_{TS,1})$
          \\
\hline
$4^{*}$
          & $\overline B{}^0\to T^0_{c s} \kappa^0$
          & $C_S-C_{ST,1}+2E_{\bar B S,1}+E_{\bar B T}$
          \\
\hline
5          
          & $\overline B{}^0\to T^+_{c}a_0^-$
          & $\frac{1}{\sqrt2}(C_S-C_{ST,1}+2E_{\bar B S,1}+E_{\bar B T})$
          \\
6         
          & $B^-\to T^0_{c} a_0^-$
          & $-\frac{1}{\sqrt2}(C_S-C_{T}+C_{ST,2}-C_{TS,1})$
          \\
6       
          & $\overline B{}^0\to T^0_{c} a_0^0$
          & $-\frac{1}{2} (C_T-C_{ST,1}-C_{ST,2}+C_{TS,1}+2E_{\bar B S,1}+E_{\bar B T})$
          \\                    
6         
          & $\overline B{}^0\to T^0_{c} f_0$
          & $-\frac{c_{\phi}}{2} (C_T-C_{ST,1}-C_{ST,2}-C_{TS,1}-2E_{\bar B S,1}-E_{\bar B T})$
          \\   
          & 
          & $   -\frac{s_{\phi}}{\sqrt2} (C_S+C_T-C_{ST,1}-C_{TS,1}-2C_{TS,2}+2E_{\bar B S,2}+E_{\bar B T})$
          \\  
6         
          & $\overline B{}^0\to T^0_{c} \sigma$
          & $\frac{s_{\phi}}{2} (C_T-C_{ST,1}-C_{ST,2}-C_{TS,1}-2E_{\bar B S,1}-E_{\bar B T})$
          \\   
          & 
          & $  -\frac{c_{\phi}}{\sqrt2} (C_S+C_T-C_{ST,1}-C_{TS,1}-2C_{TS,2}+2E_{\bar B S,2}+E_{\bar B T})$
          \\  
6 
          & $\overline B{}_s^0\to T^0_{c} \kappa^0$
          & $\frac{1}{\sqrt2} (C_T-C_S+C_{ST,1}-2C_{TS,2})$
          \\
\hline 
\hline         
7       
          & $\overline B{}^0\to T^{\prime\prime +}_{c} a_0^-$
          & $-\frac{1}{\sqrt2} (2T_S+C_S+C_{ST,1}+2E_{\bar B S,1}+2E_{\bar B S,2}-E_{\bar B T})$
          \\
8         
          & $B^-\to T^{\prime\prime 0}_{c} a_0^-$
          & $-\frac{1}{\sqrt2}(2T_S+C_S+C_T+C_{ST,2}+2C_{ST,3}+C_{TS,1})$
          \\
8          
          & $\overline B{}^0\to T^{\prime \prime 0}_{c} a_0^0$
          & $\frac{1}{2}(C_T-C_{ST,1}+C_{ST,2}+2C_{ST,3}+C_{TS,1}-2E_{\bar B S,1}-2E_{\bar B S,2}+E_{\bar B T})$
          \\
8         
          & $\overline B{}^0\to T^{\prime\prime 0}_{c} f_0$
          & $\frac{c_{\phi}}{2}(C_T-C_{ST,1}+C_{ST,2}+2C_{ST,3}-C_{TS,1}+2E_{\bar B S,1}+2E_{\bar B S,2}+3E_{\bar B T})$
          \\
          & 
          & $+\frac{s_\phi}{\sqrt2} (C_S+C_T-C_{ST,1}+2C_{ST,3}-C_{TS,1}-2 C_{TS,2}+2E_{\bar B S,1}+2E_{\bar B S,2}+E_{\bar B T})$
          \\
8         
          & $\overline B{}^0\to T^{\prime\prime 0}_{c} \sigma$
          & $-\frac{s_{\phi}}{2}(C_T-C_{ST,1}+C_{ST,2}+2C_{ST,3}-C_{TS,1}+2E_{\bar B S,1}+2E_{\bar B S,2}+3E_{\bar B T})$
          \\
          & 
          & $+\frac{c_\phi}{\sqrt2} (C_S+C_T-C_{ST,1}+2C_{ST,3}-C_{TS,1}-2 C_{TS,2}+2E_{\bar B S,1}+2E_{\bar B S,2}+E_{\bar B T})$
          \\
8
          & $\overline B{}_s^0\to T^{\prime\prime 0}_{c} \kappa^0$
          & $-\frac{1}{\sqrt2} (C_T+C_S-C_{ST,1}+2C_{ST,3}-2C_{TS,2})$
          \\                        
\hline
9          
          & $\overline B{}^0\to T^{\prime\prime +}_{c \bar s} \kappa^-$
          & $-\frac{1}{\sqrt2} (C_S-C_{ST,2}-C_{TS,1}-2 C_{TS,2}+2E_{\bar B S,1}+2E_{\bar B S,2}-E_{\bar B T})$
          \\
9
          & $\overline B{}_s^0\to T^{\prime\prime +}_{c \bar s} a_0^-$
          & $-\frac{1}{\sqrt2} (2T_S+C_{ST,1}+C_{ST,2}+C_{TS,1}+2C_{TS,2})$
          \\ 
\end{tabular}
\end{ruledtabular}
}
\end{table}

\begin{table}[t!]
\caption{\label{tab: BtoTS I -1}
$\overline B_q\to T_{c q'[ \bar q'' \bar q''']} S$ decay amplitudes in $\Delta S=-1$ transition in scenario I.}
\scriptsize{
\begin{ruledtabular}
\begin{tabular}{llccccr}
\#
&Mode
          & $A' (\overline B_q\to T_{c q[ \bar q' \bar q'']} S)$
          \\
\hline
$2^*$
          & $\overline B{}_s^0\to T^+_{c \bar s} \kappa^-$
          & $\frac{1}{\sqrt2}(C'_S-C'_{ST,1}+2E'_{\bar BS,1}+E'_{\bar B T})$
          \\
$3^{*}$
          & $\overline B{}_s^0\to T^0_{c \bar s} \overline \kappa{}^0$
          & $C'_S-C'_{ST,1}+2 E'_{\bar BS,1}+E'_{\bar B T}$
          \\
\hline
$4^{*}$ 
          & $B^-\to T^0_{c s} a^-$
          & $C'_{T}-C'_S-C'_{ST,2}+C'_{TS,1}$
          \\
$4^{*}$
          & $\overline B{}^0\to T^0_{c s} a_0^0$
          & $-\frac{1}{\sqrt2}(C'_T-C'_S-C'_{ST,2}+C'_{TS,1})$
          \\
$4^{*}$
          & $\overline B{}^0\to T^0_{c s} f_0$
          & $-\frac{c_{\phi}}{\sqrt2}(C'_T+C'_S-2C'_{ST,1}-C'_{ST,2}-C'_{TS,1})$
          \\
          & 
          & $-s_{\phi} (C'_T-C'_{TS,1}-2 C'_{TS,2})$
          \\
$4^{*}$
          & $\overline B{}^0\to T^0_{c s} \sigma$
          & $\frac{s_{\phi}}{\sqrt2}(C'_T+C'_S-2C'_{ST,1}-C'_{ST,2}-C'_{TS,1})$
          \\
          & 
          & $-c_{\phi}( C'_T-C'_{TS,1}-2 C'_{TS,2})$
          \\
$4^{*}$
          & $\overline B{}_s^0\to T^0_{c s} \kappa^0$
          & $C'_T-2 C'_{TS,2}+2 E'_{\bar B S,2}+E'_{\bar B T}$
          \\
\hline
$5$        
          & $\overline B{}^0\to T^+_{c} \kappa^-$
          & $-\frac{1}{\sqrt2}(C'_{ST,1}+C'_{ST,2}-C'_{TS,1}-2C'_{TS,2})$
          \\
$5$         
          & $\overline B{}_s^0\to T^+_{c} a^-$
          & $\frac{1}{\sqrt2}(C'_S+C'_{ST,2}-C'_{TS,1}-2C'_{TS,2}+2E'_{\bar BS,1}+E'_{\bar B T})$
          \\
$6$          
          & $B^-\to T^0_{c} \kappa^-$
          & $\frac{1}{\sqrt2} (C'_S-C'_T+C'_{ST,2}-C'_{TS,1})$
          \\
$6$         
          & $\overline B{}^0\to T^0_{c} \overline \kappa{}^0$
          & $\frac{1}{\sqrt2}(C'_S-C'_T-C'_{ST,1}+2C'_{TS,2})$
          \\                    
$6$
          & $\overline B{}_s^0\to T^0_{c} a_0^0$
          & $-\frac{1}{2}(C'_S+C'_{ST,2}-C'_{TS,1}-2C'_{TS,2}+2E'_{\bar BS,1}+E'_{\bar B T})$
          \\
$6$ 
          & $\overline B{}_s^0\to T^0_{c} f_0$
          & $\frac{c_{\phi}}{2}(C'_S+2C'_T-2C'_{ST,1}-C'_{ST,2}-C'_{TS,1}-2C'_{TS,2}+2E'_{\bar BS,1}+E'_{\bar B T})$
          \\
          & 
          & $-\frac{s_{\phi}}{\sqrt2} (C'_{TS,1}+2E'_{\bar BS,1}+E'_{\bar B T})$
          \\
$6$
          & $\overline B{}_s^0\to T^0_{c} \sigma$
          & $-\frac{s_{\phi}}{2}(C'_S+2C'_T-2C'_{ST,1}-C'_{ST,2}-C'_{TS,1}-2C'_{TS,2}+2E'_{\bar BS,1}+E'_{\bar B T})$
          \\
          & 
          & $-\frac{c_{\phi}}{\sqrt2} (C'_{TS,1}+2E'_{\bar BS,1}+E'_{\bar B T})$
          \\
\hline
$7$          
          & $\overline B{}^0\to T^{\prime\prime +}_{c} \kappa^-$
          & $-\frac{1}{\sqrt2}(2 T'_S+C'_{ST,1}+C'_{ST,2}+C'_{TS,1}+2C'_{TS,2})$
          \\
$7$          
          & $\overline B{}_s^0\to T^{\prime\prime +}_{c} a^-$
          & $-\frac{1}{\sqrt2}(C'_S-C'_{ST,2}-C'_{TS,1}-2C'_{TS,2}+2E'_{\bar BS,1}+2E'_{\bar BS,2}-E'_{\bar B T})$
          \\
$8$         
          & $B^-\to T^{\prime\prime 0}_{c} \kappa^-$
          & $-\frac{1}{\sqrt2}(2T'_S+C'_S+C'_T+C'_{ST,2}+2C'_{ST,3}+C'_{TS,1})$
          \\
$8$        
          & $\overline B{}^0\to T^{\prime\prime 0}_{c} \overline \kappa{}^0$
          & $-\frac{1}{\sqrt2}(C'_S+C'_T-C'_{ST,1}+2C'_{ST,3}-2C'_{TS,2})$
          \\
$8$
          & $\overline B{}_s^0\to T^{\prime\prime 0}_{c} a_0^0$
          & $-\frac{1}{2}(C'_S-C'_{ST,2}-C'_{TS,1}-2 C'_{TS,2}+2E'_{\bar B S,1}+2E'_{\bar BS,2}-E'_{\bar B T})$
          \\                        
$8$
          & $\overline B{}_s^0\to T^{\prime\prime 0}_{c}  f_0$
          & $\frac{c_{\phi}}{2}(2 C'_T+C'_S-2C'_{ST,1}+C'_{ST,2}+4C'_{ST,3}-C'_{TS,1}-2C'_{TS,2}$
          \\
          & 
          & $+2E'_{\bar BS,1}+2E'_{\bar BS,2}+3E'_{\bar B T})+\frac{s_{\phi}}{\sqrt2} (-C'_{TS,1}+2E'_{\bar BS,1}+2E'_{\bar BS,2}+E'_{\bar B T})$
          \\
$8$
          & $\overline B{}_s^0\to T^{\prime\prime 0}_{c} \sigma$
          & $-\frac{s_{\phi}}{2}(2 C'_T+C'_S-2C'_{ST,1}+C'_{ST,2}+4C'_{ST,3}-C'_{TS,1}-2C'_{TS,2}$
          \\
          & 
          & $+2E'_{\bar BS,1}+2E'_{\bar BS,2}+3E'_{\bar B T})+\frac{c_{\phi}}{\sqrt2} (-C'_{TS,1}+2E'_{\bar BS,1}+2E'_{\bar BS,2}+E'_{\bar B T})$
          \\
\hline
$9$
          & $\overline B{}_s^0\to T^{\prime\prime +}_{c \bar s} \kappa^-$
          & $-\frac{1}{\sqrt2}(2T'_S+C'_S+C'_{ST,1}+2E'_{\bar BS,1}+2E'_{\bar BS,2}-E'_{\bar B T})$
          \\ 
\end{tabular}
\end{ruledtabular}
}
\end{table}

The $\overline B \to T S$ decay amplitudes in scenario I, decomposed in topological amplitudes, are shown in Tables \ref{tab: BtoTS I 0} and \ref{tab: BtoTS I -1} for $\Delta S=0$ and $-1$ transitions, respectively, while those in scenario II are in Tables \ref{tab: BtoTS II 0} and \ref{tab: BtoTS II -1}.

As shown in Tables \ref{tab: BtoTS I 0} and \ref{tab: BtoTS I -1}, there are eight flavor exotic modes in $\Delta S=0$ transitions and seven in $\Delta S=-1$ transitions in scenario I.
Except $T^{++}_{c\bar s}$, which is prohibited by charge conservation, all three other flavor exotic states participate in $\overline B\to TS$ decays. 
With these amplitudes, one can find relations on rates. For example, we have
\be
\Gamma(\overline B{}_s^0\to T^+_{c  \bar s} a_0^-)&=&\Gamma(\overline B{}_s^0\to T^0_{c \bar s} a_0^0),
\non\\
\Gamma(B^-\to T^0_{c \bar s } \kappa^-)
&=&
\left|\frac{V_{ud}}{V_{us}}\right|^2 \Gamma(B^-\to T^0_{c s} a^-)
=2 \left|\frac{V_{ud}}{V_{us}}\right|^2(\overline B{}^0\to T^0_{c s} a_0^0),
 \label{eq: BtoTS I a}\\
 \Gamma(\overline B{}^0\to T^0_{c s} \kappa^0)
&=&\left|\frac{V_{ud}}{V_{us}}\right|^2 \Gamma(\overline B{}_s^0\to T^0_{c \bar s} \overline \kappa{}^0)
=2 \left|\frac{V_{ud}}{V_{us}}\right|^2(\overline B{}_s^0\to T^+_{c \bar s} \kappa^-).
 \label{eq: BtoTS I b} 
\en
 In most of these cases, we need both $\Delta S= 0$ and $-1$ transitions.
Note that the last relations are interesting, because all three flavor exotic states are involved.

The above relations can be understood as in the $\overline B\to TP$ case.
From SU(3) symmetry, we see that although we have eleven topological amplitudes, there are only three independent combinations,
namely 
$C^{(\prime)}_{ST,1}+C^{(\prime)}_{ST,2}-C^{(\prime)}_{TS,1}-2C^{(\prime)}_{TS,2}$, 
$C^{(\prime)}_{T}-C^{(\prime)}_S-C^{(\prime)}_{ST,2}+C^{(\prime)}_{TS,1}$
and 
$C^{(\prime)}_S-C^{(\prime)}_{ST,1}+2E^{(\prime)}_{\bar B S,1}+E^{(\prime)}_{\bar B T}$,
that contribute to modes involving flavor exotic states and $a_0$ or $\kappa$.
These modes are related.

\begin{table}[t!]
\caption{\label{tab: BtoTS II 0}
$\overline B_q\to T_{c q'\{ \bar q'' \bar q'''\}} S$ decay amplitudes in $\Delta S=0$ transitions in scenario II.}
\scriptsize{
\begin{ruledtabular}
\begin{tabular}{llccccr}
\#
&Mode
          & $A (\overline B_q\to T_{c q\{ \bar q' \bar q''\} } S)$
          \\
\hline
$2'^*$          
          & $\overline B{}^0\to T^+_{c \bar s} \kappa^-$
          & $\frac{1}{\sqrt2} (C_S+C_{ST,2}-C_{TS,1}+E_{\bar B T})$
          \\
$2'^*$
          & $\overline B{}_s^0\to T^+_{c  \bar s} a_0^-$
          & $\frac{1}{\sqrt2} (C_{ST,1}-C_{ST,2}-C_{TS,1})$
          \\
$3'^{*}$
          & $B^-\to T^0_{c \bar s } \kappa^-$
          & $C_{T}-C_S-C_{ST,2}+C_{TS,1}$
          \\
$3'^{*}$
          & $\overline B{}^0\to T^0_{c\bar s} \overline \kappa{}^0$
          & $C_{T}+E_{\bar B T}$
          \\  
$3'^{*}$
          & $\overline B{}_s^0\to T^0_{c \bar s} a_0^0$
          & $-\frac{1}{\sqrt2}(C_{ST,1}-C_{ST,2}-C_{TS,1})$
          \\
$3'^{*}$
          & $\overline B{}_s^0\to T^0_{c  \bar s} f_0$
          & $-\frac{c_{\theta}}{\sqrt2} (2 C_T+C_{ST,1}-C_{ST,2}+C_{TS,1})+s_{\theta} (C_S-C_{ST,1}-C_{TS,1})$
          \\
$3'^{*}$
          & $\overline B{}_s^0\to T^0_{c \bar s} \sigma$
          & $\frac{s_{\theta}}{\sqrt2} (2 C_T+C_{ST,1}-C_{ST,2}+C_{TS,1})+c_{\theta} (C_S-C_{ST,1}-C_{TS,1})$
          \\
\hline
$5'^{*}$
          & $\overline B{}^0\to T^0_{c s} \kappa^0$
          & $-C_S+C_{ST,1}+E_{\bar B T}$
          \\
$6'^{*}$        
          & $B^-\to T^-_{c s} \kappa^0$
          & $-\sqrt2(C_S-C_{ST,1}+C_{ST,2})$
          \\ 
$6'^{*}$          
          & $\overline B{}^0\to T^-_{c s} \kappa^+$
          & $\sqrt 2(C_{ST,2}+E_{\bar B T})$
          \\
\hline
$8'^{*}$
          & $\overline B{}_s^0\to T^+_{c  \bar s \bar s} \kappa^-$
          & $-\sqrt2 (C_S-C_{TS,1})$
          \\
\hline
$10'^*$  
          & $\overline B{}^0\to T^+_{c}a_0^-$
          & $\sqrt\frac{2}{3}(C_{ST,1}-C_{TS,1}+E_{\bar B T})$
          \\
$11'^*$       
          & $B^-\to T^0_{c} a_0^-$
          & $-\sqrt\frac{2}{3}(C_{T}-C_{ST,1}+C_{TS,1})$
          \\
$11'^*$    
          & $\overline B{}^0\to T^0_{c} a_0^0$
          & $\frac{1}{\sqrt3} (C_T+C_{ST,1}-C_{TS,1}+2E_{\bar B T})$
          \\                    
$11'^*$       
          & $\overline B{}^0\to T^0_{c} f_0$
          & $\frac{c_{\phi}}{\sqrt3} (C_T+C_{ST,1}+C_{TS,1})
              +\sqrt\frac{2}{3} s_{\phi} (C_T-C_S+C_{ST,1}-C_{ST,2}+C_{TS,1})$
          \\   
$11'^*$       
          & $\overline B{}^0\to T^0_{c} \sigma$
          & $-\frac{s_{\phi}}{\sqrt3} (C_T+C_{ST,1}+C_{TS,1})
              +\sqrt\frac{2}{3} c_{\phi} (C_T-C_S+C_{ST,1}-C_{ST,2}+C_{TS,1})$
          \\   
$11'^*$
          & $\overline B{}_s^0\to T^0_{c} \kappa^0$
          & $-\sqrt\frac{2}{3} (C_T-C_{ST,2})$
          \\
$12'^{*}$         
          & $B^-\to T^-_{c} a_0^0$
          & $C_T-C_{ST,1}+C_{TS,1}$
          \\ 
$12'^{*}$         
          & $B^-\to T^-_{c} f_0$
          & $-c_{\theta}(C_T+C_{ST,1}+C_{TS,1})-\sqrt2 s_\phi(C_T-C_S+C_{ST,1}-C_{ST,2}+C_{TS,1})$
          \\ 
$12'^{*}$         
          & $B^-\to T^-_{c} \sigma$
          & $s_{\theta}(C_T+C_{ST,1}+C_{TS,1})-\sqrt2 c_\phi C_T-C_S+C_{ST,1}-C_{ST,2}+C_{TS,1})$
          \\ 
$12'^{*}$          
          & $\overline B{}^0\to T^-_{c} a_0^+$
          & $\sqrt 2(C_T+E_{\bar B T})$
          \\ 
$12'^{*}$
          & $\overline B{}_s^0\to T^-_{c} \kappa^+$
          & $\sqrt2 (C_{T}-C_{ST,2})$
          \\
\hline
$13'$  
          & $\overline B{}^0\to  T^{\prime +}_{c} a_0^-$
          & $\frac{1}{2\sqrt3} (-3 C_S+C_{ST,1}+2C_{TS,1}+E_{\bar B T})$
          \\ 
$14'$        
          & $B^-\to T^{\prime 0}_{c} a_0^-$
          & $\frac{1}{2\sqrt3} (C_T-3 C_S+2C_{ST,1}-3C_{ST,2}+C_{TS,1})$
          \\    
$14'$     
          & $\overline B{}^{\prime 0}\to T^0_{c} a_0^0$
          & $-\frac{1}{2\sqrt6} (C_T+C_{ST,1}-3C_{ST,2}-C_{TS,1}-E_{\bar B T})$
          \\ 
$14'$   
          & $\overline B{}^0\to T^{\prime 0}_{c} f_0$
          & $-\frac{c_{\phi}}{2\sqrt6}(C_T+C_{ST,1}-3C_{ST,2}+C_{TS,1}-3E_{\bar B T})$
          $-\frac{s_\phi}{2\sqrt3} (-C_S+C_T+C_{ST,1}+2C_{ST,2}+C_{TS,1}+3E_{\bar B T})$
          \\ 
$14'$    
          & $\overline B{}^0\to T^{\prime 0}_{c} \sigma$
          & $\frac{s_{\phi}}{2\sqrt6}(C_T+C_{ST,1}-3C_{ST,2}+C_{TS,1}-3E_{\bar B T})$
          $-\frac{c_\phi}{2\sqrt3} (-C_S+C_T+C_{ST,1}+2C_{ST,2}+C_{TS,1}+3E_{\bar B T})$
          \\ 
$14'$
          & $\overline B{}_s^0\to T^{\prime 0}_{c} \kappa^0$
          & $\frac{1}{2\sqrt3} (C_T+3C_S-3C_{ST,1}+2C_{ST,2})$
          \\
\hline
$15'$     
          & $\overline B{}^0\to  T^{\prime +}_{c \bar s} \kappa^-$
          & $\frac{1}{2}(-C_S+C_{ST,2}+C_{TS,1}+ E_{\bar B T})$
          \\ 
$15'$
          & $\overline B{}_s^0\to T^{\prime +}_{c \bar s} a_0^-$
          & $-\frac{1}{2} (2C_S-C_{ST,1}+C_{ST,2}-C_{TS,1})$
          \\                       
\hline
\hline
$16'$       
          & $\overline B{}^0\to T^{\prime\prime +}_{c} a_0^-$
          & $\frac{1}{2} (4T_S+C_S+C_{ST,1}+2C_{TS,1}+4E_{\bar B S,2}+E_{\bar B T})$
          \\
$17'$     
          & $B^-\to T^{\prime\prime 0}_{c} a_0^-$
          & $\frac{1}{2}(4T_S+C_S+C_T+2C_{ST,1}+C_{ST,2}+4 C_{ST,3}+C_{TS,1})$
          \\
$17'$ 
          & $\overline B{}^0\to T^{\prime \prime 0}_{c} a_0^0$
          & $-\frac{1}{2\sqrt2 }(C_T+C_{ST,1}+C_{ST,2}+4C_{ST,3}-C_{TS,1}-4E_{\bar B S,2}-E_{\bar B T})$
          \\
$17'$
          & $\overline B{}^0\to T^{\prime\prime 0}_{c} f_0$
          & $-\frac{c_{\phi}}{2\sqrt2 }(C_T+C_{ST,1}+C_{ST,2}+4C_{ST,3}+C_{TS,1}+4E_{\bar B S,2}+5E_{\bar B T})$
          \\
          & 
          & $+\frac{s_\phi}{2} (C_S-C_T-C_{ST,1}-2C_{ST,2}-4C_{ST,3}-C_{TS,1}-4E_{\bar B S,2}-3E_{\bar B T})$
          \\
$17'$       
          & $\overline B{}^0\to T^{\prime\prime 0}_{c} \sigma$
          & $\frac{s_{\phi}}{2\sqrt2 }(C_T+C_{ST,1}+C_{ST,2}+4C_{ST,3}+C_{TS,1}+4E_{\bar B S,2}+5E_{\bar B T})$
          \\
          & 
          & $+\frac{c_\phi}{2} (C_S-C_T-C_{ST,1}-2C_{ST,2}-4C_{ST,3}-C_{TS,1}-4E_{\bar B S,2}-3E_{\bar B T})$
          \\
$17'$
          & $\overline B{}_s^0\to T^{\prime\prime 0}_{c} \kappa^0$
          & $\frac{1}{2} (C_T-C_S+C_{ST,1}+2C_{ST,2}+4C_{ST,3})$
          \\                        
\hline
$18'$      
          & $\overline B{}^0\to T^{\prime\prime +}_{c \bar s} \kappa^-$
          & $\frac{1}{2} (-C_S+C_{ST,2}+C_{TS,1}+4E_{\bar B S,2}+E_{\bar B T})$
          \\
$18'$
          & $\overline B{}_s^0\to T^{\prime\prime +}_{c \bar s} a_0^-$
          & $\frac{1}{2} (4T_S+2C_S+C_{ST,1}-C_{ST,2}+C_{TS,1})$
          \\ 
\end{tabular}
\end{ruledtabular}
}
\end{table}

\begin{table}[t!]
\caption{\label{tab: BtoTS II -1}
$\overline B_q\to T_{c q'\{ \bar q'' \bar q'''\}} S$ decay amplitudes in $\Delta S=-1$ transition in scenario II.}
\scriptsize{
\begin{ruledtabular}
\begin{tabular}{llccccr}
\#
&Mode
          & $A' (\overline B_q\to T_{c q\{\bar q' \bar q''\}} S)$
          \\
\hline
$2'^*$
          & $\overline B{}_s^0\to T^+_{c \bar s} \kappa^-$
          & $-\frac{1}{\sqrt2}(C'_S-C'_{ST,1}-E'_{\bar B T})$
          \\
$3'^{*}$
          & $\overline B{}_s^0\to T^0_{c \bar s} \overline \kappa{}^0$
          & $-C'_S+C'_{ST,1}+E'_{\bar B T}$
          \\
\hline
$4'^{*}$      
          & $\overline B{}^0\to T^+_{c s} a^-$
          & $-\sqrt2 (C'_S-C'_{TS,1})$
          \\
$5'^{*}$
          & $B^-\to T^0_{c s} a^-$
          & $C'_{T}-C'_S-C'_{ST,2}+C'_{TS,1}$
          \\
$5'^{*}$
          & $\overline B{}^0\to T^0_{c s} a_0^0$
          & $-\frac{1}{\sqrt2}(C'_T+C'_S-C'_{ST,2}-C'_{TS,1})$
          \\
$5'^{*}$
          & $\overline B{}^0\to T^0_{c s} f_0$
          & $-\frac{c_{\phi}}{\sqrt2}(C'_T-C'_S+2C'_{ST,1}-C'_{ST,2}+C'_{TS,1})-s_{\phi} (C'_T+C'_{TS,1})$
          \\
$5'^{*}$
          & $\overline B{}^0\to T^0_{c s} \sigma$
          & $\frac{s_{\phi}}{\sqrt2}(C'_T-C'_S+2C'_{ST,1}-C'_{ST,2}+C'_{TS,1})-c_{\phi} (C'_T+C'_{TS,1})$
          \\
$5'^{*}$
          & $\overline B{}_s^0\to T^0_{c s} \kappa^0$
          & $C'_T+E'_{\bar B T}$
          \\
$6'^{*}$
          & $B^-\to T^-_{c s} a_0^0$
          & $C'_T-C'_S-C'_{ST,2}+C'_{TS,1}$
          \\ 
$6'^{*}$ 
          & $B^-\to T^-_{c s} f_0$
          & $-c_\phi(C'_T-C'_S+2C'_{ST,1}-C'_{ST,2}+C'_{TS,1})-\sqrt2 s_\phi (C'_T+C'_{TS,1})$
          \\ 
$6'^{*}$    
          & $B^-\to T^-_{c s} \sigma$
          & $s_\phi(C'_T-C'_S+2C'_{ST,1}-C'_{ST,2}+C'_{TS,1})-\sqrt2 c_\phi (C'_T+C'_{TS,1})$
          \\ 
$6'^{*}$
          & $\overline B{}^0\to T^-_{c s} a^+$
          & $\sqrt2(C'_T-C'_{ST,2})$
          \\ 
$6'^{*}$
          & $\overline B{}_s^0\to T^-_{c s} \kappa^+$
          & $\sqrt2 (C'_T+E'_{\bar B T})$
          \\
\hline
$10'^*$
          & $\overline B{}^0\to T^+_{c} \kappa^-$
          & $-\sqrt\frac{2}{3}(C'_S-C'_{ST,1}+C'_{ST,2})$
          \\
$10'^*$
          & $\overline B{}_s^0\to T^+_{c} a^-$
          & $\sqrt\frac{2}{3}(C'_{ST,2}+E'_{\bar B T})$
          \\
$11'^*$
          & $B^-\to T^0_{c} \kappa^-$
          & $-\sqrt\frac{2}{3} (C'_S-C'_{ST,1}+C'_{ST,2})$
          \\
$11'^*$ 
          & $\overline B{}^0\to T^0_{c} \overline \kappa{}^0$
          & $\sqrt\frac{2}{3}(C'_S-C'_{ST,1}+C'_{ST,2})$
          \\                    
$11'^*$
          & $\overline B{}_s^0\to T^0_{c} a_0^0$
          & $\frac{2}{\sqrt3}(C'_{ST,2}+E'_{\bar B T})$
          \\
$12'^{*}$         
          & $B^-\to T^-_{c} \overline \kappa{}^0$
          & $-\sqrt2(C'_S-C'_{ST,1}+C'_{ST,2})$
          \\ 
$12'^{*}$
          & $\overline B{}_s^0\to T^-_{c} a^+$
          & $\sqrt2 (C'_{ST,2}+E'_{\bar B T})$
          \\
\hline
$13'$  
          & $\overline B{}^0\to  T^{\prime +}_{c} \kappa^-$
          & $\frac{1}{2\sqrt3} (2 C'_S+C'_{ST,1}-C'_{ST,2}-3 C'_{TS,1})$
          \\ 
$13'$
          & $\overline B{}_s^0\to T^{\prime +}_{c} a_0^-$
          & $\frac{1}{2\sqrt3} (3C'_S+C'_{ST,2}-3C'_{TS,1}+E'_{\bar B T})$
          \\            
$14'$
          & $B^-\to T^{\prime 0}_{c} \kappa^-$
          & $-\frac{1}{2\sqrt3} (3 C'_T-C'_S-2C'_{ST,1}-C'_{ST,2}+3C'_{TS,1})$
          \\    
$14'$       
          & $\overline B{}^0\to T^{\prime 0}_{c} \overline\kappa{}^0$
          & $-\frac{1}{2\sqrt3} (3 C'_T+C'_S-C'_{ST,1}-2C'_{ST,2})$
          \\ 
$14'$
          & $\overline B{}_s^0\to T^{\prime 0}_{c} a_0^0$
          & $\frac{1}{2\sqrt6} (3C'_S+C'_{ST,2}-3C'_{TS,1}+E'_{\bar B T})$
          \\
$14'$
          & $\overline B{}_s^0\to T^{\prime 0}_{c} f_0$
          & $\frac{\sqrt3 c_\phi}{2\sqrt2} (2C'_T-C'_S+2C'_{ST,1}-C'_{ST,2}+C'_{TS,1}+E'_{\bar B T})+\frac{\sqrt3 s_\phi}{2} (C'_{T S,1}-E'_{\bar B T})$
          \\
$14'$
          & $\overline B{}_s^0\to T^{\prime 0}_{c} \sigma$
          & $-\frac{\sqrt3 s_\phi}{2\sqrt2} (2C'_T-C'_S+2C'_{ST,1}-C'_{ST,2}+C'_{TS,1}+E'_{\bar B T})+\frac{\sqrt3 c_\phi}{2} (C'_{T S,1}-E'_{\bar B T})$
          \\
\hline
$15'$
          & $\overline B{}_s^0\to T^{\prime +}_{c \bar s} \kappa^-$
          & $\frac{1}{2} (C'_S+C'_{ST,1}-2C'_{TS,1}+E'_{\bar B T})$
          \\                       
\hline
\hline
$16'$       
          & $\overline B{}^0\to T^{\prime\prime +}_{c} \kappa^-$
          & $\frac{1}{2}(4 T'_S+2C'_S+C'_{ST,1}-C'_{ST,2}+C'_{TS,1})$
          \\
$16'$     
          & $\overline B{}_s^0\to T^{\prime\prime +}_{c} a^-$
          & $-\frac{1}{2}(C'_S-C'_{ST,2}-C'_{TS,1}-4E'_{\bar BS,2}-E'_{\bar B T})$
          \\
$17'$     
          & $B^-\to T^{\prime\prime 0}_{c} \kappa^-$
          & $\frac{1}{2}(4T'_S+C'_T+C'_S+2C'_{ST,1}+C'_{ST,2}+4C'_{ST,3}+C'_{TS,1})$
          \\
$17'$  
          & $\overline B{}^0\to T^{\prime\prime 0}_{c} \overline \kappa{}^0$
          & $\frac{1}{2}(C'_T-C'_S+C'_{ST,1}+2C'_{ST,2}+4C'_{ST,3})$
          \\
$17'$
          & $\overline B{}_s^0\to T^{\prime\prime 0}_{c} a_0^0$
          & $-\frac{1}{2\sqrt2 }(C'_S-C'_{ST,2}-C'_{TS,1}-4E'_{\bar BS,2}-E'_{\bar B T})$
          \\                        
$17'$
          & $\overline B{}_s^0\to T^{\prime\prime 0}_{c}  f_0$
          & $-\frac{c_{\phi}}{2\sqrt2}(2 C'_T-C'_S+2C'_{ST,1}+3C'_{ST,2}+8C'_{ST,3}+C'_{TS,1}+4E'_{\bar BS,2}+5E'_{\bar B T})$
          \\
          & 
          & $-\frac{s_{\phi}}{2} (C'_{TS,1}+4E'_{\bar BS,2}+3E'_{\bar B T})$
          \\
$17'$
          & $\overline B{}_s^0\to T^{\prime\prime 0}_{c} \sigma$
          & $\frac{s_{\phi}}{2\sqrt2}(2 C'_T-C'_S+2C'_{ST,1}+3C'_{ST,2}+8C'_{ST,3}+C'_{TS,1}+4E'_{\bar BS,2}+5E'_{\bar B T})$
          \\
          & 
          & $-\frac{c_{\phi}}{2} (C'_{TS,1}+4E'_{\bar BS,2}+3E'_{\bar B T})$
          \\
\hline
$18'$
          & $\overline B{}_s^0\to T^{\prime\prime +}_{c \bar s} \kappa^-$
          & $\frac{1}{2}(4T'_S+C'_S+C'_{ST,1}+2C'_{TS,1}+4E'_{\bar BS,2}+E'_{\bar B T})$
          \\ 
\end{tabular}
\end{ruledtabular}
}
\end{table}

We now turn to scenario II.
From Tables \ref{tab: BtoTS II 0} and \ref{tab: BtoTS II -1}, 
we see that there are twenty-two flavor exotic modes in $\Delta S=0$ transitions and twenty in $\Delta S=-1$ transitions in scenario I.
In $\Delta S=0$ transitions, eight out of twelve flavor exotic states are involved, while in $\Delta S=-1$ transitions, we also have eight flavor exotic states involved.
Explicitly, three doubly charged states are forbidden by charge conservation, while $T^+_{c s}$ is absent from the $\Delta S=0$ transitions 
and $T^+_{c \bar s\bar s}$ from the $\Delta S=-1$ transitions. The situation is similar to the $\overline B\to TP$ case.

With these amplitudes, one can find relations on rates. For example, we have
\be
\Gamma(\overline B{}_s^0\to T^+_{c  \bar s} a_0^-)&=&\Gamma(\overline B{}_s^0\to T^0_{c \bar s} a_0^0),
\non\\
%
\Gamma(B^-\to T^0_{c \bar s } \kappa^-)
&=&
\left|\frac{V_{ud}}{V_{us}}\right|^2 \Gamma(B^-\to T^0_{c s} a^-)
\neq 2 \left|\frac{V_{ud}}{V_{us}}\right|^2\Gamma(\overline B{}^0\to T^0_{c s} a_0^0),
\label{eq: BtoTS II a}
\\
\Gamma(B^-\to T^0_{c \bar s } \kappa^-)
&=&\left|\frac{V_{ud}}{V_{us}}\right|^2 \Gamma( B^-\to T^-_{c s} a_0^0),
\non\\
\Gamma(\overline B{}^0\to T^0_{c s} \kappa^0)
&=&\left|\frac{V_{ud}}{V_{us}}\right|^2 \Gamma(\overline B{}_s^0\to T^0_{c \bar s} \overline \kappa{}^0)
=2 \left|\frac{V_{ud}}{V_{us}}\right|^2 \Gamma(\overline B{}_s^0\to T^+_{c \bar s} \kappa^-),
\label{eq: BtoTS II b}\\
2\Gamma(\overline B{}^0\to T^0_{c\bar s} \overline \kappa{}^0)
&=&\Gamma(\overline B{}^0\to T^-_{c} a_0^+)
= 2 \left|\frac{V_{ud}}{V_{us}}\right|^2 \Gamma(\overline B{}_s^0\to T^0_{c s} \kappa^0)
\non\\
&=& \left|\frac{V_{ud}}{V_{us}}\right|^2 \Gamma(\overline B{}_s^0\to T^-_{c s} \kappa^+).
\en
Note that some of the relations are similar to those in scenario I, 
such as Eqs. (\ref{eq: BtoTS II a}) and (\ref{eq: BtoTS II b}) and Eqs. (\ref{eq: BtoTS I a}) and (\ref{eq: BtoTS I b}).
However, in Eq. (\ref{eq: BtoTS I a}), we have $\Gamma(B^-\to T^0_{c s} a^-)= 2 \Gamma(\overline B{}^0\to T^0_{c s} a_0^0)$,
while in Eq. (\ref{eq: BtoTS II a}), we have $\Gamma(B^-\to T^0_{c s} a^-)\neq 2 \Gamma(\overline B{}^0\to T^0_{c s} a_0^0)$.
Note that the former amplitudes are governed by $C'_{T}-C'_S-C'_{ST,2}+C'_{TS,1}$, while the latter by $C'_{T}+C'_S-C'_{ST,2}-C'_{TS,1}$.
As explained previously, the difference in the isospin quantum numbers of $T^0_{cs}$ in different scenarios leads to this discrepancy.
One can work out other relations from Tables \ref{tab: BtoTS II 0} and \ref{tab: BtoTS II -1}.

The above relations can be understood in the same manner as in previous cases.
From SU(3) symmetry, we see that although we have nine topological amplitudes, there are only four independent combinations,
namely 
$C^{(\prime)}_{ST,1}-C^{(\prime)}_{ST,2}-C^{(\prime)}_{TS,1}$, 
$C^{(\prime)}_{T}-C^{(\prime)}_S-C^{(\prime)}_{ST,2}+C^{(\prime)}_{TS,1}$,
$C^{(\prime)}_{T}+C^{(\prime)}_S-C^{(\prime)}_{ST,2}-C^{(\prime)}_{TS,1}$
and 
$C^{(\prime)}_{T}+E^{(\prime)}_{\bar B T}$
that contribute to modes involving flavor exotic states and $a_0$ or $\kappa$.
These modes are related. 

The final state $T$ can further decay to $DP$ and $DS$, or they can connect to these states virtually. 
One can obtain their full amplitudes using results in the section and those in Sec.~\ref{sec: T to DP, DS}.

\section{$B^-_c\to T \overline T$ decays}

\begin{figure}[t]
\centering
\subfigure[]{
  \includegraphics[width=0.35\textwidth]{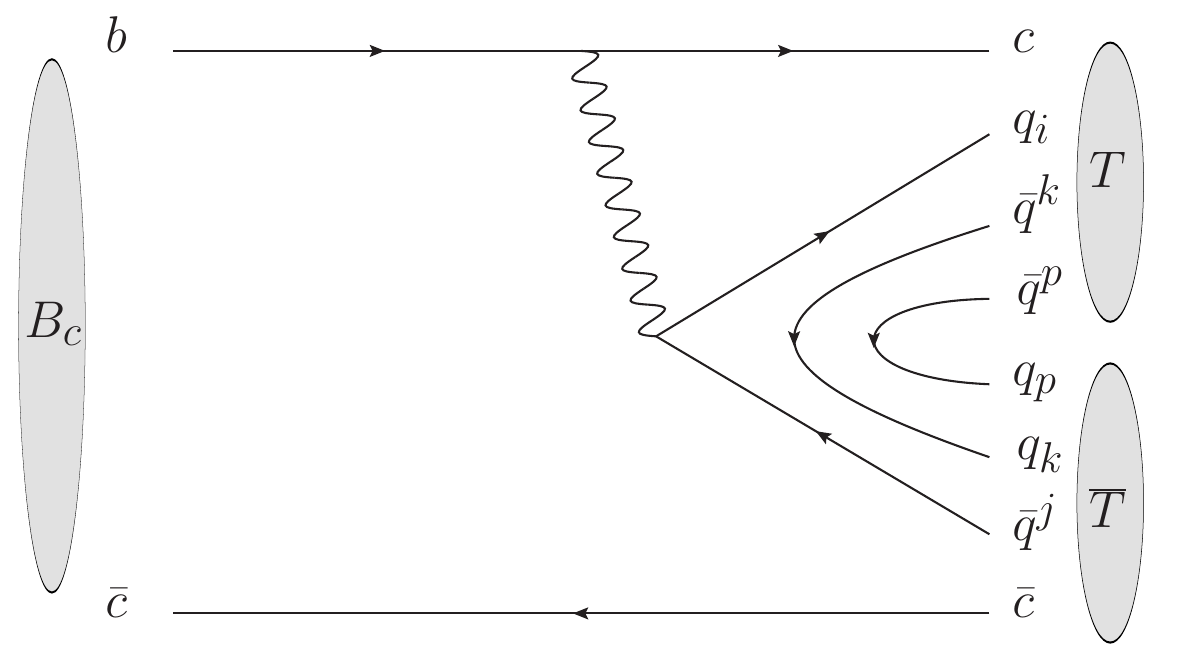}
}
\hspace{24pt}
\subfigure[]{
  \includegraphics[width=0.35\textwidth]{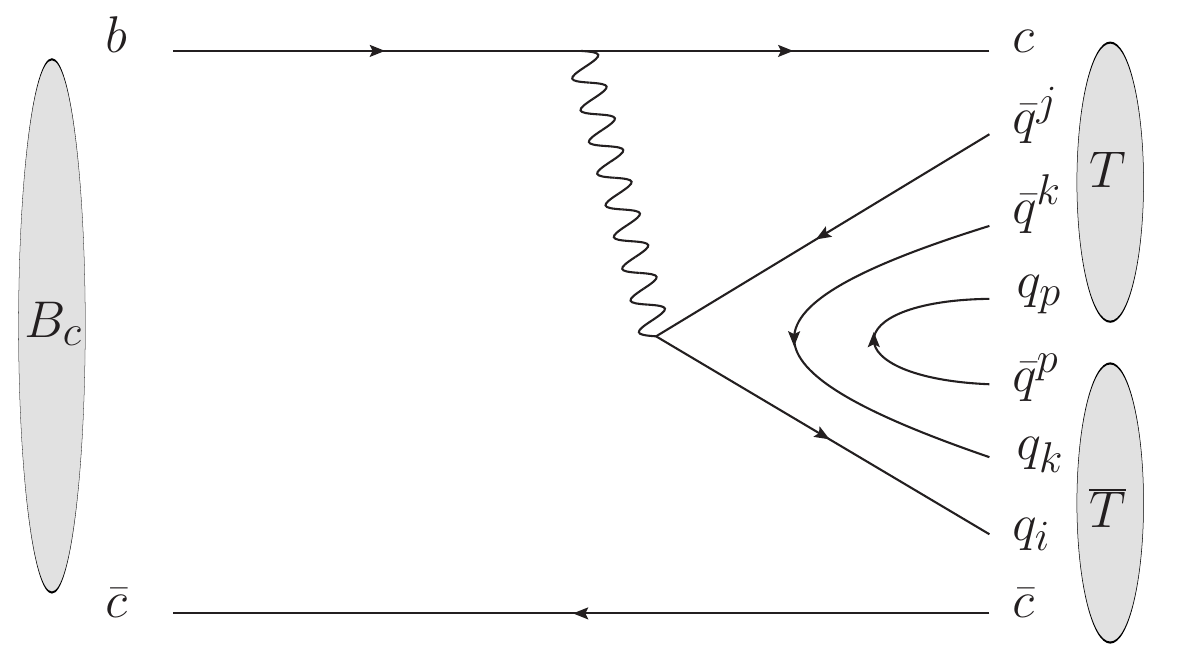}
}
\subfigure[]{
  \includegraphics[width=0.35\textwidth]{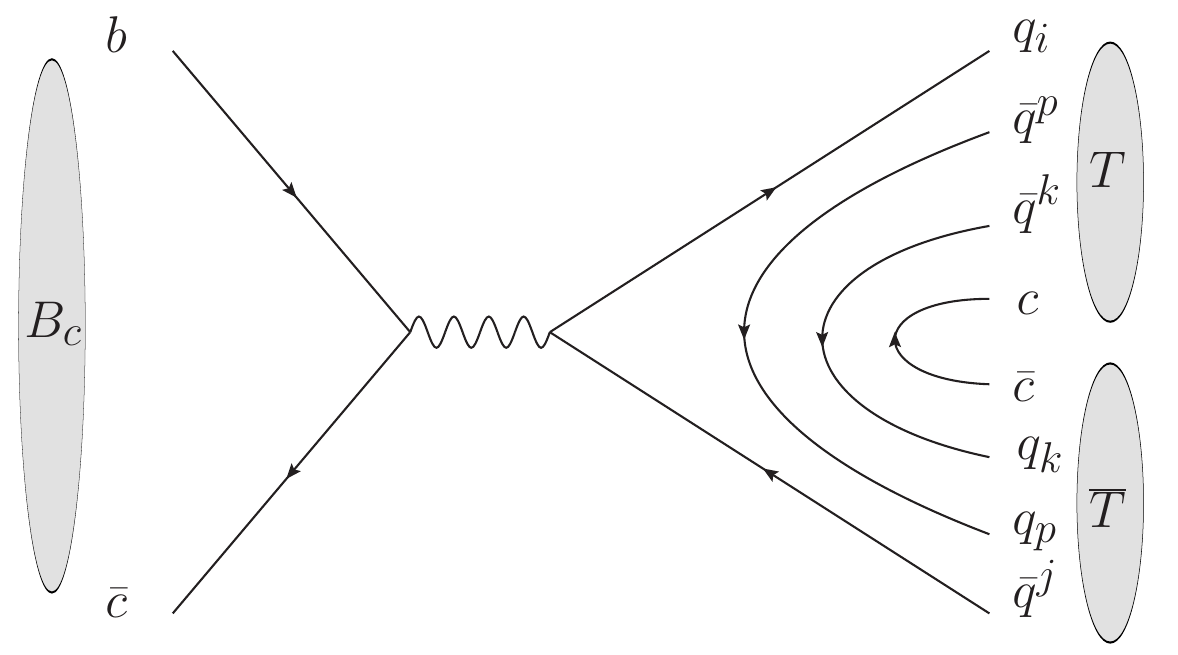}
}
\hspace{24pt}
\subfigure[]{
  \includegraphics[width=0.35\textwidth]{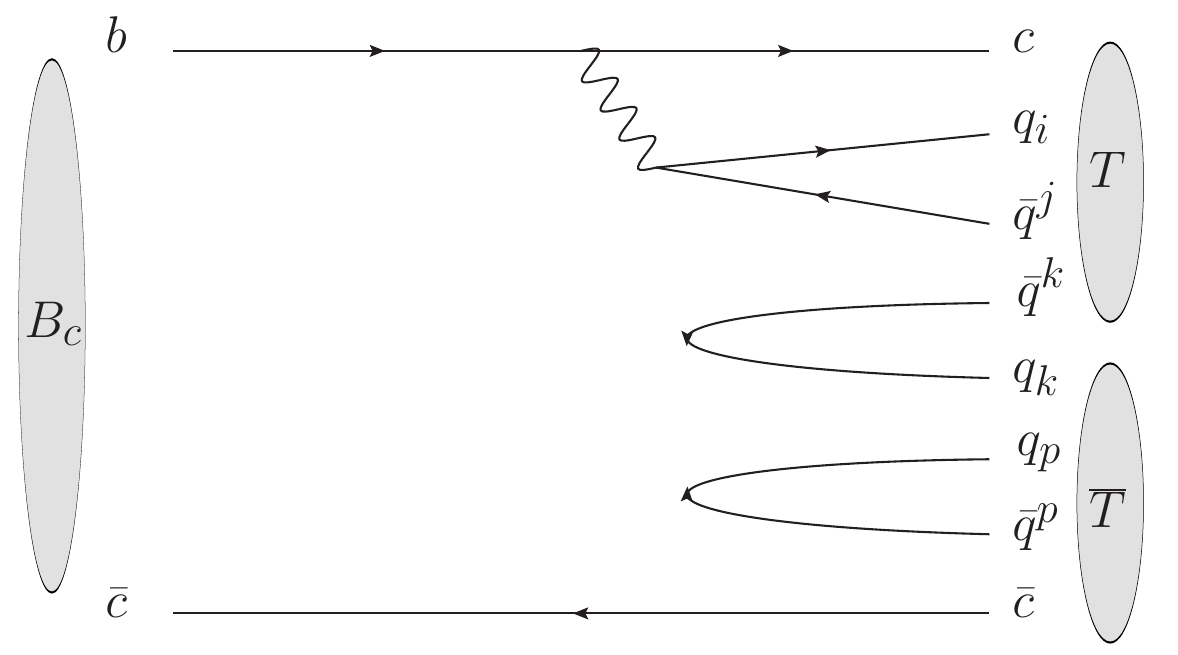}
}
\subfigure[]{
  \includegraphics[width=0.35\textwidth]{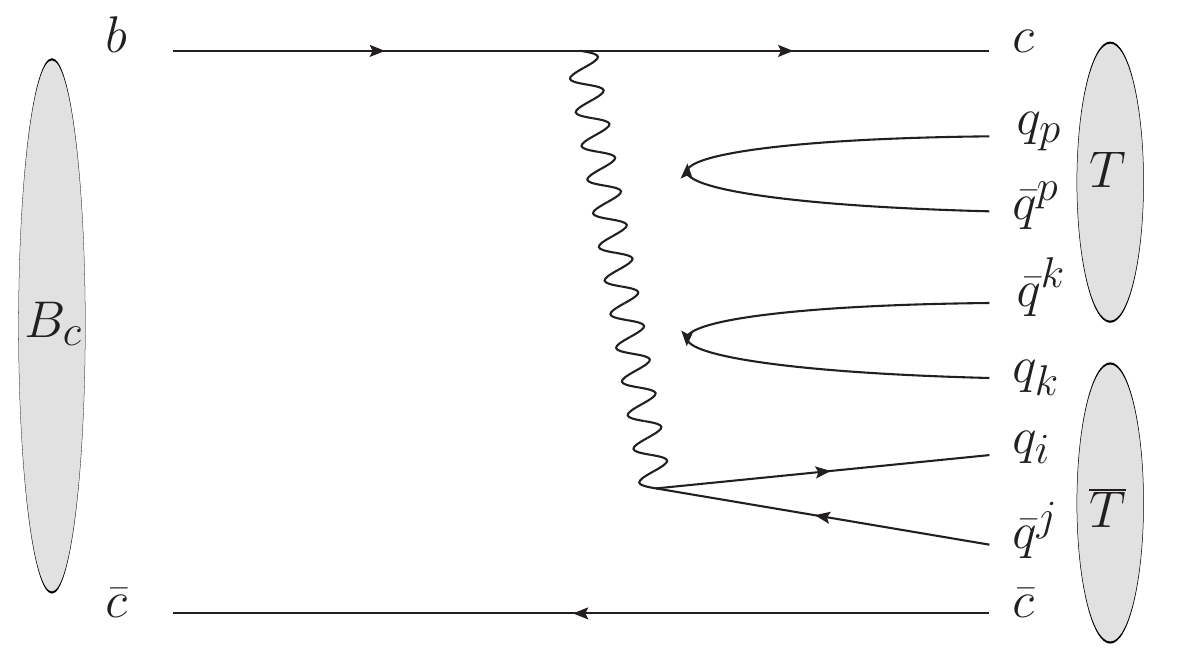}
}
\hspace{24pt}
\subfigure[]{
  \includegraphics[width=0.35\textwidth]{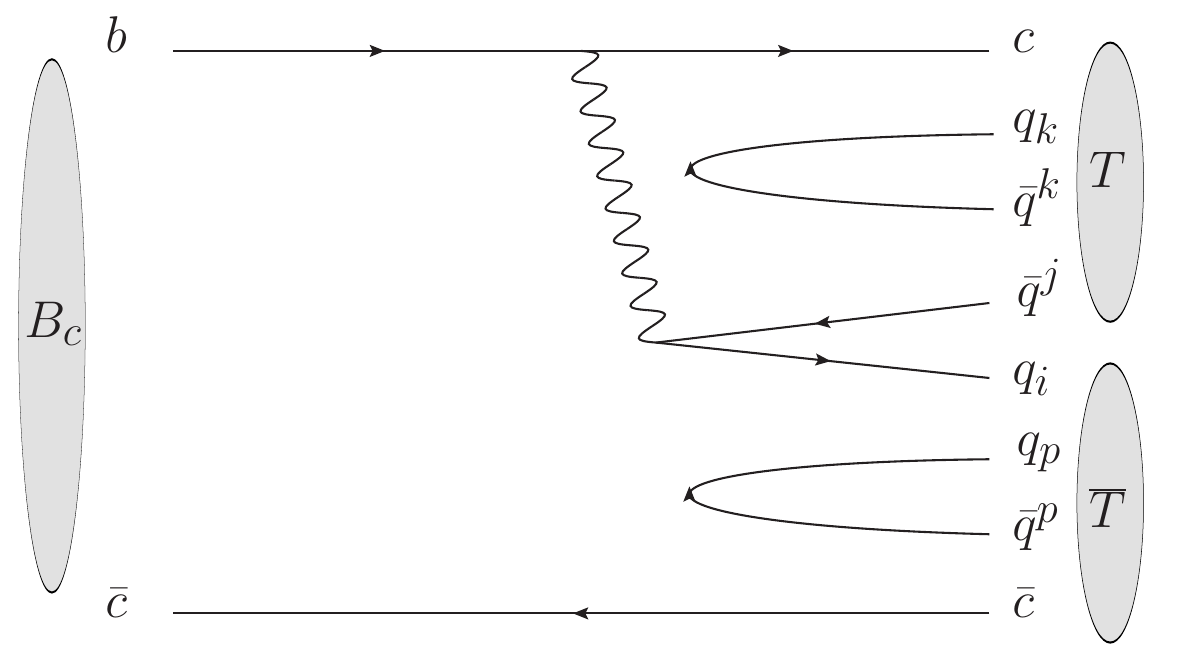}
}
\hspace{24pt}
\subfigure[]{
  \includegraphics[width=0.35\textwidth]{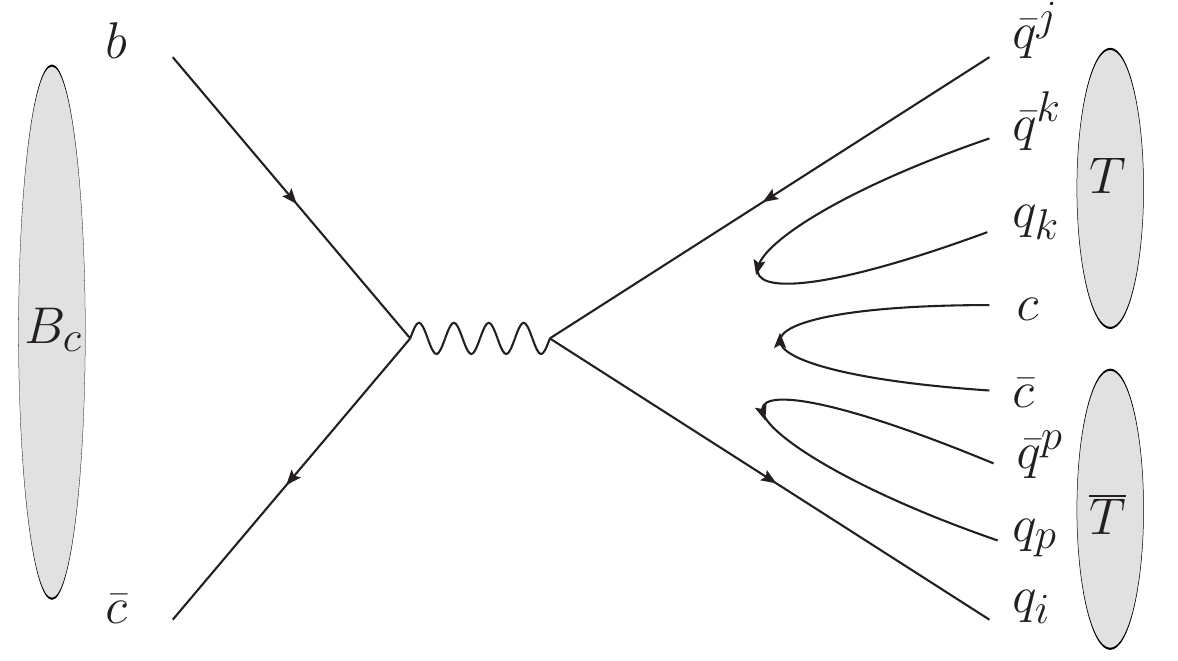}
}
\caption{Topological diagrams of
(a) $C_{T\bar T}$,
(b) $C_{\bar T T,1}$,
(c) $A_{T \bar T}$,
(d) $C_{T}$,
(e) $C_{\bar T}$,
(f) $C_{\bar T T,2}$ and 
(g)~$A_{\bar T T}$
amplitudes in $B^-_c\to T \overline T$ decays, where $C$ and $A$ are internal $W$ and annihilation diagrams, respectively.
Diagrams (a) to (c) contribute to modes with flavor exotic $T$ and $\overline T$.
} \label{fig: TA Bc2TT}
\end{figure}

From the above discussion, we note that $T^{++}_{c\bar s}$ and other doubly charged states do not participate in $\overline B\to \overline D T$, $\overline B\to TP$ and $\overline B\to T S$ decays,
as required by charge conservation. 
It is preferable to have other modes to include them, besides $\overline B\to D \overline T$ decays, where $T^{++}_{c\bar s}$ was discovered.  
In this section, we consider $B^-_c$ decays, which are kinematically allowed to produce a $T \overline T$ pair.
As we shall see $T^{++}_{c\bar s}$ and other doubly charged states can be produced in $B^-_c\to T \overline T$ decays.

\subsection{Topological amplitudes in $B^-_c\to T \overline T$ decays}

For $B^-_c\to T \overline T$ decays in $\Delta S=0$ transitions, with $T=T_{cq[q' q'']}$ in scenario I, we have
\be
H_{\rm eff}
&=&
        C_{\bar T T,1}\, B_c^-  \overline T_l^{[jk]}\, H^i_j\, T^l_{[ik]}
        + C_{\bar T T,2}\, B_c^-  \overline T_k^{[jk]}\, H^i_j\, T^l_{[il]}
\non\\
         && +C_{T\bar T}\, B_c^-  \overline T_i^{[kl]}\, H^i_j\, T^j_{[lk]}
              +C_T\, B_c^-  \overline T_i^{[jk]}\, H^i_j\, T^l_{[kl]}
\non\\
         &&+C_{\bar T}\, B_c^-  \overline T_l^{[kl]}\, H^i_j\, T^j_{[ik]}
         +A_{T \bar T} B^-_c \overline T^{[kl]}_i H^i_j T^j_{[lk]}
         +A_{\bar T T} B^-_c \overline T^{[jk]}_k H^i_j T^l_{[il]},
\label{eq: H B to TTbar I}
\en
while for $B^-_c\to T \overline T$ decays with $T=T_{cq\{q' q''\}}$ in scenario II, 
we have
\be
H_{\rm eff}
&=&
         C_{\bar T T,1}\, B_c^-  \overline T_l^{\{jk\}}\, H^i_j\, T^l_{\{ik\}}
         +C_{\bar T T,2}\, B_c^-  \overline T_k^{\{jk\}}\, H^i_j\, T^l_{\{il\}}
\non\\
         && +C_{T\bar T}\, B_c^-  \overline T_i^{\{kl\}}\, H^i_j\, T^j_{\{lk\}}
              +C_T\, B_c^-  \overline T_i^{\{jk\}}\, H^i_j\, T^l_{\{kl\}}
\non\\
         &&+C_{\bar T}\, B_c^-  \overline T_l^{\{kl\}}\, H^i_j\, T^j_{\{ik\}}
              +A_{T \bar T} B^-_c \overline T^{\{kl\}}_i H^i_j T^j_{\{lk\}}
              +A_{\bar T T} B^-_c \overline T^{\{jk\}}_k H^i_j T^l_{\{il\}}. 
\label{eq: H B to TTbar II}
\en
For notational simplicity, we do not distinguish the topological amplitudes in scenarios I and II.
We have seven topological amplitudes, namely
$C_{T\bar T}$,
$C_{\bar T T,1}$,
$A_{T \bar T}$,
$C_{T}$,
$C_{\bar T}$,
$C_{\bar T T,2}$ and 
$A_{\bar T T}$, in $B^-_c\to T \overline T$ decays, where $C$ and $A$ are internal $W$ and annihilation diagrams, respectively.
The subscripts in the topological amplitudes denote the open-charmed tetraquark(s), where the $q^i \bar q_j$, from the $b\to c W^-\to c q^i \bar q_j$ transition, contribute to.
The topological amplitudes are depicted in Fig.~\ref{fig: TA Bc2TT}.
Diagrams (a) to (c) contribute to modes with flavor exotic $T$ and $\overline T$.

The above effective Hamiltonian can be easily transformed to those for $\Delta S=-1$ transition. 
To achieve that, one simply replaces the $H$ in Eqs. (\ref{eq: H B to TTbar I}) and (\ref{eq: H B to TTbar II}) with $H'$ as shown in Eq. (\ref{eq: H' 3x3}), and replace
$C_{T\bar T}$,
$C_{\bar T T,1}$,
$A_{T \bar T}$,
$C_{T}$,
$C_{\bar T}$,
$C_{\bar T T,2}$ and 
$A_{\bar T T}$
with
$C'_{T\bar T}$,
$C'_{\bar T T,1}$,
$A'_{T \bar T}$,
$C'_{T}$,
$C'_{\bar T}$,
$C'_{\bar T T,2}$ and 
$A'_{\bar T T}$, respectively,
and they are related by CKM factors. 
For example, we have $C'_{T\bar T}=\frac{V^*_{us}}{V^*_{ud}} C_{T\bar T}$ and so on.

Comparing Fig.~\ref{fig: TA Bc2TT} to Figs.~ \ref{fig: TA B2TP}, 
we see that the topological diagrams in $B^-_c\to T\overline T$ decay have an additional $q\bar q$ line produced. 
According to the OZI rule, one expects some suppression factors.
But the diagrams in $B^-_c\to T\overline T$, see Fig.~\ref{fig: TA Bc2TT}, and $\overline B\to T P$ decays, see Figs.~ \ref{fig: TA B2TP}, are rather different, 
we cannot offer a reliable comparison of these rates on this basis.

\begin{table}[t!]
\caption{\label{tab: BctoTT I}
$B^-_c\to T \overline T$ decay amplitudes with $T=T_{c q[\bar q' \bar q'']}$ in $\Delta S=0$ and $\Delta S=-1$ transitions in scenario I.}
\footnotesize{
\begin{ruledtabular}
\begin{tabular}{llccccr}
\#
&Mode
          & $A (B^-_c\to T\overline T)$
          \\
\hline
$(2^*, \bar 1^{*})$  
          & $B^-_c\to T^+_{c \bar s}  \overline T^{\,--}_{\bar c  s}$
          & $\frac{1}{\sqrt2}(C_{\bar T T,1}+2C_{T\bar T}+2 A_{T\bar T})$
          \\ 
$(3^{*}, \bar 2^*)$ 
          & $B^-_c\to T^0_{c\bar s} \overline T^{\,-}_{\bar c s }$
          & $-\frac{1}{\sqrt2}(C_{\bar T T,1}+2C_{T\bar T}+2 A_{T\bar T})$
          \\
$(6, \bar 5)$  
          & $B^-_c\to T^0_{c} \overline T^{\,-}_{\bar c}$
          & $-\frac{1}{2}(C_{\bar T T,1}+2C_{T\bar T}+2A_{T\bar T})$
          \\
\hline          
$(3^{*}, \bar 9)$ 
          & $B^-_c\to T^0_{c \bar s} \overline T^{\,\prime\prime -}_{\bar c s}$
          & $-\frac{1}{\sqrt2}(2C_T-C_{\bar T T,1}+2C_{T\bar T}+2 A_{T\bar T})$
          \\
$(6, \bar 7)$ 
          & $B^-_c\to T^0_{c} \overline T^{\,\prime\prime -}_{\bar c}$
          & $-\frac{1}{2}(2 C_T-C_{\bar T T,1}+2 C_{T\bar T}+2A_{\bar T T})$
          \\ 
$(8, \bar 5)$ 
          & $B^-_c\to T^{\prime\prime 0}_{c} \overline T^{\,-}_{\bar c}$
          & $\frac{1}{2}(2 C_{\bar T}-C_{\bar T T,1}+2 C_{T\bar T}+2 A_{\bar T T})$
          \\ 
$(9, \bar 1^{*})$  
          & $B^-_c\to T^{\prime\prime +}_{c \bar s} \overline T^{\,--}_{\bar c s}$
          & $-\frac{1}{\sqrt2}(2C_{\bar T}-C_{\bar T T,1}+2C_{T\bar T}+2 A_{T\bar T})$
          \\    
\hline
$(8, \bar 7)$  
          & $B^-_c\to T^{\prime\prime 0}_{c} \overline T^{\,\prime\prime -}_{\bar c}$
          & $\frac{1}{2}(2C_T+2 C_{\bar T}+C_{\bar T T,1}+4C_{\bar T T,2}+2C_{T\bar T}+4 A_{\bar T T}+2A_{T\bar T})$
          \\ 
\hline
\hline
\#
&Mode
          & $A' (B^-_c\to T\overline T)$
          \\
\hline
$(4^{*}, \bar {5})$ 
          & $B^-_c\to T^0_{c s} \overline T^{\,-}_{\bar c}$
          & $-\frac{1}{\sqrt2}(C'_{\bar T T,1}+2 C'_{T\bar T}+2A'_{T\bar T})$
          \\                      
$(5, \bar {1}^{*})$  
          & $B^-_c\to T^+_{c} \overline  T^{\,--}_{\bar c s}$
          & $-\frac{1}{\sqrt2} (C'_{\bar T T,1}+2C'_{T \bar T}+2 A'_{T \bar T})$
          \\          
$(6, \bar 2^*)$  
          & $B^-_c\to T^0_{c} \overline T^{\,-}_{\bar c s }$
          & $\frac{1}{2}(C'_{\bar T T,1}+2C'_{T\bar T}+2A'_{T\bar T})$
          \\
\hline
$(4^{*}, \bar {7})$  
          & $B^-_c\to T^0_{c s} \overline T^{\,\prime\prime -}_{\bar c}$
          & $-\frac{1}{\sqrt2}(C'_T-C'_{\bar T T,1}+2 C'_{T\bar T}+2A'_{T\bar T})$
          \\    
$(6, \bar 9)$  
          & $B^-_c\to T^0_{c} \overline T^{\,\prime\prime -}_{\bar c s}$
          & $\frac{1}{2}(2C'_T-C'_{\bar T T,1}+2C'_{T\bar T}+2A'_{T \bar T})$
          \\ 
$(8, \bar 2^*)$  
          & $B^-_c\to T^{\prime\prime 0}_{c} \overline T^{\,-}_{\bar c s}$
          & $\frac{1}{2}(2C'_{\bar T}-C'_{\bar T T,1}+2C'_{T\bar T }+2A'_{T\bar T})$
          \\ 
$(7, \bar {1}^*)$  
          & $B^-_c\to T^{\prime\prime +}_{c}  \overline T^{\,--}_{\bar c s}$
          & $\frac{1}{\sqrt2} (2C'_{\bar T}-C'_{\bar T T,1}+2C'_{T \bar T}+2 A'_{T \bar T})$
          \\
\hline  
$(8, \bar 9)$  
          & $B^-_c\to T^{\prime\prime 0}_{c} \overline T^{\,\prime\prime -}_{\bar c s}$
          & $\frac{1}{2} (2C'_T+2C'_{\bar T}+C'_{\bar T T,1}+4 C'_{\bar T T,2}+2 C'_{T\bar T}+4 A'_{\bar T T}+2 A'_{T\bar T})$
          \\ 
\end{tabular}
\end{ruledtabular}
}
\end{table}

\begin{table}[t!]
\caption{\label{tab: BctoTT II 0}
$B^-_c\to T \overline T$ decay amplitudes with $T=T_{c q\{\bar q' \bar q''\}}$ in $\Delta S=0$ transition in scenario II}.}
\footnotesize{
\begin{ruledtabular}
\begin{tabular}{llccccr}
\#
&Mode
          & $A (B^-_c\to T\overline T)$
          \\
\hline
$(5'^{*}, \overline {4}'^{*})$  
          & $B^-_c\to T^0_{c s} \overline T^{\,-}_{\bar c \bar s}$
          & $\sqrt2 C_{\bar TT,1}$
          \\                      
$(6'^{*}, \bar{5}'^{*})$ 
          & $B^-_c\to T^-_{c s} \overline T^0_{\bar c \bar s}$
          & $\sqrt2 C_{\bar T T,1}$
          \\  
$(8'^{*}, \bar{7}'^{*})$ 
          & $B^-_c\to T^+_{c \bar s \bar s} \overline T^{\,--}_{\bar cs s}$
          & $2(C_{T\bar T}+A_{T\bar T})$
          \\                        
$(2'^*, \bar 1'^{*})$  
          & $B^-_c\to T^+_{c \bar s}  \overline T^{\,--}_{\bar c  s}$
          & $\frac{1}{\sqrt2}(C_{\bar T T,1}-2C_{T\bar T}-2 A_{T\bar T})$
          \\ 
$(3'^{*}, \bar 2'^*)$ 
          & $B^-_c\to T^0_{c\bar s} \overline T^{\,-}_{\bar c s }$
          & $-\frac{1}{\sqrt2}(C_{\bar T T,1}-2C_{T\bar T}-2 A_{T\bar T})$
          \\
$(10'^*, \bar 9'^{*})$ 
          & $B^-_c\to T^+_{c}  \overline T^{\,--}_{\bar c}$
          & $\frac{2}{\sqrt3}(C_{\bar T T,1}-C_{T\bar T}-A_{T\bar T})$
          \\          
$(11'^*, \overline{10}'^*)$  
          & $B^-_c\to T^0_{c} \overline T^{\,-}_{\bar c}$
          & $\frac{4}{3}(C_{\bar T T,1}-C_{T\bar T}-A_{T\bar T})$
          \\
$(12'^{*}, \overline{11}'^*)$ 
          & $B^-_c\to T^{-}_{c} \overline T^0_{\bar c}$
          & $-\frac{2}{\sqrt3}(C_{\bar T T,1}-C_{T\bar T}-A_{T\bar T})$
          \\   
$(3'^{*}, \overline {15}')$ 
          & $B^-_c\to T^0_{c \bar s} \overline T^{\,\prime -}_{\bar c s}$
          & $\frac{1}{2}(C_{\bar T T,1}+2C_{T\bar T}+2 A_{T\bar T})$
          \\  
$(11'^*, \overline{13}')$ 
          & $B^-_c\to T^0_{c} \overline T^{\,\prime -}_{\bar c}$
          & $- \frac{1}{3\sqrt2} (C_{\bar T T,1}+2C_{T \bar T}+2A_{T\bar T})$
          \\           
$(12'^{*}, \overline{14}')$ 
          & $B^-_c\to T^-_{c} \overline T^{\prime 0}_{\bar c}$
          & $\frac{1}{\sqrt6} (C_{\bar T T,1}+2 C_{T\bar T}+2 A_{T\bar T})$
          \\  
$(13', \bar 9'^{*})$  
          & $B^-_c\to T^+_{c} \overline T^{\,--}_{\bar c}$
          & $\frac{1}{\sqrt6} (C_{\bar T T,1}+2 C_{T\bar T}+2 A_{T\bar T})$
          \\  
$(14', \overline{10}'^*)$ 
          & $B^-_c\to T^{\prime 0}_{c} \overline T^{\,-}_{\bar c}$
          & $\frac{1}{3\sqrt2} (C_{\bar T T,1}+2 C_{T\bar T}+2 A_{T\bar T})$
          \\  
$(15', \bar 1'^{*})$  
          & $B^-_c\to T^+_{c \bar s} \overline T^{\,--}_{\bar c s}$
          & $\frac{1}{2} (C_{\bar T T,1}+2 C_{T\bar T}+2 A_{T\bar T})$
          \\
$(14', \overline{13}')$  
          & $B^-_c\to T^{\prime 0}_{c} \overline T^{\,\prime-}_{\bar c}$
          & $\frac{1}{12} (13C_{\bar T T,1}+2 C_{T\bar T}+2 A_{T\bar T})$
          \\  
\hline             
$(3'^{*}, \bar 18')$  
          & $B^-_c\to T^0_{c \bar s} \overline T^{\,\prime\prime -}_{\bar c s}$
          & $\frac{1}{2}(4C_T+C_{\bar T T,1}+2C_{T\bar T}+2 A_{T\bar T})$
          \\
$(12'^{*}, \overline{17}')$  
          & $B^-_c\to T^-_{c} \overline T^{\prime\prime 0}_{\bar c}$
          & $\frac{1}{\sqrt2}(4C_T+C_{\bar T T,1}+2C_{T\bar T}+2A_{T\bar T})$
          \\  
$(14', \overline{16}')$  
          & $B^-_c\to T^{\prime 0}_{c} \overline T^{\,\prime\prime -}_{\bar c}$
          & $\frac{1}{4\sqrt3} (4C_T+C_{\bar T T,1}+2 C_{T\bar T}+2 A_{T\bar T})$
          \\  
$(11'^*, \overline{16}')$  
          & $B^-_c\to T^0_{c} \overline T^{\,\prime\prime -}_{\bar c}$
          & $- \frac{1}{\sqrt6} (4C_T+C_{\bar T T,1}+2C_{T \bar T}+2A_{T\bar T})$
          \\ 
$(16', \bar 9'^{*})$  
          & $B^-_c\to T^{\prime\prime +}_{c}  \overline T^{\,--}_{\bar c}$
          & $\frac{1}{\sqrt2}(4 C_{\bar T}+C_{\bar T T,1}+2C_{T\bar T}+2A_{T\bar T})$
          \\  
$(17', \overline{10}'^*)$  
          & $B^-_c\to T^{\prime\prime 0}_{c} \overline T^{\,-}_{\bar c}$
          & $ \frac{1}{\sqrt6} (4C_{\bar T}+C_{\bar T T,1}+2C_{T \bar T}+2A_{T\bar T})$
          \\ 
$(17', \overline{13}')$ 
          & $B^-_c\to T^{\prime\prime 0}_{c}  \overline T^{\,\prime -}_{\bar c}$
          & $\frac{1}{4\sqrt3}(4 C_{\bar T}+C_{\bar T T,1}+2C_{T\bar T}+2A_{T\bar T})$
          \\
$(18', \bar 1'^{*})$  
          & $B^-_c\to T^{\prime\prime +}_{c \bar s} \overline T^{\,--}_{\bar c s}$
          & $\frac{1}{2}(4C_{\bar T}+C_{\bar T T,1}+2C_{T\bar T}+2 A_{T\bar T})$
          \\
\hline              
$(17', \overline{16}')$  
          & $B^-_c\to T^{\prime\prime 0}_{c} \overline T^{\,\prime\prime -}_{\bar c}$
          & $\frac{1}{4}(4C_T+4 C_{\bar T}+5C_{\bar T T,1}+16C_{\bar T T,2}+2C_{T\bar T}+16 A_{\bar T T}+2A_{T\bar T})$
          \\ 
\end{tabular}
\end{ruledtabular}
\end{table}

\begin{table}[t!]
\caption{\label{tab: BctoTT II -1}
$B^-_c\to T \overline T$ decay amplitudes with $T=T_{c q\{\bar q' \bar q''\}}$ in $\Delta S=-1$ transition in scenario II.}
\footnotesize{
\begin{ruledtabular}
\begin{tabular}{llccccr}
\#
&Mode
          & $A' (B^-_c\to T\overline T)$
          \\
\hline
$(2'^*, \bar {7}'^*)$ 
          & $B^-_c\to T^+_{c \bar s} \overline T^{\,--}_{\bar c s s}$
          & $C'_{\bar T T,1}$
          \\ 
$(3'^{*}, \bar{8}'^*)$ 
          & $B^-_c\to T^0_{c \bar s} \overline T^{\,-}_{\bar c s s}$
          & $\sqrt2 C'_{\bar T T, 1}$
          \\
$(10'^*, \bar {1}'^*)$ 
          & $B^-_c\to T^+_{c} \overline  T^{\,--}_{\bar c s}$
          & $\sqrt\frac{2}{3} C'_{\bar T T,1}$
          \\          
$(11'^*, \bar {2}'^*)$ 
          & $B^-_c\to T^0_{c} \overline T^{\,-}_{\bar c s }$
          & $\frac{2}{\sqrt3} C'_{\bar T T,1}$
          \\
$(12'^{*}, \bar 3'^{*})$  
          & $B^-_c\to T^-_{c} \overline T^0_{\bar c s}$
          & $\sqrt2 C'_{\bar T T,1}$
          \\                    
$(4'^{*}, \bar {9}'^*)$  
          & $B^-_c\to T^+_{c s} \overline T^{\,--}_{\bar c}$
          & $2(C'_{T\bar T}+A'_{T\bar T})$
          \\ 
$(5'^{*}, \overline{10}'^*)$ 
          & $B^-_c\to T^0_{c s} \overline T^{\,-}_{\bar c}$
          & $2\sqrt\frac{2}{3} (C'_{T\bar T}+A'_{T\bar T})$
          \\                      
$(6'^{*}, \overline {11}'^*)$ 
          & $B^-_c\to T^-_{c s} \overline T^0_{\bar c}$
          & $\frac{2}{\sqrt3} (C'_{T\bar T}+A_{T\bar T})$
          \\  
$(5'^{*}, \overline{13}')$ 
          & $B^-_c\to T^0_{c s} \overline T^{\,\prime -}_{\bar c}$
          & $-\frac{1}{2\sqrt3}(3 C'_{\bar T T,1} - 2 C'_{T\bar T} - 2 A'_{T\bar T})$
          \\ 
$(6'^{*}, \overline {14}')$ 
          & $B^-_c\to T^-_{c s} \overline T^{\prime 0}_{\bar c}$
          & $-\frac{1}{\sqrt6} (3C'_{\bar T T,1}-2 C'_{T\bar T}- 2 A'_{T\bar T})$
          \\ 
$(14', \overline {15}')$ 
          & $B^-_c\to T^{\prime 0}_{c} \overline T^{\,\prime -}_{\bar c s}$
          & $\frac{\sqrt 3}{4} (3 C'_{\bar T T,1}-2 C'_{T \bar T} - 2 A'_{T \bar T})$
          \\  
$(15', \bar {7}'^*)$ 
          & $B^-_c\to T^{\prime +}_{c \bar s} \overline T^{\,--}_{\bar c s s}$
          & $\frac{1}{\sqrt2} (C'_{\bar T T,1}-2 C'_{T\bar T}-2 A'_{T\bar T})$
          \\   
$(13', \bar {1}'^*)$ 
          & $B^-_c\to T^{\prime +}_{c} \overline T^{\,--}_{\bar c s}$
          & $\frac{1}{2\sqrt3} (C'_{\bar T T,1}-6 C'_{T\bar T}-6 A'_{T\bar T})$
          \\  
$(14', \bar {2}'^*)$  
          & $B^-_c\to T^{\prime 0}_{c} \overline T^{\,-}_{\bar c s}$
          & $\frac{1}{2\sqrt6} (C'_{\bar T T,1}-6 C'_{T \bar T}-6 A'_{T \bar T})$
          \\  
$(5'^{*}, \overline{16}')$ 
          & $B^-_c\to T^0_{c s} \overline T^{\,\prime\prime -}_{\bar c}$
          & $\frac{1}{2}(4 C'_T +C'_{\bar T T,1} +2C'_{T\bar T}+2A'_{T\bar T})$
          \\    
$(6'^{*}, \overline {17}')$ 
          & $B^-_c\to T^-_{c s} \overline T^{\prime\prime 0}_{\bar c}$
          & $\frac{1}{\sqrt2} (4C'_{T}+C'_{\bar T T,1}+2 C'_{T\bar T}+2 A'_{T\bar T})$
          \\  
$(14', \overline {18}')$ 
          & $B^-_c\to T^{\prime 0}_{c} \overline T^{\,\prime\prime-}_{\bar c s}$
          & $-\frac{\sqrt3}{4} (4C'_T + C'_{\bar T T,1} + 2 C'_{T\bar T}+2 A'_{T\bar T})$
          \\  
$(16', \bar {1}'^*)$ 
          & $B^-_c\to T^{\prime\prime +}_{c}  \overline T^{\,--}_{\bar c s}$
          & $\frac{1}{2} (4C'_{\bar T}+C'_{\bar T T,1}+2C'_{T \bar T}+2 A'_{T \bar T})$
          \\  
$(17', \bar {2}'^*)$ 
          & $B^-_c\to T^{\prime\prime 0}_{c} \overline T^{\,-}_{\bar c s}$
          & $\frac{1}{2\sqrt2 }(4 C'_{\bar T} + C'_{\bar T T,1} + 2C'_{T\bar T } + 2A'_{T\bar T})$
          \\ 
$(17', \overline {15}')$ 
          & $B^-_c\to T^{\prime\prime 0}_{c}  \overline T^{\,\prime -}_{\bar c s}$
          & $\frac{1}{4} (4 C'_{\bar T} + C'_{\bar T T,1} + 2C'_{T\bar T}+2A'_{T\bar T})$
          \\
$(18', \bar {7}'^*)$ 
          & $B^-_c\to T^{\prime\prime +}_{c \bar s} \overline T^{\,--}_{\bar c s s}$
          & $\frac{1}{\sqrt2} (4C'_{\bar T}+C'_{\bar T T,1}+2 C'_{T\bar T}+2 A'_{T\bar T})$
          \\    
\hline
$(17', \overline {18}')$ 
          & $B^-_c\to T^{\prime\prime 0}_{c} \overline T^{\,\prime\prime -}_{\bar c s}$
          & $\frac{1}{4} (4C'_T+4C'_{\bar T}+5C'_{\bar T T,1}+16 C'_{\bar T T,2}+2 C'_{T \bar T}+16 A'_{\bar T T}+2 A'_{T\bar T})$
          \\ 
\end{tabular}
\end{ruledtabular}
}
\end{table}

\subsection{$B^-_c\to T \overline T$ decay amplitudes}

The $B^-_c\to T \overline T$ decay amplitudes in $\Delta S=0$ and $-1$ transitions are shown in Table \ref{tab: BctoTT I} for scenario I, and in Table \ref{tab: BctoTT II 0} and \ref{tab: BctoTT II -1} for scenario II.
We shall concentrate on modes where both $T$ and $\overline T$ are flavor exotic states.

From Table \ref{tab: BctoTT I}, we see that, for scenario I, 
none of modes with both flavor exotic $T$ and $\overline T$ involved can be found in $\Delta S=-1$ transitions,
but there are two such modes in $\Delta S=0$ transitions, namely
$B^-_c\to T^+_{c \bar s}  \overline T^{\,--}_{\bar c  s}$ and $B^-_c\to T^0_{c\bar s} \overline T^{\,-}_{\bar c s }$ decays.
In fact, not just the doubly charged state, $T^{++}_{c\bar s}$, but all three members in the $S=+1$ isotriplet, $(T^{++}_{c\bar s}, T^{+}_{c\bar s}, T^{0}_{c\bar s})$, participate in these two modes. 

Although there are three topological amplitudes $C_{T\bar T}$, $C_{\bar T T,1}$,
and $A_{T \bar T}$ that can contribute to these modes, they occur in a unique combination, namely 
$C_{\bar T T,1}+2C_{T\bar T}+2 A_{T\bar T}$. Hence, the rates of the above two modes are highly related; in fact, they are identical,
\be
\Gamma(B^-_c\to T^+_{c \bar s}  \overline T^{\,--}_{\bar c  s})=\Gamma(B^-_c\to T^0_{c\bar s} \overline T^{\,-}_{\bar c s }).
\en 
This can be easily understood. 
$B^-_c$ is an SU(3) singlet, the operator $(\bar c b)(\bar d u)$ belongs to an eight ${\bf 8}$. For flavor exotic $T$ and $\overline T$, they form ${\bf 6}\otimes\bar{\bf 6}$, in scenario I.
There is only one ${\bf 8}$ from ${\bf 6}\otimes\bar{\bf 6}={\bf 27}\oplus{\bf 8}\oplus {\bf 1}$, see Eq.~(\ref{eq: SU(3) decompositions 1}),
and, hence, there is only one combination that can match the SU(3) quantum numbers of the decaying particle $B^-_c$, the operator $(\bar c b)(\bar d u)$, and the flavor exotic final state $T$, $\overline T$.

If one of the $T$ or $\overline T$ is allowed to be non-flavor exotic, we have two additional modes in $\Delta S=0$ transitions and six modes in $\Delta S=-1$ transitions.

We now turn to scenario II.
This situation is rather different.
From Tables \ref{tab: BctoTT II 0} and \ref{tab: BctoTT II -1}, we see that, we now have eight modes in $\Delta S=0$ transitions with both flavor exotic $T$ and $\overline T$ involved,
and eight other modes in $\Delta S=-1$ transitions.
Interestingly, all flavor exotic states are involved in these decays.
Furthermore, the amplitudes of all these modes are linear combinations of $C^{(\prime)}_{\bar T T,1}$ and  $C^{(\prime)}_{T\bar T} + A^{(\prime)}_{T\bar T}$.
This is reasonable. 
Now the $T$ and $\overline T$ form $\overline{\bf 15}\otimes{\bf 15}$.
As there are two ${\bf 8}$ from 
$\overline{\bf 15}\otimes{\bf 15}$, see Eq.~(\ref{eq: SU(3) decompositions 1})
and, consequently, 
there are two combinations to match the SU(3) quantum numbers of the decaying particle $B^-_c$, the operator $(\bar c b)(\bar d u)$ or $(\bar c b)(\bar s u)$, and the flavor exotic final state $T$, $\overline T$.

The rates of these sixteen modes are related as follows,
\be
2\Gamma(B^-_c\to T^0_{c s} \overline T^{\,-}_{\bar c \bar s})
&=&2\Gamma(B^-_c\to T^-_{c s} \overline T^0_{\bar c \bar s})
=4\left|\frac{V_{ud}}{V_{us}}\right|^2\Gamma(B^-_c\to T^+_{c \bar s} \overline T^{\,--}_{\bar c s s})
\non\\
&=&2\left|\frac{V_{ud}}{V_{us}}\right|^2\Gamma(B^-_c\to T^0_{c \bar s} \overline T^{\,-}_{\bar c s s})
=6\left|\frac{V_{ud}}{V_{us}}\right|^2\Gamma(B^-_c\to T^+_{c} \overline  T^{\,--}_{\bar c s})
\non\\
&=&3 \left|\frac{V_{ud}}{V_{us}}\right|^2\Gamma(B^-_c\to T^0_{c} \overline T^{\,-}_{\bar c s })
=2 \left|\frac{V_{ud}}{V_{us}}\right|^2\Gamma(B^-_c\to T^-_{c} \overline T^0_{\bar c s}),
\en
\be
2\Gamma(B^-_c\to T^+_{c \bar s \bar s} \overline T^{\,--}_{\bar cs s})
&=&2\left|\frac{V_{ud}}{V_{us}}\right|^2\Gamma(B^-_c\to T^+_{c s} \overline T^{\,--}_{\bar c})
=3\left|\frac{V_{ud}}{V_{us}}\right|^2\Gamma(B^-_c\to T^0_{c s} \overline T^{\,-}_{\bar c})
\non\\
&=&6\left|\frac{V_{ud}}{V_{us}}\right|^2\Gamma(B^-_c\to T^-_{c s} \overline T^0_{\bar c}),
\en
\be
\Gamma(B^-_c\to T^+_{c \bar s}  \overline T^{\,--}_{\bar c  s})
=\Gamma(B^-_c\to T^0_{c\bar s} \overline T^{\,-}_{\bar c s }),
\en
and
\be
4\Gamma(B^-_c\to T^+_{c}  \overline T^{\,--}_{\bar c})
=3\Gamma(B^-_c\to T^0_{c} \overline T^{\,-}_{\bar c})
=4\Gamma(B^-_c\to T^{-}_{c} \overline T^0_{\bar c}).
\en
One can also work with triangle relations, as their amplitudes are highly related.
It will be interesting to search for these modes and verify the above relations, especially since all twelve flavor exotic states are involved in these decays.

Note that there is no ${\bf 8}$ in ${\bf 6}\times\overline{\bf 15}$ or $\overline{\bf 15}\times\bar{\bf 6}$, see Eq. (\ref{eq: SU(3) decompositions 3}), as they cannot have real irreducible representations.
Hence, even if flavor exotic states in scenarios I and II can mix, their $B^-_c\to T\overline T$ amplitudes can be obtained readily using those shown in this section.

The final state $T$ or $\overline T$ can further decay to $DP$ and $DS$ or $\bar DP$ and $\bar DS$, or they can connect to these states virtually. 
One can obtain their full amplitudes using results in this section and those in Sec.~\ref{sec: T to DP, DS}.

\section{Conclusion}

In this work, we study the decays of heavy mesons to these open-charmed tetraquark states ($T$) using a topological amplitude approach. We first obtain the $T\to DP$ and $DS$ strong decay amplitudes by decomposing them into several topological amplitudes, where $P$ is a light psedo-scalar particle and $S$ is a low-lying scalar particle. 
Subsequently, weak decay amplitudes of $\overline B\to D\overline T$, $\overline D T$ and $\overline B\to TP$, $TS$ decays are decomposed topologically.  
In addition, we also discuss $B^-_c\to T\overline T$ decays.
Our results can be extended to other modes by replacing $D$ and $P$ with $D^*$ and $V$, respectively. 
Furthermore, as our analysis is based on flavor, our result can also be applied to other open-charmed tetraquark states having the same flavor structure.
Our main results are:

\begin{table}[t!]
\caption{\label{tab: participation I}
Participation of flavor exotic states in scenario I in $\overline B\to D\overline T$, $\overline B\to \overline T$, $\overline B\to T P$, $\overline B\to T S$ and $B^-_c\to T \overline T$ decays in $\Delta S=0$ and $\Delta S=-1$ transitions. Those with parentheses are modes also involving non-flavor exotic states. These results and those in the Tables can be generalized with $D$ and $P$ replaced with $D^*$ and $V$, respectively.}
{
\begin{ruledtabular}
\begin{tabular}{lccccc}
Decays
          & $T^{++}_{c\bar s}$
          & $T^+_{c\bar s}$
          & $T^0_{c\bar s}$
          & $T^0_{c s}$
          & Tables
          \\
\hline
$\overline B\to D \overline T$, $\Delta S=-1$
          & $\checkmark$
          & $\checkmark$
          & $\checkmark$
          & {\sffamily X}
          & \ref{tab: BtoDTbar I}
          \\
$\overline B\to D \overline T$, $\Delta S=0$
          & $\checkmark$
          & $\checkmark$
          & {\sffamily X}
          & $\checkmark$
          & \ref{tab: BtoDTbar I}
          \\
\hline
$\overline B\to  \overline D T$, $\Delta S=-1$
          & {\sffamily X}
          & {\sffamily X}
          & {\sffamily X}
          & $\checkmark$
          & \ref{tab: BtoDbarT I}
          \\
$\overline B\to  \overline D T$, $\Delta S=0$
          &  {\sffamily X}
          & $\checkmark$
          & $\checkmark$
          & {\sffamily X}
          & \ref{tab: BtoDbarT I}
          \\                  
\hline
$\overline B\to T P$, $\Delta S=0$
          & {\sffamily X}
          & $\checkmark$
          & $\checkmark$
          & $\checkmark$
          & \ref{tab: BtoTP I}
          \\
$\overline B\to T P$, $\Delta S=-1$
          & {\sffamily X}
          & $\checkmark$
          & $\checkmark$
          & $\checkmark$
          & \ref{tab: BtoTP I}
          \\
\hline           
$\overline B\to  T S$, $\Delta S=0$
          & {\sffamily X}
          & $\checkmark$
          & $\checkmark$
          & $\checkmark$
          &  \ref{tab: BtoTS I 0}
          \\          
$\overline B\to T S$, $\Delta S=-1$
          &  {\sffamily X}
          & $\checkmark$
          & $\checkmark$
          & $\checkmark$
          &  \ref{tab: BtoTS I -1}
          \\                  
\hline           
$B^-_c\to T \overline T$, $\Delta S=0$
          & $\checkmark$
          & $\checkmark$
          & $\checkmark$
          & {\sffamily X}
          & \ref{tab: BctoTT I}
          \\
$B^-_c\to T \overline T$, $\Delta S=-1$
          &  $(\checkmark)$
          & $(\checkmark)$
          & {\sffamily X}
          & $(\checkmark)$
          & \ref{tab: BctoTT I}
          \\           
\end{tabular}
\end{ruledtabular}
}
\end{table}

\begin{table}[t!]
\caption{\label{tab: participation II}
Participation of flavor exotic states in scenario II in $\overline B\to D\overline T$, $\overline B\to \overline T$, $\overline B\to T P$, $\overline B\to T S$ and $B^-_c\to T \overline T$ decays in $\Delta S=0$ and $\Delta S=-1$ transitions. These results and those in the Tables can be generalized with $D$ and $P$ replaced with $D^*$ and $V$, respectively.}
\begin{ruledtabular}
\begin{tabular}{lccccccc}
Decays
          & $T^{++}_{c\bar s}$
          & $T^+_{c\bar s}$
          & $T^0_{c\bar s}$
          & $T^+_{c s}$
          & $T^0_{c s}$
          & $T^-_{c s}$
          & Tables
          \\
\hline
$\overline B\to D \overline T$, $\Delta S=-1$
          & $\checkmark$
          & $\checkmark$
          & $\checkmark$
          & {\sffamily X}
          & {\sffamily X}
          & {\sffamily X}
          & \ref{tab: BtoDTbar II}
          \\
$\overline B\to D \overline T$, $\Delta S=0$
          & $\checkmark$
          & $\checkmark$
          & {\sffamily X}
          & $\checkmark$
          & $\checkmark$
          & {\sffamily X}
          & \ref{tab: BtoDTbar II}
          \\
\hline
$\overline B\to  \overline D T$, $\Delta S=-1$
          & {\sffamily X}
          & {\sffamily X}
          & {\sffamily X}
          & $\checkmark$
          & $\checkmark$
          & $\checkmark$
          & \ref{tab: BtoDbarT II}
          \\
$\overline B\to  \overline D T$, $\Delta S=0$
          &  {\sffamily X}
          & $\checkmark$
          & $\checkmark$
          & {\sffamily X}
          & {\sffamily X}
          & {\sffamily X}
          & \ref{tab: BtoDbarT II}
          \\                  
\hline
$\overline B\to T P$, $\Delta S=0$
          & {\sffamily X}
          & $\checkmark$
          & $\checkmark$
          &  {\sffamily X}
          & $\checkmark$
          & $\checkmark$
          & \ref{tab: BtoTP II 0}
          \\
$\overline B\to T P$, $\Delta S=-1$
          & {\sffamily X}
          & $\checkmark$
          & $\checkmark$
          & $\checkmark$
          & $\checkmark$
          & $\checkmark$
          & \ref{tab: BtoTP II -1}
          \\
\hline           
$\overline B\to  T S$, $\Delta S=0$
          & {\sffamily X}
          & $\checkmark$
          & $\checkmark$
          & {\sffamily X}
          & $\checkmark$
          & $\checkmark$
          &  \ref{tab: BtoTS II 0}
          \\          
$\overline B\to T S$, $\Delta S=-1$
          &  {\sffamily X}
          & $\checkmark$
          & $\checkmark$
          & $\checkmark$
          & $\checkmark$
          & $\checkmark$
          &  \ref{tab: BtoTS II -1}
          \\                  
\hline           
$B^-_c\to T \overline T$, $\Delta S=0$
          & $\checkmark$
          & $\checkmark$
          & $\checkmark$
          & $\checkmark$
          & $\checkmark$
          & $\checkmark$
          & \ref{tab: BctoTT II 0}
          \\
$B^-_c\to T \overline T$, $\Delta S=-1$
          & $\checkmark$
          & $\checkmark$
          & $\checkmark$
          & $\checkmark$
          & $\checkmark$
          & $\checkmark$
          & \ref{tab: BctoTT II -1}
          \\           
\hline
\hline
Decays
          & $T^{++}_{c\bar s\bar s}$
          & $T^+_{c\bar s\bar s}$
          & $T^{++}_{c}$
          & $T^+_{c}$
          & $T^0_{c}$
          & $T^-_{c}$
          & Tables
          \\
\hline
$\overline B\to D \overline T$, $\Delta S=-1$
          & $\checkmark$
          & $\checkmark$
          & {\sffamily X}
          & {\sffamily X}
          & {\sffamily X}
          & {\sffamily X}
          & \ref{tab: BtoDTbar II}
          \\
$\overline B\to D \overline T$, $\Delta S=0$
          & {\sffamily X} 
          & {\sffamily X} 
          & $\checkmark$
          & $\checkmark$
          & $\checkmark$
          & {\sffamily X}
          & \ref{tab: BtoDTbar II}
          \\
\hline
$\overline B\to  \overline D T$, $\Delta S=-1$
          & {\sffamily X}
          & {\sffamily X}
          & {\sffamily X}
          & {\sffamily X}
          & {\sffamily X}
          & {\sffamily X}
          & \ref{tab: BtoDbarT II}
          \\
$\overline B\to  \overline D T$, $\Delta S=0$
          &  {\sffamily X}
          & $\checkmark$
          & {\sffamily X}
          & $\checkmark$
          & $\checkmark$
          & $\checkmark$
          & \ref{tab: BtoDbarT II}
          \\                  
\hline
$\overline B\to T P$, $\Delta S=0$
          & {\sffamily X}
          & $\checkmark$
          &  {\sffamily X}
          & $\checkmark$
          & $\checkmark$
          & $\checkmark$
          & \ref{tab: BtoTP II 0}
          \\
$\overline B\to T P$, $\Delta S=-1$
          & {\sffamily X}
          & {\sffamily X}
          & {\sffamily X}
          & $\checkmark$
          & $\checkmark$
          & $\checkmark$
          & \ref{tab: BtoTP II -1}
          \\
\hline           
$\overline B\to  T S$, $\Delta S=0$
          & {\sffamily X}
          & $\checkmark$
          &  {\sffamily X}
          & $\checkmark$
          & $\checkmark$
          & $\checkmark$
          &  \ref{tab: BtoTS II 0}
          \\          
$\overline B\to T S$, $\Delta S=-1$
          & {\sffamily X}
          & {\sffamily X}
          & {\sffamily X}
          & $\checkmark$
          & $\checkmark$
          & $\checkmark$
          &  \ref{tab: BtoTS II -1}
          \\                  
\hline           
$B^-_c\to T \overline T$, $\Delta S=0$
          & $\checkmark$
          & $\checkmark$
          & $\checkmark$
          & $\checkmark$
          & $\checkmark$
          & $\checkmark$
          & \ref{tab: BctoTT II 0}
          \\
$B^-_c\to T \overline T$, $\Delta S=-1$
          & $\checkmark$
          & $\checkmark$
          & $\checkmark$
          & $\checkmark$
          & $\checkmark$
          & $\checkmark$
          & \ref{tab: BctoTT II -1}
          \\           
\end{tabular}
\end{ruledtabular}
\end{table}

\begin{itemize}

\item We introduce two different scenarios of open-charmed tetraquarks $T_{cq\bar q\bar q}$, where in scenario~I the light antiquarks are antisymmetric, while in scenario II they are symmetric. 
Scenario I consists of a ${\bf 6}$ and a $\bar {\bf 3}$, while scenario II consists of a $\overline{\bf 15}$ and another $\bar {\bf 3}$.
Furthermore, $T_{cs 0} (2870)^0$, $T^*_{c\bar s 0}(2900)^0$ and $T^*_{c\bar s 0}(2900)^{++}$ can be in ${\bf 6}$ or $\overline {\bf 15}$, 
but can only in one of them at the same time.

\item In scenario I, we identify four flavor exotic states, namely $T^{++}_{c\bar s}$, $T^{+}_{c\bar s}$, $T^{0}_{c\bar s}$ and $T^{0}_{c s}$, where the first three are in the $S=+1$ isotriplet, while the last one is an $S=-1$ isosinglet, see Fig. \ref{fig: SU(3)} (a).
They are in ${\bf 6}$. 
 These four states are protected by their isospin and $S$ quantum numbers from mixing with other states in the scenario. 
 Note that $T^{+}_{c\bar s}$ cannot be identify with $T_{cs 0} (2870)^0$, $T^*_{c\bar s 0}(2900)^0$ or $T^*_{c\bar s 0}(2900)^{++}$ and it has not been considered a flavor exotic state before.
 It is new.

\item In scenario II, we identify twelve flavor exotic states, which are in 
the $S=+2$ isodoublet, $(T^{++}_{c\bar s\bar s},T^{+}_{c\bar s\bar s})$,
the $S=+1$ isotriplet, $(T^{++}_{c\bar s},T^{+}_{c\bar s},T^{0}_{c\bar s})$, 
the $S=0$ iso-quarplet, $(T_c^{++},T_c^{+},T_c^{0},T_c^{-})$,
and
the $S=-1$ isotriplet, $(T^{+}_{c s},T^{0}_{c s},T^{-}_{c s})$,  
see Fig.~\ref{fig: SU(3)}~(b). 
They are in $\overline{\bf 15}$.
These twelve states are protected by their isospin and $S$ quantum numbers from mixing with other states in the scenario. 
Except $T^{++}_{c\bar s}$, $T^{0}_{c\bar s}$ and $T^{0}_{c s}$, all other states cannot be identify with $T_{cs 0} (2870)^0$, $T^*_{c\bar s 0}(2900)^0$ or $T^*_{c\bar s 0}(2900)^{++}$.
Furthermore, $T^{+}_{c\bar s}$,  and $T_c^{+}$ and $T_c^{0}$ have not been considered as flavor exotic states before.

\item We decompose $T\to DP$ and $DS$ decay amplitudes into topological amplitudes, where three to four topological amplitudes are needed.
However, only one topological amplitude, namely the fall-apart amplitude $F_{DP}$, is needed in $T\to DP$ decays for flavor exotic $T$ in scenarios I or II. Hence, their amplitudes are highly related. The situation in $T\to DS$ decays is similar, but with $F_{DP}$ replaced with $2C_S+C_{DS}$ in scenario I and $C_{DS}$ in scenario II.
Although some of the modes may be kinematically forbidden, the information about their amplitudes is still useful, as the process can be a part of a larger diagram.
The decomposition of decay amplitudes in terms of topological amplitudes is unaffected by rescattering. 
For illustration, we consider the effect of quasi-elastic rescattering on $T\to DP$ decays.

\item We decompose $\overline B\to D\overline T$ and $\overline B\to \overline D T$ decay amplitudes in $\Delta S=-1$ and $0$ transitions into topological amplitudes, where six topological amplitudes are needed in each case.
However, only $C_{\overline T}+P_{\bar T,1}$ or $C_{\overline D T}+P_{T}$, are needed in $\overline B\to D\overline T$ or $\overline B\to \overline D T$ decays for modes involving flavor exotic $T$. Hence, their amplitudes are highly related. 
As there are many flavor exotic states in scenario II and the amplitudes of their production are highly related, the continued unobservation of these flavor exotic states in $\overline B\to D\overline T$ and $\overline B\to \overline D T$ decays poses tension in this scenario.

\item
A doubly charged open-charmed tetraquark $T^{++}$ cannot be produced in $\overline B\to\overline  D T$ decays, as such a decay violates charge conservation.
In scenario I, one needs to consider both $\Delta S=-1$ and $\Delta S=0$ transitions in $\overline B\to D\overline T$ decays to have all four flavor exotic states participate. 
This is important as we need to verify whether $T_{cs 0} (2870)^0$, $T^*_{c\bar s 0}(2900)^0$ and $T^*_{c\bar s 0}(2900)^{++}$ are in the same multiplet or not.
In scenario II, we also need to consider both $\Delta S=-1$ and $\Delta S=0$ transitions in $\overline B\to D\overline T$ decays.
Note that the $\overline B{}_s^0\to D^0\overline {T} ^0_{\bar c\bar s}$ decay involving $\overline{T^*_{c s 0}}(2870)^0$ has not been observed yet, as it is a CKM suppressed $\Delta S=-1$ transition.
It will be interesting and useful to search for this mode to complete the above verification.
It should be noted that there is penguin pollution in the above programs. 
Although the effects on rate ratios are estimated to be mild, further works are needed for verification.
We also note that the $\overline B^0\to \overline D^0 T^0_{c s}$ decay has not been observed yet. 
It will be useful to search for this mode and check the relations on rates. 

\item The amplitudes of $\overline B\to T P$ and $\overline B\to T S$ decays in $\Delta S=0$ and $-1$ transitions are decomposed into topological amplitudes, where we need seven (seven) and eleven (nine) independent topological amplitudes, respectively, in scenario I (II).
The number of topological amplitudes contributing to modes with flavor exotic states is reduced considerably.
In $\overline B\to T P$ with flavor exotic $T$ and $P=\pi, K$, only three (four) independent combinations of topological amplitudes are needed in scenario I (II), resulting in several relations on rates of these modes.
Similarly, in $\overline B\to T S$ with flavor exotic $T$ and $S=a_0, \kappa$, only three (four) independent combinations of topological amplitudes are needed in scenario I (II), giving several rate relations. 
Although the doubly charged $T^*_{c\bar s 0}(2900)^{++}$ cannot be produced in $\overline B\to T P$ nor $\overline B\to T S$ decays,
there are relations that can be used to verify whether $T_{cs 0} (2870)^0$ and $T^*_{c\bar s 0}(2900)^0$ are in the same multiplets or not.
According to the OZI rule, the $\overline B\to T S$ topological amplitudes are expected to receive some suppressions.

\item We note that $B^-_c\to T\overline T$ decays are kinematically allowed. Interestingly, doubly charged $T$ can be produced in these decays.
The $B^-_c\to T\overline T$ decay amplitudes can be decomposed into seven topological amplitudes, while only three of them contribute to modes with flavor exotic $T$ and $\overline T$.
In scenario I, these three topological amplitudes only occur in one combination for modes with flavor exotic $T$ and $\overline T$ ; these amplitudes are all related.
In scenario II, all flavor exotic states participate in these decays.
According to the OZI rule, the topological amplitudes in $B_c^-\to T \overline T$ decays are expected to have some suppressions.

\item A summary of the participation of flavor exotic states in scenario I in $\overline B\to D\overline T$, $\overline B\to \overline T$, $\overline B\to T P$, $\overline B\to T S$ and $B^-_c\to T \overline T$ decays in $\Delta S=0$ and $\Delta S=-1$ transitions is shown in Table~\ref{tab: participation I}, while the summary for scenario II is in Table~\ref{tab: participation II}.

\end{itemize}


\begin{acknowledgments}
This work is supported in part by
the National Science and Technology Council of R.O.C.
under Grant No NSTC-112-2112-M-033-006 and NSTC-113-2112-M-033-004.
\end{acknowledgments}

\appendix

\section{SU(3) transformation of $q\bar q'\bar q''$ states and SU(3) decompositions}\label{app: SU(3)}

It is well known that $q$ is the fundamental representation of SU(3), while $\bar q$ is the complex representation.
Hence, the generators of SU(3) transformation in $q\bar q'\bar q''$ space is given by
\be
({\cal T}_a)_i^{i'} {}^j_{j'} {}^k_{k'}
=\left(\frac{{\lambda}_a}{2}\right)_i^{i'}
   \otimes {\mathbb {I}}^j_{j'}
   \otimes{\mathbb {I}}^k_{k'}
-{\mathbb {I}}_i^{i'} 
   \otimes\left(\frac{{\lambda}^*_a}{2}\right)^j_{j'}
   \otimes {\mathbb {I}}^k_{k'}
-{\mathbb {I}}_i^{i'} 
   \otimes {\mathbb {I}}^j_{j'}
   \otimes \left(\frac{{\lambda}^*_a}{2}\right)^k_{k'},
\label{eq: generator}
\en
with $a=1,2,\dots, 8$, ${\lambda}_a$ the Gell-mann matrix and ${\mathbb {I}}$ a $3\times 3$ unity matrix, see \cite{Lee:1981mf, Georgi:2000vve} for example.
Note that we use a field convention, where transformations act on quark or anti-quark fields. 
Equivalently, we consider transformations on states in bra instead of in ket.
For simplicity and convenience, we do not always distinguish carefully between these two statements, even though when the state is in ket.
One should be able to make the necessary ``translation" if needed. 

From the above equation, it is evident that we have
\be
({\cal T}_a)_i^{i'} {}^j_{j'} {}^k_{k'}=({\cal T}_a)_i^{i'} {}^k_{k'} {}^j_{j'},
\en
and, for a traceless combination $V_i{}^{j k}$, satisfying $V_i{}^{i k}=V_{i}{}^{k i}=0$, 
\be
V_{i'}{}^{i' k}=({\cal T}_a)^i_{i'} {}_j^{i'} {}_k^{k'} V_i{}^{j k}=0,
\quad
V_{i'}{}^{k' i'}=({\cal T}_a)^i_{i'} {}_j^{j'} {}_k^{i'} V_i{}^{j k}=0.
\en
Consequently, SU(3) transformation does not mix up $\bar q'\bar q''$ in different permutation symmetry, 
nor does it mix up traceless $q\bar q'\bar q''$ combinations with the traceful combinations.
Therefore, we can decompose $\bar q'\bar q''$ into $[\bar q'\bar q'']$ and $\{\bar q'\bar q''\}$ at the outset.
Hence,
we have scenario I and scenario II, where scenario I has $T=T_{cq[\bar q'\bar q'']}$, while scenario II has $T=T_{cq\{\bar q'\bar q''\}}$,
and they can be further separated into traceless and traceful parts.

Explicitly, for the generators, we have
\be
2{\cal T}_3
&=&
\left(
\begin{array}{ccc}
1 &0 &0\\
0 &-1 & 0\\
0 &0 & 0\\
\end{array}
\right)
\otimes {\mathbb {I}}\otimes {\mathbb {I}}
- {\mathbb {I}}
\otimes
\left(
\begin{array}{ccc}
1 &0 &0\\
0 &-1 & 0\\
0 &0 & 0\\
\end{array}
\right)
\otimes {\mathbb {I}}
- {\mathbb {I}}
\otimes {\mathbb {I}}
\otimes
\left(
\begin{array}{ccc}
1 &0 &0\\
0 &-1 & 0\\
0 &0 & 0\\
\end{array}
\right),
\non\\
2\sqrt3 {\cal T}_8
&=&
\left(
\begin{array}{ccc}
1 &0 &0\\
0 &1 & 0\\
0 &0 & -2\\
\end{array}
\right)
\otimes {\mathbb {I}}\otimes {\mathbb {I}}
- {\mathbb {I}}
\otimes
\left(
\begin{array}{ccc}
1 &0 &0\\
0 &1 & 0\\
0 &0 & -2\\
\end{array}
\right)
\otimes {\mathbb {I}}
- {\mathbb {I}}
\otimes {\mathbb {I}}
\otimes
\left(
\begin{array}{ccc}
1 &0 &0\\
0 &1 & 0\\
0 &0 & -2\\
\end{array}
\right),
\en
\begin{table}[t!]

\caption{\label{tab: SU(3) trans scenario I}
The results of SU(3) generators acting on states in scenario I with quark contents given in Table \ref{tab: 6+3bar}.  
They are in their own irreducible representations after these transformations.
Comparing this table with Fig. \ref{fig: SU(3)},
one can verify that these transformations indeed follow Eq.~(\ref{eq: Ti pm iTj}).
}
\footnotesize{
\begin{ruledtabular}
\begin{tabular}{lccccccccc}
$T_c$
          & ${\cal I}_3$
          & ${\cal S}$
          & ${\cal T}_1+i {\cal T}_2$
          & ${\cal T}_1-i {\cal T}_2$
          & ${\cal T}_4+i {\cal T}_5$
          & ${\cal T}_4-i {\cal T}_5$
          & ${\cal T}_6+i {\cal T}_7$
          & ${\cal T}_6-i {\cal T}_7$
          \\
\hline 
$T^{++}_{c\bar s}$
          & $+T^{++}_{c\bar s}$
          & $+T^{++}_{c\bar s}$
          & 0
          & $-\sqrt 2 T^+_{c\bar s}$
          & 0
          & $\sqrt2 T_c^+$
          & 0
          & 0
          \\
$T^{+}_{c\bar s}$
          & $0$ 
          & $+T^{+}_{c\bar s}$ 
          & $-\sqrt2 T^{++}_{c\bar s}$
          & $\sqrt2 T^0_{c\bar s}$
          & 0
          & $-T^0_c$
          & 0
          & $- T^+_c$
          \\          
$T^{0}_{c\bar s}$
          & $-T^0_{c\bar s}$
          & $+T^0_{c\bar s}$
          & $\sqrt 2 T_{c\bar s}^+$
          & 0
          & 0
          & $0$
          & 0
          & $-\sqrt2 T^0_c$
          \\     
\hline
$T^{0}_{cs}$
          & $0$
          &$-T^{0}_{cs}$
          & 0
          & 0
          & $\sqrt 2 T^+_c$ 
          & 0
          & $\sqrt 2 T^0_c$
          & 0
          \\
\hline
$T^{+}_c$
          & $+\frac{1}{2}T^{+}_c$
          & 0
          & 0
          & $T_c^0$
          & $\sqrt2 T^{++}_{c\bar s}$
          & $\sqrt2 T^0_{c s}$
          & $-T_{c\bar s}^+$
          & 0
          \\ 
$T^{0}_c$
          & $-\frac{1}{2}T^{0}_c$
          & 0
          & $T^{+}_c$
          & 0
          & $-T^{+}_{c\bar s}$
          & 0
          & $-\sqrt{2} T_{c\bar s}^0$
          & $\sqrt{2}T_{c s}^0$
          \\       
\hline
\hline
$T^{\prime\prime +}_{c}$
          & $+\frac{1}{2}T^{\prime\prime +}_{c}$
          & 0
          & 0
          & $-T^{\prime\prime 0}_c$
          & 0
          & 0
          & $-T^{\prime\prime +}_{c\bar s}$
          & 0
          \\                                                                                  
$T^{\prime \prime 0}_{c}$
          & $-\frac{1}{2} T^{\prime \prime 0}_{c}$
          & 0
          & $-T^{\prime\prime +}_c$
          & 0
          & $-T^{\prime\prime +}_{c\bar s}$
          & 0
          & 0
          & 0
          \\ 
\hline                                       
$T^{\prime \prime+}_{c\bar s}$
          & 0
          & $+T^{\prime \prime+}_{c\bar s}$
          & 0
          & 0
          & 0
          & $-T^{\prime\prime 0}_c$
          & 0
          & $-T^{\prime\prime +}_c$
          \\  
\end{tabular}
\end{ruledtabular}
}
\end{table}
\be
{\cal T}_1+i {\cal T}_2
&=&
\left(
\begin{array}{ccc}
0 &1 &0\\
0 &0 & 0\\
0 &0 & 0\\
\end{array}
\right)
\otimes {\mathbb {I}}\otimes {\mathbb {I}}
- {\mathbb {I}}
\otimes
\left(
\begin{array}{ccc}
0 &0 &0\\
1 &0 & 0\\
0 &0 & 0\\
\end{array}
\right)
\otimes {\mathbb {I}}
- {\mathbb {I}}
\otimes {\mathbb {I}}
\otimes
\left(
\begin{array}{ccc}
0 &0 &0\\
1 &0 & 0\\
0 &0 & 0\\
\end{array}
\right),
\non\\
{\cal T}_4+i {\cal T}_5
&=&
\left(
\begin{array}{ccc}
0 &0 &1\\
0 &0 & 0\\
0 &0 & 0\\
\end{array}
\right)
\otimes {\mathbb {I}}\otimes {\mathbb {I}}
- {\mathbb {I}}
\otimes
\left(
\begin{array}{ccc}
0 &0 &0\\
0 &0 & 0\\
1 &0 & 0\\
\end{array}
\right)
\otimes {\mathbb {I}}
- {\mathbb {I}}
\otimes {\mathbb {I}}
\otimes
\left(
\begin{array}{ccc}
0 &0 &0\\
0 &0 & 0\\
1 &0 & 0\\
\end{array}
\right),
\non\\
{\cal T}_6+i {\cal T}_7
&=&
\left(
\begin{array}{ccc}
0 &0 & 0\\
0 &0 &1\\
0 &0 & 0\\
\end{array}
\right)
\otimes {\mathbb {I}}\otimes {\mathbb {I}}
- {\mathbb {I}}
\otimes
\left(
\begin{array}{ccc}
0 &0 &0\\
0 &0 & 0\\
0 &1 & 0\\
\end{array}
\right)
\otimes {\mathbb {I}}
- {\mathbb {I}}
\otimes {\mathbb {I}}
\otimes
\left(
\begin{array}{ccc}
0 &0 &0\\
0 &0 & 0\\
0 &1 & 0\\
\end{array}
\right),
\en
and their hermitian conjugations,
while generators for isospin ${\cal I}_3$ and stangeness ${\cal S}$ can be defined as 
\be
{\cal I}&=&{\cal T}_3,
\non\\
{\cal S}&\equiv& {\mathbb {I}}\otimes {\mathbb {I}} \otimes{\mathbb {I}}+2\sqrt3 {\cal T}_8
\non\\
&=& -{\mathbb {I}}\otimes {\mathbb {I}} \otimes{\mathbb {I}}
+{\mathbb {I}}\otimes {\mathbb {I}} \otimes{\mathbb {I}}
+{\mathbb {I}}\otimes {\mathbb {I}} \otimes{\mathbb {I}}+2\sqrt3 {\cal T}_8
\non\\
&=&
\left(
\begin{array}{ccc}
0 &0 &0\\
0 &0 & 0\\
0 &0 & -1\\
\end{array}
\right)
\otimes {\mathbb {I}}\otimes {\mathbb {I}}
- {\mathbb {I}}
\otimes
\left(
\begin{array}{ccc}
0 &0 &0\\
0 &0 & 0\\
0 &0 & -1\\
\end{array}
\right)
\otimes {\mathbb {I}}
- {\mathbb {I}}
\otimes {\mathbb {I}}
\otimes
\left(
\begin{array}{ccc}
0 &0 &0\\
0 &0 & 0\\
0 &0 & -1\\
\end{array}
\right).
\en  
These generators are responsible for generating SU(3) transformation.
In particular, they will move states with quantum numbers $(I_3, S)$ in an irreducible multiplet to other states in the same multiplet in the following way:
\be
{\cal T}_1\pm i {\cal T}_2&:& (I_3, S)\to (I_3\pm 1, S),
\non\\
{\cal T}_4\pm i {\cal T}_5&:& (I_3, S)\to (I_3\pm 1, S\pm 1),
\non\\
{\cal T}_6\pm i {\cal T}_7&:& (I_3, S)\to (I_3\mp 1, S\pm 1).
\label{eq: Ti pm iTj} 
\en
These generators are known as the raising and lowering operators of isospin, $V$-spin and $U$-spin, respectively.
Note that if the transformed state does not exist in the multiplet, the result of that transformation will end up in a null state.

\begin{table}[t!]

\caption{\label{tab: SU(3) trans scenario II}
Same as Table. \ref{tab: SU(3) trans scenario I} but for scenario II with states defined in Table \ref{tab: 15bar+3bar}.
}
\footnotesize{
\begin{ruledtabular}
\begin{tabular}{lccccccccc}
$T_c$
          & ${\cal I}_3$
          & ${\cal S}$
          & ${\cal T}_1+i {\cal T}_2$
          & ${\cal T}_1-i {\cal T}_2$
          & ${\cal T}_4+i {\cal T}_5$
          & ${\cal T}_4-i {\cal T}_5$
          & ${\cal T}_6+i {\cal T}_7$
          & ${\cal T}_6-i {\cal T}_7$
          \\
\hline 
$T^{++}_{c\bar s}$
          & $+T^{++}_{c\bar s}$
          & $+T^{++}_{c\bar s}$
          & 0
          & $-\sqrt 2 T^+_{c\bar s}$
          & 0
          & $-\frac{\sqrt2 T_c^++2 T^{\prime +}_c}{\sqrt3}$
          & $-\sqrt2 T^{++}_{c\bar s\bar s}$
          & $-\sqrt2 T_c^{++}$
          \\
$T^{+}_{c\bar s}$
          & $0$ 
          & $+T^{+}_{c\bar s}$ 
          & $-\sqrt2 T^{++}_{c\bar s}$
          & $\sqrt2 T^0_{c\bar s}$
          & $-T^{++}_{c\bar s\bar s}$
          & $-\frac{2 T^0_c+\sqrt2 T^{\prime 0}_c}{\sqrt3}$
          & $T_{c\bar s\bar s}^+$
          & $\frac{-2 T^+_c+\sqrt2 T^{\prime +}_c}{\sqrt3}$
          \\          
$T^{0}_{c\bar s}$
          & $-T^0_{c\bar s}$
          & $+T^0_{c\bar s}$
          & $\sqrt 2 T_{c\bar s}^+$
          & 0
          & $-\sqrt 2 T_{c\bar s\bar s}^+$
          & $-\sqrt 2 T_{c}^-$
          & 0
          & $\frac{\sqrt 2 T^0_c-2 T^{\prime 0}_c}{\sqrt3}$
          \\     
\hline
$T^{+}_{cs}$
          & $+T^{+}_{cs}$
          & $-T^{+}_{cs}$
          & $0$
          & $-\sqrt2 T^0_{c s}$
          & $T^{++}_c$
          & 0           
          & $\frac{-T^+_c+2\sqrt2 T^{\prime +}_c}{\sqrt 3}$
          & 0
          \\
$T^{0}_{cs}$
          & $0$
          &$-T^{0}_{cs}$
          & $-\sqrt2 T^+_{cs}$
          & $-\sqrt2 T^-_{cs}$
          & $\frac{\sqrt 2 T^+_c+2T^{\prime +}_c}{\sqrt3} $ 
          & 0
          & $\frac{-\sqrt 2 T^0_c+2 T^{\prime 0}_c}{\sqrt3}$
          & 0
          \\
$T^{-}_{cs}$
          & $-T^{-}_{cs}$
          & $-T^{-}_{cs}$
          & $-\sqrt2 T^0_{c s}$
          & $0$
          & $\frac{T^0_c+2\sqrt2 T^{\prime 0}_c}{\sqrt3}$
          & $0$
          & $T^-_c$
          & 0 
          \\
\hline
$T^{++}_{c\bar s\bar s}$
          & $+\frac{1}{2} T^{++}_{c\bar s\bar s}$
          & $+2 T^{++}_{c\bar s\bar s}$
          & 0
          & $T^{+}_{c\bar s\bar s}$
          & 0
          & $-T^+_{c\bar s}-\sqrt2 T^{\prime +}_{c\bar s}$
          & 0
          & $-\sqrt2 T^{++}_{c\bar s}$
          \\
$T^{+}_{c\bar s\bar s}$
          & $-\frac{1}{2}T^{+}_{c\bar s\bar s}$
          & $+2 T^{+}_{c\bar s\bar s}$
          & $T^{++}_{c\bar s\bar s}$
          & 0
          & 0
          & $-\sqrt2 T^0_{c\bar s}$
          & 0
          & $T^+_{c\bar s}-\sqrt2 T^{\prime +}_{c\bar s}$
          \\
\hline
$T^{++}_c$
          & $+\frac{3}{2}T^{++}_c$
          & 0
          & 0
          & $-\sqrt3 T^+_c$
          & 0
          & $T^+_{cs}$
          & $-\sqrt2 T^{++}_{c\bar s}$
          & 0
          \\
$T^{+}_c$
          & $+\frac{1}{2}T^{+}_c$
          & 0
          & $-\sqrt3 T^{++}_c$
          & $-2 T_c^0$
          & $-\sqrt{\frac{2}{3}}T^{++}_{c\bar s}$
          & $\sqrt{\frac{2}{3}}T^0_{c s}$
          & $-\frac{2 T_{c\bar s}^+}{\sqrt3}$
          & $-\frac{T_{c s}^+}{\sqrt3}$
          \\ 
$T^{0}_c$
          & $-\frac{1}{2}T^{0}_c$
          & 0
          & $-2 T^{+}_c$
          & $\sqrt3 T_c^-$
          & $-\frac{2T^{+}_{c\bar s}}{\sqrt3}$
          & $\frac{T^-_{c s}}{\sqrt3}$
          & $\sqrt{\frac{2}{3}} T_{c\bar s}^0$
          & $-\sqrt{\frac{2}{3}}T_{c s}^0$
          \\       
$T^{-}_c$
          & $-\frac{3}{2} T^{-}_c$
          & 0
          & $\sqrt3 T^0_c$
          & 0
          & $-\sqrt2 T^0_{c\bar s}$
          & 0
          & 0
          & $T^-_{cs}$
          \\  
\hline  
$T^{\prime +}_c$
          & $+\frac{1}{2}T^{\prime +}_c$
          & 0
          & 0
          & $-T^{\prime 0}_c$
          & $-\frac{2 T_{c\bar s}^{++}}{\sqrt3}$
          & $\frac{2 T_{cs}^0}{\sqrt3}$
          & $\frac{\sqrt2 T_{c\bar s}^+-3 T^{\prime +}_{c\bar s}}{\sqrt3}$
          & $2\sqrt{\frac{2}{3}}T_{cs}^+$
          \\   
$T^{\prime 0}_c$
          & $-\frac{1}{2}T^{\prime 0}_c$
          & 0
          & $-T^{\prime +}_c$
          & 0
          & $-\frac{\sqrt2 T_{c\bar s}^+ +3 T^{\prime +}_{c\bar s}}{\sqrt3}$
          & $2\sqrt{\frac{2}{3}}T_{cs}^-$
          & $-\frac{2 T_{c\bar s}^{0}}{\sqrt3}$
          & $\frac{2 T_{cs}^0}{\sqrt3}$
           \\ 
\hline
$T^{\prime +}_{c\bar s}$
          & 0
          & $+T^{\prime +}_{c\bar s}$
          & 0
          & 0
          & $-\sqrt2 T_{c\bar s\bar s}^{++}$
          & $-\sqrt3 T^{\prime 0}_c$
          & $-\sqrt2 T_{c\bar s\bar s}^+$
          & $-\sqrt3 T^{\prime +}_c$
          \\ 
\hline
\hline
$T^{\prime\prime +}_{c}$
          & $+\frac{1}{2}T^{\prime\prime +}_{c}$
          & 0
          & 0
          & $-T^{\prime\prime 0}_c$
          & 0
          & 0
          & $-T^{\prime\prime +}_{c\bar s}$
          & 0
          \\                                                                                  
$T^{\prime \prime 0}_{c}$
          & $-\frac{1}{2} T^{\prime \prime 0}_{c}$
          & 0
          & $-T^{\prime\prime +}_c$
          & 0
          & $-T^{\prime\prime +}_{c\bar s}$
          & 0
          & 0
          & 0
          \\ 
\hline                                       
$T^{\prime \prime+}_{c\bar s}$
          & 0
          & $+T^{\prime \prime+}_{c\bar s}$
          & 0
          & 0
          & 0
          & $-T^{\prime\prime 0}_c$
          & 0
          & $-T^{\prime\prime +}_c$
          \\  
\end{tabular}
\end{ruledtabular}
}
\end{table}

As shown in Tables~\ref{tab: SU(3) trans scenario I} and \ref{tab: SU(3) trans scenario II},
using the above generators the states in ${\bf 6}$ and $\bar{\bf 3}$ in scenario I, 
given in Table~\ref{tab: 6+3bar}, and states in $\overline{\bf 15}$ and $\bar{\bf 3}'$ in scenario II, 
given in Table~\ref{tab: 15bar+3bar}, are in their own irreducible representations after SU(3) transformations. 
By comparing Tables~\ref{tab: SU(3) trans scenario I} and \ref{tab: SU(3) trans scenario II} and Fig. \ref{fig: SU(3)},
one can also verify that these transformations indeed follow Eq. (\ref{eq: Ti pm iTj} ).
Furthermore, the isospin multiplets within those irreducible representations can be verified using ${\cal I}_\pm={\bf \cal T}_1\pm i {\bf \cal T}_2$ and ${\cal I}^2$ with
\be
{\cal I}^2=({\cal I}_3)^2+\frac{1}{2} ({\cal T}_1+i {\cal T}_2)({\cal T}_1-i {\cal T}_2)+\frac{1}{2} ({\cal T}_1-i {\cal T}_2)({\cal T}_1+i {\cal T}_2).
\en 
This quantum number can be obtained easily using the results in Tables~\ref{tab: SU(3) trans scenario I} and \ref{tab: SU(3) trans scenario II}. 
The final results are shown in Tables~\ref{tab: 6+3bar} and~\ref{tab: 15bar+3bar}.


When an SU(3) decomposition is needed, we follow the method in ref.~\cite{Coleman:1965afp}.
For example, it is easy to obtain
\be
\bar {\bf 3}\otimes \bar{\bf 3}=\bar {\bf 6}\oplus{\bf 3},
\quad
{\bf 3}\otimes\bar{\bf 6}=\overline{\bf 15}\oplus\bar{\bf 3},
\quad
{\bf 3}\otimes{\bf 3}={\bf 6}\oplus\bar{\bf 3},
\label{eq: SU(3) decompositions 0}
\en
\be
\bar {\bf 3}\otimes {\bf 8}
&=&\overline{\bf 15}\oplus{\bf 6}\oplus\bar{\bf 3},
\quad
\bar{\bf 3}\otimes {\bf 1}=\bar{\bf 3},
\non\\
{\bf 6}\otimes\bar{\bf 6}&=&{\bf 27}\oplus{\bf 8}\oplus{\bf 1},
\non\\
\overline{\bf 15}\otimes{\bf 15}
&=&{\bf 64}\oplus{\bf 35}\oplus\overline{\bf 35}\oplus{\bf 27}\oplus{\bf 27}
\non\\
&&\oplus{\bf 10}\oplus\overline{\bf 10}\oplus{\bf 8}\oplus{\bf 8}\oplus {\bf 1},
\label{eq: SU(3) decompositions 1}
\en
\be
{\bf 6}\otimes{\bf 8}&=&{\bf 24}\oplus\overline {\bf 15}\oplus{\bf 6}\oplus\bar{\bf 3},
\non\\
\overline{\bf 15}\otimes{\bf 8}
&=&\overline {\bf 42}\oplus{\bf 24}\oplus\overline{\bf 15}'\oplus\overline {\bf 15}\oplus\overline {\bf 15}\oplus{\bf 6}\oplus\bar {\bf 3},
\label{eq: SU(3) decompositions 2}
\en
and
\be
{\bf 6}\otimes\overline{\bf 15}
&=&{\bf 42}\oplus\overline{\bf 24}\oplus{\bf 15}\oplus\bar{\bf 6}\oplus{\bf 3},
\non\\
\bar{\bf 6}\otimes{\bf 15}
&=&\overline{\bf 42}\oplus{\bf 24}\oplus\overline{\bf 15}\oplus{\bf 6}\oplus\bar{\bf 3},
\label{eq: SU(3) decompositions 3}
\en
which are used in the text. Note that the $\overline {\bf 15}'$ in Eq. (\ref{eq: SU(3) decompositions 2}) is a $D(0,4)$, which is different from the $D(1,2)$, the $\overline {\bf 15}$ used extensively in this work.

\section{Quasi-elastic strong rescattering in the $DP$ system}\label{sec: FSI}

In this Appendix, we want to demonstrate that the form of $T\to DP$ decay amplitudes in terms of topological amplitudes does not change in the presence of quasi-elastic strong rescattering, whereas the topological amplitudes may contain the effect of rescattering. 

In the presence of quasi-elastic strong rescattering, the $T\to DP$ decay amplitude, $A_i$, is related to the amplitude before rescattering, $A^0_j$, in the following way \cite{Chua:2001br, Chua:2005dt,Chua:2007qw},
\be
A_i={\cal S}^{1/2}_{ij} A^0_j,
\en
where ${\cal S}^{1/2}$ is given by
\be
{\cal S}^{1/2}&=&e^{i\delta_{\overline{15}}}\sum_{a=1}^{15} |DP(\overline {\bf 15});a\rangle\langle DP(\overline {\bf 15});a|
+e^{i\delta_{{6}}}\sum_{b=1}^{6} |DP({\bf 6});b\rangle\langle DP({\bf 6});b|
\non\\
&&+\sum_{m,n=\overline{\bf 3},\overline{\bf 3}{}^\prime}\sum_{c=1}^{3} 
|DP(m);c\rangle
\,
U^{1/2}_{mn}
\,
\langle DP(n);c|,
\label{eq: S1half}
\en
with
\begin{equation}
U^{1/2}=\left(
\begin{array}{cc}
\cos\tau
       &\sin\tau
       \\
-\sin\tau
       &\cos\tau
\end{array}
\right)
\left(
\begin{array}{cc}
e^{i\delta_{\bar 3}}
       &0
       \\
0
       &e^{i\delta_{{\bar 3}'}}
\end{array}
\right)
\left(
\begin{array}{cc}
\cos\tau
       &-\sin\tau
       \\
\sin\tau
       &\cos\tau
\end{array}
\right). \label{eq:U}
\end{equation}
In the above equation, 
$|DP(\overline{\bf 15}),a\ra$, $|DP({\bf 6}),b\rangle$, $|DP(\overline {\bf 3}),c\rangle$ and 
$|DP(\overline {\bf 3}'),c\rangle$
are $DP$ final state in SU(3) $\overline{\bf 15}$, ${\bf 6}$,  ${\bf 3}$ and ${\bf 3}'$ multiplets, respectively.

Explicitly, for $|DP(\overline{\bf 15}),a\ra$ we have
\begin{eqnarray}
(S=1,\,I=1)
&:&
\frac{|D^+K^+\rangle+|D_s^+\pi^+\rangle}{\sqrt2},
\quad
\frac{|D^+K^0\rangle-|D^0K^+\rangle-\sqrt2 |D^+_s\pi^0\rangle}{2},
\non\\
&&\frac{|D^0K^0\rangle+|D_s^+\pi^-\rangle}{\sqrt2};
\non\\
(S=1,\,I={1\over2})
&:&|D_s^+K^+\rangle,\quad|D_s^+K^0\rangle;
\non\\
(S=1,\,I=0)
&:&
\frac{|D^+K^0\rangle+|D^0K^+\rangle+\sqrt6 |D_s^+\eta_8\rangle}{2\sqrt2};
\\
(S=0,\,I={3\over2})
&:&|D^+\pi^+\rangle,\quad
{1\over\sq3}|D^0\pi^+\rangle+\sqrt{2\over3}|D^+\pi^0\rangle,\quad
{1\over\sqrt3}|D^+\pi^-\rangle-\sqrt{2\over3}|D^0\pi^0\rangle,
\non\\
&&
|D^0\pi^-\rangle;
\non\\
(S=0,\,I={1\over2})
&:&\frac{2|D^0\pi^+\rangle-\sqrt2 |D^+\pi^0\rangle+3\sqrt6|D^+\eta_8\rangle
     -6|D^+_s\overline K{}^0\rangle}{4\sqrt6},
\non\\
&&\frac{2|D^+\pi^-\rangle+\sqrt2 |D^0\pi^0\rangle+3\sqrt6|D^0\eta_8\rangle
     -6|D^+_sK^-\rangle}{4\sqrt6};
\non\\
(S=-1,\,I=1)&:&|D^+\overline K{}^0\rangle,
\quad
\frac{|D^+K^-\rangle+|D^0\overline K{}^0\rangle}{\sqrt2},
\quad
|D^0K^-\rangle.
\end{eqnarray}
Similarly $|DP({\bf 6}),b\rangle$ are
\begin{eqnarray}
(S=1,\,I=1)
&:&\frac{|D^+K^+\rangle-|D_s^+\pi^+\rangle}{\sqrt2},\quad
   \frac{|D^+K^0\rangle-|D^0K^+\rangle+\sqrt2|D^+_s\pi^0\rangle}{2},
   \non\\
&&
\frac{|D^0K^0\rangle-|D_s^+\pi^-\rangle}{\sqrt2};
\non\\
(S=0,\,I={1\over2})
&:&
\frac{2|D^0\pi^+\rangle-\sqrt2 |D^+\pi^0\rangle-\sqrt6|D^+\eta_8\rangle
     -2|D^+_s\overline K{}^0\rangle}{4},
\nonumber\\
&&
\frac{2|D^+\pi^-\rangle+\sqrt2 |D^0\pi^0\rangle-\sqrt6|D^0\eta_8\rangle
     -2|D^+_sK^-\rangle}{4};
\non\\
(S=-1,\,I=0)
&:&
\frac{|D^+K^-\rangle-|D^0\overline K{}^0\rangle}{\sqrt2}.
\end{eqnarray}
The $|DP(\overline {\bf 3}),c\rangle$ are 
\begin{eqnarray}
(S=0,\,I=1/2)
&:&
\frac{6|D^0\pi^+\rangle-3\sqrt2 |D^+\pi^0\rangle+\sqrt6|D^+\eta_8\rangle
     +6|D^+_s\overline K{}^0\rangle}{4\sqrt6},
\nonumber\\
&&\frac{6|D^+\pi^-\rangle+3\sqrt2 |D^0\pi^0\rangle+\sqrt6|D^0\eta_8\rangle
     +6|D^+_sK^-\rangle}{4\sqrt6};
\non\\
(S=1,\,I=0)
&:& 
\frac{3|D^+K^0\rangle+3|D^0K^+\rangle-\sqrt6|D^+_s\eta_8\rangle}{2\sqrt6}.
\end{eqnarray}
Finally, the $|DP(\overline {\bf 3}'),c\rangle$ are given by
\begin{eqnarray}
(S=0,\,I=1/2)
&:&
|D^+\eta_1\rangle,\quad
|D^0\eta_1\rangle;
\non\\
(S=1,\,I=0)
&:& 
|D^+_s\eta_1\rangle.
\end{eqnarray}

We give some illustrations on the effect of quasi-elastic strong rescattering on $T\to DP$ decays.
Using the above equations, the quasi-elastic strong rescattering for $S = +1$, $Q=+2$ and $S = +1$, $Q=+1$ $DP$ final
states can be written as
\begin{equation}
\left(
\begin{array}{l}
A_{D^+ K^+}\\
A_{D^+_s \pi^+}
\end{array}
\right) ={\cal S}^{1/2}_{(+1,+2)}\, \left(
\begin{array}{l}
A^0_{D^+ K^+}\\
A^0_{D^+_s \pi^+}
\end{array}
\right),
\quad
\left(
\begin{array}{l}
A_{D^0 K^+}\\
A_{D^+ K^0}\\
A_{D^+_s \pi^0}\\
A_{D^+_s\eta_8}\\
A_{D^+_s \eta_1}
\end{array}
\right) ={\cal S}^{1/2}_{(+1,+1)}\, \left(
\begin{array}{l}
A^0_{D^0 K^+}\\
A^0_{D^+ K^0}\\
A^0_{D^+_s \pi^0}\\
A^0_{D^+_s\eta_8}\\
A^0_{D^+_s \eta_1}
\end{array}
\right),
 \label{eq:FSIDpi}
\end{equation}
where ${\cal S}^{1/2}_{(S=+1,Q=+2)}$ and ${\cal S}^{1/2}_{(S=+1,Q=+1)}$ are given by
\be
{\cal S}^{1/2}_{(+1,+2)}
&=&\left(
\begin{array}{ccccc}
\frac{e^{i\delta_{\overline {15}}}+e^{i\delta_6}}{2}
       &\frac{e^{i\delta_{\overline {15}}} -e^{i\delta_6}}{2}
       \\
\frac{e^{i\delta_{\overline {15}}} -e^{i\delta_6}}{2}
       &\frac{e^{i\delta_{\overline {15}}}+e^{i\delta_6}}{2}
\end{array}
\right),
\non\\
{\cal S}^{1/2}_{(+1,+1)}
&=&\left(
\begin{array}{ccccc}
\frac{3 e^{i\delta_{\overline {15}}}+2 e^{i\delta_6}+3 U^{1/2}_{\bar 3\bar 3}}{8}
       &\frac{- e^{i\delta_{\overline {15}}} -2 e^{i\delta_6} +3 U^{1/2}_{\bar 3\bar 3}}{8}
       &\frac{e^{i\delta_{\overline {15}}} - e^{i\delta_6}}{2\sqrt2}
       &\sqrt{\frac{3}{2}}\frac{e^{i\delta_{\overline {15}}} -U^{1/2}_{\bar 3\bar 3}}{4}
       &\sqrt{\frac{3}{2}}\frac{U^{1/2}_{\bar 3\bar 3'}}{2}
       \\
\frac{- e^{i\delta_{\overline {15}}} -2 e^{i\delta_6} +3 U^{1/2}_{\bar 3\bar 3}}{8}
       &\frac{3e^{i\delta_{\overline {15}}} +2 e^{i\delta_6} +3 U^{1/2}_{\bar 3\bar 3}}{8}
       &-\frac{e^{i\delta_{\overline {15}}} - e^{i\delta_6}}{2\sqrt2}
       &\sqrt{\frac{3}{2}}\frac{e^{i\delta_{\overline {15}}} -U^{1/2}_{\bar 3\bar 3}}{4}
       &\sqrt{\frac{3}{2}}\frac{U^{1/2}_{\bar 3\bar 3'}}{2}
       \\
\frac{e^{i\delta_{\overline {15}}} - e^{i\delta_6}}{2\sqrt2}
       &-\frac{e^{i\delta_{\overline {15}}} - e^{i\delta_6}}{2\sqrt2}
       &\frac{e^{i\delta_{\overline {15}}} + e^{i\delta_6}}{2}
       &0
       &0
       \\
\sqrt{\frac{3}{2}}\frac{e^{i\delta_{\overline {15}}} -U^{1/2}_{\bar 3\bar 3}}{4}
       &\sqrt{\frac{3}{2}}\frac{e^{i\delta_{\overline {15}}} -U^{1/2}_{\bar 3\bar 3}}{4}
       &0
       &\frac{3 e^{i\delta_{\overline {15}}} + U^{1/2}_{\bar 3\bar 3}}{4}
       &-\frac{U^{1/2}_{\bar 3\bar 3'}}{2}
       \\
\sqrt{\frac{3}{2}}\frac{U^{1/2}_{\bar 3'\bar 3}}{2}
       &\sqrt{\frac{3}{2}}\frac{U^{1/2}_{\bar 3'\bar 3}}{2}
       &0
       &-\frac{U^{1/2}_{\bar 3'\bar 3}}{2}
       & U^{1/2}_{\bar 3'\bar 3'}
\end{array}
\right), 
\non\\
\label{eq:S1halfDpi}
\en
with
\be
U^{1/2}_{\bar 3\bar 3}
&=&e^{i\delta_{\bar 3}}\cos^2\tau+e^{i\delta_{\bar 3'}}\sin^2\tau,
\non\\
U^{1/2}_{\bar 3\bar 3'}=U^{1/2}_{\bar 3'\bar 3}
&=&-(e^{i\delta_{\bar 3}}-e^{i\delta_{\bar 3'}})\sin \tau\cos\tau,
\non\\
U^{1/2}_{\bar 3'\bar 3'}
&=&e^{i\delta_{\bar 3'}}\cos^2\tau+e^{i\delta_{\bar 3}}\sin^2\tau.
\en

One can check that these ${\cal S}^{1/2}_{S, Q}$ are unitary and reduce to unity matrices when the phases vanish.
Furthermore $({\cal S}^{1/2}_{S, Q})^N$ has the same form of ${\cal S}^{1/2}_{S, Q}$ but with phases multiplied by a common factor $N$. 
These properties on ${\cal S}^{1/2}_{S, Q}$ follow directly from Eq. (\ref{eq: S1half}).

Note that in this appendix we will consider rescattering within four groups of modes (3 in scenario I and 1 in scenario II), but only the above two ${\cal S}^{1/2}_{S, Q}$, namely ${\cal S}^{1/2}_{(S=+1,Q=+2)}$ and ${\cal S}^{1/2}_{(S=+1,Q=+1)}$, are employed. Interestingly, different results will be obtained even with the same ${\cal S}^{1/2}_{S, Q}$.

Now we apply the above equations on $T^{++}_{c\bar s}\to DP$, $T^{+}_{c\bar s}\to DP$ and $T^{\prime\prime +}_{c\bar s}\to DP$ decays in scenario I.
Using Table \ref{tab: TtoDP1}, but before rescattering, we have
\be
&&\left(
\begin{array}{l}
A^0_{T^{++}_{c\bar s}\to D^+ K^+}\\
A^0_{T^{++}_{c\bar s}\to D^+_s \pi^+}
\end{array}
\right)
=
\left(
\begin{array}{c}
F^0_{DP}\\
-F^0_{DP}
\end{array}
\right),
\qquad
\left(
\begin{array}{l}
A^0_{T^{+}_{c\bar s}\to D^0 K^+}\\
A^0_{T^{+}_{c\bar s}\to D^+ K^0}\\
A^0_{T^{+}_{c\bar s}\to D^+_s \pi^0}\\
A^0_{T^{+}_{c\bar s}\to D^+_s\eta_8}\\
A^0_{T^{+}_{c\bar s}\to D^+_s \eta_1}
\end{array}
\right)
=
\left(
\begin{array}{c}
\frac{1}{\sqrt2} F^0_{DP}\\
-\frac{1}{\sqrt2} F^0_{DP}\\
F^0_{DP}\\
0\\
0
\end{array}
\right),
\non\\
&&\left(
\begin{array}{l}
A^0_{T^{\prime\prime +}_{c\bar s}\to D^0 K^+}\\
A^0_{T^{\prime\prime+}_{c\bar s}\to D^+ K^0}\\
A^0_{T^{\prime\prime+}_{c\bar s}\to D^+_s \pi^0}\\
A^0_{T^{\prime\prime+}_{c\bar s}\to D^+_s\eta_8}\\
A^0_{T^{\prime\prime+}_{c\bar s}\to D^+_s \eta_1}
\end{array}
\right)
=
\left(
\begin{array}{c}
\frac{1}{\sqrt2} (F^0_{DP}+2 AC^0_P)\\
-\frac{1}{\sqrt2} (F^0_{DP}+2 AC^0_P)\\
0\\
-\frac{1}{\sqrt3} (F^0_{DP}+2 AC^0_P)\\
-\sqrt{\frac{2}{3}} (F^0_{DP}-AC^0_P-3 AC^0_D)
\end{array}
\right).
\label{eq: A0}
\en
In the presence of quasi-elastic strong rescattering, using Eqs. (\ref{eq:FSIDpi}) and (\ref{eq:S1halfDpi}), we obtain
\be
\left(
\begin{array}{l}
A_{T^{++}_{c\bar s}\to D^+ K^+}\\
A_{T^{++}_{c\bar s}\to D^+_s \pi^+}
\end{array}
\right)
={\cal S}^{1/2}_{(+1,+2)}
\left(
\begin{array}{c}
F^0_{DP}\\
-F^0_{DP}
\end{array}
\right)
=\left(
\begin{array}{c}
e^{i\delta_6} F^0_{DP}\\
-e^{i\delta_6} F^0_{DP}
\end{array}
\right),
\non\\
\left(
\begin{array}{l}
A_{T^{+}_{c\bar s}\to D^0 K^+}\\
A_{T^{+}_{c\bar s}\to D^+ K^0}\\
A_{T^{+}_{c\bar s}\to D^+_s \pi^0}\\
A_{T^{+}_{c\bar s}\to D^+_s\eta_8}\\
A_{T^{+}_{c\bar s}\to D^+_s \eta_1}
\end{array}
\right)
={\cal S}^{1/2}_{(+1,+1)}
\left(
\begin{array}{c}
\frac{1}{\sqrt2} F^0_{DP}\\
-\frac{1}{\sqrt2} F^0_{DP}\\
F^0_{DP}\\
0\\
0
\end{array}
\right)
=\left(
\begin{array}{c}
\frac{1}{\sqrt2} e^{i\delta_6} F^0_{DP}\\
-\frac{1}{\sqrt2} e^{i\delta_6} F^0_{DP}\\
e^{i\delta_6} F^0_{DP}\\
0\\
0
\end{array}
\right),
\label{eq: AFSI1}
\en
and
\be
\left(
\begin{array}{l}
A_{T^{\prime\prime +}_{c\bar s}\to D^0 K^+}\\
A_{T^{\prime\prime+}_{c\bar s}\to D^+ K^0}\\
A_{T^{\prime\prime+}_{c\bar s}\to D^+_s \pi^0}\\
A_{T^{\prime\prime+}_{c\bar s}\to D^+_s\eta_8}\\
A_{T^{\prime\prime+}_{c\bar s}\to D^+_s \eta_1}
\end{array}
\right)
&=&
{\cal S}^{1/2}_{(+1,+1)}
\left(
\begin{array}{c}
\frac{1}{\sqrt2} (F^0_{DP}+2 AC^0_P)\\
-\frac{1}{\sqrt2} (F^0_{DP}+2 AC^0_P)\\
0\\
-\frac{1}{\sqrt3} (F^0_{DP}+2 AC^0_P)\\
-\sqrt{\frac{2}{3}} (F^0_{DP}-AC^0_P-3 AC^0_D)
\end{array}
\right)
\non\\
&=&
\left(
\begin{array}{c}
\frac{1}{\sqrt2} \left[U^{1/2}_{\bar 3 \bar 3} (F^0_{DP}+2 AC^0_P)-\frac{U^{1/2}_{\bar 3\bar 3'}}{\sqrt2}  (F^0_{DP}-AC^0_P-3 AC^0_D)\right]\\
-\frac{1}{\sqrt2} \left[U^{1/2}_{\bar 3 \bar 3} (F^0_{DP}+2 AC^0_P)-\frac{U^{1/2}_{\bar 3\bar 3'}}{\sqrt2}  (F^0_{DP}-AC^0_P-3 AC^0_D)\right]\\
0\\
-\frac{1}{\sqrt3} \left[U^{1/2}_{\bar 3 \bar 3} (F^0_{DP}+2 AC^0_P)-\frac{U^{1/2}_{\bar 3\bar 3'}}{\sqrt2}  (F^0_{DP}-AC^0_P-3 AC^0_D)\right]\\
-\sqrt{\frac{2}{3}} \left[-\sqrt2 U^{1/2}_{\bar 3'\bar 3}(F^0_{DP}+2 AC^0_P) +U^{1/2}_{\bar 3' \bar 3'}  (F^0_{DP}-AC^0_P-3 AC^0_D)\right]
\end{array}
\right).
\non\\
\label{eq: AFSI2}
\en
Comparing $A^0_{T\to DP}$ in Eq. (\ref{eq: A0}) and $A_{T\to DP}$ in Eqs. (\ref{eq: AFSI1}) and (\ref{eq: AFSI2}), it is evident that $A^0_{T\to DP}$ and $A_{T\to DP}$ take the same form, but with the following replacements
\be
F^0_{DP}&\to& e^{i\delta_6} F^0_{DP},
\non\\
F^0_{DP}+2 AC^0_P
&\to& U^{1/2}_{\bar 3 \bar 3} (F^0_{DP}+2 AC^0_P)-\frac{U^{1/2}_{\bar 3\bar 3'}}{\sqrt2}  (F^0_{DP}-AC^0_P-3 AC^0_D),
\non\\
F^0_{DP}-AC^0_P-3 AC^0_D
&\to& -\sqrt2 U^{1/2}_{\bar 3'\bar 3}(F^0_{DP}+2 AC^0_P) +U^{1/2}_{\bar 3' \bar 3'}  (F^0_{DP}-AC^0_P-3 AC^0_D).
\en
One can simply redefine those on the right-hand sides as $F_{DP}$, $F_{DP}+2 AC_P$ and $F_{DP}-AC_P-3 AC_D$, respectively, as in Eq. (\ref{eq: FSI I}).
Hence, the form of $T\to DP$ decay amplitudes in terms of topological amplitudes, as shown in Table \ref{tab: TtoDP1}, does not change in the presence of quasi-elastic strong rescattering, whereas the topological amplitudes may contain the effect of rescattering. 
Note that if one reiterates the above equations $N$ times, the final results will be identical to the above equation, or equivalently Eq. (\ref{eq: FSI I}), but with $\delta_{6}$, $\delta_{\bar 3}$, and $\delta_{\bar 3'}$ multiplied by a common factor $N$.

Similar results also hold for the amplitudes given in scenario II.
For example, according to Table \ref{tab: TtoDP1}, before rescattering the $T^{++}_{c\bar s}\to D^+ K^+$ and $D_s^+\pi^+$ amplitudes are 
\be
\left(
\begin{array}{l}
A^0_{T^{++}_{c\bar s}\to D^+ K^+}\\
A^0_{T^{++}_{c\bar s}\to D^+_s \pi^+}
\end{array}
\right)
=
\left(
\begin{array}{c}
F^0_{DP}\\
F^0_{DP}
\end{array}
\right).
\en
Using the same ${\cal S}^{1/2}_{(+1,+2)}$ given in Eq. (\ref{eq:S1halfDpi}), we obtain, in the presence of quasi-elastic scattering,
\be
\left(
\begin{array}{l}
A_{T^{++}_{c\bar s}\to D^+ K^+}\\
A_{T^{++}_{c\bar s}\to D^+_s \pi^+}
\end{array}
\right)
={\cal S}^{1/2}_{(+1,+2)}
\left(
\begin{array}{c}
F^0_{DP}\\
F^0_{DP}
\end{array}
\right)
=\left(
\begin{array}{c}
e^{i\delta_{\overline {15}}} F^0_{DP}\\
e^{i\delta_{\overline {15}}} F^0_{DP}
\end{array}
\right),
\en
which agree to those shown in Table \ref{tab: TtoDP2} by taking $F_{DP}=e^{i\delta_{\overline {15}}} F^0_{DP}$.
Therefore, the form of the amplitudes in terms of topological amplitudes is unchanged, but the topological amplitude itself is affected by rescattering.
Complete results of the effect of the rescattering on topological amplitudes in scenario II are given in Eq. (\ref{eq: FSI II}).  

\section{The flavor structure of the low lying scalar particles}\label{sec: scalar}

In this appendix, we give explicitly the flavor structure of the low-lying scalar particles as follows.
Their quark contents are given by~\cite{Jaffe:1976ig, Jaffe:1976ih} 
\begin{equation}
\begin{gathered}
a^+_0(980)=S([su][\bar d\bar s]),\,\,
a^0_0(980)=\frac{S([su][\bar s\bar u])-S([d s][\bar d \bar s])}{\sqrt 2},\,\,
a^-_0(980)=S([ds][\bar s \bar u]),
\\
f_0([ud] [\bar u \bar d])=S([u d][\bar u\bar d]),\,\,
f_0([ns] [\bar n \bar s])=\frac{S([su][\bar s\bar u])+S([d s][\bar d \bar s])}{\sqrt 2},
\\
\kappa^+=S([ud][\bar d\bar s]),\,\,
\kappa^0=S([ud][\bar s \bar u]),\,\,
\bar\kappa^0=S([su][\bar u\bar d]),\,\,
\kappa^-=S([ds][\bar u\bar d]).
\end{gathered}
\end{equation}
The physical $f_0(980)$ and $\sigma(500)$ are linear combinations of the above $f_0([ns] [\bar n \bar s])$ and $f_0([ud] [\bar u \bar d])$,
\be
f_0(980)&=&\cos\phi\, f_0([ns] [\bar n \bar s])+\sin\phi \,f_0([ud] [\bar u \bar d]),
\non\\
\sigma(500)&=&-\sin\phi\, f_0([ns] [\bar n \bar s])+\cos\phi\, f_0([ud] [\bar u \bar d]),
\en
or equivalently, we have
\be
f_0([ns] [\bar n \bar s]))&=&\cos\phi\, f_0(980)-\sin\phi \,\sigma(500),
\non\\
f_0([ud] [\bar u \bar d])&=&\sin\phi\, f_0(980)+\cos\phi\, \sigma(500).
\en


\begin{thebibliography}{99}
\bibitem{Zweig:1964ruk}
G.~Zweig,
``An SU(3) model for strong interaction symmetry and its breaking. Version 1,''
doi:10.17181/CERN-TH-401
\end{thebibliography}
\end{document}